\documentclass[12pt,epsf]{book}
\usepackage{latexsym,amsfonts,amsmath,amscd,amssymb,epsf}
\usepackage[utf8]{inputenc}
\usepackage[american]{babel}
\usepackage[dvips]{graphicx}
\usepackage{epsfig}
\usepackage{graphics,color} 

\usepackage{graphicx}
\usepackage{hyperref}

\usepackage{mathrsfs}
\usepackage{color}
\usepackage{graphicx}
\usepackage{caption}

\usepackage{subcaption}
\usepackage{cancel}
\usepackage{mathtools}
\usepackage{slashed}

\usepackage[toc,page]{appendix}

\usepackage{xspace}
\usepackage{amsthm}
\usepackage{amsmath,amssymb,amsthm,amscd}
\usepackage{epsfig,comment}
\usepackage{graphicx,caption,subcaption}
\usepackage{graphics,graphicx,epsfig}
\usepackage{amsfonts,psfrag,verbatim,color}

\advance \topmargin by -\headheight
\advance \topmargin by -\headsep

\evensidemargin \oddsidemargin
\def\del{{\delta^{\hbox{\sevenrm B}}}}

\def\im{{\hbox{\rm Im}}}  \def\tr{{\hbox{\rm Tr}}}

\def\ie{{\em i.e.}}
\def\ie{\hbox{\it i.e.}}

\def\CC{{\mathchoice
{\rm C\mkern-8mu\vrule height1.45ex depth-.05ex
width.05em\mkern9mu\kern-.05em}
{\rm C\mkern-8mu\vrule height1.45ex depth-.05ex
width.05em\mkern9mu\kern-.05em}
{\rm C\mkern-8mu\vrule height1ex depth-.07ex
width.035em\mkern9mu\kern-.035em}
{\rm C\mkern-8mu\vrule height.65ex depth-.1ex
width.025em\mkern8mu\kern-.025em}}}

\def\RR{{\rm I\kern-1.6pt {\rm R}}}

\def\ZZ{{\rm Z}\kern-3.8pt {\rm Z} \kern2pt}
\def\IB{\relax{\rm I\kern-.18em B}}
\def\ID{\relax{\rm I\kern-.18em D}}
\def\II{\relax{\rm I\kern-.18em I}}
\def\IP{\relax{\rm I\kern-.18em P}}

\def\im{Invent. Math.}

\newcommand{\beq}{\begin{equation}}
\newcommand{\eeq}{\end{equation}}

\newcommand{\bear}{\begin{eqnarray}}
\newcommand{\eear}{\end{eqnarray}}
\newcommand{\W}{{\cal W}}

\def\to{\rightarrow}

\def\tr{{\rm Tr}}
\def\to{\rightarrow}

\newfont{\namefont}{cmr10}
\newfont{\addfont}{cmti7 scaled 1440}
\newfont{\boldmathfont}{cmbx10}
\newfont{\headfontb}{cmbx10 scaled 1728}
\def\ie{{\it i.e.}}

\def\revise#1       {\raisebox{-0em}{\rule{3pt}{1em}}%
                     \marginpar{\raisebox{.5em}{\vrule width3pt\
                     \vrule width0pt height 0pt depth0.5em
                     \hbox to 0cm{\hspace{0cm}{%
                     \parbox[t]{4em}{\raggedright\footnotesize{#1}}}\hss}}}}

\def\del          {\partial}

\def\ee           {{\rm e}}

\def\tr           {\mathop{\rm Tr}}

\def\half{{\frac12}}

\def\sqr#1#2{{\vcenter{\vbox{\hrule height.#2pt
 \hbox{\vrule width.#2pt height#1pt \kern#1pt
 \vrule width.#2pt}\hrule height.#2pt}}}}

\def\a{\alpha}
\def\b{\beta}
\def\r{\rho}

\def\be{\begin{equation}}
\def\ee{\end{equation}}

\def\m{\mu}
\def\g{\gamma}
\def\l{\lambda}
\def\n{\nu}

\def \ov{\over}

\def \del { \partial}

\def\e{\epsilon}
\def\rt{\rightarrow}
\def\tr{{\tilde\rho}}
\newcommand{\eel}[1]{\label{#1}\end{equation}}
\newcommand{\bea}{\begin{eqnarray}}
\newcommand{\eea}{\end{eqnarray}}
\newcommand{\eeal}[1]{\label{#1}\end{eqnarray}}

\newcommand{\nn}{\nonumber}

\def \ov {\over}

\def \del { \partial}

\def\o{\omega}
\def\O{\Omega}
\def\e{\epsilon}

\def\e{\epsilon}
\def\m{\mu}
\def\n{\nu}
\def\a{\alpha}
\def\b{\beta}
\def\g{\gamma}

\def\d{\delta}
\def\r{\rho}

\def\s{\sigma}

\def\l{\lambda}

\def\be{\begin{equation}}
\def\ee{\end{equation}}

\def \g {\gamma}

\def \l {\lambda}

\def \m {\mu}
\def \n {\nu}
\def \W{{\cal W}}
\def \eps {\epsilon}

\def\l{\lambda}

\def\g{\gamma}

\def\hd{\hat{d}}

\def\CC{\mathcal{C}}

\def\tr{{\rm Tr}}

\def\href#1#2{#2}
\def\half{{1 \over 2}}

\definecolor{M_Beige}         {rgb}{0.96 , 0.96 , 0.86}

\definecolor{M_Brown}         {rgb}{0.65 , 0.16 , 0.16}
\definecolor{M_Gold}          {rgb}{0.12 , 0.84 , 0.30}

\definecolor{M_LemonChiffon}  {rgb}{1.00 , 0.98 , 0.80}

\definecolor{M_Orange}        {rgb}{1.00 , 0.60 , 0.00}

\definecolor{M_Pink}          {rgb}{1.00 , 0.75 , 0.80}

\definecolor{M_Gre}          {rgb}{0.05 ,0.46 , 0.00}

\def\a{\alpha}
\def\b{\beta}

\def\d{\delta}
\def\e{\epsilon}           
\def\f{\phi}               
  
\def\vp{\varphi}
\def\g{\gamma}

\def\l{\lambda}
\def\m{\mu}
\def\n{\nu}
\def\o{\omega}  
  \def\th{\theta}                  
\def\r{\rho}                                     
\def\s{\sigma}                                   

\def\u{\upsilon}

\def\pt{\tilde{\varphi}}

\def\6{\partial}
\def\wg{\wedge}

\def\bt{\bar{\theta}}

\usepackage[innercaption]{sidecap} 

	\def\IB {\mathbb{B}}	
\def\ID {\mathbb{D}}		
		\def\II {\mathbb{I}}

\def\IP {\mathbb{P}}

		\def\CC {\mathcal{C}}

\def\a {\alpha}			\def\b {\beta}			\def\d {\delta}
\def\e {\epsilon}		\def\g {\gamma}			
				\def\l {\lambda}
		\def\m {\mu}			\def\n {\nu}
\def\o {\omega}						
\def\s {\sigma}						\def\O {\Omega}

				\def\dd {\dot{\delta}}

\def\tilde{\widetilde}

\def\sfU{\mathsf{U}}

\def\ImO{\mathrm{Im}\mathcal{O}}
\def\ImOt{\mathrm{Im}\tilde{\mathcal{O}}}

\catcode`@=12

\def\dd{\mathrm{d}}

\def\eq{&=&}

\def\rt{\tilde{\rho}_0}
\def\pt{\tilde{\pi}_0}
\newcommand{\prt}[1]{{\left( {#1} \right)}}

\def\eq{&=&}
\def\f2{\phi_2}

\newcommand{\wh}[1]{\widehat{#1}}

\def\W2{\omega^2}

\def \calI {\mathcal I}

\def\ac{action\xspace}

\def\smod{$\sigma$-model\xspace}

\def\st{string theory\xspace}
\def\sc{strong coupling\xspace}
\def\lN{large $N$\xspace}

\def\holo{holography\xspace}
\def\hoc{holographic\xspace}
\def\hocy{holographically\xspace}
\def \bdry{boundary\xspace}

\def\dft{dual field theory\xspace}

\def\sts{strings theories\xspace}
\def\bg{background\xspace}  \def\bgs{backgrounds\xspace}
\def\dbg{dual background\xspace} 
\def\hd{higher dimension\xspace}
\def\hdal{higher dimensional\xspace}
\def\fn{function\xspace}   \def\fns{functions\xspace}

\def\qns{quantities\xspace}

\def\td{T-duality\xspace}
\def\dua{duality\xspace}

\def\natd{NATD\xspace}
\def\brs{Buscher rules\xspace}

\def\susy{SUSY\xspace}
\def\sugra{SUGRA\xspace}
\def\su2{$\mathrm{SU(2)}$\xspace}

\def\wi{Ward identities\xspace}

\def\sb{symmetry breaking\xspace}
\def\csb{concomitant symmetry breaking\xspace}

\def\hren{holographic renormalization\xspace}

\def\esb{ESB\xspace}
\def\ssb{SSB\xspace}
\def\cc{conserved current\xspace}
\def\gb{GB\xspace}
\def\aq{alternative quantization\xspace}
\def \vp{variational principle\xspace}
\def\bt{boundary terms\xspace}
\def\sc{scalar\xspace}  
\def\bc{boundary conditions\xspace}

\def\fl{fluctuations\xspace} 
\def\corr{correlator\xspace}

\def\uv{UV\xspace}

\def\u1{$\mathrm{U(1)}$\xspace}

\newcommand{\esp}[1]{\langle {#1} \rangle}
\def\o{O_\phi} 
\def\reo{\mathrm{Re}O_\phi}
\def\imo{\mathrm{Im}O_\phi}
\def\dmu{\partial_\mu}
\def\dj{\partial_\mu J^\mu}
\def\djx{\partial_\mu J^\mu(x)}

\def\imx{\mathrm{Im}O_\phi(x)}
\def\dj0{\partial_\mu J^\mu(0)}
\def\re0{\mathrm{Re}O_\phi(0)}
\def\im0{\mathrm{Im}O_\phi(0)}

\def\imoimo {\esp{\imo \imo}}
\def\djimo  {\esp{\partial_\mu J^\mu  \imo}}
\def \ro{\rho_0}
\def \ru {\rho_1}
\def \po {\pi_0} 
\def \pu {\pi_1}
\def \lo {l_0}
\def \lu {l_1}

\begin{document}


\begin{titlepage}

	\begin{center} 
		
		\vskip.5cm
		

		\vspace{.3cm}

		\hspace{0.5cm}
		\large Departamento de F\'\i sica de Part\'\i culas 
		
	\end{center}

	\vspace{6cm}

	\begin{center} 
		
		\LARGE  \bf On gravitational Phase Transitions, T-duality and Symmetry Breaking in AdS/CFT
		
	\end{center}

	\vspace{3cm}

	\vspace{5cm}

	\begin{center} 
		
{\sf\bf \large Jes\'us An\'ibal Sierra Garc\'ia}

\sf Santiago de Compostela, Junio de 2017.
		
	\end{center}

\end{titlepage}

\pagestyle{empty}


\begin{titlepage}

	\begin{center} 
		
		\large \sf  UNIVERSIDADE DE SANTIAGO DE COMPOSTELA

		\vspace{.3cm}

		\large Departamento de F\'\i sica de Part\'\i culas 
		
	\end{center}

	\vspace{6cm}

	\begin{center} 
		
		\LARGE  \bf On gravitational Phase Transitions, T-duality and Symmetry Breaking in AdS/CFT	
		
	\end{center}

	\vspace{3cm}

	\vspace{5cm}

	\begin{center} 
		
{\sf\bf \large Jes\'us An\'ibal Sierra Garc\'ia}

\sf Santiago de Compostela, Junio de 2017.
		
	\end{center}

\end{titlepage}


\pagestyle{empty}
\
\newpage

\begin{center}

	\large \sf  UNIVERSIDADE DE SANTIAGO DE COMPOSTELA

	\vspace{.3cm}

	\large Departamento de F\'\i sica de Part\'\i culas 
	
\end{center}

\vspace{4cm}

\begin{center} 
	
\Large \textsc{ On gravitational Phase Transitions, T-duality and Symmetry Breaking in A}\MakeLowercase{d}\textsc{S/CFT} 
	
\end{center}

\vspace{5cm}

\hspace{6.5cm}
Tesis presentada para optar al grado

\hspace{6.5cm}
de Doctor en F\'\i sica por:

\vspace{.5cm}
\hspace{6.5cm}
{\bf Jes\'us An\'ibal Sierra Garc\'ia}

\vspace{.4cm}
\hspace{6.5cm}
Junio, 2017


\newpage
\
\newpage

\begin{center} 
	
	\large \sf  UNIVERSIDADE DE SANTIAGO DE COMPOSTELA

	\vspace{.3cm}

	\large Departamento de F\'\i sica de Part\'\i culas 
	
\end{center}

\vspace{5cm}
Jos\'e D. Edelstein Glaubach, Profesor Titular de F\'\i sica Te\'orica
de la Universidad de Santiago de Compostela,

\vspace{1cm}

\textbf{ CERTIFICO:} que la memoria titulada \textit{On gravitational Phase Transitions, T-duality and Symmetry Breaking in $\mathrm{AdS/CFT}$} fue realizada bajo mi direcci\'on por Jes\'us An\'ibal Sierra Garc\'ia en el Departamento de F\'\i sica de Part\'iculas de esta Universidad y constituye el trabajo de Tesis que presenta para optar al grado de Doctor en F\'\i sica.

\vspace{2.5cm}

\begin{center} 
	\begin{displaymath}
	\begin{matrix}
	&&\textrm{Firmado:}\cr\cr
	\cr\cr\cr\cr
	&&\textsf{Jos\'e D. Edelstein Glaubach.}\cr
	&&\textsf{Santiago de Compostela, Junio de 2017.}
	\end{matrix}
	\end{displaymath}
\end{center}
\setcounter{footnote}{0}

\pagestyle{empty}

\setcounter{page}{0}

\newpage
\
\newpage


\pagestyle{empty}

\

\vspace{9cm}

\hspace{13cm}
\textsl{A mi familia}

\setcounter{page}{0}

\newpage
\
\newpage



\pagestyle{empty}

{\LARGE \bf {\centerline{Agradecimientos}}}

\vspace{1cm}

En primer lugar corresponde recordar que nada de este trabajo habría sido posible sin la beca FPI BES 2012-062365 del Ministerio de Econom\'ia y Competitividad, adem\'as de las siguientes fuentes de financi\'on: proyectos MINECO (FPA2014-52218-P), Xunta de Galicia (GRC2013-024) y fondos FEDER.

Me gustaría agradecer a mi director de tesis José Edelstein, por guiar mis primeros pasos en el extraordinario campo de la Física Teórica, por su paciencia y su atención en todo momento. Quiero remarcar además la gran libertad que me dio a la hora de trabajar. Por otra parte, tengo que reconocer su cuidadosa lectura de esta tesis, cuya presentación y redacción sin duda contribuyó a mejorar. Una mención distinguida les corresponde a Josiño Sánchez de Santos y Xián Camanho, que me ayudaron en la etapa inicial del doctorado. También quiero recordar a mis coautores: Riccardo Argurio, Yago Bea, Gastón Giribet, Andy Gomberoff, Karta Kooner, Andrea Marzolla, Daniel Naegels, Carlos Núñez, Daniel Schofield y Kostas Sfetsos: a todos ellos debo también gratitud por su entusiasmo por la Física.

No puedo olvidarme tampoco de los compañeros que hicieron más llevadero el día a día, tanto en los lugares de estancia como en Santiago. De manera especial quiero señalar a aquellos que tuve el placer de conocer mejor: Alex, Daniele, Georgios y Yago. También es justo acordarme de los otros profesores del grupo, Alfonso y Javier.

Finalmente aprovecho la ocasión para reconocer el importante servicio que arXiv.org e inSPIRE hacen a toda la comunidad de físicos teóricos.

\setcounter{page}{0}

\newpage
\
\newpage


\pagestyle{headings}

\setcounter{page}{1}

\tableofcontents


\chapter[\texorpdfstring{Overview of A\MakeLowercase{d}S/CFT, strings and gravitation }{Overview of AdS/CFT, strings and gravitation }]{Overview of AdS/CFT, strings and gravitation } \label{motivation}

Everything in this thesis is directly or indirectly motivated by the AdS/CFT correspondence. In its broadest, non rigorous form, it asserts that string theory in AdS spaces can be described in terms of Quantum Field Theory, and vice versa. We will provide now a historical introduction to AdS/CFT and afterwards introduce the main problems studied throughout the present work.

The history of AdS/CFT began with the attempt to unify Quantum Theory and General Relativity. Let's elaborate briefly on these two theories before. Quantum Theory describes our most essential principles of the very small scales, roughly ranging from molecular to known elementary particles. Although it is fairly counterintuitive and abstract, its success in explaining and predicting physical phenomena remains unparalleled both in precision and applicability breadth. Apart from explaining phenomena that classical physics could not (atom stability, blackbody radiation, discreteness of spectra, photoelectric effect, ... ), it became the cornerstone in the following great domains:
\begin{itemize}
\item Molecules and condensed matter.
\item Atoms.
\item Nuclei.
\item Elementary particles.
\end{itemize}
That short enumeration encompasses an absolutely huge variety of physical effects tested in experiments of all sorts. We will be more concerned in this introduction with advances in the high energy regime. In particular, its description requires of the other great revolution of the 20th century Fundamental Physics: Relativity. The principles of Quantum Mechanics and Special Relativity were successfully combined in Quantum Field Theory, and the so called electromagnetic, weak and strong interactions were explained using it. 

Nevertheless, the Special Theory of Relativity needed extension to be consistent with non-inertial frames and gravity; this was the birth of Einstein's theory of General Relativity. Apart from solving theoretical inconsistencies and although less thoroughly tested than Quantum Theory, it has its own explained non-Newtonian phenomena, including: 
\begin{itemize}
\item Anomalous precession of orbits.
\item Light deflection near very massive bodies, gravitational lensing and Shapiro time delay.
\item Gravitational frequency shift.
\item Gravitational waves (measured recently by the LIGO collaboration \cite{TheLIGOScientific:2016wfe}),
\item Universe expansion. \footnote{The theory allows a simple mathematical model of expansion through the so-called cosmological constant. Nevertheless, astronomical observations indicate that the value of the constant is many orders of magnitude smaller than the estimated using modern theories of elementary particles. This is known as the cosmological constant problem, and it remains one of the main unsolved puzzles of contemporary fundamental Physics \cite{Weinberg:1988cp}. }
\end{itemize}
Despite this success of General Relativity, we know it cannot the ultimate theory of gravity, as it generically predicting singularities of spacetime in the interior of BHs and at the beginning of Big Bang. When we try to combine it with Quantum Field Theory for very short distances, it becomes not renormalizable, losing its predictive power. This is one of the main open problems in contemporary Fundamental Physics: the quantization of gravity. In this situation is where string theory enters our discussion, and ends up leading to AdS/CFT.

In early 1970's String Theory changed fundamental point particles for extended strings to reproduce the Veneziano amplitudes, a proposed description of strongly interacting mesons. Such a radical program was nearly abandoned when Quantum Chromodynamics explained the strong force; however strings would be thoroughly reconsidered. In 1974 Scherk and Schwarz found a spin 2 particle and the Einstein equations of motion in the low energy limit of the theory: strings implied gravity! At the same time it contained other lower spin particles.  Meanwhile, Supersymmetry was a popular idea trying to unify fermions and bosons. Its application to strings brought Superstring theory, that eliminated some inconsistencies of bosonic case while demanding a 10-dimensional spacetime. Superstrings showed potential way to quantize gravity and unify interactions. \\
With such exciting panorama, the consistent possibilities were quickly explored, and the first theories given the unpretentious names of Type I, Type IIA and Type IIB. In 1984 anomaly cancellation through Green-Schwarz mechanism ignited the so called 'first Superstring revolution', which unveiled heterotic $SO(32)$ and heterotic $E_8 \times E_8$ as new string theories. Moreover, it was possible to reproduce the gauge groups $\mathrm{U(1)}\times \mathrm{SU(2)}\times SU(3)$ of the Standard Model. To carry experimental agreement further, the 10 spacetime dimensions had to be lowered down to 4. The number of possible ways to do so was estimated of order $10^{500}$, with no clear reason to discard many of them. No surprisingly, enthusiasm diminished due to this enormous "vacua landscape". Furthermore, five string theories were already known at the time, somewhat in contradiction with the pretensions of being a final theory of interactions. 

And yet, strings multiplicity would not last for long, since Witten alleviated it while prompting the 'second Superstring revolution' in the mid 90's: he and others shown that the so called S,T,U 'dualities' relate the five Superstring theories! They are five perturbative expansions of an underlying M-theory in different regimes. For example, S-duality relates the strong coupling regime of Type I to the weakly coupled of $SO(32)$; T-duality reveals the equivalence of type IIA on a cylinder of radius $R$ and type IIB in radius $\alpha'/R$; and U-duality is a combination of both T and S. Although M-theory reduces to 11D Supergravity at low energy, the full theory is not yet completely understood.

The dualities not only cured strings multiplicity but also shown D-branes necessity, eventually leading to AdS/CFT.\footnote{They also allow extremal BH microstate counting in subsequent agreement with Bekenstein-Hawking entropy, among other virtues. } Branes in general are non-perturbative dynamical hypersurfaces that join strings as dynamical objects in the theory. In particular D-branes (where D stands for Dirichlet) are hypersurfaces to which open strings attach their ends. Non-trivially, the massless modes of the open strings realize gauge theories. From another point of view, D-branes are also sources to closed strings and modify the spacetime. That means that the gauge theory from the open string description is related to the spacetime of the closed one. This observation by Maldacena in 1997 starts the ongoing, AdS/CFT-motivated era of string theory. 

The canonical example of the correspondence happens between strings on AdS$_5\times$S$^5$ that is related to D3 branes and $\mathcal N=4$ SYM at  $N$ in 3+1 dimensions. The D3 brane description is weakly coupled and the computations are feasible, while the CFT is strongly coupled. Despite it, the maximal Supersymmetry makes the computations possible in the gauge theory, and both sides yield the same result for explicit calculations. This was the first and so far best established instance of AdS/CFT correspondence. 

In rough terms, the interest on the correspondence is easy to justify. While the field theory is strongly coupled and no perturbative treatment applies, its gravitational dual interacts weakly and perturbation theory is readily available. 

The gauge/gravity correspondence has been extended to much more general cases. Some of these extensions and their intended applications (along with a few representative works or reviews) are:  \begin{itemize}
\item Other integrable cases, partly to test the explicit matching of both side results \cite{Beisert:2010jr}.
\item Reduction of the amount of SUSY (notably $ \mathcal N=1$ in Klebanov-Witten field theory) \cite{Edelstein:2006kw}, RG flows and different representations and gauge groups to approach more realistic phenomena;  for example for QCD one can check \cite{Sakai:2004cn,Pons:2004dk,Bigazzi:2015bna,Faedo:2015urf,Attems:2016tby,Casalderrey-Solana:2016hrm}, 
\item Higher and lower spacetime dimensions, 
\item Higher derivative corrections in the gravity action \cite{Brigante:2007nu,Hofman:2008ar,Camanho:2013pda},
\item Higher spin theories \cite{Vasiliev:1990en,Giombi:2009wh},
\item Non-equilibrium situations, for example for thermalization in QFT \cite{Chesler:2008hg,Chesler:2010bi},
\item Non-stringy actions that allows to handle simpler dynamics in the gravity side and to tune them for more specific purposes. This is called the called bottom-up approach, and it has become very popular for all kind of purposes,
\item Non-relativistic Physics to tackle condensed matter problems \cite{Hartnoll:2008kx,Hartnoll:2008vx,Herzog:2009xv,Faulkner:2009wj,Charmousis:2010zz}. This typically involves  a change of spacetime boundary from AdS to Lifshitz spaces\cite{Kachru:2008yh,Taylor:2015glc}),
\item Low energy limit of the QFT, giving rise to the fluid/gravity correspondence \cite{Hubeny:2011hd}, 
\item Going beyond classical strings in the AdS side, for example for applications to black hole Physics \cite{Berkowitz:2016muc},
\item Quantum information theory for many body systems and spacetime emergence from it \cite{Rangamani:2016dms},
\item ...
\end{itemize}
This list is not even complete due to extensive research during the last 20 years. The original work on AdS/CFT by Maldacena \cite{Maldacena:1997re} is the most cited article in inSPIRE with nearly 13000 citations, even more than collaborations like WMAP, RPP and the Higgs discovery by ATLAS and CMS. Even though the latter are more recent, it clearly signals the importance reached by the correspondence in contemporary High Energy Physics.

In addition, the number of QFT quantities computable from the AdS side has also grown in time, leading to a respectable maturity: amount of SUSY, n-point functions, thermodynamic quantities, transport coefficients, Wilson loops, central charges, entanglement entropy and other information measures, etcetera.

At this point it is also fitting to remark the main limitations of the correspondence:
\begin{itemize}
\item The dual field content and action are almost never known. This may cast some doubts on whether the results correspond to any field theory, although typically they satisfy general physical expectations even for bottom-up approach. 
\item Due to to the previous issue, extracting numerical predictions for a realistic given system seems far from feasible at the moment \footnote{One notable exception is the value of $\eta/s=\frac 1 {4\pi}$ of the quark-gluon plasma, where $\eta, s$ are the shear viscosity and entropy density.}. As stated before, the main goal so far is to obtain qualitative insights about strongly coupled phenomena. 
\item Finite $N$ or finite coupling effects are normally difficult or impossible to compute.
\end{itemize}
The standard approach is to start with a gravity theory and compute the dual holographic quantities corresponding to strongly coupled phenomena. Any result from them is of interest as in general little is known about strongly coupled Physics.
In most of this presentation we have emphasized the applications for strongly coupled systems, but it is not its only interest. On the one hand, AdS/CFT can also be used to learn new things about gravity. On the other hand this provides an additional strong motivation to further investigate GR and its modifications, as any progress will be followed by a question about its implications in the dual CFT. We will approach the study of gravity using stringy effective actions. There are several reasons to do so:
\begin{itemize}
\item String theory already provides a candidate description of microscopic gravitational degrees of freedom. Among other, the explicit reproduction of Bekenstein-Hawking entropy for an extremal black hole stands out.  \cite{Strominger:1996sh}.  
\item One does not deal directly with hypothetical components of spacetime, but with effective actions than can be more easily compared to General Relativity. Furthermore, it is much simpler to study classical gravity than its quantum processes. Finally, classical gravity is almost always used in the AdS side of the correspondence.
\item Consistency of stringy effective actions can be explored (causality for example), \cite{Camanho:2014apa}. 
\item The bosonic string gravity truncation to order $\alpha'$ is the LGB gravity theory, that belongs like GR to the Lovelock family. The latter is a special set of theories with a great deal of previous literature, even from the holographic point of view. 
\item Stringy dualities like T-duality can be applied to generate new solutions. 
\end{itemize}  
In the following page, we will summarize how the present thesis contributed to the topics above.

\section{Scope and organization of this thesis}
The goal of this thesis is to make progress in AdS/CFT concrete problems, with the common motivation of testing and extending its domain of applicability. The main text of it is based on the articles \cite{Camanho:2015zqa, Bea:2015fja, Argurio:2016xih}. The organization is the following: 
\vskip5mm
\begin{itemize}
\item Chapter \ref{motivation} contains the present motivation and overview of AdS/CFT and gravity.
\item Chapter \ref{intro} contains a few pages of non-formal introduction to the AdS/CFT topics discussed in the following chapters alongside a brief overview of the main results.
\item In chapter \ref{gb} we find phase transitions between AdS and dS geometries in the Gauss-Bonnet theory of gravity. They are mediated by bubble nucleation and do not require addition of any matter. The phenomenon is strongly expected to be generic in higher curvature theories of gravity. As dS/CFT correspondence is not as well developed as its AdS counterpart, the holographic interpretation of the final state and the transition as a whole remains unclear.
\item In chapter \ref{natd} non-Abelian T-duality (NATD) is applied to generate new Supergravity backgrounds and study their holographic duals. The solutions upon which NATD is used are $\mathcal N=1$ supersymmetric deformations of Klebanov-Witten background flowing from AdS$_5$ to AdS$_3$ in the IR. In particular, new regular $\mathcal N=1$ supersymmetric AdS$_3$ fixed points are constructed. The new solutions seem to be dual to long linear quiver gauge theories. 
\item In chapter \ref{ward} we \hocy derive the \wi of a 1+1 QFT. This is achieved by holographic renormalization with appropriate boundary conditions. The resulting Ward identities are the same as in higher dimensions and they show the expected spontaneous symmetry breaking for large $N$. 

\item Chapter \ref{sac} contains the summary and conclusions of the previous results.
\end{itemize}
\vskip2mm
 \newpage

\chapter{Introduction} \label{intro}


This thesis is intended to contribute to the very general goal of understanding and extending the applicability of the AdS/CFT correspondence. Although many new physical situations and observables have been described in its 20 years of development, there is still much ongoing research oriented to a better comprehension of the correspondence itself (specially including quantum gravity effects) and to realistic modeling of strongly coupled systems. We have contributed with the investigation of three different topics:
\begin{itemize}
	\item In the first we studied the role of higher curvature gravity corrections on thermal phase transitions between AdS and dS geometries. This research may also be helpful to clarify the correspondence in the case of dS geometries, which is not as well developed as its AdS counterpart.
	\item  The second concerns new solutions of Supergravity found using Non-Abelian T-duality, a generating technique based on string theory. Such new solutions are dual to supersymmetric RG flows, and we investigate the effects of the Non-Abelian T-duality on the holographic observables. There is also a (super)gravity motivation, because the solutions generated are highly non-trivial, and can fall outside existing classifications of backgrounds. Apart from being potentially interesting in themselves, the new field theories may help to understand the effect of Non-Abelian duality on the string theory sigma model, its supergravity approximation and its generic interplay with AdS/CFT.
	
	\item In the third we extend holographic renormalization to reproduce the \wi of spontaneous symmetry breaking in a 1+1 holographic superconductor. The model could find application in AdS/CMT. Furthermore, the use of a bottom up model correctly reproduced the Ward Identities, confirming for this the validity of the conjecture outside SUGRA and in this peculiar dimension. 
\end{itemize}
In the following few pages we introduce these problems in more detail and mention the main results.
\section[\texorpdfstring{A\MakeLowercase{d}S to \MakeLowercase{d}S phase transitions in higher curvature gravity}{AdS to dS phase transitions in higher curvature gravity}]{AdS to dS phase transitions in higher curvature gravity}
We have analyzed the role of higher curvature gravity corrections on thermal phase transitions between AdS and dS geometries. Apart from the gravitational interest, this research may also be helpful to clarify the correspondence in the case of dS geometries, that is less developed than it AdS counterpart. The details can be found in chapter \ref{gb}.

 Higher curvature corrections to gravity are studied for a variety of reasons: as possible corrections to GR coming from some UV completion candidate (like string theory), to examine their cosmological effects, to explore their holographic meaning, etcetera. Among such theories, Lovelock family of actions have some unique properties, as the absence of ghosts around flat space and second order EoM in general. We will focus on the simplest non-GR case of Lovelock family, known as Lanczos-Gauss-Bonnet gravity (\textit{cf}. \ref{gbintrogb}), the unique quadratic action with no ghosts around flat space as well as the (heterotic) string theory effective action correction to GR. EoM differ from those of GR only in more than four bulk dimensions, bringing another piece of interest for higher dimensional gravity besides holography and a richer phase diagram.  
 
Despite this interest in higher curvature theories, gravitational phase transitions already occur in GR. One of the best known examples is the Hawking-Page transition \cite{Hawking:1982dh}, in which thermal AdS decays to a BH above a certain temperature. Furthermore, it is certainly possible for a phase transition in GR to change spacetime asymptotics \textit{ when matter is added}. A simple way is the quantum or thermal fluctuation between the different non-degenerated minima of a scalar potential. After spending some time in a metastable minimum, the field nucleates to a more stable one \cite{Coleman:1980aw}. Transitions changing the value of asymptotic curvature have been studied in relation to the cosmological constant problem for decades. 

Due to the effect of higher curvature corrections, these transitions can happen between AdS geometries of different curvature radius \textit{even in the absence of matter} \cite{Camanho:2013uda}. The transitions are mediated by a thermalon, i.e. the euclidean section of a bubble (which we describe in the next paragraph). To understand the phenomenon it is important to know that higher curvature gravities have several different maximally symmetric spacetimes (vacua), instead of the unique one in GR. The endpoints of the transitions are metrics asymptoting to different vacua. In the particular case of LGB there are only two possible metrics for spherical black hole ansatz, and each of them asymptotes to one of the two possible vacua. These two different type of solutions for BH ansatz are called branches.

The interior of the bubble is given by the branch of the final vacuum and it is glued to an exterior solution with the initial asymptotics. For the whole configuration to be a solution of the Lanczos-Gauss-Bonnet EoM, the so called junction conditions must be satisfied. They are similar in spirit to the Israel-Darmois matching equations, but the physical implications are very different.

The thermalon forms when it has lower free energy than thermal AdS, like any other thermodynamical phase. In our case, it is unstable against small metric perturbations and will either collapse or expand. The latter possibility gives rise to the desired change of geometry from AdS to dS. Let us remark that the metric is continuous at the junction but the derivatives are not; despite it such solution must be considered as a legitimate competing saddle in the path integral approximation to the free energy \cite{Camanho:2013uda}.

We can summarize our procedure and final results saying that we found a phase transition from AdS to dS geometry in Lanczos-Gauss-Bonnet gravity. The transition happens through the formation of a bubble with a dS  black hole in the interior and an AdS metric in the exterior. Afterwards it can expand filling the spacetime and eventually changing the geometry. No matter fields are required to match both sides of the bubble due to the Lanczos-Gauss-Bonnet junction conditions. For other higher curvature corrections we generically expect the same phenomenon. 

The precise holographic meaning of the geometry change from AdS to dS, if any, is unknown to us. Our work was the continuation of previous studies of transitions between different AdS vacua in the Lovelock theory, which in turn are thought to be dual to quantum quenches. Holographically, the peculiar interest of our case is that a physical process is changing the geometry from AdS to dS \textit{dynamically}. A proper interpretation may help to advance the dS/CFT correspondence. 
\section[\texorpdfstring{Deformations of Klebanov-Witten CFT and A\MakeLowercase{d}S$_3 $ backgrounds via NATD}{Deformations of Klebanov-Witten CFT and AdS$_3 $ backgrounds via NATD}]{Deformations of Klebanov-Witten CFT and AdS$_\textbf{3} $ backgrounds via NATD} \label{natdsummary}
We found new solutions of Supergravity using Non-Abelian T-duality (NATD), a generating technique based on string theory. Such new solutions represent supersymmetric RG flows, and we investigate the effects of the Non-Abelian T-duality on the holographic observables. The problem also present a pure (super)gravity interest, as the generated SUSY solutions can fall outside known classifications. Apart from their intrinsic interest, the new field theories may help to understand the effect of Non-Abelian T-duality on the string theory $\sigma $- model, its supergravity approximation and its interplay with AdS/CFT. 

Non-Abelian T-duality (or NATD) can be seen as a SUGRA solution generating technique. As such, it is a powerful instance in which knowledge on (super)gravity leads to novelty in QFT through holography. It is not known if the $ \sigma $-model partition function is invariant under NATD, as it is in the case of Abelian duality. The effect on holographic observables may help to clarify this issue. \cite{Itsios:2016ooc}.

We can make a one paragraph summary of the results presented in chapter \ref{natd} in the following terms: there are deformations of Klebanov-Witten background AdS$_5\times$T$^{1,1}$ flowing to an AdS$_3$ factor in the IR. We generate new examples applying Non-Abelian T-duality to the previous ones. We finally compare the holographic observables of the generated solutions with those of the original RG flows.

Let us explain general aspects of the transformation, that will be extended in section \ref{natdintronatd}. \natd is the non-Abelian version of standard Abelian T-duality, and when applied to a background with non-Abelian isometry the resulting one is guaranteed to be a solution of SUGRA. When the non-Abelian isometry is $ \mathrm{SU(2)} $ (as in our case), it relates backgrounds in IIA with backgrounds in IIB. The $ \mathrm{SU(2)} $ symmetry is always destroyed and it cannot be used a second time. In this regard it is very different of Abelian T-duality, that yields the same result when applied twice. Not surprisingly, the reduction of spacetime symmetry can lead to SUSY reduction. Another important difference is that the range the dual coordinates is not fixed by the generating technique. This is turns implies the the dual field theories are not fully determined.

For the Neveu-Schwarz sector of SUGRA non-Abelian T-duality was discovered in 1993 \cite{delaOssa:1992vci} as a symmetry of the string sigma model. However, the implications for AdS/CFT were not examined until 2010, when Sfetsos and Thompson \cite{Sfetsos:2010uq} extended the transformation to the Ramond-Ramond sector. Although we do not know yet if it is a duality for the full (perturbative) string theory, the use as a generating technique does not rely on it. 

Typically the solutions produced are highly non-trivial although they always have closed expressions; one cannot reasonably expect to find them via an ansatz as there is dependence on several coordinates. When applied to AdS$_5\times$S$^5$, the generated background is dual to a $ \mathcal N =2 $ known field theory \cite{Sfetsos:2010uq}; however, in less symmetric seed backgrounds it is usually impossible to fully identify the dual field theory of the new solution. Nevertheless, one can compare the holographic observables of the initial background and those of its non-Abelian T-dual to get relevant information of the generated field theory. 

Having already summarized the technique, we introduce the backgrounds on which we will apply it (seed backgrounds). All of them are holographically dual to compactifications of Klebanov-Witten (KW) CFT \cite{Klebanov:1998hh} on a spacelike 2-surface $ \Sigma_2 $. Let us remember that KW CFT it is a $ d=4 $ $ \mathcal N =1 $ SUSY field theory with gauge group $ \mathrm{SU(N)} \times \mathrm{SU(N)} $, and one of the very few instances in which both sides of the correspondence are known explicitly. The holographic dual is given by a type IIB background with geometry AdS$_5\times$T$^{1,1}$, where  T$^{1,1} $ is a Sasaki-Einstein manifold given by $ \mathrm T^{1,1} = \mathrm{SU(2)}\times \mathrm{SU(2)}/\mathrm{U(1)} $, supported by an $ F_5 $ RR field. Its non-Abelian T-dual was already known \cite{Macpherson:2014eza}, although the holographic analysis was not complete.

The KW deformation duals on which NATD is applied can be separated into two groups:
(see table \ref{natdsolutionstable} for a summary and section \ref{natdprevioussolutions} for details). The first is given by the duals of some 4d QFT when placed on a manifold $\Sigma_2$ of constant curvature; in particular an $\mathcal N=1$ SUSY flow with $\Sigma=\mathrm{H_2}$ and non-SUSY fixed points with $\mathrm{\Sigma_2=\mathrm H_2, T^2, \mathrm{S^2}}$; we called them twisted solutions. The second kind is the deformation of KW found in \cite{Donos:2014eua} by Donos and Gauntlett (DG), we present it in section \ref{natdDG}. This last \bg was chosen as a seed of the generating technique for several reasons, that include: 
\begin{enumerate}
	\item It interpolates between KW in the UV and AdS$_3\times\mathbb R^2\times$S$^2\times$S$^3$ in the IR. It is a type IIB solution where $B_2$ and $F_3$ are turned on. (They vanish in the twisted solutions).
	\item It can be embedded in $\mathcal N=4$ $D=5$ gauged \sugra. 
	\item (0,2) \susy is preserved all along the flow; in the IR it enhances to (4,2) superconformal \susy.  The solution is regular everywhere. 
	\item It is related by two Abelian dualities to AdS$_3\times$S$^3\times$S$^3\times$S$^1$. The presence of additional $\mathrm{U(1)}$ symmetries allows the combination of \natd with further Abelian \td. 
\end{enumerate}
For the flowing backgrounds above ($ \Sigma = \mathrm{H_2} $ and DG solution), the Non-Abelian T-dual preserves \susy and smoothness of the seed solutions; the generated backgrounds are shown in sections \ref{natdtwistedduals} and \ref{natddualDG}. 
 
We have also analyzed the Page charges, c-function, entanglement entropy (EE) on a strip and a rectangular Wilson loop (WL), before and after applying \natd. This calculation suggests that the new backgrounds' dual field theories is related to an infinitely long linear quiver gauge theory, as the dependence of central charge is cubic in the number of NS5 branes.  The c-function, EE and Wilson loop change by a constant factor along the RG flow under the effect of the duality.  See \ref{natdcentralcharge} and \ref{natdEEWilson} for more details.

\section{Holographic Ward Identities in 1+1 QFT}
 \label{wardsummary}
We have studied the extension of the AdS/CFT correspondence, in particular the holographic renormalization procedure, to reproduce the \wi of spontaneous symmetry breaking in a 1+1 holographic superconductor, which is a bottom up model. The result could be of interest for one-dimensional condensed matter applications. A peripheral motivation is to know if the bottom up model we use behaves in physically sensible manner for this low dimension. A final goal is to explore what is the role of peculiarities of three dimensional gravity in the bulk. This content is presented in chapter \ref{ward}.

Spontaneous symmetry breaking (SSB) is important in High Energy Physics as well as in Condensed Matter. A key result about it is the Goldstone theorem: there is a massless boson (called Goldstone boson) for every given generator whose global symmetry is spontaneously broken. The properties of \sb\ are reflected on the Ward identities: the one point value signals SSB whereas the 2 point correlator contains the Goldstone boson as a massless pole.  The Goldstone theorem does not apply directly to QFT in 1+1 dimensions: at finite $N$ SSB and Goldstone bosons are not possible (Coleman theorem); this restriction disappears in the strict \lN limit. 

Holographically, \sb\ has been described in several previous works with the double motivation of writing the \hoc\ dictionary and applying it to the less understood case of strong coupling. In particular, the authors of  \cite{Argurio:2015wgr} found the Ward Identities and an analytic pseudo-Goldstone bosons in 2+1. An introduction of the relevant aspects is in sections \ref{introwardsbwi} and \ref{introwardpreviouspaper}. 

In this chapter, we extend \cite{Argurio:2015wgr} and \hocy\ derive \wi\ of \csb in a 1+1 QFT. The gravity side is a (bottom up) \hoc\ superconductor (eq (\ref{wardrenormchargedscalar})); its $\mathrm{U(1)}$ symmetry is broken. In the 2+1 dimensional bulk, the \hoc description of \cc\ requires an \aq\ for the $\mathrm{U(1)}$ vector field, as explained in sections \ref{wardordynary} and \ref{wardalternative}. The essential reason can be summarized as follows. In any dimension, the current couples to the constant, gauge-dependent term in the near horizon expansion of $A_\m$. Therefore, that mode must be a source in our on-shell action. In \hd\ such mode is the leading one, whereas in 2+1 bulk it is the subleading. In this last case, the \hren\ needs different \bt for that mode to become a source. Such choice of the subleading mode as a source, and the addition of the required different \bt is what we mean by alternative quantization.

Once the whole procedure is carried out, the standard \hdal field-theoretic \wi \eqref{wardwardeq} are recovered. Furthermore, a Goldstone boson pole appears analytically for the spontaneous symmetry breaking case, see section \ref{wardaltscalar}. This is not in contradiction with the Coleman theorem because holographically we are in the strict \lN limit. 
\chapter[\texorpdfstring{A\MakeLowercase{d}S to \MakeLowercase{d}S phase transitions in Lanczos-Gauss-Bonnet gravity}{AdS to dS phase transitions in Lanczos-Gauss-Bonnet gravity}]{AdS to dS phase transitions in Lanczos-Gauss-Bonnet gravity} \label{gb}
\newpage
\vskip-15mm

This chapter covers the phase transitions between  AdS  and dS thermalons than can happen in Lanczos-Gauss-Bonnet (LGB) gravity. We begin introducing the LGB action as a member of the Lovelock family. In the second section we define the thermalon configurations that mediates the transition, as well as its dynamics and Thermodynamics. The next sections contain the original results. Finally we have added an appendix with some generalities of euclidean on-shell action formalism for BH Thermodynamics.
\section{Lovelock and Lanczos-Gauss-Bonnet gravity}  \label{gbintrogb}
\subsection*{Bulk action} 
In this section we introduce the Thermodynamics of Lanczos-Gauss-Bonnet (LGB) theory of gravity, the simplest member of the Lovelock family different from GR. We will show the action with the appropriate boundary terms, the branches of vacua and black holes and finally some general remarks about its Thermodynamics. 

Let us start with the motivation: why to consider higher curvature corrections to gravity at all? The covariance under general changes of coordinates is a fundamental principle behind the description of gravity, but it can be satisfied with an infinite number of gravity actions. In fact, any general tensorial expression that is fully contracted will be invariant by diffeomorphisms, and higher curvature corrections are a distinguished particular case. In the following we give some more specific motivations to study the role of higher curvature corrections:
\begin{enumerate}
	\item We know that GR is not complete. Nevertheless, given its theoretical and experimental successes it is natural to consider perturbative corrections around it.
	\item Any higher energy gravity proposal, no matter how different from the geometric setup of GR, will have to reproduce it in the classical limit and also give some higher derivatives terms as leading corrections. 
	\item For some Superstring theories, such (leading order) corrections are known. It is interesting to test the consistency of those truncated approximations. In particular, for heterotic and bosonic string theories, the pure gravity first correction in $ \alpha' $  to GR is given by the LGB term \eqref{gbintrogbaction} \cite{Metsaev:1987zx}.
	\item Do the corrected actions satisfy basic physical requirements like: causality, absence of ghosts, uniqueness of initial value problem, sensible BH Thermodynamics, cosmic censorship, ... ?
	\item We desire to understand the implications on the AdS/CFT dual field theory.
\end{enumerate}
Among the modifications of Einstein-Hilbert (EH) action without introduction of matter fields (what we call pure gravity theories), the Lovelock family have some unique properties. Let us begin explaining the defining one. If one demands second order equations of motion \textit{in four dimensions} to be of the form
\bea
\mathcal{F}_{\m\n} (g_{\a\b},\partial_\g, \partial^2_{\gamma,\delta} g_{ab}) \eq T_{\m\n},
\eea
where $\mathcal{F}$ is a local, symmetric and conserved tensor (i.e. $\mathcal{F^{\m\n}_{\;\;,\n}}=0$), Lovelock \cite{Lovelock:1971yv} established that the only possibility are the standard Einstein equations of motion: 
\bea
R_{\m\n} - \half g_{\m\n} R + \Lambda g_{\m\n} \eq 8 \pi G_N T_{\m\n} .
\eea
As it is well known, these EoM can be derived from the Einstein-Hilbert action:
\bea
S \eq \int d ^4 x \sqrt{-g}\; \prt{R - 2\Lambda} + S_{matter}.
\eea
Another way to single out EH gravity in any dimension is to demand linearity of the equations in the second derivatives of the metric. This can be slightly relaxed to quasi-linearity: there are not squared or higher order terms in second derivatives, more details on this condition can be found in \cite{Deruelle:2003ck}. If in five dimensions we require again having second order EoMs with local, symmetric and conserved $\mathcal{F}$, we end up with the Lanczos-Gauss-Bonnet theory in five dimensions \cite{Lanczos:1932zz,Lanczos:1938sf}\footnote{This work was performed by Cornelius Lanczos, although the name Lanczos-Gauss-Bonnet is not found often in the literature. }:
\begin{equation}
	S = \frac{1}{16\pi G_d} \int d^d x\sqrt{-g}\,\left( R -\frac{\tilde \Lambda (d-1)(d-2)}{L^2} + \frac{\lambda L^2}{(d-3)(d-4)}\mathcal{R}^2 \right),
	\label{gbintrogbaction}
\end{equation}
where $\mathcal{R}^2 = R^{2} - 4 R_{\mu\nu} R^{\mu\nu} + R_{\mu\nu\lambda\rho} R^{\mu\nu\lambda\rho}$, which guarantees that the equations of motion are second order in $d\geq 5$ spacetime dimensions. A normalized characteristic length $L$  is introduced defining the \textit{bare} cosmological constant $ \Lambda $ as $\Lambda =\tilde \Lambda (d-1)(d-2)/2 L^2$, and we are left with a single dimensionless coupling, $\lambda$. The adjective bare is used because the asymptotic curvature of the vacuum will no longer be proportional to $\Lambda$ as in GR. Actually, there are in general two different vacua and not just one. $ \tilde \Lambda $ is the sign of the bare cosmological constant $ \Lambda $. We will always use $\lambda > 0$ which is the natural sign inherited, for instance, from string theory embeddings of \eqref{gbintrogbaction}.  

Generalizing for any dimensions and quasi-linear second order EoMs, Lovelock \cite{Lovelock:1971yv} found all the possibilities, that are henceforth known as the Lovelock family. For any given order in the curvature, there is a new Lovelock term, and the sum of them also fulfills the special properties demanded above. The quadratic Lovelock (Lanczos-Gauss-Bonnet) is topological for dimension 4, the cubic Lovelock for dimension 6, and the $k$-th for dimension $2k$. The particular case of GR is of this kind for $k=1$, and it is indeed topological in 2 dimensions as follows from the 2 dimensional Gauss-Bonnet theorem. 

An additional important property of the Lovelock family is the absence of higher derivative ghosts \cite{Lovelock:1971yv,Zumino:1985dp}. Indeed, when one consider gravitons around flat space, the LGB term \eqref{gbintrogbaction} is the only quadratic in Riemann action with that property. It must be stressed, however, that the theory can present ghosts around non-flat solutions despite having second order equations, as it will appear later under the name of Boulware-Deser instability. Finally, LGB requires at least dimension five to not be topological, what is a motivation for higher dimensions in addition to AdS/CFT. 
\subsection*{Boundary terms} 
\label{gbintrogbbt}
As explained in the appendix \ref{gbintrothermo}, the Euclidean path integral requires an action that is stationary under metric variations with a certain restriction on the boundary; in our case that we will fix the induced metric on the boundary. Now we will introduce the terms corresponding to that condition that are required for the Lovelock family and for LGB in particular.

To discuss boundary terms, it is convenient to write the LGB action \eqref{gbintrogbaction} using the vierbein formalism. \footnote{We are not going to consider connections with torsion, and therefore the results in this section will be the same found with the metric formalism.}. This language allows to quote the form of the entire Lovelock family of actions, as the expression is very systematic for the different Lovelock orders:
\begin{equation}
	S =\frac{1}{16\pi G_N (d-3)!}\, \sum_{k=0}^{K} {\frac{c_k}{d-2k}} \mathop\int \mathcal{L}_{k} ~,
	\label{gbintrogbgenerallovelock}
\end{equation}
$G_N$ being the Newton constant in $d$ spacetime dimensions. $\{c_k\}$ are the couplings with length dimensions $L^{2(k-1)}$, $L$ is a length scale related to the (bare) cosmological constant (as explained after eq. \eqref{gbintrogbaction}) and $K$ is the maximum value of $ k $ for which $ \mathcal{L}_k $ is not a boundary term, given by:  %
\begin{equation}
	K\leq \left[\frac{d-1}{2}\right] ~,
	\label{maximalK}
\end{equation}
labeling the highest non-vanishing coefficient, {\it i.e.}, $c_{k>K} = 0$.   $\mathcal{L}_{k}$ is simply:  %
\begin{equation}
	\mathcal{L}_{k} = \e\!\left(R^k e^{d-2k}\right)=\epsilon_{f_1 \cdots f_{d}}\; R^{f_1 f_2\ldots f_{2k-1} f_{2k}} \wedge e^{f_{2k+1}\ldots f_d} ~.
\end{equation}
where $ \epsilon $ is the Levi-Civita symbol. The zeroth and first term in \eqref{gbintrogbgenerallovelock} correspond, respectively, to the cosmological term and the Einstein-Hilbert action. The first Lovelock term beyond GR only contributes for dimensions larger than four and corresponds to the LGB coupling $c_2=\lambda L^2$ as given in \eqref{gbintrogbaction}.
The Lovelock actions defined by \eqref{gbintrogbgenerallovelock} are not stationary under Dirichlet boundary conditions. The situation can be more clearly understood with a one-particle analogy taken from \cite{Camanho:2015ysa}. Consider the following free particle action:
\begin{equation}\label{introgbgbparticledirichlet}
	S=\frac{1}{2}\int_{t_1}^{t_2} \dot{x}^2dt ~.
\end{equation}
This action is stationary under Dirichlet boundary conditions \footnote{In a single particle system, fields only depend on one coordinate $ t $. The boundary is the initial and final values of it, not a spatial region.}, namely, fixing $x$ at initial and final times, $t_{1,2}$,\ \ie \ $\delta x(t_{1,2})=0$:
\begin{eqnarray} \label{gbintrogbparticledvariation}
	\delta S &=& \int_{t_1}^{t_2} \dot{x}\delta \dot{x}  \ dt=   \int_{t_1}^{t_2} \left[ - \ddot{x} \delta x +  \partial_t (\dot{x} \delta x )\right] dt \\
	&=&  \int_{t_1}^{t_2} (-\ddot{x})\delta x \ dt  + \dot{x}(t_2) \delta x(t_2) -   \dot{x}(t_1) \delta x(t_1)~. \nonumber 
\end{eqnarray}
Therefore for a solution of the equations of motion $\ddot{x}=0$ the action is stationary, $\delta S=0$.
But we can also consider the action: 
\beq\label{gbintrogbparticlemixed}
S'= - \frac{1}{2}\int_{t_1}^{t_2} x\ddot{x}dt ~,
\eeq
that still has the same equations of motion of \eqref{introgbgbparticledirichlet}. Nevertheless its variations no longer fix the position in the extremes of the time interval:
\begin{eqnarray} \label{gbintrogbactionparticlemixedvariation}
	\delta S'&=&  \int_{t_1}^{t_2} (-\ddot{x})\delta x \ dt   \\ &-&\frac{1}{2} {x}(t_2) \delta \dot{x}(t_2) + \frac{1}{2} {x}(t_1) \delta \dot{x}(t_1)
	+ \frac{1}{2} {\dot x}(t_2) \delta x (t_2) - \frac{1}{2} {\dot x}(t_1) \delta x (t_1)\ \ . \nonumber
\end{eqnarray}
We need to add a boundary term, the analogue of Gibbons-Hawking term, in order to fix the metric in GR. In this example, it is not difficult to see that the $ S' $ is related to $ S $ by integration by parts, the difference being a total derivative (i.e. a boundary term in 0+1 case):
\bea
S \eq  S' + \frac{d}{d t}\prt{\frac{x \dot x}{2}}.
\eea
The variation of $ S-S' $cancels the first two terms in \eqref{gbintrogbactionparticlemixedvariation} and adds up to the other two such that we get the original result \eqref{gbintrogbparticledvariation}. In our Lovelock case, the bulk action \eqref{gbintrogbaction} behaves like \eqref{gbintrogbparticlemixed}. In the following we introduce the necessary total derivatives (analog to $ \frac{d}{d t}\prt{\frac{x \dot x}{2}} $) necessary to have Dirichlet boundary conditions, following the results of \cite{Myers:1987yn} .

To introduce the necessary boundary terms that make the action stationary under Dirichlet boundary conditions, we define first the spin connection difference $\theta$ with respect to a reference connection $ \omega_0 $:
\bea
\theta \eq \omega - \omega_0,
\eea
$ \omega_0 $ is typically taken to correspond to the product metric that agrees with $ g $ in the boundary. The result is independent of the $ \omega_0 $ choice. For more details see \cite{Myers:1987yn}.

With $ \theta $ definition we can already quote the boundary term for any general Lovelock action with coupling constants $c_k $ as defined by \eqref{gbintrogbgenerallovelock}:
\begin{equation}
	S_{\partial} = \frac1 {16\pi G_N (d-3)!}\sum_{k=1}^{K} {\frac{c_k}{d-2k}\int_{\partial\cal M}\mathcal{Q}_k} ~,
\end{equation}
where
\begin{equation}
	\mathcal{Q}_k =  k \int_0^1 d\xi \, \epsilon_{a_1\cdots a_d}\,\theta^{a_1 a_2}\wedge \mathcal{R}_\xi^{a_3 a_4}\wedge\ldots\wedge \mathcal{R}_\xi^{a_{2k-1} a_{2k}} e^{a_{2k+1}\cdots a_d} ~,
	\label{genGH0}
\end{equation}
and 
\be
\mathcal{R}_\xi^{ab}= \xi R^{ab}+(1-\xi)R_0^{ab}-\xi(1-\xi)(\theta\wedge \theta)^{ab}    ~.
\ee
The total action with Dirichlet boundary conditions is just:
\bea
\widetilde{S}=S-S_\partial
\eea (notice the $-$ sign). For the GR and LGB cases ($ k=1,2 $ respectively), the boundary terms can be simplified to: 
\bea
\mathcal{Q}_{1}\eq \theta^{a_1 a_2}\wedge e^{a_3\cdots a_d}\, \epsilon_{a_1\cdots a_d} ~,\\
\mathcal{Q}_2 \eq 2\theta^{a_1 a_2}\wedge (R^{a_3 a_4}-\frac{2}{3}\theta^{a_3}_{\ \;c}\wedge\theta^{c a_4}) \wedge e^{a_5\cdots a_d}\, \epsilon_{a_1\cdots a_d}~. \label{gbintroLGBbterms}
\eea
\subsection*{LGB black holes and their Thermodynamics} 
\paragraph{Branches of vacua and black holes}
One of the central aspects of interest for any gravity theory are the black hole solutions. We restrict ourselves to  spherically symmetric ansatz:
\begin{equation}
	d s^2 = -f(r)\,d t^2 + \frac{d r^2}{f(r)} + r^2 d\Omega^2_{d-2} ~,
	\label{gbsBHansatz}
\end{equation}
where $d\Omega^2_{d-2}$ is the metric of a unit sphere. Notice that this ansatz is of the form \eqref{gbintrothermobhansatz} we calculated the temperature for, and it is given by eq. \eqref{gbintrothermotemperaturespherical}. The equations of motion of  \eqref{gbintrogbaction} can be readily solved and two branches of solutions characteristic of LGB emerge  \cite{Boulware:1985wk,Wheeler:1985nh,Wheeler:1985qd}:
\begin{equation}
	f_{\pm}(r) = 1 + \frac{r^2}{2\lambda L^{2}} \left( 1\pm \sqrt{1 + 4\lambda \left( 1  + \frac{{\rm \tilde \Lambda M}_\pm}{r^{d-1}}\right)} \right) ~,
	\label{gbintrogbholes}
\end{equation}
where M$_\pm$ is the (properly normalized) mass parameter of the spacetime \cite{Boulware:1985wk}. The dS case under consideration has been analyzed in detail in \cite{Cai:2003gr}. Notice that the only branch admitting a black hole solution ({\it i.e.}, a smooth event horizon) is that with the minus sign. Each of the two branches in \eqref{gbintrogbholes} is associated with a different value of the effective cosmological constant,
\begin{equation} \label{gbintrovacua}
	\Lambda_{\pm} = - \frac{1\pm \sqrt{1 +  \tilde \Lambda 4\lambda}}{2\lambda L^{2}} ~.
\end{equation}
It is very important to keep in mind the existence of several vacua in higher curvature theories of pure gravity. This property allows new physical processes, in particular the phase transition between different vacua. For $ \tilde \Lambda = 1 $, that will be our case of interest, observe how $ \Lambda_\pm $ behave in the GR limit $ \lambda \rightarrow 0 $: $ \Lambda_+ \longrightarrow - \infty $ while $ \Lambda_- \longrightarrow \frac{1}{L^2} $. So we see how the $ - $ branch of vacua is continuously connected with the EH one. This also happens for the black holes in eq. \eqref{gbintrogbholes}. 

The AdS vacuum $\Lambda_+$ is afflicted by the so-called Boulware-Deser (BD) instability \cite{Boulware:1985wk}. This means that the gravitational perturbations around it present ghosts, and therefore $ \Lambda_+ $ is not a healthy vacuum.\footnote{We will elaborate on this point when needed in our original results presentation.}

\paragraph{General aspects of LGB BH Thermodynamics} \label{gbintrogeneralgbthermo}
The laws of BH Thermodynamics for GR are established in all generality (\textit{cf.} section \ref{gbintrothermohistory}). But some of the proofs depend on the form of the action, and even perturbative corrections may induce unphysical Thermodynamics. Given the remarkable properties of LGB, the study of this issue in it has been more profound than in generic theories of gravity. Nevertheless, the results are not as sound as in GR, as explained in the following paragraphs. 

The zeroth law is satisfied for matter respecting the dominant energy condition \cite{Sarkar:2012wy}, the same as in the Einstein-Hilbert case. The first law is also verified, as expected on general grounds at least since the work of Wald in \cite{Wald:1993nt}. Indeed, it was established before for Lovelock in \cite{Myers:1988ze} and further investigated in \cite{Jacobson:1993xs}. As it happens generically in gravity with higher curvature corrections, the entropy is no longer given by the area of the horizon; instead for  particular case of LGB we have:
\bea
S_{BH} \eq \frac{1}{4G_{N}} \int_H d A \; \prt{1+\lambda R_{(d-2)}},
\eea
where $ \lambda $ is the higher order coupling as defined in \eqref{gbintrogbaction} and $ R_{(d-2)} $ is the \textit{intrinsic} scalar curvature of the horizon.\ Generic corrections to area law  do not depend only on intrinsic geometry (notice that GR is a member of the Lovelock family and also fulfills this property). If the horizon is compact \footnote{This is not always the case, for example in black funnels \cite{Hubeny:2009kz}.}, the correction to area law in four dimensions will depend only of the horizon topology.  A particular problem in LGB is that the entropy can become negative in some cases, therefore incompatible with the entropy of any physical system. This cannot happen in GR, though.  

Another key result is the second law. It has been shown that it also holds for general Lovelock theories in a number of cases \cite{Akbar:2008vz,Sarkar:2010xp}. An interesting physical process version has been analyzed in \cite{Kolekar:2012tq}. Nevertheless the present results are not as general as Hawking area theorem for GR. Some study on the third law was performed in \cite{Torii:2006gu}.

Given that state of affairs, we will take the usual thermodynamic interpretation for granted but always keeping in mind that the second law is not fully proven. Our completely solid identification of the Noether charge with entropy is based only in the first law.
\section{Gravitational phase transitions mediated by thermalon} \label{gbintrobubbles}
\subsubsection{Previous bibliography} 
Concerning the macroscopical properties of any system, it is of  the  greatest relevance to know the different possible phases and the transitions among them. This applies for black holes as well. For the asymptotically flat BHs of GR in 4D, the possible phases allowed by the Einstein-Maxwell equations are well known. An stationary BH is specified uniquely by the three extensive parameters of the first law $M,J,Q$ \cite{Israel:1967za,Israel:1967wq,Carter:1971zc}. For asymptotically AdS solutions in four dimensions, thermal AdS can decay to an AdS black hole at a certain temperature. This is known as the Hawking-Page transition \cite{Hawking:1982dh}. Another important factor that makes the phase diagram richer is the number of spacetime dimensions. When equal to five, the BH uniqueness turns into multiplicity. For example, for $Q=J=0$ and a given mass there are black holes and black rings solutions, that are different even in the topology of the horizon \cite{Emparan:2008eg}.

As we are interested in LGB we have to consider $ d>5 $, otherwise the LGB term does not contribute to the EoM. Furthermore, a new dimensionful quantities is introduced as a coupling constants. We therefore expect in general rather non-trivial phases diagram in this context, and indeed we do not intend an exhaustive investigation. We are exclusively concerned with a transition between pure thermal AdS and a dS BH in the absence of matter. In this section we are going to present the mechanism that makes it possible, namely the nucleation of spacetime bubbles.\\
Before describing in some detail the bubbles and their formation, dynamics and Thermodynamics, let us clarify a point that may confuse the reader familiar with gravitational phase transitions. Once we choose the canonical ensemble, the competing phases must have the same temperature, charges ($ J,Q,... $) and asymptotics. But we said that the transition would lead from AdS to dS  asymptotics.Then, how can we compare their free energies? The answer is that we will not compare them directly. Thermal AdS will compete with a bubble that contains dS BH in its interior with AdS asymptotics; only in that way their free energies can be compared. After the bubble is formed, it can expand in a purely dynamical manner dictated by the EoM and eventually change the boundary to dS.

In the following we expose the essentials of previous literature dealing with gravitational phase transitions mediated by a thermalon. Due to the cosmological constant problem, some attention has been devoted to phase transitions that can change its value and therefore may justify its abnormally large value. In this instances, the existence of several possible values of asymptotic curvature in EH gravity comes from the different vacua of matter. Although we do not have a cosmological motivation, we will rely on the bubble nucleations mechanism they proposed, but we will not require the addition of matter.

The first step is to illustrate bubble nucleation in everyday systems, say water. When the temperature gets close to evaporation threshold, bubbles of the new phase (steam) appear. The same can happen in quantum field theory: bubbles of true vacuum can appear inside a metastable one. The actual mechanism can includes thermal and quantum fluctuations, and we will focus on the former, that was studied in gravitational context by \cite{Klinkhamer:1984di,Gomberoff:2003zh}. They found that the probability of bubble nucleation goes as
\bea
P \propto e^{-I_E}
\eea 
where $ I_E $ is the euclidean action difference between initial and bubble state. The euclidean section of that bubble is what we call thermalon, and we say that the transition is thermalon mediated. Notice the peculiarity that this transition can happen even when $I_E>0$. Therefore we can say that the initial state is indeed \textit{metastable}. 
 Other examples related to instanton mediated phase transitions in gravity are \cite{Brown:1987dd,Brown:1988kg}, in relation to the cosmological constant problem. In the case of \cite{Gomberoff:2003zh}, pure de Sitter ends up in a dS black hole after the thermalon decay, somewhat reminiscent of Hawking-Page transition. Normally the change is from dS (higher vacuum energy) to AdS (lower vacuum energy), but there are instances where the opposite happens as in \cite{Cvetic:2001bk,Nojiri:2001pm,Kim:2007ix,Gupt:2013poa}. 

Now that we summarized the previous works using EH actions, let us introduce the higher curvature corrections. The difference with the previous cases is that in LGB the thermalon can be the result of matching the two different black hole branches \eqref{gbintrogbholes}, \textit{without any matter in the junction} \cite{Gravanis:2007ei,Garraffo:2007fi}; this is impossible in EH gravity. One of the branches is in the interior and the other in the exterior of the bubble; we will sometimes say junction to refer the bubble's surface. The existence of several branches and the the possibility of gluing them together without matter is not specific of LGB, but fairly generic; therefore we strongly expect that this kind of phase transition is generic. For the LGB case, the Thermodynamics for the two asymptotically AdS branches was investigated in detail in \cite{Camanho:2013uda} with the main results: 
\begin{itemize}
	\item Generic Lovelock matching conditions, dynamics and Thermodynamics are analyzed for spherical thermalons.
	\item Thermalon is indeed the dominant phase in LGB and cubic Lovelock at a certain temperature. 
	\item The change of asymptotics takes place through the expansion of the bubble, when it is dynamically unstable and can reach the boundary. 
	\item Lovelock parameters constraints to avoid some pathological configurations.
	\item Discussion of bubble collapse and naked singularities formation.
\end{itemize}
In the present chapter we extend their work for LGB to include similar transitions from AdS to dS asymptotics. There are several motivations. The first is that we consider the problem is potentially interesting holographically, as a physical process would be dynamically substituting the AdS boundary for dS. Moreover, the most frequently studied transitions are from dS to AdS and not vice versa; finally there is the additional interest of using higher curvature corrections and not matter. In the following subsection we describe the thermalon configuration formally.  

\subsubsection{Thermalon configuration and dynamics} \label{gbintrobubblesthermalonconfiguration}
Before going into details, we make a clarification about terminology. A source of possible confusion to the reader is the somewhat exchangeable use of the words bubble and thermalon. Being precise, \textit{the thermalon is the euclidean section of a static bubble}.  In strict sense, when we talk about dynamics (expansion or contraction, normally) we are always referring to the bubble. When we consider the bubble as a thermodynamic phase, we are really meaning the thermalon.  With this in mind, let us formally characterize the thermalon, following \cite{Camanho:2013uda}.
\begin{figure}	
	\centering
	\includegraphics[width=.65\textwidth]{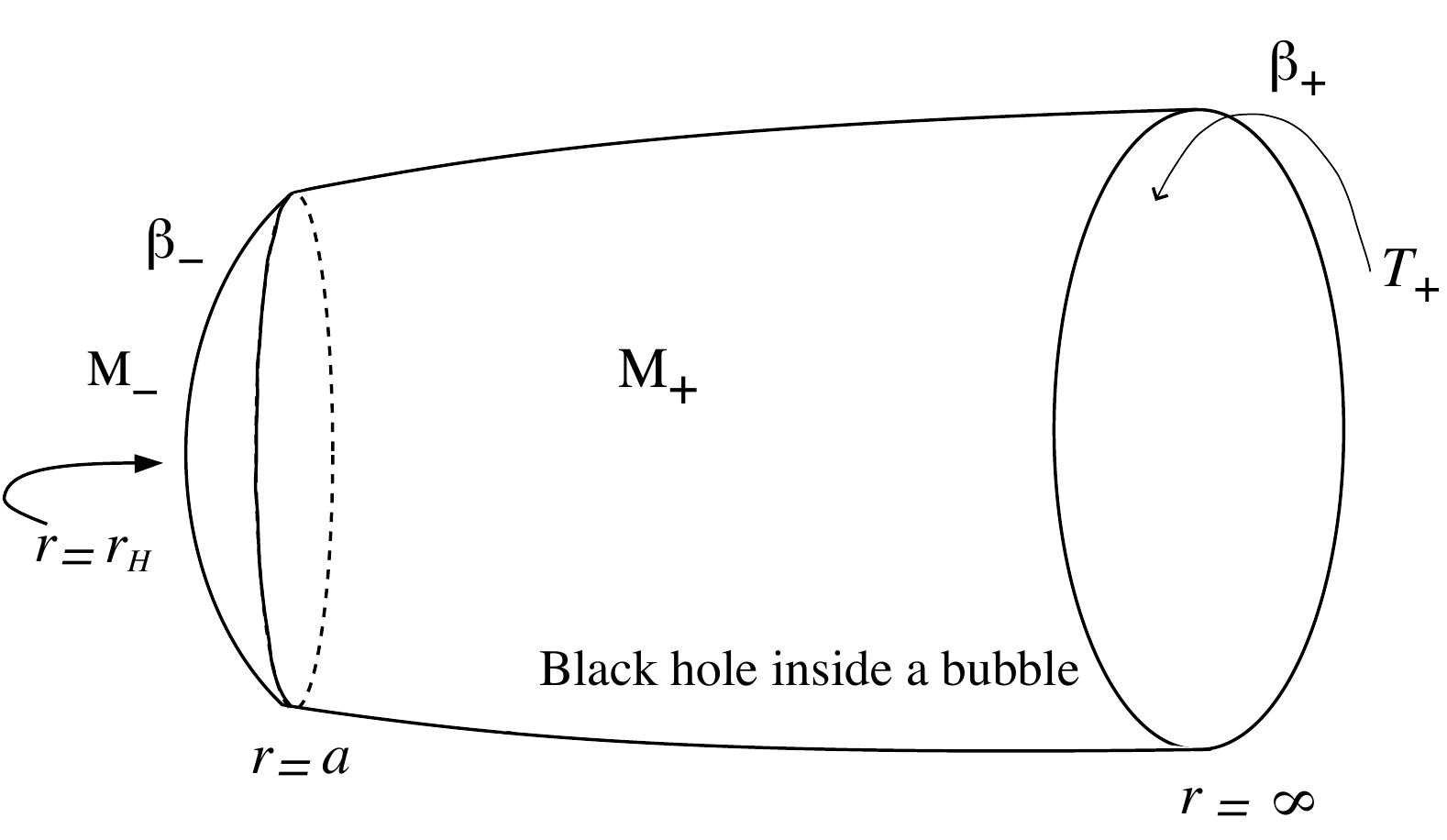}
	\caption{ Thermalon configuration. In the interior $ r<a_\star $, the thermalon hosts the - branch that is a dS BH corresponding to mass and temperature ($M_-,\beta_-$), and the exterior $ r>a_\star $ with the + branch with parameters ($ M_+,\beta_+ $) presents AdS asymptotics (the exact expression of the branches is in eq. \eqref{gbintrogbholes}). The decay of thermal AdS into the thermalon, and its later expansion causes the change of asymptotics. Both solutions are glued together at $ r=a_\star $ without matter due to the higher curvature corrections of Lanczos-Gauss-Bonnet gravity theory.} 
	\label{gbintrobubblethermalonfig}
\end{figure}
First we start with the physical intuition for the static case, a depiction of the setup can be found in figure \ref{gbintrobubblethermalonfig}. We will have a timelike closed surface, that in coordinates will be given by $ r=a_\star $. For $ r<a_\star $ we choose as metric one the branches explained in \eqref{gbintrogbholes}, and the other for $ r>a_\star $. The equations of motion are satisfied for all $ r \neq a_\star $ \footnote{Of course, the solutions \eqref{gbintrogbholes} are only defined for $ r>r_h $ when they have horizons, but let us forget about that subtlety for the sake of simplicity.}, because both branches are solutions. For the whole metric to be a distributional solution of LGB action, we must impose the so called \textit{junction conditions}. Those will constraint the three free parameters $ M_\pm, a_\star $, and the match may not even be possible in some cases. If we success in satisfying the junction conditions, we ended up with a new solution by matching two previous solutions together. For the non-static bubbles, the idea is the same except that now $a$ is time dependent. 

Now we describe the thermalon configuration in slightly more precise terms. Let us consider a manifold that at least outside horizons will be globally covered by one chart with coordinates $ \tau, r, \phi_i $. We work in euclidean signature so $ \tau $ is spacelike but its Lorentzian counterpart is timelike. We divide the manifold into two disjoint regions $ r<a_\star $ and $ r>a_\star $, that we call interior and exterior respectively. Here comes the difference with the previous paragraph: we will not associate exactly the solutions in \eqref{gbintrogbholes} to the interior/exterior. Instead, we will associate the following metrics diffeomorphic to them:%
\bea
	ds_\pm^2 \eq  f_{\pm}(r) \dot T^2_\pm d\tau^2 + \frac{dr^2}{f_{\pm}(r)} + r^2\ d\Omega_{d-2}^2 ~,
	\label{gbintrobubblesthermalonmetric}
\eea
where $ t_\pm= T_\pm (\tau)$ must be determined afterwards. The functions $ T_\pm $ should not be confused with temperatures, and are nothing but the effect of time coordinate redefinition in each of them. The position of the bubble at a given time corresponds to $ r=a(\tau)$. Using the same $ \tau,\phi_i $ coordinates to cover the bubble, the metric pullback on it becomes:
\begin{equation}
	ds^2_\Sigma = \prt{f_{\pm}(a)\,\dot{T}^2_{\pm}+ \frac{\dot{a}^2}{f_{\pm}(a)}}d\tau^2 + a(\tau)^2\ d\Omega_{d-2}^{2} ~,
	\label{indmetric}
\end{equation}
The first junction condition is to have continuous $ ds^2_\Sigma $ across the bubble, and that is the reason why we introduced the functions $ T_\pm $.  If we had taken simply $ T_\pm =\tau $, the metric would be discontinuous. We avoid that problem with the requirement:
\begin{equation}
	f_{\pm}(a)\,\dot{T}^2_{\pm}+ \frac{\dot{a}^2}{f_{\pm}(a)} = 1 ~, \qquad \forall \tau ~.
	\label{gbintrobubbleTglue}
\end{equation}
that guarantees the desired continuity of $ ds^2_\Sigma $ across the bubble. These choice must be explicitly checked once the $ a(\tau) $ is found to verify its consistency. 

In addition to the continuity of induced metric across the junction \eqref{gbintrobubbleTglue}, there is another junction condition in the absence of matter fields in the bubble, namely the continuity of momenta: 
\bea \label{gbintrobubblepis}
\Pi^+_{ab} \eq \Pi^-_{ab} 
\eea
where the $ \Pi_{ab}$ are the canonical gravitational momenta $\Pi_{ab} = \frac{\partial L}{\partial \dot g_{ab}} $, and $ a,b $ indices refer to the directions of  the junction surface (not necessarily flat indices): $ - $,$ + $ refer to the interior,exterior side of the bubble.

\begin{figure}[h!]	
	\centering
	\includegraphics[width=0.65\textwidth]{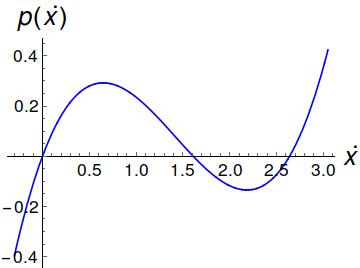}
	\caption{Momentum  $ p $ of the free particle as a function of the velocity $ \dot x $. $ \dot x $ is multivalued as a function of $ p $; this is due to the higher derivative corrections in \eqref{gbintrobubblepaction}. Analogous gravity higher curvature corrections allow to glue two solutions with different derivatives in each side, \textit{without the need of matter in the junction surface} (see paragraph after eq. \eqref{gbintrobubblepis}).} 
	\label{gbintrobubblepofdotx}
\end{figure}

Although in GR the second condition can also be phrased as momentum continuity\footnote{The matching equations of Einstein-Hilbert action are the so-called Israel-Darmois conditions \cite{Israel:1966rt}. They imply the continuity of induced metric and momenta in both sides of the junction.} \eqref{gbintrobubblepis}, the different expression of $ \Pi $ in LGB leads to very different implications. In particular, it allows to glue solutions with different metric derivatives in each side with no matter in the junction. Indeed, in Einstein-Hilbert action the momenta $ \Pi_{ab} $ are in one-to-one correspondence with the velocities $\dot g_{ab} $. If the momenta are continuous in the junction, so are the velocities.  But with higher curvature corrections, for the same momenta in both sides of the junction there may be two or more different velocities. In this way, the discontinuity of metric derivative is compatible with the momenta continuity without matter, while it is not in Einstein-Hilbert. Let us illustrate with an explicit one-particle example how the higher order corrections can induce multivaluedness of the velocities \cite{Henneaux:1987zz}. Consider a free particle lagrangian containing higher powers of velocities such as:
\vskip-1mm
\bea\label{gbintrobubblepaction}
L(\dot{x})\eq \frac{1}{2}\dot{x}^2-\frac13\dot{x}^3+\frac1{17}\dot{x}^4
,\\
p(\dot x) \eq \dot x-\dot x^2+\frac{4 \dot x^3}{17}.
\eea
The EoM is given by $ \frac{d p}{d t} = 0 $.  $ \dot x (p)$ is a multivalued function, as can be seen in figure \ref{gbintrobubblepofdotx}:\\
Even further, the solution to the variational principle is not unique. If we fix boundary conditions $x(t_{1,2})=x_{1,2}$, an obvious solution would be constant speed 
$
\dot{x}=(x_2-x_1)/(t_2-t_1)\equiv v.
$
For this lagrangian it is possible to have two velocities $ v_1,v_2 $ with $ v_1 < v < v_2 $ and  $p(v_1)= p(v_2) $. As the momentum is the same for both velocities, we can glue them despite the discontinuous velocity, forming the trajectory:
\bea \label{gbintrobubblediscontsolution}
\dot x \eq v_1, \qquad \qquad t_1 \leq t <  t_1 + \Delta t,\\
\dot x \eq v_2, \qquad \qquad t_1+ \Delta t  < t  \leq t_2.\nn
\eea 
If we choose $ \Delta t $ such that $ v_1 \Delta t + v_2 (t_2 -\Delta t ) = v (t_2-t_1)$, we have found another solution to the variational problem because the boundary conditions are fulfilled and the momentum $ p(v_1) = p(v_2) $ is conserved. Then, as we have two solutions fulfilling the EoM and boundary conditions, what is the actual physical solution of the variational problem? The answer is given by the absolute minimum of the action, and it can be the discontinuous one in some cases \cite{Camanho:2015ysa}.

Now we compare back the free particle with the gravitational case. The interior/exterior of the bubble $ f_{\mp} $ are the analog of the different velocity pieces of \eqref{gbintrobubblediscontsolution}. The bubble sides can be matched despite having different velocities because  $ \Pi^-_{ab}=\Pi^+_{ab} $ \eqref{gbintrobubblepis}, this is analog to equality of momenta $ p(v_1) = p(v_2) $ in eq. \eqref{gbintrobubblediscontsolution}. In both gravity and particle case, different velocities can be matched because they have the same momentum. This is possible due the higher curvature corrections in \eqref{gbintrobubblepaction}, that make $ \dot x(p) $ multivalued.  

So we conclude that a priori non-continuous solutions in the absence of matter like the thermalon \eqref{gbintrobubblesthermalonmetric} could exist; and actually they do \cite{Gravanis:2007ei,Garraffo:2007fi}. It is important to realize that they are another competing saddle and may become the dominant one as in \cite{Camanho:2013uda}. Indeed that is our mechanism of phase transition from AdS to dS we present in this chapter. The phenomenon of solution with discontinuous derivatives is expected generically in higher curvature theories in addition to LGB gravity.

The junction conditions \eqref{gbintrobubblepis} for our spherical bubble \eqref{gbintrobubblesthermalonmetric} turn out to have only one independent component $ \Pi^\pm_{\tau\tau} $. The rest are related to that one as \cite{Davis:2002gn,Gravanis:2007ei}:
\begin{equation}
	\frac{d}{d\tau}\left(a^{d-2}\, \Pi^\pm_{\tau\tau}\right) = (d-2)\, a^2 \dot{a}\, \Pi^\pm_{\varphi_i\varphi_i} ~, \qquad \forall i ~.
	\label{Bianchi}
\end{equation}
Therefore if $\Pi^\pm_{\tau\tau}$ verifies \eqref{gbintrobubblepis}, all the components automatically do. We will therefore focus on  $\Pi^\pm_{\tau\tau} $, with expression given by \cite{Camanho:2013uda}:
\begin{equation} \label{gbintropidefinition}
	\Pi^\pm_{(\tau\tau)} \equiv \frac{\sqrt{\dot{a}^2+f_\pm(a)}}{a} \int_0^1\! d\xi\ \Upsilon'\left[\frac{1-\xi^2\,f_\pm(a)+ (1 - \xi^2)\,\dot{a}^2}{a^2}\right] ~,
\end{equation}
where we ignored some factors as they are the same for both $ \Pi^- $ and $\Pi^+$.  $\Upsilon$ is an important polynomial defined for any Lovelock action\footnote{This quantity appears often in many Lovelock applications, including Thermodynamics, presence of ghosts, dynamics, etc..., \textit{cf}. \cite{Camanho:2015ysa}.} (see \eqref{gbintrogbgenerallovelock}): 
\bea \label{gbintroupsilon}
\Upsilon[x] := \sum_{k=0}^{K} c_k x^k =  \lambda L^2  x^2 + x - \frac{ \tilde \Lambda}{L^2}  ,
\eea  
where $  K$ is the highest Lovelock order present in the action and the expression at the right is the value for our LGB choice of parameters. From now on we avoid the use of indices i.e. $\Pi^{\pm} \equiv \Pi^\pm_{\tau\tau}$. Let us define $\widetilde{\Pi}=\Pi^+-\Pi^-$; the junction equations are rewritten as $\widetilde{\Pi}=\partial_\tau\widetilde{\Pi}=0$. After we introduce the auxiliary quantities:
\begin{equation}
	g_\pm \equiv g_\pm(a) = \frac{1-f_{\pm}(a)}{a^2} ~, \qquad H \equiv H(a,\dot{a}) = \frac{1 + \dot{a}^2}{a^2} ~,
\end{equation}
we can recast the momenta as:
\begin{equation}
	\Pi^\pm\,[g_\pm, H] = \sqrt{H-g_\pm}\int_0^1\! d\xi\ \Upsilon'\left[\xi^2\,g_\pm + (1 - \xi^2)\,H \right] ~,
	\label{Sjunc}
\end{equation}
where it becomes clear that all the information about the branches is contained only in $g_\pm$, that can be found using \eqref{gbintrogbholes}. 

%
As mentioned before, it is possible to rewrite the condition \eqref{gbintrobubblepis} as the dynamics of one particle with an effective potential $ V_{\mathrm{th}} (a)$ \cite{Camanho:2015ysa} constrained to have a vanishing Hamiltonian, namely:
\begin{equation}
	{\Pi^{+}}^2 = {\Pi^-}^2 \qquad \Longleftrightarrow \qquad \frac12\dot{a}^2+V_{th}(a) = 0, ~\ddot{a} = -V'_{th}(a) ,
	\label{gbintrobubbleVgeneral}
\end{equation}
%
Here $ \dot a $ is the derivative with respect to euclidean time $ \tau $, i.e. $ \dot a \equiv \frac{d a(\tau)}{d\tau} $. The precise form of $ V_{th} $ clearly must depend on $ g_\pm $ and it will be given for our problem later in eq.  \eqref{gbvform}. 

Now that we have specified the dynamics of time dependent bubbles, we particularize for the thermalon, that is static. It has seven unknowns: $T_\pm, M_\pm, \beta_\pm$ and $a_\star$. In the previous,  $ M_\pm $ are the mass parameters implicit in $ f_\pm $, $\beta_{-}$ is the usual inverse Hawking temperature of the inner solution while $\beta_{+}$ is the one seen by the asymptotic observer (that it is in the outer region); $a_\star =a(\tau)$ is the location of the thermalon (the subindex $ \star $ means that it is the static value). Four constraints are given by \eqref{gbintrobubbleTglue} and the r.h.s. of  \eqref{gbintrobubbleVgeneral}. As we are in static configuration, they can be simplified to:
\bea
\sqrt{f_-(a_\star)}\,T_- =\sqrt{f_+(a_\star)}\,T_+=\tau,\\
V_{\mathrm{th}} (a_\star) = V'_{\mathrm{th}} (a_\star)=0 . 
\eea
A fifth constraint is given by the absence of conical singularity in the horizon, fixing $ \beta_- $ in the usual way. The sixth and last is enforced by the continuity of the total length of the thermal circle  across the junction:
\begin{equation}
	\sqrt{f_{-}(a_\star)}\ \beta_{-} = \sqrt{f_{+}(a_\star)}\ \beta_{+} ~,
	\label{gbintrobubblebetaglue}
\end{equation}
Only one degree of freedom remains, and we choose it to be the temperature at infinity $ \beta_+ $. Namely, if we consider the canonical ensemble at temperature $ \beta_e $, the corresponding thermalon phase has $ \beta_+ = \beta_e $. It must be stressed that not any choice of $ \beta_+ $ will allow the existence of a thermalon. In our case they shall exist only below a certain temperature (see figure \ref{gbft}). 

We have so far described the thermalon to which our initial thermal AdS will decay. To finish the transition to dS geometry, the bubble must expand. Such expansion is a purely dynamical process determined by the right hand side of \eqref{gbintrobubbleVgeneral}. For the expansion to happen, the thermalon must sit in the maximum of the real time potential $ V(a) = -V_{th}(a) $ \footnote{Notice that the euclidean evolution has the opposite potential of real time.}. Once it has been formed, the bubble will either expand or contract due to small metric fluctuations. If it contracts, nothing will happen at large scale\footnote{For a discussion on possible naked singularity formation see \cite{Camanho:2013uda}.}, as the bubbles are formed all throughout space very much like they do in boiling water. If it expands, the dS phase in the interior will expand eventually reaching the cosmological horizon. At that point, an observer in the interior of the cosmological horizon will measure the Thermodynamics of a dS space.  

\paragraph{Thermodynamics of the thermalon}  \label{gbintrobubblethermodynamics}
In this subsection we address how to compute free energy of the thermalon in LGB gravity. A more detailed treatment for the whole Lovelock family and more general thermalon configurations can be found in \cite{Camanho:2013uda}. 

We will use the Euclidean on-shell action method described in the appendix of this chapter, section \ref{gbintrothermo}.The main difference with standard BH Thermodynamics is that  \textit{surface terms of the action must be included at the junction}. The action for the whole thermalon can be split as:
\be
\widehat{S}=\widehat{S}_-+\widehat{S}_{\Sigma}+\widehat{S}_+   ~.
\ee
where $ \widehat{S}_-  $ is the standard bulk action to be integrated in $ r_h<r<a_\star $.  $  \widehat{S}_+    $  is integrated in $ r>a_\star $. $S_+ $ includes on the one hand the boundary terms necessary for good variational principle \eqref{gbintroLGBbterms} and the other the subtraction of the $ \Lambda_+ $ vacuum, necessary to make the action finite. The non-standard piece $ \wh I_\Sigma $ is given by:
\bea
\wh S_\Sigma \eq -\wh S_\partial^- +  \wh S_\partial^+, \label{gbintrobubblethermoIsigma}
\eea
this is, the boundary terms \eqref{gbintroLGBbterms} evaluated in both sides of the junction $r=a_\star $. Such term is necessary for the thermalon to be a distributional solution satisfying the variational principle, and therefore it is not added ad-hoc only to compute free energy. Notice that it must be included despite the absence of matter in the junction. 

Before we continue, let us define 
\bea
\wh \calI_{BH} \eq \wh S - \wh S_{bub} = \beta_- M_- -S_{BH},
\eea
that is the familiar contribution to $F$ of the interior black hole. The evaluation of $\wh S_{bub} $ is possible for generic Lovelock gravity \cite{Camanho:2013uda}, and the result is remarkably neat:
\bea
\wh S_{bub} \eq \beta_+ M_+ - \beta_- M_-. 
\eea 
With it, the whole action becomes simply:
\bea
\wh S \eq \wh \calI_{BH} + \wh S_{bub} = \beta_+ M_+ - S_{BH} 
\label{gbintrobubblefreeenergy}
\eea
meaning that the junction contributes as mass but not as entropy. Both $ \beta_+, M_+ $ are the quantities corresponding to the exterior observer.

\section[\texorpdfstring{Thermalon configuration and $\mathrm{A \MakeLowercase{d}S}$ to $\mathrm{\MakeLowercase{d}S}$ phase transition}{Thermalon configuration and AdS to dS phase transition}]{Thermalon configuration and AdS to dS phase transition}

From now on, we introduce the original results based on the previous work summarized above. As mentioned before, we are interested in studying AdS to dS phase transitions. It turns out \cite{Camanho:2012da} that thermalons connecting AdS  to  dS  are not possible with $\tilde \Lambda =-1$ (remember that $\tilde \Lambda = \mathrm{sign}(\Lambda)$), therefore we have to consider $ \tilde \Lambda =1 $, i.e. positive bare cosmological constant. To make things more concrete, we explicitly state now all the parameters we are choosing in the action \eqref{gbintrogbaction}:
\begin{equation}
	\tilde \Lambda =1, \qquad d = 5,\qquad \lambda > 0 , \qquad L = 1. 
\end{equation}
With these choices $ \lambda $ is the only remaining tunable parameter in the theory. It is important to note that the $ \Lambda_+ $ vacuum (given by \eqref{gbintrovacua}) is Boulware-Deser unstable. $ \Lambda_+ $ is our initial state, and therefore strictly speaking our results are unphysical. The reason why we overlook that fact is because we want to show that a thermalon configuration that expands yielding dS asymptotic can be thermodynamically preferred. We strongly expect the existence of similar transitions between healthy vacua in higher order Lovelock theories. Indeed, for Ads to AdS transition, it was found to be possible in higher order Lovelock in \cite{Camanho:2013uda}, becoming clear that the transition is not caused by the unhealthy vacuum. Furthermore, the vacua curvature are given by the real roots of $ \Upsilon $ polynomial defined in \eqref{gbintroupsilon}. If $ \Upsilon' $ is positive on them, they are Boulware-Deser stable. In LGB, as $ \Upsilon $ is quadratic, both solutions cannot have positive derivative. But in higher order Lovelock, $ \Upsilon $ is a higher order polynomial and it is possible to find two different vacua with positive derivatives, and therefore both are Boulware-Deser stable. We think that more general higher order gravity theories will also display the desired transitions AdS to dS transition.

The thermalon we are considering is exactly of the form explained in the previous sections. To determine its dynamics using the r.h.s. of \eqref{gbintrobubbleVgeneral}, we compute the momenta $\Pi_{\tau\tau}^{\pm}(a,\dot{a})$ using \eqref{gbintropidefinition}. The result in Lorentzian signature is:
\begin{equation}
	\Pi_{\tau\tau}^{\pm}(a,\dot{a}) = \frac{\sqrt{\dot{a}^2+f_\pm(a)}}{a} \int_0^1\! d\xi \left[ 1 + 2\lambda \frac{L^2}{a^2} \left( \dot{a}^2 + 1 - \xi^2 (f_\pm(a) + \dot{a}^2) \right) \right] ~.
\end{equation}
We can work out now the explicit form of $V_{\rm th}(a)$:
\begin{equation}
	V_{\rm th}(a) = \frac{1+4\lambda}{24\lambda} \left[ a^{d-1} \frac{f_-(a) - f_+(a)}{{\rm M}_+-{\rm M}_-} + \frac{4\lambda}{1+4\lambda} \frac{{\rm M}_- f_-(a) - {\rm M}_+ f_+(a)}{{\rm M}_+-{\rm M}_-} + \frac{8 (a^2 + 2\lambda)}{1+4\lambda} \right] ~.
	\label{gbvform}
\end{equation}
As explained in the thermalon configuration section, there is one degree of freedom in the thermalon configurations. We take it to be $ \beta_+ $ as we are going to work in the canonical ensemble. Following the discussion of subsection \ref{gbintrobubblethermodynamics} we can compute the thermodynamic free energy at temperature $ \beta_+ $. 
We also have checked that the relation between the various quantities do satisfy the first law of Thermodynamics. 

\begin{figure}[ht]
	\begin{center} 
		\includegraphics[width=0.58\textwidth]{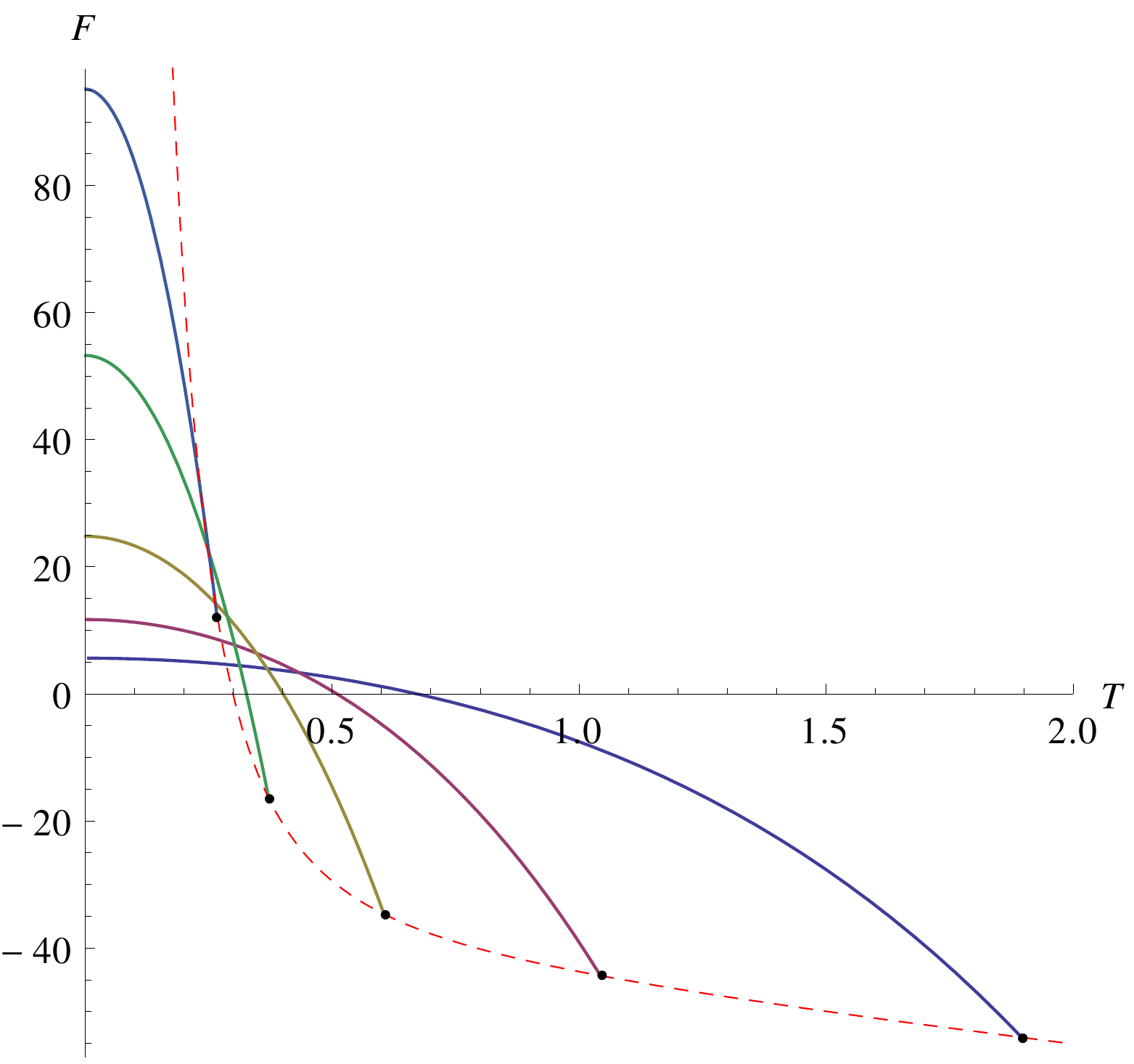}
	\end{center}
	\caption{Free energy of the bubble configuration compared to the thermal vacuum $\Lambda_+$ ($F=0$) as a function of the temperature $T = \beta_+^{-1}$. From bottom to top in $y$-intercept we have the LGB couplings $\lambda = 0.1, 0.2, 0.4, 0.8, 1.35$. The results are for five dimensions and $L=1$. The dashed line indicated the locus of the points where the different curves end (black dots). Beyond that temperature the thermalon configuration is no longer allowed by the equations of motion.} 
	\label{Freefig}
	\label{gbft}
\end{figure}

The free energy as a function of the temperature is depicted in figure \ref{gbft}, where we can analyze the global and local stability of the different solutions.
For each value of the temperature, the saddle with the lowest value of the free energy will be the dominant one. The first thing we notice is that the thermalon configuration exists only for a limited range of temperatures for each $\lambda$. Above some threshold temperature, $T_\star$, which decreases with $\lambda$, the thermal AdS vacuum would be the only available static solution, thus being in this respect stable. We again should remind the reader that in our toy LGB action the AdS vacuum $\Lambda_+$ is Boulware-Deser unstable. 

Maximal temperature for the thermalon $ T_\star $ happens when the mass of the interior spherically symmetric black hole with dS asymptotics reaches its upper bound given by the Nariai threshold \cite{Camanho:2011rj}. This phenomenon only occurs in the $ \tilde \Lambda >0 $ case we explored in our original results.


\begin{figure}[ht]
	\begin{center} 
		\includegraphics[width=0.58\textwidth]{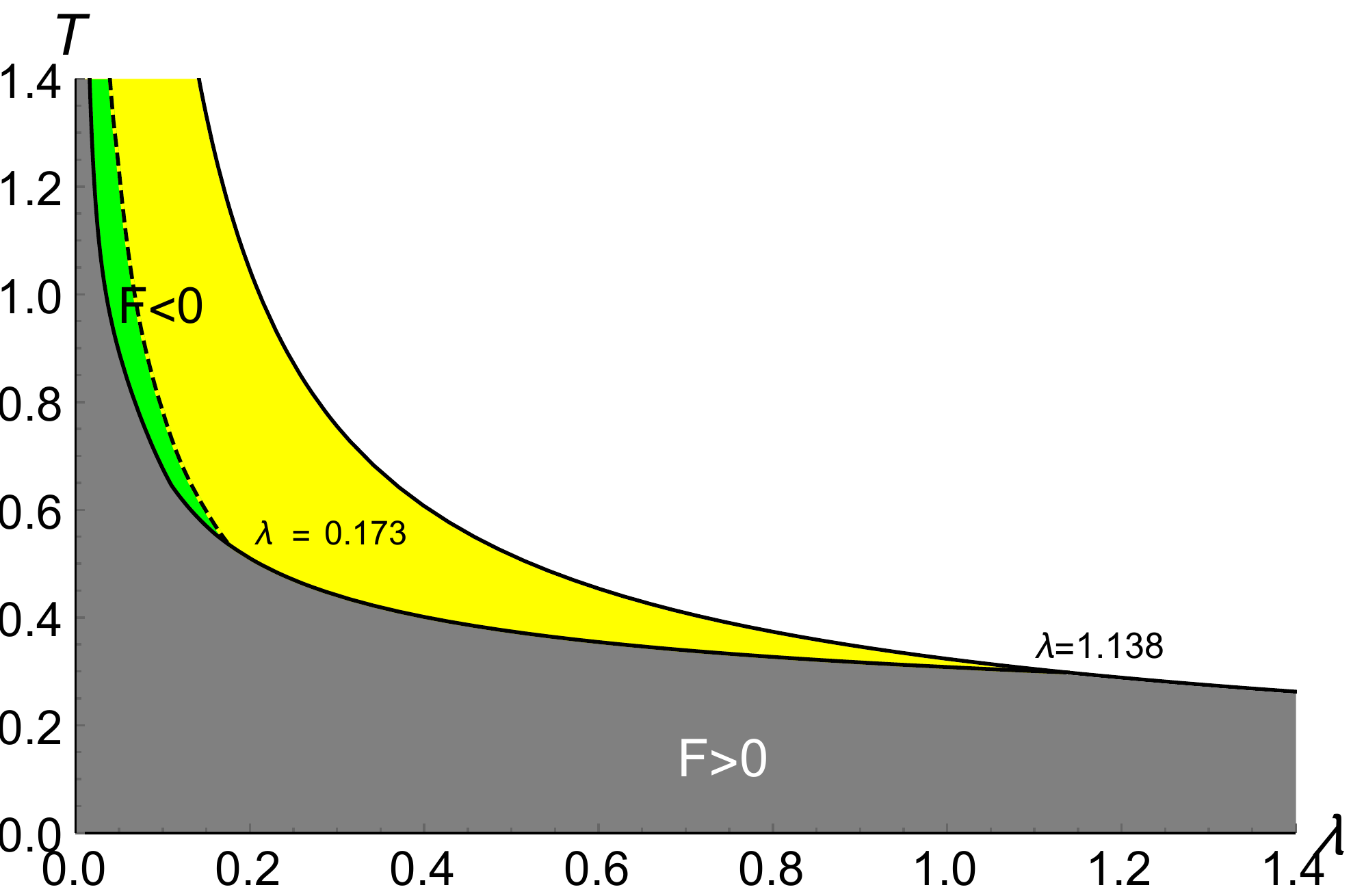}
	\end{center}
	\caption{Phase diagram, where T stands for $\beta_+^{-1}$ and $\lambda$ for the LGB coupling. The upper solid curve corresponds to the maximal temperature $T_\star(\lambda)$ for which the thermalon exists. The lower one is the critical temperature $T_c(\lambda)$ for which thermalon configuration is thermodynamically preferred. Both curves merge at $\lambda_\star = 1.13821$. Beyond $\lambda_\star$ the thermalon exists but is never the dominant phase. Between both lines there is the region where the thermalon free energy is negative and the transition is favored. In gray at the bottom we have the region of positive free energy of the thermalon. The dashed line separates the (upper--yellow/lower--green) regions with negative/positive black hole's specific heat. In white at the top the region where no thermalon is dynamically possible.}
	\label{gbphasediagram}
\end{figure}

Regarding the sign of the free energy, we can distinguish two different regimes depending on the value of the LGB coupling (see Figure \ref{gbphasediagram}). For small enough values, $\lambda < \lambda_\star = 1.13821$, the free energy of the thermalon is positive at low temperatures, while it changes sign as we go to higher temperatures. For higher values of $\lambda$, on the other hand, the free energy remains positive for the whole range of temperatures of the thermalon. In the latter case, then, the dominant saddle is always the $\Lambda_+$ vacuum, whereas in the former it is so except for a limited range of intermediate temperatures, $T \in (T_c, T_\star)$. In this range the thermalon is the dominant saddle and the nucleation process is favored.
Notice that despite having $ F>0 $, the thermalon phase may form at low temperature. The probability of bubble formation in this situation is not zero but roughly $e^{-\hat{\mathcal{I}}}$ 
\cite{Langer:1969bc,Linde:1977mm,Linde:1981zj,Linde:1981zj,Affleck:1980ac}. Waiting long enough the bubble will form and mediate a transition to the healthy branch of solutions. In this respect, we may say that thermal AdS is metastable.

The only range of temperatures for which there is no transition is that of high temperatures for which the thermalon configuration does not even exist. The threshold temperature for this regime diverges as we approach small values of $\lambda$. At the same time, the free energy becomes lower and lower in the range in which it is positive. This means that the thermalon is more easily excited (the nucleation probability grows) as we go to lower values of $\lambda$, and that this configuration is available for all temperatures in the limit of very small $\lambda$. This is a strong indication that the AdS vacuum is more and more unstable as $\lambda \to 0$.

When reaching the critical temperature the thermalon will form, but it will not remain in equilibrium for long. Even though thermodynamically stable, this configuration is dynamically unstable. Our bubble sits at a maximum of the potential $ V_{th} $ given by \eqref{gbvform}. Eventually expands reaching the asymptotic region in finite proper time, thus changing the effective cosmological constant of the whole spacetime.


We may wonder what happens when the bubble expansion reaches the boundary. In asymptotically AdS spacetimes, boundary conditions are of paramount importance. In the case of transitions between two AdS vacua of different radii, one may argue that fixing the asymptotics (by means of reflecting boundary conditions) would make the bubble bounce back and collapse. In the case of AdS to dS transitions, though, this is actually impossible since the formation of a dS horizon makes the expansion of the bubble irreversible from that moment onwards \cite{Camanho:2013uda}.

We may summarize that the thermalon effectively effectively changes from branch of solutions to another. This is important because the final (dS) branch does not suffer from the pathologies characterizing the unstable LGB AdS branch that was our initial asymptotics. 

As an aside comment figure \ref{gbphasediagram} seems to entail the occurrence of so-called {\it reentrant phase transitions},\footnote{We thank David Kubiznak and Robert Mann for pointing out this interesting possibility to us.} a familiar phenomenon in chemical physics that was earlier observed in the context of black hole Thermodynamics \cite{Frassino:2014pha,Altamirano:2013ane}. There is, however, an important subtlety hidden in the fact that Figure \ref{gbphasediagram} is not comparing two different stable thermodynamical configurations. Instead, one of them --the {\it thermalon}-- is a finite temperature instanton describing an intermediate state, a bubble of true-vacuum that grows after popping-up, filling space with the new phase that changes the asymptotics to dS. Once this happens, we cannot refer any longer to the same diagram. Therefore, we cannot reverse the process by changing the temperature, as required to have reentrant phase transitions.

\section{Discussion and outlook}
Higher curvature gravity, such as GB, do typically posses several vacua. Gravitational phase transition among them is in principle possible despite the change of boundary curvature. In particular, we have shown a case where there a sing change in these curvature. 

We have described a novel scenario for transitions between AdS and dS asymptotics in higher-curvature gravity, without involving any additional matter fields. The phenomenon discussed in the present chapter is expected to take place in any higher-curvature theory as long as AdS and dS branches are allowed at the same time. The framework described here is general enough as to accommodate any series of higher-curvature terms, we would only need to adequately generalize the junction conditions.

For illustrative purposes we have chosen LGB gravity as the minimal model where this kind of transition takes place. For this theory we have all the necessary ingredients without the usual complications of higher-derivative terms. In particular, we have explicit junction conditions (given by {\it momenta} conservation in the Hamiltonian formalism) that enable us to construct the thermalon configuration. Nonetheless, this example has to be taken with a grain of salt, as a simple toy model, given that the theory has well known issues; namely, it generically violates causality\footnote{This signals the existence of an infinite tower of higher-spin particles whose presence does not necessarily affect the gravitational phase transitions discussed in this chapter.} \cite{Camanho:2014apa}.

We are also truncating the effective string theory action and analyzing the regime where the higher-curvature corrections become of the same order as the Einstein-Hilbert term, precisely where the rest of the higher-curvature series becomes relevant as well. In particular, the unstable vacuum curvature diverges as we take the LGB coupling to zero. All these concerns would be dealt with in a brane setup in string theory, where a consistent field theory limit is taken. In that respect, the LGB action is a promising candidate given that it arises both in heterotic string theory and in type II Superstrings in the presence of wrapped probe D-branes.

This model might have as well interesting applications in the context of the gauge/gravity duality. In addition to the well known AdS/CFT correspondence there is a proposal for a dS/CFT duality \cite{Strominger:2001pn}. The transition mechanism presented in this chapter suggests a possible avenue for a deeper understanding of both formulations by means of a framework where both types of asymptotics appear on equal footing in a given gravitational theory. Notice, nonetheless, that in all cases studied so far one of the vacua is unstable for some reason and we can always argue that we have to stick to the other asymptotics. There may be more general cases of transitions involving two perfectly healthy vacua in Lovelock theory.

The approach considered here is very different to other AdS/CFT descriptions of similar transitions (see, for instance, \cite{Freivogel:2005qh}) where the boundary in which the CFT lives is always unchanged, the transition taking place in the other asymptotic region of an eternal black hole. Our description is completely insensitive to this {\it other side} and even if we try to describe it, the same transition would happen there as well.
\begin{subappendices}
	\setcounter{equation}{0}
	\section{Introduction to black hole Thermodynamics} \label{gbintrothermo}
	In this appendix we introduce the main concepts of BH black hole Thermodynamics we have used in the chapter. The topic of black hole Thermodynamics is so far the main window to quantum gravity, and its importance cannot be overemphasized. Furthermore, it is in this context where the holographic principle by t'Hooft \cite{tHooft:1993dmi} appeared, of which   AdS/CFT is a much more concrete and developed realization. Any Quantum Theory of Gravity must be able to reproduce the laws of BH Thermodynamics. Finally, it is crucial in holographic applications involving thermal properties in the dual field theory.
	
	\subsection{History} \label{gbintrothermohistory}
	Back in the 1970s, Bekenstein \cite{Bekenstein:1973ur} conjectured that black holes should have entropy. The basic reasoning is as follows: consider matter before and after falling into a black hole. At the beginning of the process, matter has some entropy, after it, the matter is no longer there and therefore the total entropy of the universe would be reduced. To avoid this violation of the second law of Thermodynamics, black hole entropy must be taken into account. But what is the actual amount of entropy that the BH should have? The right answer turns out to be proportional to the area of the horizon, not the volume as in standard Thermodynamics. Motivated by this proposal, Hawking proved that the total area of horizons cannot decrease over time and the area-entropy identification became very strong.  

	This was in itself an impressive feat, but it raised an important question: if BHs are thermodynamical objects, they should also have a temperature and radiate; ironically they would no longer be black. Using quantum field theory on curved spacetime, Hawking \cite{Hawking:1974sw} showed a microscopic description of BH radiation linking the temperature $T_H$ with the purely geometrical surface gravity $\kappa$; indeed Hawking radiation is closely related to the Unruh effect, in which accelerated observers perceive a non-vanishing temperature \cite{Unruh:1976db}\cite{Davies:1976ei}. Surprisingly enough, using this temperature one could proof the first law of Thermodynamics for BHs \cite{Bardeen:1973gs}:
	\bea
	\delta M = \frac{\kappa}{8\pi G_N} \delta A + \Omega_H \delta J + \phi_H \delta Q
	\eea
	where $M$ is the mass, $\Omega_H, J$ is the angular velocity at the horizon and the angular momentum , $\phi_H, Q$ are the electric potential and electric charge.  

	From this point on, the thermodynamical properties of the horizon allow to ask thermodynamical questions like in more down to earth systems. In particular, Nernst version of third law was also proved to hold \cite{Bardeen:1973gs}. It must be honestly remarked that until less than a year ago there was no experimental evidences concerning BH Thermodynamics. Recently, the LIGO collaboration made the first direct observation of gravitational waves of a black hole merger \cite{TheLIGOScientific:2016wfe}, and with it, the area law can be checked in that particular BH merging, being fulfilled.  

	This description of Thermodynamics explained above can be extended to more general spacetime dynamics: higher dimensions, different actions and/or matter fields. If we change the action, some the expressions entering the first law of Thermodynamics will also be different: this is the case in particular  for the extensive quantities $M,S,J$. For example, the entropy will no longer be given by the horizon area. Despite this, Iyer and Wald \cite{Wald:1993nt, Iyer:1994ys} found a  generalization of the entropy formula still satisfying the first law. The key is the identification of the entropy as a Noether charge. With it, one can in principle compute the  thermodynamical quantities for very general actions. An even more general approach is \cite{Fursaev:1995ef}, based on conical singularity method, although it is more difficult to use.

	It must be stressed that the second law is not guaranteed in general theories of gravitation, like the first is after Iyer-Wald work. In particular, it is not fully established for the Lanczos-Gauss-Bonnet (\textit{cf.} section \ref{gbintrogeneralgbthermo}). In what follows we will introduce the euclidean formalism to compute the partition function of a BH in the canonical ensemble. 
	%
	%
	\newcommand{\cD}{\mathcal{D}}
	%
	%
\subsection{Euclidean path integral formalism. Temperature and free energy} 
In the previous section we introduced the elementary notions of black hole Thermodynamics. At the beginning, area law for the entropy was an heuristic conjecture. Nevertheless, it is possible to derive it as well as the free energy of the canonical ensemble from the so-called euclidean path integral formalism \cite{Gibbons:1976ue}, which is also a formal approach to quantum gravity. After its introduction, we will in particular emphasize how to compute the temperature and the free energy, as they will be essential for the phase transitions in the present chapter. We will be exclusively in the canonical ensemble.

As the name already explains, the method is based in the path integral approach to Quantum Field Theories. For real Lorentzian metrics $g$ and real matter fields the action $\mathcal{I}[g,\phi]$ will be real. The real action path integral (whose integrand is essentially $ e^{iS} $) will then oscillate and it is not clear whether it converges. To describe systems at a fixed temperature (like in the canonical ensemble), imaginary time $ t= -i \tau $ is introduced. We can write the canonical partition function as  \cite{Gibbons:1976ue}:
	\bea
	\label{gbintrothermozcanonical}
	Z\eq \sum_{E_n}e^{-\b E_n}=\int{\cD[g,\phi]\, e^{-S_E[g,\phi]}}~,
	\eea
	where $S_E = -i S$ is called the {\it Euclidean} action, $E_n$ is the energy of the $n$-th eigenstate, and $g$ and $\phi$ are the Wick rotated metric and matter fields.  There are several important remarks concerning this formula:
	\begin{enumerate}
		\item The first equality is essentially a formal definition, impossible to fully compute in practice. The second is just a path integral representation of the former.
		\item The action variations must yield the desired boundary conditions of the variational principle, typically of Dirichlet type. To achieve it, extra boundary terms in the action might be needed. This is the reason why the Gibbons-Hawking term was added to the EH term in GR in the first place. \footnote{Modified theories of gravity do often require modified GH-like terms. For our case of Lanczos-Gauss-Bonnet and Lovelock gravity, we introduced them section \ref{gbintrogbbt}. } In all the cases in this thesis the asymptotic geometry will be of constant curvature: AdS or dS. 
		\item The path integral is taken over all fields $\phi$ in the Euclidean section and periodic in imaginary time with period $\b$. We explain how to compute the range of $ \beta $ for a spherical BH in the next paragraph. 
	\end{enumerate}
	
	Let us explain how to relate the range of $ \tau $, $ \Delta \tau $, with the temperature $ T $ of the canonical ensemble in the case of a spherical black hole ansatz:
	\bea \label{gbintrothermobhansatz}
	d s^2 \eq f(r) d \tau^2 + \frac{d r^2}{f(r)} + r^2 d \Omega^2_{d-2}.
	\eea
	For asymptotically flat spacetime, one takes the null vector generating the horizon $ \xi $, normalized so that $ \xi^2 = -1 $ at infinity. Let's remark (this vector is null only at the horizon). For metrics of the form \eqref{gbintrothermobhansatz} with $ f(r=\infty) =1 $, $ \xi = \partial_t $.

	To begin with, we take the euclidean section of \eqref{gbintrothermobhansatz}. The first step is replacing $ t \rightarrow i \tau $:
	\bea
	d s^2 \eq f(r) d \tau^2 + \frac{d r^2}{f(r)} + r^2 d \Omega^ 2_{d-2}.
	\eea 
	If we allow any range of $ \tau $, the metric will generically display conical singularities in the horizon. Indeed the second step is to expand the metric near the horizon:
	\be
	d s^2\approx\frac{4}{f'(r_h)}\left(d\rho^2+\rho^2\frac{d\tau^2}{4/f'(r_h)^2}\right)+r_h^2 d\Omega_{d-2}^2~,
	\ee
	where $\rho^2=r-r_h$ and $'$ is the derivative with respect to the radial coordinate $r$. Then, to avoid conical singularity at $\rho=0$ we must impose:
	\be
	\frac{\beta}{2/f'(r_h)} = 2 \pi,
	\ee 
	as $ \Delta \tau = \beta =1/T$ as implied by the euclidean path integral formalism. From it we immediately identify the temperature:
	\bea \label{gbintrothermotemperaturespherical}
	T \eq \frac{f'(r_h)}{4\pi}.
	\eea
	For asymptotically AdS spacetimes, it is customary to normalize as $ \xi^2 = -\frac{r^2}{l^2}$. We will be concerned with asymptotically  AdS  or  dS  geometries.

	Now that we specified the range of $ \tau $ in terms of the temperature, the question is how to obtain physical quantities from \eqref{gbintrothermozcanonical}. As stressed before the expression in the right of \eqref{gbintrothermozcanonical} is a formal one: we do not have any hope to compute it. Instead, we use the so called saddle point approximation restricting our integration to the classical solution of the EoMs instead of all possible metrics:
	\bea
	Z \approx e^{-S_E[g_{cl},\phi_{cl}]}  
	\eea
	where the subscript $cl$ represents the classical (euclidean) solution of temperature $\beta$ and desired boundary conditions. When one does this in practice, the integral of $S_E$ is typically divergent and some renormalization is necessary. Often, the criterion is to compute the difference with respect to some reference solution, typically the maximally symmetric solution with the desired boundary conditions (for example thermal AdS at the temperature fixed by the canonical ensemble). We will indeed follow this procedure.

	The relation of $ Z $ with Thermodynamics is given by the standard canonical ensemble statistical mechanics:
	\bea
	F=M-TS_{BH}=-T\log Z .
	\eea 
	Once a finite expression for $ Z $ has been achieved, it must be written as a function of $ T $ and the conserved charges fixed by the canonical ensemble \footnote{The most common ones are the electric charge and the angular momentum, but others are possible: magnetic charge, winding charge, extra angular momenta for dimensions higher than four, etc...}.  Then we can obtain thermodynamic quantities from $ F $ in the standard manner, for instance the mass $ M $ and the entropy $ S $: 
	\bear
	M &=& \frac 1 Z\sum_{E_n}{E_n e^{-\b E_n}}=-\frac{\partial}{\partial\b} \log Z \label{Emass}\\
	S_{BH}&=&-\left( \frac{\partial F}{\partial T}\right)=\b M+\log 
	Z\label{Eentropy}~.
	\eear
	
\end{subappendices}

\chapter[\texorpdfstring{Deformations of KW CFT and new A\MakeLowercase{d}S$_\textbf{3}$ backgrounds via NATD}{Deformations of KW CFT and new AdS$_3$ backgrounds via NATD}]{Deformations of KW CFT and new AdS$_\textbf{3}$ backgrounds via non-Abelian T-duality} \label{natd}
\newpage



We start by briefly reviewing the topic of T-duality (both Abelian and non-Abelian) that will become central in our study. After it we include basic review of the concepts of Wilson loop, entanglement entropy and central charge. Furthermore, a brief review of AdS/CFT and NATD research is included. The original results will start in section \ref{natdcontext}.
\section{T-duality as a generating technique}
\label{natdintrotduality}

\subsection{Abelian T-duality} \label{natdintroabelian}

In this section we briefly state the basics of Abelian T-duality as a \sugra solution generating technique. As mentioned in the motivation of chapter \ref{motivation}, \td was introduced in \st in the late eighties. Consider a particular \st (say for example type IIA) \smod on a background with a $\mathrm{U(1)}$ symmetry. \td relates it with another \smod on a different \bg. The dual \smod belonging to another (potentially the same) different string theory (for example type IIA related with IIB). To be more concrete, consider as \smod the Polyakov action defined on a purely geometrical ($B_{\m\n}=0,\Phi=0$) Neveu-Schwarz background with a compact direction of radius $R$ (in units $\a'=1$). Via T-duality, it is equivalent to the same \smod on a background of radius $1/R$ and $e^{-2\Phi} = R^2$. This is the simplest instance of the duality. What is common to all forms of Abelian \td is the key requirement of having a global $\mathrm{U(1)}$ symmetry for all the fields in the background. \footnote{For Non-Abelian T-duality the requirement will be a non-Abelian isometry, and we will take $ \mathrm{SU(2)} $.}

So far we elaborated on the equivalence of different \sts in different \bgs. But it turns out that \td applies also to the low energy description of \sts given by SUGRA. In the present work we will be concerned mostly with \td relating \bgs of type IIA and IIB SUGRA, we will say the \bgs are dual to each other. This is made precise in the subsequent paragraphs.

Let us start from a type IIA/IIB \sugra \bg with a global $\mathrm{U(1)}$-isometry, which leaves invariant not only the metric but all the fields of the solution. We will employ coordinates $(x^\m,\psi)$, $\m=0,1,2,...,8$ adapted to the $\mathrm{U(1)}$ isometry; therefore the Killing vector is just $\partial_\psi$. The form of the \dbg is given by the \brs  \cite{Buscher:1987sk,Buscher:1987qj,Bergshoeff:1995as}. For the NSNS sector they read:
\begin{eqnarray}
&  & \wh{g}_{\psi\psi}=1/g_{\psi\psi}~, ~~~~~~~ \wh{g}_{\psi\m}=B_{\psi \m} / g_{\psi\psi}  ~,   \nonumber \\
& & \wh{g}_{\m \nu}= g_{\m \nu} -(g_{\psi\m}g_{\psi\nu} - B_{\psi\m} B_{\psi \nu})/g_{\psi\psi} ~, \nonumber \\
& & \wh{B}_{\psi \m} =g_{\psi \m} /g_{\psi\psi} ~,   \nonumber \\
& & \wh{B}_{\m \nu}=B_{\m \nu}  - (g_{\psi\m} B_{\psi \nu} - g_{\psi\nu} B_{\psi \m})/g_{\psi\psi}    ~,    \nonumber \\
& & \wh{\phi}=\phi - \frac{1}{2}  \log g_{\psi\psi}~.
\label{natdintroabeliannsrules}
\end{eqnarray}
where the hatted fields are those of the \dbg; such notation will apply not only to fields but to any kind of quantities related to the \dbg. For the RR sector let us define:
\begin{equation}
\text{IIB}  ~~~~  P:=\frac{e^{\Phi}}{2} \sum_{n=0}^4 \slashed{F}_{2n+1} ~, ~~~~ \text{IIA}  ~~~~ P:=\frac{e^{\Phi}}{2} \sum_{n=0}^5 \slashed{F}_{2n}  ~,
\label{natdintroabelianppolynomial}
\end{equation}
where $\slashed{F}_{i}:=1/i! \Gamma_{\mu_1...\mu_i} F_i^{\mu_1...\mu_i} $. Then, the \td rule is given by:
\begin{equation}
\hat{P}=P \ \Omega^{-1}~,
\label{natdintroabelianprule}
\end{equation}
where:
\begin{equation}
\Omega= \frac{1}{\sqrt{g_{\psi\psi}}} \Gamma_{11} \Gamma_9 ~.
\label{natdintroabelianpomega}
\end{equation}
~\newline
$\Gamma_9$ refers to a vierbein defined by:
\bea
d s^2 \eq g_{\m\n} d x^\m d x^\n + e^{2\s}(d y + V_\m d x^\m)^2,\\
g_{\m\n} \eq \eta_{ab} e^a_\m e^b_\n d x^\m d x^\n,\\
e^9\eq e^\s(d y + V_\m d x^\m).
\eea 
with $a,b=0,...,8$.  \\
The NS \brs can be derived dealing directly with the Polyakov action, via a gauging procedure (for more details on this as well as a general review of NS Abelian and non-Abelian T-duality, see for example \cite{Alvarez:1994dn}). The idea is to start from the sigma model action for the NSNS fields, and gauge the $\mathrm{U(1)}$ isometry: $\partial \rightarrow D= \partial +A$ introducing a gauge field $A$.  Then, a Lagrange multiplier term is added, ensuring that the field strength vanishes. Finally, integrating out the gauge field we are left with the dual action, from which the dual metric and KR 2-form $\wh g, \wh B$ are read. The Lagrange multiplier acts as the new coordinate. Notice in \eqref{natdintroabeliannsrules} that also the dilaton $\Phi$ is also modified. Its proof in the general case is much more involved and can be found in \cite{Picos:2009pc}. 

Two relevant features of Abelian \dua will appear in chapter \ref{natd}:
\begin{itemize}
	\item If the initial \bg is in type IIA, the dual is in IIB and viceversa. The \dbg is guaranteed to be a solution of its corresponding SUGRA. Therefore the \brs above \eqref{natdintroabeliannsrules}, \eqref{natdintroabelianprule} are a solution generating technique. This also happens for \natd, and it is what makes the present chapter possible.
	\item To completely specify the dual \bg it is necessary to fix the \textit{range of the new dual coordinate $\Delta \wh \psi$}, guaranteeing in that manner that the \smod on the new \bg is equivalent the one in the original \bg; this is achieved when \cite{Rocek:1991ps} : 
	\bea
	\Delta \psi\, \Delta \wh {\psi} = (2\pi)^2 . \label{natdintroabelianperiodicities}
	\eea
\end{itemize}
Having introduced the Abelian \td, especially as a solution generating technique in \sugra, we proceed now with its non-Abelian generalization, that will focus most of our attention.
\subsection{Non-Abelian T-duality (\natd)} \label{natdintronatd}
When the isometry of the \bg is given by some non-Abelian Lie group, there exists an analog solution generating technique called Non-Abelian T-duality (\natd). The whole present chapter is based upon this. This section will introduce the transformation rules and some relevant properties. As in the Abelian case, all the fields must be invariant under the symmetry, that we will always restrict to be \su2 for simplicity. The non-Abelian T-duality for the Neveu-Schwarz sector was originally presented in \cite{delaOssa:1992vci}.

We begin with a brief overview of the procedure yielding \natd transformation rules (we will follow the notation and conventions of \cite{Itsios:2013wd}). Let us consider a IIA/IIB \sugra \bg 
with \su2 isometry. If $L^1, L^2, L^3$ are the $\mathrm{SU(2)}$ left invariant Maurer-Cartan one-forms, we can write the metric and NSNS $B_2$ form as:
\begin{align}
ds^2&=G_{\mu\nu} dx^{\mu} dx^{\nu} + 2 G_{ \mu i} dx^{\mu} L^i + g_{ij} L^i L^j ~, \\
B_2&=\frac{1}{2} B_{\mu \nu} dx^{\mu} \wedge dx^{\nu} + B_{\mu i} dx^{\mu}\wedge L^i + \frac{1}{2} b_{ij} L^i \wedge L^j ~,
\label{natdintronatdansatz}
\end{align}
where $\mu, \nu \in \{1,2,...,7\}$. Let us define:
\begin{equation}
Q_{\mu \nu}:=G_{\mu \nu}+B_{\mu \nu} ~, ~~ Q_{\mu i}:=G_{\mu i} + B_{\mu i} ~, ~~  Q_{i \mu}:=G_{i \mu} + B_{i \mu}      ~, ~~  E_{ij}:=g_{ij} + b_{ij}   ~,
\label{natdintronatddefinitionsQAB}
\end{equation}
and from them the following block matrix:
\begin{equation}
Q_{AB}:=\left( \begin{array}{ccc}
Q_{\mu \nu} &   Q_{\mu i}  \\
Q_{\nu j} &  E_{ij}  \end{array} \right) ~,
\label{natdintronatddefinitionsQAB2}
\end{equation}
where $A,B \in \{1,2,...,10\}$. We can identify the dual metric and $B_2$ field as the symmetric and antisymmetric components, respectively,  of:
\begin{equation}
\hat{Q}_{AB}:=\left( \begin{array}{ccc}
Q_{\mu \nu}- Q_{\mu i } M_{ij}^{-1} Q_{j \nu} ~~ &   Q_{\mu j}M_{ji}^{-1}  \\
-M_{ij}^{-1} Q_{j \mu}  ~~ &  M_{ij}^{-1}  \end{array} \right) ~,
\label{natdintronatdqrule}
\end{equation}
where:
\begin{equation}
M_{ij}:=E_{ij}+ \epsilon_{ij}^k v_k ~,
\label{natdintromatrixMij}
\end{equation}
and $v_k$ with $(k=1,2,3)$ are the new dual coordinates. Moreover, one finds that the dilaton receives a contribution at the quantum level just as in Abelian case:
\begin{equation}
\hat{\Phi}=\Phi - \frac{1}{2} \ln \left( \det M \right) ~.
\label{natdintronatddilatonrule}
\end{equation}
Again the dual fields are denoted with a hat. \eqref{natdintronatdqrule} and \eqref{natdintronatddilatonrule} yield the dual NS sector but the transformation of RR fluxes is also needed. This was discovered much more recently in \cite{Sfetsos:2010uq}, and reignited interest in this technique for holographic applications. Surprisingly, the solution starts in very much the same way as the Abelian case, by defining the polyforms $P,\wh P$:
\begin{equation}
\text{IIB}  ~~~~  P:=\frac{e^{\Phi}}{2} \sum_{n=0}^4 \slashed{F}_{2n+1} ~, ~~~~ \text{IIA}  ~~~~ P:=\frac{e^{\Phi}}{2} \sum_{n=0}^5 \slashed{F}_{2n}~,
\label{natdintrobispinortypeIIBIIA}
\end{equation}
where $\slashed{F}_{i}:= \frac{1}{i!} \Gamma_{\mu_1...\mu_i} F_i^{\mu_1...\mu_i} $. Then, the dual RR fluxes are obtained from:
\begin{equation}
\hat{P}=P \ \Omega^{-1} ~,
\label{natdintronatdrrrules}
\end{equation}
where $\Omega=( \Gamma^1 \Gamma^2 \Gamma^3 + \zeta_{a} \Gamma^{a}) \Gamma_{11} / \sqrt{1+ \zeta^2}$, $\zeta^2=\zeta_a \zeta^a$,  $\zeta^a= \kappa^a_i z^i$, $\kappa^a_i \kappa^a_j=g_{ij}$ and $z_i=(b_i+v_i)/\det \kappa$, $b_{ij}=\epsilon_{ijk} b_k$. The only difference with the Abelian transformation \eqref{natdintroabelianprule} is that $\O$ matrix becomes more involved. With \eqref{natdintronatdqrule}, \eqref{natdintronatddilatonrule} and \eqref{natdintronatdrrrules} we can calculate the non-Abelian T-dual of a given \bg. 

Before discussing some essential properties, let us sketch how to obtain the NSNS Buscher rules \footnote{The rules of NATD transformation are sometimes referred to as Buscher rules, due to the original transformation rules discovered by Buscher in the Abelian case.}above from the T-dual action. They are similar in spirit, although the non-Abelian isometry introduces more indices and other complications. The lagrangian density for the NSNS sector reads:
\begin{equation}
{\cal L}=Q_{AB} \partial_+ X^A \partial_- X^B  ~,
\label{natdinitiallagrangian}
\end{equation}
where $\partial_{\pm} X^A=(\partial_{\pm} X^{\mu}, L^i_\pm)$. We then gauge the $\mathrm{SU(2)}$ isometry by replacing the derivatives by covariant derivatives $\partial_{\pm} g \rightarrow D_{\pm} g=\partial_{\pm} g - A_{\pm} g$. Afterwards we add a Lagrange multiplier term to constrain the gauge fields to be pure gauge:
\begin{equation}
-i \tr (v F_{\pm}) ~, ~~~~ F_{\pm}=\partial_+ A_- - \partial_- A_+ - \left[ A_+, A_- \right]~.
\label{natdlagrangemultipliertermNATD}
\end{equation}
After a gauge fixing, we can choose the 3 Lagrange multipliers $v_i$ as the new coordinates. The last step is to integrate out the gauge fields, obtaining the dual lagrangian density:
\begin{equation}
\hat{{\cal L}}=\hat{Q}_{AB} \partial_+ \hat{X}^A \partial_- \hat{X}^B  ~,
\label{natdfinallagrangian}
\end{equation}
where $\partial_{\pm} \hat{X}^B = ( \partial_{\pm} X^{\mu} , \partial_{\pm} v^i)$. 
Despite some obvious similarities, there are some differences with the previous Abelian transformation of section \ref{natdintroabelian}:
\begin{itemize}
	\item When Abelian \td is applied twice, the original \bg is recovered. Nevertheless, when NATD is applied for the first time, the \su2 isometry is destroyed, and therefore cannot be applied again. 
	\item For compact commuting isometries \td is a perturbative (in both $\alpha'$ and $g_s$) duality of the string partition function \cite{Rocek:1991ps}; meanwhile \natd may well be a symmetry only for \sugra. 
	\item The range of the dual coordinates that makes the transformation a string duality is known for the Abelian case \eqref{natdintroabelianperiodicities}. For \natd is not even known if it is a string duality.
\end{itemize}
The last point is relevant for holographic application, as several physical quantities depend on the range of dual coordinates as it, entanglement entropy and c-function described in sections \ref{natdintroentanglemententropy}, \ref{natdintrocfunction}. Such slight limitation will become manifest in our results, and eventually prompted further understanding of the field theory effect of NATD in later papers \cite{Lozano:2016kum,Lozano:2016wrs,Lozano:2017ole,Itsios:2017cew}. 

There are more aspects of \natd that we ignored in this short introduction. The interested reader might check \cite{Giveon:1993ai,Borlaf:1996na,Quevedo:1997jb} for some reference works and lectures. For more holography oriented application, \cite{Itsios:2013wd} has a very good balance of \natd fundamentals, previous literature, and citations. 
\section{Selected holographic observables}
To learn about Quantum Field Theory using AdS/CFT we have to relate gravitational quantities with boundary observables. In the following three sections we elaborate on three remarkable field theory side quantities that can be computed without any need of correlators and holographic renormalization: Wilson loop, entanglement entropy and c-function. 

\subsection{Wilson loop} \label{natdintrowilsonloop}
The Wilson loop is one of the oldest but most relevant gauge invariant operators in the study of non-Abelian gauge theories. It was introduced in \cite{Wilson:1974sk} in connection with the crucial phenomenon of confinement. Here we give a review of its most basic aspects that will be enough for the present thesis. Let's start with its definition in the field theory side. For a non-Abelian gauge theory with gauge field $A^a_\m$, the Wilson loop is defined as:
\bea
W(\mathcal{C}) = \frac{1}{N} \mathrm{Tr}\; P \mathrm{exp} \prt{i \oint A_\m\, d x^\m},
\eea
where $A_\m=A^a_\m T^a$, and $P$ is the path ordering operator along $\mathcal C$. $ \mathcal C $ is a closed contour in the (field theory) spacetime. 

In this work, the contour of integration will always be a rectangle, with length $T$ in the time direction and $d$ in some spatial one. In more physical terms, it represents the amplitude of propagation of a $q\bar{q}$ pair separated by distance $d$. In the limit of $T\rightarrow \infty$:
\bea
\lim_{T\to\infty} \langle W(\mathcal C) \rangle \approx e^{-T E(d)} 
\eea
where $E(d)$ is the energy of the $q \bar q$ pair at distance $d$. When the field theory is confining (meaning that quarks cannot be isolated at long enough distances), the energy grows linearly:
\bea
E(d) \sim \rho d,\; \r \approx \mathrm{constant}.
\eea
Using this in the previous equation, we quickly find:
\bea
\lim_{T\to\infty} \langle W(\mathcal C) \rangle \approx e^{-T E(d)} \approx e^{-T d\, \r} \approx e^{-\r\, \mathrm{area}} ,
\eea
because $T d$ is the area of the rectangle enclosed by $ \mathcal C $.  This fact is referred as the Wilson loop area law. It must be emphasized that determining whether a gauge theory is or not confining is of great physical interest.
 
On the other, AdS/CFT allows its computation (as usual for the correspondence, for large $N$ and strong coupling) \cite{Maldacena:1998im,Rey:1998ik}. A key point to keep in mind  is that the open string ending on a D-brane is dual to a quark \cite{Maldacena:1998im,Rey:1998ik}. This allows to conjecture that the Wilson loop is dual to an open string with worldsheet boundary $\partial \Sigma$ given by $\mathcal C$:
\bea
\langle W(\mathcal{C}) \rangle = Z_{string}(\partial \Sigma = \mathcal{C}).
\eea
For the pair of quarks to be non-dynamical, the D-brane to which the string is attached must lie on the boundary, thus fixing the boundary conditions for $Z_{string}$:
\bea
Z_{string}(\partial \Sigma = \mathcal{C}) = e^{-S(\mathcal{C})} ,
\eea
where $S(\mathcal{C})$ is the on-shell string action (that we will compute using the Nambu-Goto action). Therefore, we end up with the relation:
\bea
\langle W(\mathcal{C)} \rangle = e^{-S(\mathcal{C})}. \label{natdintrowilsonloopprescription}
\eea
If one computes $S(\mathcal{C})$ it will diverge in the UV. To make physical sense of it, it is necessary to renormalize it, normally by subtracting the quarks infinite mass (as they are non-dynamical). In practice, the on-shell embedding problem is straightforward to pose yet it may need numerical methods to solve, particularly in time dependent backgrounds. For some simple detailed examples, see \cite{Ramallo:2013bua} and references therein. A last basic but relevant case is the conformal field theory, in which:
\bea
E \sim 1/d
\eea 
due to conformal invariance. The precise coefficient depends on spacetime dimensionality.  We will use that dependence to probe the effective dimension in a holographic situation where it is not obvious, as shown in eq. \eqref{natdwilsonloop}. In that case, two dimensions of spacetime become very small along the RG flow, so the actual CFT dimension is not totally obvious. The Wilson loop and entanglement entropy results are consistent with a reduction of field theory dimension from four to two. A notable holographic application of Wilson loop to jet-quenching was discovered \cite{Liu:2006ug}. 
\subsection{Holographic entanglement entropy} \label{natdintroentanglemententropy}
Before introducing the entanglement entropy (EE), let's define the Shannon entropy $S_{Shannon}$ of a discrete probability distribution\footnote{Normally the Shannon entropy is defined with logarithms to the base 2 as it is counting bits.}
\bea
S_{Shannon} \eq - \Sigma_i p_i \ln p_i ,
\eea
where the $p_i$ are the probabilities of each outcome. The Shannon entropy is a key measure appearing in classical information theory, and equals the natural logarithm of the effective number of possible outcomes of a probability distribution. In this regard, it is totally analogous to Statistical Mechanics Gibbs' entropy. 

Similarly, the von-Neumann entropy of a finite density matrix $\rho$ is defined by:
\bea
S_{VN} \eq - \mathrm{Tr}\,\prt{ \rho \ln \rho}. \label{natdintroeevnentropy}
\eea
Notice that when the matrix is normalized, the eigenvalues are all non-negative and sum up to 1, thus forming a discrete probability distribution. Therefore, it equals the Shannon entropy computed with the eigenvalues of $\rho$. If the $\rho$ correspond to a pure state, $S_{VN}=0$, and $S_{VN}>0$ otherwise.

So far we discussed discrete quantum systems, but in an holographic context, we are almost always concerned about quantum fields that are spatially extended. Let us clarify how we define the entanglement entropy in those systems for time independent situations. \footnote{Time dependent covariant formulation for holographic entanglement entropy is also available \cite{Hubeny:2007xt}, but we are not going to consider this more involved situation.} We choose a region with some closed contour in space at fixed time, defining in this way a subset of the degrees of freedom.  After that, we define the entanglement entropy of the region bounded by the contour to be the von-Neumann entropy of its reduced matrix. The calculation of EE in QFT is normally very difficult, and typically achieved through the so-called replica trick. 
 
A relevant generic fact of entanglement entropy in QFT is the so-called area law: for local interactions, the EE is proportional to the boundary of the region, not its volume. This obviously resembles the area law of BH Thermodynamics with one crucial difference: it is fairly counterintuitive in gravity, but relatively intuitive in field theory.  

A very important application of entanglement \footnote{Another crucial one is the entire field of Quantum Information.} is to the theory of quantum many body systems appearing in Condensed Matter Physics. is the comprehension of different phases. Even being quantum mechanical systems, some of the phases can be easily understood with classical descriptions and order parameters; this is known as the Landau-Ginsburg paradigm, originally developed for superconductivity. Nevertheless this theory is not adequate when strong quantum effects are present, for example in fractional quantum Hall effect and quantum magnets. The key insight here is that entanglement entropy can be used as a meaningful \textit{parameter order} to distinguish these very non-classical phases. This effect is particularly dramatic in fractional quantum Hall, where at low energy the theory behaves as a topological field theory, and the correlations are not only not useful as order parameter but actually trivial.

Related to this, and more quantitative, is the amount of numbers required to specify the state of a quantum system, and to evolve it on a computer. This will justify the relevance of the von-Neumann \eqref{natdintroeevnentropy} and EE definitions. As the dimension of the Hilbert space of a system grows exponentially with the number of particles, brute force diagonalization of the Hamiltonian is impossible even for moderate number of particles. But it was eventually recognized that for not very excited states, the scaling of entanglement entropy with size of the region is directly related to the number of effective d.o.f. necessary to characterize the phase. This is for example applied in the density matrix renormalization group (DMRG) method. This somehow endows the EE definition with quantitative physical relevance. It is also at the core of another group of closely related methods collectively called tensor networks, which sharpened the connection to entanglement in AdS/CFT. Being quantitatively relevant and fairly universal as it does not refer to a particular observable but to the quantum state itself, we expect the same for the holographic description of field theories at large $N$ and strong coupling.

Now we turn again to AdS/CFT. How can we compute EE in the holographic description of the field theory? To do this, consider a minimal surface attached to the contour defining the space region at the boundary. This surface is typically called entangling surface. The surface minimizes the value of the Ryu-Takayanagi formula \cite{Ryu:2006bv,Ryu:2006ef}, which in string frame is given in \cite{Nishioka:2006gr}: 
\bea
S_{EE} \eq \frac{1}{4 G_N^D}\int_\Sigma d^{D-2}\s\; e^{-2\Phi} \sqrt{G_{ind}^{(8)}} \label{natdintroeeryutakayanagi}
\eea
where $D$ is the total spacetime dimension, $G_{ind}$ is the metric pull backed to the minimal surface $\Sigma$. There are several similarities with Wilson loops: we have to find a minimal surface attached to a contour on the boundary, renormalization in the UV is needed, and it can be used to signal confinement/deconfinement transition \cite{Nishioka:2006gr,Klebanov:2007ws}. Some of this similarities will show up in our examples in section \ref{natdEEWilson}. Like in the Wilson loop, the main practical difficulty is to solve the differential equation of the extremization.

The holographic entanglement entropy proposal is very firmly established at this stage, with many non-trivial tests satisfied: general quantum mechanical properties of (strong) subadditivity \cite{Headrick:2007km}, fixed point behaviour and area law derivation \cite{Ryu:2006bv}, recovery the 2d CFT result \cite{Ryu:2006bv}, EE equality in complementary regions of globally pure state\cite{Ryu:2006bv}, consistent relation to central charges in even dimensions \cite{Ryu:2006ef}..., and also a number of equally working extensions: Renyi entropies \cite{Headrick:2010zt}, time dependent situations \cite{Hubeny:2007xt}, relative entropy \cite{Blanco:2013joa}, mutual information \cite{Hayden:2011ag}, higher curvatures gravity corrections \cite{Dong:2013qoa}...  Finally, the Ryu-Takayanagi prescription was derived in \cite{Lewkowycz:2013nqa}.

Indeed, quantum many-body systems are not the only motivation of EE in QFT. It is expected that entanglement between inside and outside of the BH is relevant to understand the BH entropy, and of course there are some formal similarities regarding the area law. This motivations of course get even stronger in the presence of AdS/CFT. This important subfield of holographic EE is under strong development at the moment, but we are not going to use it in this thesis. The reader interested in general aspects of holographic EE may check \cite{Nishioka:2009un} for a middled-size review. A longer one, more recent and more oriented to quantum gravity is \cite{Rangamani:2016dms}. 

In this thesis, the use of entanglement entropy will be rather modest, both in extension and depth (see section \ref{natdEEWilson}). But the wealth of past literature, physical applications and future potential more than justifies this lengthy introduction about it. 
\subsection{Central charge and c-function} \label{natdintrocfunction}
In the forthcoming section, we will introduce the basic ideas behind this important observable of the field theory. Roughly, it is intended to count the effective volume of the phase space diminishing along the RG flow.

In the case of CFTs in 1+1 dimensions, the d.o.f. in the theory are proportional to the central charge of the Virasoro algebra. Consider a flowing theory (perhaps trivially as in the conformal case), and call $t=-\ln \mu$ with $\mu$ an energy scale. Therefore the renormalization group flow corresponds to increasing values of $t$. When we are at a conformal point, the couplings do not depend on $t$ $\beta_i = -\frac{d \lambda_i}{d t}=0$. 
Hence, the beta functions $\beta_i(\lambda)$ are the 'velocities' of the motion towards the IR. It is interesting to define a 'c-function' $c(t)$, constant for CFT's and equal to its central charge, and decreasing along the flow to lower energies:
\begin{equation}
\frac{d c(t)}{d t} \leq 0. \label{natdintrocfunctioncdefinition}
\end{equation}
This $c(t)$ is somehow intended at counting the d.o.f. of the theory at different values of $t$. The idea that it will decrease is essentially motivated by the fact that relevant deformations will lift part the massless modes and the RG flow will coarse grain some of them without creating new ones. These intuitions have been made precise and proved to be true in  two different situations. The first is the two dimensional case, with Zamolodchikov's definition of $c(g)$ \cite{Zamolodchikov:1986gt}. The second is the four dimensional case, where Cardy's conjecture for the 'a-theorem' \cite{Cardy:1988cwa} was proven by Komargodski and Schwimmer in \cite{Komargodski:2011vj}, although that proposal does not apply to odd dimensional cases. There are different versions of the c-theorem, varying in 'strength' and generality. A basic summary can be found in \cite{Barnes:2004jj}.

Now let us turn to its calculation in the bulk. For conformal field theories there exists a well-established formalism to calculate central charges using its relation with the Weyl anomaly $\mathcal{A}$ in the QFT when placed on a curved space.  This was first discovered in \cite{Brown:1986nw} and a complete holographic understanding was developed in \cite{Henningson:1998gx}. For conformal field theories in two and four dimensions holographically described by SUGRA, the relations are respectively as follows:

\begin{equation}
\label{natdintrocfunctionweylanomaly2d}
\hskip -4.2cm \textrm{2 dimensions:}\  \qquad \mathcal{A}=-\frac{L \bar{R}}{16 \pi G_N^{(3)}}=-\frac{c}{24 \pi}\bar{R} , ~~~~~ ~~~ c=\frac{3}{2} \frac{L}{G_N^{(3)}} \ ,
\end{equation}
\begin{equation}
\label{natdintrocfunctionweylanomaly4d}
\begin{aligned}
& \textrm{4 dimensions:} \, \quad   \mathcal{A}=-\frac{L^3 }{8 \pi G_N^{(5)}}  \left( - \frac{1}{8}\bar{R}^{ij}\bar{R}_{ij} +\frac{1}{24} \bar{R}^2 \right)  =\frac{1 }{16 \pi^2 }\Bigg[c W^2 - a E^2   \Bigg]  , 
\end{aligned}
\end{equation}
with $ W,E,c $ defined by:
\bea
W &:=&  \bar{R}^{ijkl}\bar{R}_{ijkl} -2 \bar{R}^{ij}\bar{R}_{ij} +\frac{1}{3}\bar{R}^2  ,\\
E &:=&  \bar{R}^{ijkl}\bar{R}_{ijkl} -4 \bar{R}^{ij}\bar{R}_{ij} + \bar{R}^2,\\
c&=&a=\frac{\pi}{8}\frac{L^3}{G_N^{(5)}} , G_N^{(10)}=8 \pi^6  \alpha'^4,  ~~~~~ G_N^{(10-d)}=\frac{G_N^{(10)}}{vol (M_d)} . 
\label{natdintrocfunctionnewton}
\eea
where  $\bar{R}_{ijkl}$, $\bar{R}_{ij}$ and $\bar{R}$ are the Riemann and Ricci tensors and scalar of the boundary metric.  The coefficients $c,a$ are equal in the large $N$ approximation captured by the leading order supergravity, but this is not true in holographic field theories described by higher derivative gravities in the bulk. 

Now we finally introduce the holographic c-function for flowing field theories in the spirit of eq. \eqref{natdintrocfunctioncdefinition}. Surprisingly, it is possible to define it satisfying the correct monotonicity on-shell (under some very generic conditions of positivity of energy \cite{Freedman:1999gp}). First, we introduce the result for a very simple yet relevant case is a metric of the form:
\begin{equation}
d s_{D}^2= e^{2A(r)}d x_{1,D-2}^2 + d r^2 \ ,
\label{manaza}
\end{equation}
where the proposed c-function with the correct monotonicity is:
\bea
c \eq \frac{1}{(A')^{D-2}}.
\label{natdintrocfunctionaprime}
\eea
Another step forward was taken by Klebanov, Kutasov and Murugan \cite{Klebanov:2007ws}, who found a similar result for the following very general ansatz in SUGRA:
\bea
d s^2 \eq \alpha_0 (r) \prt{d x^2_{d,1} + \beta_0(r) d r} + g_{ij}(r,\vec{\th})\, d\th_i d\th_j ,\\
\Phi \eq \Phi(r),
\label{natdintrocfunctionmiddle}
\eea
where $\Phi$ is the dilaton. Finally, when the $\alpha_0,\Phi$ depend also in the internal coordinates $\th_i$, the same proposal was also used with sucess in \cite{Macpherson:2014eza}:
\bea \label{natdintrocfunctiongeneraldef}
c\eq d^d \frac{\beta_0(r)^\frac{d}{2} \wh{H}^{\frac{2d+1}{2}} }{\pi G_N^{(10)} (\wh H')^d},\\ 
\wh H \eq \prt{\int d \vec \th \sqrt{ e^{-4\Phi} g_{int} \a_0^d } }^2.
\eea
%
\section{Some aspects of NATD in AdS/CFT} \label{natdcontext}
In this section we briefly mention the main conclusions of some previous papers applying NATD to SUGRA and AdS/CFT, and the outline of how our work extends them. The general notions of Abelian and \natd have been explained in sections \ref{natdintroabelian}, \ref{natdintronatd}. 

The organization of this chapter is as follows. In the first part
of this work, covered in sections \ref{natdprevioussolutions} to \ref{natdlift}, we present  type IIB backgrounds that are already 
known and the new ones (in type IIA, IIB and M-theory)
that we construct using NATD. Table \ref{natdsolutionstable} summarizes these solutions. In the second part of this work, starting in section \ref{natdcomments},
we begin the study of  the
field theoretical aspects encoded by the backgrounds presented in the first part.
Section \ref{natdpagecharges}  
deals with the quantized charges, defining ranks of the gauge groups.
Section \ref{natdcentralcharge}
studies the central charge computed holographically, either at the 
fixed points or along the anisotropic flows 
(where a proposal for a c-function is analyzed).
This observable  presents many clues
towards the understanding of the associated  QFTs. 
Section \ref{natdEEWilson}
presents a detailed study of Wilson loops and entanglement entropy of the QFT
at the fixed points and along the flow. We discuss the results in section \ref{natdconclusions}. The computations of the supersymmetry preserved by the solutions is relegated to appendix  \ref{natdappendixsusy2}.

NATD was used for the first time in the context of the gauge/gravity duality in the paper \cite{Sfetsos:2010uq}. Indeed, Sfetsos and Thompson applied NATD to the maximally supersymmetric example of AdS$_5\times$S$^5$, finding a metric and RR-fields that preserved ${\cal N}=2$ SUSY. When lifted to eleven dimensions, the background fits (not surprisingly) into the classification of \cite{Lin:2004nb}. What is interesting is that Sfetsos and Thompson \cite{Sfetsos:2010uq} {\it generated} a new solution to the 
Gaiotto-Maldacena differential equation \cite{Gaiotto:2009gz}, describing ${\cal N}=2$ SUSY CFTs of the Gaiotto-type \cite{Gaiotto:2009we}, \cite{Tachikawa:2013kta}.
This logic was profusely applied to less supersymmetric cases in 
\cite{Itsios:2012zv,Itsios:2012dc}; finding new metrics and defining new QFTs by
the calculation of their observables.

An important aspect of the generated solution is the supersymmetry; it can be partially broken even for Abelian \td \cite{Duff:1998us}. For \natd, the transformation may also lower it (for example the case of AdS$_5\times$S$^5$ goes from $\mathcal N=4$ to $\mathcal N=2$.\cite{Sfetsos:2010uq} ) \footnote{Or break it completely. This happens for example when the symmetry is $\mathrm{S^2}$ instead of \su2, as \natd can be generalized for cosets, \textit{cf.} \cite{Lozano:2011kb}.}, as mentioned in the paragraph above. It is possible to know the amount of \susy preserved after NATD without explicitly computing the supersymmetric variations \eqref{natdintrosugraiiavariations} or \eqref{natdintrosugraiibvariations} \cite{Sfetsos:2010uq,Kelekci:2014ima}. To do it, it is necessary to compute the Lie-Lorentz-Kosmann derivative \cite{Kosmann:1972kd,Ortin:2002qb} of the Killing spinor along the Killing vector that generates the isometry of the NATD transformation. We will also rely on this technique, as the \susy of the \sugra \bg is extremely useful to identify candidates for the \dft, as well as to determine the stability of the generated solution. 

Nevertheless, various puzzles associated with NATD remain. As explained in \ref{natdintronatd}, the global properties of the new manifold remain still elusive. In more concrete terms, the ranges of the dual coordinates (the Lagrange multipliers in the $\s$-model) are not known. For this reason, the precise field theory dual to the backgrounds generated by NATD is not fully determined by the transformation rules. 

\section{Summary of results}
In this chapter we present possible solutions to these puzzles---
at least in particular examples. Our study begins with type IIB backgrounds dual to a compactification of the
Klebanov-Witten CFT \cite{Klebanov:1998hh} to a 2-d CFT. We will present these backgrounds for different compactifications and perform a NATD transformation on them, hence generating new smooth and SUSY 
solutions with AdS$_3$-factors in type IIA and M-theory. Application of a further T-duality generates new backgrounds in type IIB with an AdS$_3$-factor which are also smooth and preserve the same amount of SUSY. We will make a proposal for the dual QFT and interpret the range of the dual coordinates in terms of a field theoretic operation.

The picture that emerges is that our geometries describe QFTs that become conformal at low energies. These CFTs live on the intersection of D2- and D6-branes suspended
between NS5-branes. While crossing the NS5-branes, 
charge for D4-branes is induced and new nodes of the quiver appear. We will present the calculation of different observables of the associated QFTs that support the proposal made above. These calculations are performed in smooth supergravity solutions, hence they are trustable and capture the strong dynamics of the associated 2-d CFTs.

The work performed in this chapter is a continuation of  the previous works of Sfetsos-Thompson
\cite{Sfetsos:2010uq} and Itsios-Nunez-Sfetsos-Thompson (INST) \cite{Itsios:2012zv,Itsios:2013wd}. The connection between the material in this work
and those papers is depicted in Fig. \ref{RoadMap} below.
\begin{figure}[h]
	\centering
	\includegraphics[scale=0.6]{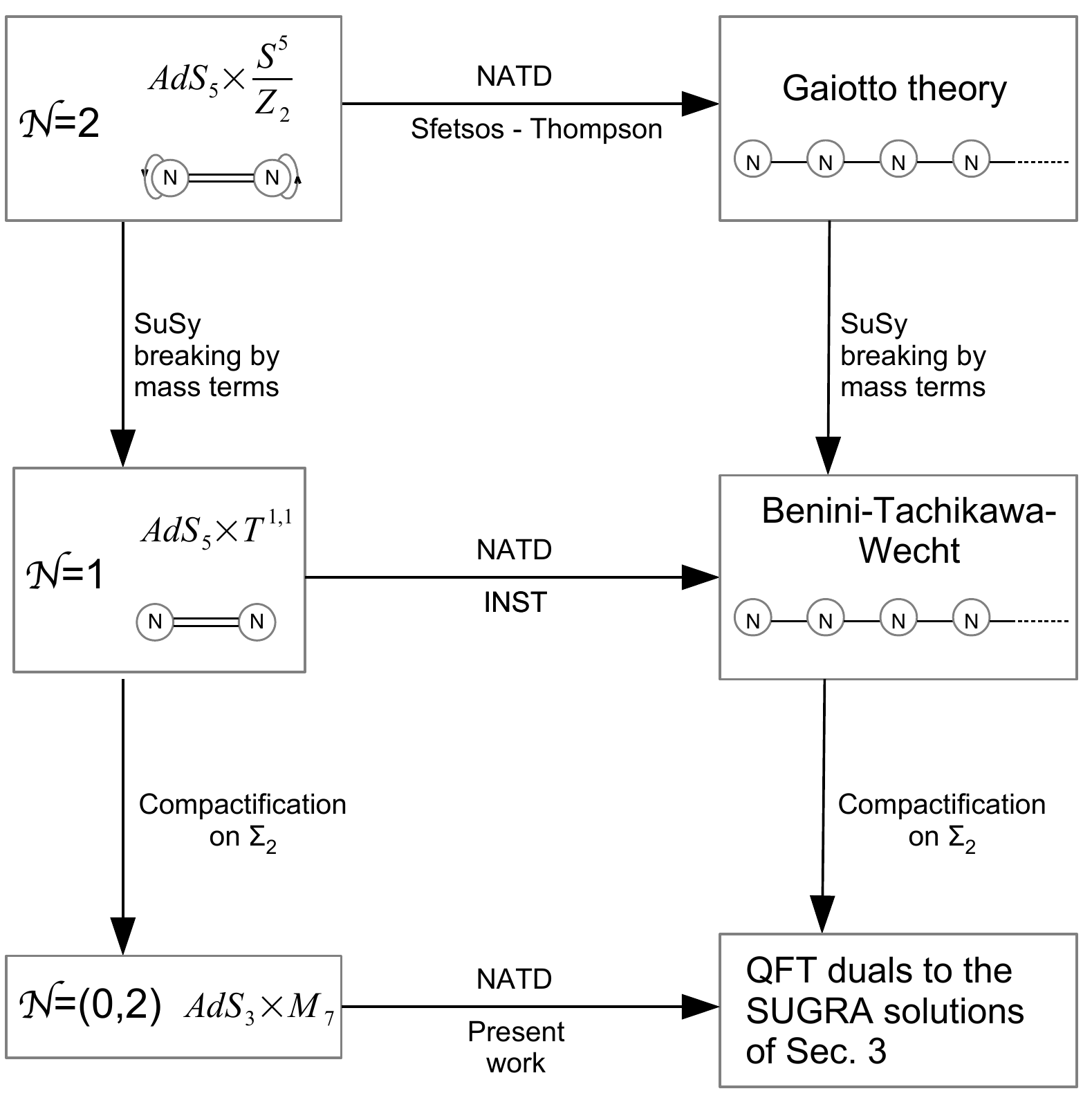}
	\caption{On the left: known solutions on which NATD is performed. On the right: QFT's that correspond to the NATD SUGRA solutions.}
	\label{RoadMap}
\end{figure}
\vskip 20pt
\begin{table}[h]
	\begin{tabular}{|c|c|c|c|c|}
		\hline
		\textrm{\textbf{Solutions}} & \textrm{\textbf{IIB}} &  \textrm{\textbf{NATD}} & \textrm{\textbf{NATD-T}} &\textrm{\textbf{Uplift}}
		\\
		\hline
		$
		\begin{array}{l}
		\textrm{Flow from} \;\mathrm{ AdS_5\times T^{1,1}} \; \textrm{to} \; \mathrm{AdS_3 \times \mathrm H_2\times T^{1,1}}  \; \big ( \mathcal{N} = 1\big)
		\\[5pt]
		\textrm{Fixed points} \;\mathrm{ AdS_3\times\Sigma_2 \times T^{1,1}}:  \Sigma_2 = \mathrm{S^2}, \mathrm{T^2}, \mathrm H_2 
		\end{array}
		$
		&  \ref{natdtwistedsolutions} & \ref{natdtwistedduals}  & \ref{natdt} & \ref{uplifteds2h2natd} 
		\\
		\hline
		\textrm{The Donos-Gauntlett solution} &  \ref{natdDG} &  \ref{natddualDG} & \textrm{-}  &  \ref{upliftnatdDG}\\
		\hline
	\end{tabular}
	\caption{Summary of solutions contained in sections \ref{natdprevioussolutions}, \ref{natdduals}, \ref{natdt} and \ref{natdlift}. The numbers indicate  the section where they are discussed.}
	\label{natdsolutionstable}
\end{table}
In the following (sections \ref{natdprevioussolutions} to \ref{natdlift}) we will exhaustively present a large set  of backgrounds solving the type IIB  or type IIA supergravity equations. Most of them are new, but some are already present in the bibliography. New solutions in eleven-dimensional
supergravity will also be discussed. These geometries, for the most part, preserve some amount of SUSY.

The common denominator of these backgrounds will be the presence of an AdS$_3$ sub-manifold in the ten or eleven-dimensional metric. This will be interpreted as the dual description of strongly coupled two dimensional conformal dynamics. In most of the cases, there is also a flow, connecting from an AdS$_3$ fixed point to an AdS$_5$, with boundary $\mathbb R^{1,1}\times \Sigma_2$. The manifold $\Sigma_2$ will be a constant curvature Riemann surface. As a consequence we conjecture that the full geometry is describing the strongly coupled dynamics of a four dimensional QFT, that is conformal at high energies and gets compactified on $\Sigma_2$ while flowing to the IR (typically preserving some amount of SUSY). The QFT flows at low energies to a  2-d CFT that is also strongly coupled.

The solutions that we are going to present in the next section, can be found by inspection of 
the type II equations, but requires a quite inspired ansatz. 
More practical is to search for solutions of this kind in 
five dimensional gauged supergravity, see the papers \cite{Buchel:2006gb,Gauntlett:2009zw}
for a detailed account of the lagrangians. 
Some other solutions are efficiently obtained by the use 
of generating techniques, for example a combination of
Abelian and  non-Abelian T-duality, 
that are applied to known (or new) backgrounds, as we show below. 

\section[\texorpdfstring{Simple flows from A\MakeLowercase{d}S$_5 \times$T$^{1,1}$ to A\MakeLowercase{d}S$_3\times$M$_7$ in type IIB}{Simple flows from $AdS_5 \times T^{1,1}$ to $AdS_3 \times M_7$ in type IIB}]{Simple flows from AdS$_\textbf 5 \times$T$^\textbf{1,1}$ to AdS$_\textbf{3}\times$M$_\textbf{7}$ in type IIB}
\label{natdprevioussolutions}
We start this section by proposing a 
simple background in type IIB. In the sense of the Maldacena duality,
this describes the strongly coupled dynamics of an   
${\cal N}=1$ SUSY QFT in four dimensions, that 
is compactified to two dimensions on a manifold $\Sigma_2$.
In order to allow such a compactification we turn on a 1-form field, $A_1$, on the Riemann surface
$\Sigma_2$. Motivated by the works \cite{Buchel:2006gb,Gauntlett:2009zw}, where the authors consider dimensional reductions to five dimensions of type IIB supergravity backgrounds on any Sasaki-Einstein manifold, we propose the following ansatz,
\begin{equation}
\begin{aligned}
& \frac{ds^{2}}{L^2}=e^{2 A} \left(- d y_0^2+d  y_1^2\right)+e^{2 B} d s^2_{\Sigma_{{2}}}+ d r^2+e^{2U} d s_{KE} ^2+e^{2V} \left( \eta + z A_1 \right)^2, 
\\[5pt]
& \frac{F_5}{L^4}=4 e^{-4U-V} \textrm{Vol}_5+2 J \wedge J \wedge \left( \eta + z A_1 \right) - z \textrm{Vol}_{\Sigma_2} \wedge J \wedge \left( \eta + z A_1 \right)
\\[5pt]
&   ~~~~~ - ze^{-2B-V} \textrm{Vol}_\mathrm{AdS_3} \wedge J, \;\;\; 
\\[5pt]
&    \Phi=0, ~~~~~~ C_0=0, ~~~~~~ F_3=0, ~~~~~~ B_2=0.
\end{aligned}
\label{NN02}
\end{equation}
%
%
We will focus on the case in which the Sasaki-Einstein space is $T^{1,1}$, hence the K\"ahler-Einstein manifold is
\begin{equation}
ds_{KE}^2=\frac{1}{6}(\sigma_{1}^2+\sigma_{2}^2+\omega_1^2+\omega_2^2) \ ,
\label{zazar}
\end{equation}
where we have defined,
\begin{equation}
\label{gaga}
\begin{array}{lll}
\sigma_{1}= d \theta_1,  &\hspace{-3mm} \sigma_{2}= \sin \theta_1 d\phi_1,   &\hspace{-6mm} \sigma_{3}= \cos \theta_1 d\phi_1,
\\[10pt]
\omega_1=\cos\psi \sin\theta_2 d\phi_2 -\sin\psi d\theta_2  , &\hspace{-3mm} \omega_2=\sin\psi \sin\theta_2 d\phi_2  + \cos\psi d\theta_2 , &\hspace{-6mm} \omega_3=d\psi +\cos\theta_2 d\phi_2  , 
\\[10pt]
\textrm{Vol}_{AdS_3}=e^{2A} dy_0 \wedge dy_1 \wedge dr, &\hspace{-3mm} \textrm{Vol}_5= e^{2B} \textrm{Vol}_{AdS_3} \wedge \textrm{Vol}_{\Sigma_{{2}}} , &\hspace{-6mm} z \in \mathbb{R} ,
\\[10pt]
\eta =\frac{1}{3}\left( d\psi + \cos \theta_1 d \phi_1 + \cos \theta_2 d \phi_2 \right), & \hspace{-3mm} J=\frac{-1}{6}\left( \sin \theta_1 d \theta_1 \wedge d \phi_1 + \sin \theta_2 d \theta_2\wedge d \phi_2  \right). & \textrm{}
\end{array}
\end{equation}
The forms $\eta$ and $J$ verify the relation $d\eta = 2 \ J$.  The 
range of the angles in the $\sigma_{i}'s$ and the $\omega_i\textrm{'}s$ 
---the left invariant forms of $\mathrm{SU(2)}$---  
is given by $0\leq\theta_{1,2}<\pi$, 
$0\leq\phi_{1,2}<2\pi$ and $0\leq\psi<4\pi$. {The $\omega_i$
	satisfy $d \omega_i = \frac{1}{2} \ \epsilon_{ijk} \ \omega_j \wedge \omega_k$.}
We also defined a one form $A_1$, that verifies $d C_1= \textrm{Vol}_{\Sigma_2}$. As usual $ds_{\Sigma_2}^2$ is the metric of the two dimensional surface of curvature $\kappa=(1,-1,0)$, denoting a sphere, hyperbolic plane\footnote{To be precise, we do not consider the hyperbolic plane $\mathrm H_2$, as it has infinite volume. What we consider is a compact space $\mathrm H_2/\Gamma$ obtained by quotient by a proper Fuchsian group \cite{Kehagias:2000dga}, and its volume is given by $4\pi (g-1)$, where $g$ is the genus of $\mathrm H_2/\Gamma$.} or a torus 
respectively. In local coordinates these read,
\begin{equation}
\label{NN0}
\begin{array}{llll}
A_1= - \cos \alpha \ d \beta,  &  \textrm{Vol}_{\Sigma_{{2}}}= \sin \alpha \ d\alpha \wedge d \beta,     &  ds^2_{\Sigma_{{2}}}= d\alpha^2 + \sin^2 \alpha d \beta^2, & (\kappa=1) \ ,
\\[10pt]
A_1= \cosh \alpha \ d \beta,  & \textrm{Vol}_{\Sigma_{{2}}}= \sinh \alpha \ d\alpha 
\wedge d \beta,  &  ds^2_{\Sigma_{{2}}}= d\alpha^2 + \sinh^2 \alpha 
d \beta^2, & (\kappa=-1) \ ,
\\[10pt]
A_1 =  \alpha d \beta,  & \textrm{Vol}_{\Sigma_{{2}}}= d\alpha \wedge d \beta,  &  d s^2_{\Sigma_{{2}}}= d\alpha^2 + d \beta^2, & (\kappa=0) \ .
\end{array}
\end{equation}
%
%
A natural vielbein for the metric (\ref{NN02}) is,
\begin{equation}
\begin{aligned}
& e^{y_0}=L e^A d y_0 \ ,   ~~~~~  e^{y_1}=L e^A d y_1 \ ,  ~~~~~  e^{\alpha}=L e^B d\alpha \ ,~~~~~~~  e^{\beta}=L e^B A_0 d\beta \ ,  ~~~~~~~  e^{r}=L r \ , ~~~~~~~~~~~~
\\[5pt]
& e^{\sigma_{1}}= L \frac{e^{U}}{\sqrt{6}} \sigma_{1} \ ,  ~~~~  e^{\sigma_{2}}= L \frac{e^{U}}{\sqrt{6}} \sigma_{2} \ ,    ~~~~  e^{1}= L \frac{e^{U}}{\sqrt{6}} \omega_{1} \ ,  ~~~~ e^{2}=  L \frac{e^{U}}{\sqrt{6}} \omega_{2} \ ,   ~~~~ e^{3}=L e^V (\eta + z A_1) \ ,
\label{vielbein00}
\end{aligned}
\end{equation}
with $A_0=\sinh \alpha$ for $\mathrm H_2$, $A_0=\sin \alpha$ for $\mathrm{S^2}$  and $A_0=1$ for $\mathrm{T^2}$. 

As anticipated, the background above describes 
the strong dynamics for a compactification of a 
four dimensional QFT to two dimensions. 
{In the case that we are interested in this work} --- in which the 
K\"ahler-Einstein manifold is the one in eq. \eqref{zazar}--- 
the four dimensional QFT at high energy asymptotes to  
the Klebanov-Witten quiver 
\cite{Klebanov:1998hh} on $\mathbb R^{1,1}\times \Sigma_2$. 
As it will be clear, most of our results will be valid
for the case of a general $Y^{p,q}$ or any other Sasaki-Einstein 
manifold and their associated QFT. Indeed, these solutions 
can be obtained by lifting to type IIB, simpler backgrounds of
the five-dimensional supergravity in \cite{Gauntlett:2009zw}.
In fact, the 5-d supergravity lagrangian was written for any Sasaki-Einstein 
internal space.

Assuming that the functions $A,B,U,V$ depend only on the radial 
coordinate $r$, we can calculate the BPS equations describing 
the SUSY preserving flow from  AdS$_5\times$T$^{1,1}$  
at large values of the radial coordinate to AdS$_3\times$M$_7$. 
The end-point of the flow will be dual to a  
2-d CFT obtained after taking the low energy limit of
a twisted KK compactification of the Klebanov-Witten QFT
on $\Sigma_2$. Imposing a set of projections on the 
SUSY spinors of type IIB---see appendix \ref{natdappendixsusy2} for details--- 
we find,
\begin{eqnarray}
& & A'-e^{-V-4U}\pm \frac{z}{2}e^{-2B-2U-V}=0,\nonumber\\[5pt]
& & B'-e^{-V-4U}\mp \frac{z}{2}e^{-2B-2U-V} \mp \frac{z}{2}e^{-2B+V}=0,\nonumber\\[5pt]
& & U'+e^{-V-4U}-e^{V-2U}=0,\label{O17}
\\[5pt]
& & V'-3e^{-V}+2e^{V-2U}+e^{-V-4U} \mp \frac{z}{2}e^{-2B-2U-V} \pm \frac{z}{2}e^{-2B+V}=0 \ ,\nonumber
\end{eqnarray}
where the upper signs are for $\mathrm H_2$, the lower signs for $\mathrm{S^2}$,  and $z=-\frac{1}{3}$ for both cases.
In the case of the torus the variation of the gravitino will force $z=0$, obtaining $A'=B'$, which does not permit an AdS$_3$ solution.
We now attempt {to} find simple solutions to the eqs.(\ref{O17}).
\subsection[Solution of the form AdS$_3 \times  \Sigma_2$, with 'twisting']{Solution of the form AdS$_\textbf{3} \times  \Sigma_{\textbf{2}}$, with 'twisting'}
\label{natdtwistedsolutions}
{At this point we are going to} construct a flow between AdS$_5\times$T$^{1,1}$ 
and AdS$_3\times\Sigma_2\times$M$_5$.
To simplify the task, we  propose that the functions 
$U,V$ are constant, then the BPS equations imply that $U=V=0$, 
leaving us---in the case 
of $\mathrm H_2$ with,
\begin{equation}
A'= 1 + \frac{e^{-2B}}{6},\;\;\; B'=1-\frac{e^{-2B}}{3} \ ,
\label{BPSH2}
\end{equation}
that can be immediately integrated,
\begin{equation}
A=\frac{3}{2}r-\frac{1}{4} \ln \left( 1+e^{2r} \right) +a_0 \ ,    ~~~~~~ B=\ln\frac{1}{\sqrt{3}} + \frac{1}{2} \ln \left( 1+e^{2r} \right).
\label{O19}
\end{equation}
One of the  integration constants associated with these solutions 
corresponds to a choice of the origin 
of the holographic variable and the other constant, 
$e^{2a_0}$, sets the size of the 
three dimensional space ($y_0,y_1,r$). We will choose $a_0=0$ 
in what follows.

In the limit $r\rightarrow \infty$ (capturing the UV-dynamics of the QFT) we recover a Klebanov-Witten metric
\begin{equation}
A\sim r -\frac{1}{4}e^{-2r} +\frac{1}{8}e^{-4r} \ ,   ~~~~~ 
B \sim r -\frac{1}{2}\ln 3 +\frac{1}{2}e^{-2r}-\frac{1}{4}e^{-4r} \ ,
\label{O20x}
\end{equation}
whilst in the limit $r\rightarrow -\infty$ (that is dual to  the IR 
in the dual QFT) we obtain a supersymmetric solution of 
the form AdS$_3\times$H$_2$ ,
\begin{equation}
A\sim\frac{3}{2}r -\frac{1}{4}e^{2r}+\frac{1}{8}e^{4r} \ ,  ~~~~~ 
B\sim\ln\frac{1}{\sqrt{3}} +\frac{1}{2}e^{2r} -\frac{1}{4}e^{4r}.
\label{O21}
\end{equation}
Let us write explicitly this background using eq. \eqref{zazar},
\begin{eqnarray}
& &      \frac{ds_{10}^2}{L^2}=\frac{e^{3r}}{\sqrt{1+e^{2r}}}\left( -dy_0^2+dy_1^2 \right)+\frac{1+e^{2r}}{3}\left( d\alpha^2+\sinh^2\alpha d\beta^2 \right)+dr^2+ {ds^2_{KE}} +\left( \eta - \frac{1}{3} A_1 \right)^2 \ ,\nonumber\\[5pt]
& & \frac{F_5}{L^4}=\frac{4}{3}e^{3r}\sqrt{1+e^{2r}}\sinh \alpha \ dy_0 \wedge dy_1 \wedge d\alpha \wedge d\beta \wedge dr +2 J \wedge J \wedge \left( \eta  - \frac{1}{3} A_1 \right) \nonumber\\[5pt]
& &  
\qquad +\frac{1}{3} \sinh \alpha \ d \alpha \wedge d \beta \wedge J \wedge \left( \eta - \frac{1}{3}  A_1 \right) + \frac{e^{3r}}{\left(1+e^{2r}\right)^{\frac{3}{2}}} dy_0 \wedge dy_1 \wedge dr  \wedge J \ ,
\\[5pt]
& &       A_1= \cosh \alpha \ d \beta,~~~~~~\Phi=0, ~~~~~~ C_0=0 ,~~~~~~ F_3=0, ~~~~~~ B_2=0 \ .
\nonumber
\label{O20}
\end{eqnarray}
The solution above was originally presented in \cite{Gauntlett:2006af}.

We consider now the case of $\mathrm{S^2}$. With the same assumptions 
about the functions $U,V$, the BPS equations (\ref{O17}) read:
\begin{equation}
A'=1-\frac{e^{-2B}}{6},\;\;\;     B'=1+\frac{e^{-2B}}{3}.
\label{O25}
\end{equation}
These can also be immediately integrated (with a suitable 
choice of integration constants), 
\begin{equation}
A=\frac{3}{2}r-\frac{1}{4} \ln \left( e^{2r} -1 \right) \ ,   ~~~~~~ B=\ln\frac{1}{\sqrt{3}} + \frac{1}{2} \ln \left( e^{2r}-1 \right) \ , ~~~~~~ r>0 \ .
\label{O26}
\end{equation}
This solution seems to be problematic close to $r=0$. 
Indeed, if we compute the Ricci scalar we obtain $R=0$, 
nevertheless, $R_{\mu\nu}R^{\mu\nu}\sim \frac{3}{32 L^4 r^4}+.... $ 
close to $r=0$. The solution is singular and we will not
study it any further.

It is interesting to notice that  a family of {\it non-SUSY} 
fixed point solutions exists. Indeed, we can consider 
the situation where $B,U,V$  and $A'(r)=a_1$ are constant. 
For $\mathrm{S^2}$ and $\mathrm H_2$, we find then that the Einstein equations impose $U=V=0$ and 
\begin{equation}
8+e^{-4B}z^2-4 \ a_1^2=0 \ ,  ~~~~~~ 4-z^2 e^{-4B}+\kappa \ e^{-2B}=0.
\label{O12}
\end{equation}
Where $\kappa=+1$ for $\mathrm{S^2}$ and $\kappa=-1$ for $\mathrm H_2$. The solution is,
\begin{equation}
z^2= e^{2B} \left( 4 e^{2B}  +  \kappa \right) \ ,   ~~~~~~  a_1^2=3  +  \kappa \ \frac{e^{-2B}}{4}~.
\label{O13}
\end{equation}
For the $\mathrm{S^2}$ case the range of parameters is,
\begin{equation}
B \in \mathbb{R} \ , ~~~~~~ z \in \mathbb{R}-\{0\} \ , ~~~~~~ a_1 \in \left( -\infty,\sqrt{3}\right) \cup \left( \sqrt{3}, +\infty \right) \ ,
\label{O14}
\end{equation}
while for ${\mathrm H_2}$ we find,
\begin{equation}
B \in \Big[-\ln 2, +\infty \Big) \ , ~~~~~~ z \in \mathbb{R} \ , ~~~~~~ a_1 \in \Big( -\sqrt{3},-\sqrt{2} \ \Big] \cup \Big[ \ \sqrt{2}, \sqrt{3} \ \Big) \ .
\label{O15}
\end{equation}
Notice that in the case of $\mathrm H_2$, there is a non-SUSY solution with $z=0$, hence no fibration
between the hyperbolic plane and the Reeb 
vector $\eta$. The SUSY fixed point  
in eq. \eqref{O20} is part of the family 
in eq. \eqref{O15}, with $z=-\frac{1}{3}$.

For the $\mathrm{T^2}$ we also find an AdS$_3$ solution,
\begin{equation}
a_1^2=3 \ , ~~~~~~ z ^2=4e^{4B} \ , ~~~~~~ U=0 \ , ~~~~~~ V=0 \ .
\label{T7}
\end{equation}

Up to this point, we have set the stage for our study, 
but most of the solutions discussed above have already been stated in the literature. {Here} we just added some new backgrounds and technical elaborations on the known ones.
Below, we will present genuinely new type IIA/B solutions. 
The technical tool {that we have used is} non-Abelian T-duality. 
This technique, when applied to an $\mathrm{SU(2)}$
isometry of the previously discussed solutions will generate new type IIA configurations. {In our examples these new solutions are nonsingular}. Their lift to eleven dimensions will 
produce new, smooth, AdS$_3$ configurations in M-theory. {Moreover, performing an additional Abelian T-duality transformation
	we generate new type IIB  backgrounds} with all fluxes turned on and an AdS$_3$ fixed point at the IR.
We move on to describe these.

\subsection{The Donos-Gauntlett(-Kim) background}
\label{natdDG}

In this section we revisit a beautiful solution  written in 
\cite{Donos:2014eua}---this type of solution
was first studied in \cite{Donos:2008ug}. Due to the more
detailed study of \cite{Donos:2014eua}, we will refer to it as
the Donos-Gauntlett solution in the rest of this chapter.  

The background in \cite{Donos:2014eua} 
describes a flow  in the radial coordinate, from 
AdS$_5\times$T$^{1,1}$ to AdS$_3\times$M$_7$. 
The solution is very original. While the boundary of AdS$_5$ 
is of the form $\mathbb R^{1,1}\times$T$^2$, the compactification on the
Riemann surface (a torus) does not use a 'twist'
of the 4d-QFT. This is reflected by the absence of a 
fibration of the Riemann surface on the R-symmetry direction $\eta$. 
Still, the background preserves SUSY
\footnote{The solutions in Eqs.(63)-(80) of the paper \cite{Nunez:2001pt}, 
	can be thought
	as an 'ugly' ancestor of the Donos-Gauntlett background.}. 
The {solution} contains 
an active NS-three form $H_3$ that together with 
the RR five form $F_5$ implies the presence of a 
RR-three form $F_3$. The authors of \cite{Donos:2014eua} found this configuration
by using a very inspired ansatz. 
We review this solution below, adding new information to complement that in
\cite{Donos:2014eua}.

The metric ansatz is given by,
\begin{align}
& &  \frac{ds^2}{L^2}   = e^{2A}\left(-dy_0^2+dy_1^2\right)+e^{2B}\left(d\alpha^2+d\beta^2\right)+dr^2+ e^{2U}ds_{KE}^{2}+e^{2V} \eta^2 \ , 
\label{metric-bef}
\end{align}
where $A,B,U,V$ are functions of the radial coordinate $r$ only. 
The line element
$ds_{KE}^{2}$ is defined in eq. \eqref{zazar} 
and $\sigma_{i}$, $\omega_{i}$, $\eta$ are given in eq. \eqref{gaga}.
%
The natural vielbein is,
\begin{equation}
\begin{aligned}
& e^{y_0}=L e^{A} dy_0, \quad e^{y_1}=L e^{A} dy_1, \quad e^{\alpha}= Le^{B} d\alpha, \quad e^{\beta}= L e^{B} d\beta, \quad e^{r} = L dr,		
\\[5pt]
& e^{\sigma_{1}}= L \frac{e^{U}}{\sqrt{6}}\sigma_{1},  
\quad e^{\sigma_{2}}=L \frac{e^{U}}{\sqrt{6}}\sigma_{2},   \quad e^{1}= L \frac{e^{U}}{\sqrt{6}} \omega_1, \quad e^{2}= L \frac{e^{U}}{\sqrt{6}} \omega_2, \quad e^{3}=L e^V \eta\ .
\end{aligned}
\label{veil-bef}
\end{equation}
%
Note that compared to \cite{Donos:2014eua} 
we have relabeled $\phi_i \rightarrow -\phi_i$.
To complete the definition of the background, we also need 
\begin{align}
{{\textrm{Vol}}_1= - \sigma_{1} \wedge \sigma_{2}, 
	\qquad  {\textrm{Vol}}_2=\omega_1\wedge\omega_2}, 
\label{bdefn-bef}
\end{align}
and the fluxes,
\begin{equation}
\begin{aligned}
	& \frac{1}{L^{4}}F_{5} =  4e^{2A+2B-V-4U}dy_0 \wedge dy_1 \wedge d\alpha \wedge d\beta \wedge d r+
	\frac{1}{9}\eta \wedge {\textrm{Vol}}_1 \wedge {\textrm{Vol}}_2		
	\\[5pt]
	& \qquad \quad  {+\frac{\lambda^{2}}{12} \Big[ d\alpha \wedge d\beta \wedge \eta \wedge ( \textrm{Vol}_1+{\textrm{Vol}}_2) + e^{2A-2B-V} dy_0 \wedge dy_1 \wedge d r \wedge ({\textrm{Vol}}_1+{\textrm{Vol}}_2)\Big]}, 
	\\[5pt]
	&  {\frac{1}{L^{2}} F_3=\frac{\lambda}{6} d\beta \wedge ({\textrm{Vol}}_1-{\textrm{Vol}}_2), \qquad \frac{1}{L^{2}}  B_2 = -\frac{\lambda }{6}\alpha\ ({\textrm{Vol}}_1-{\textrm{Vol}}_2)},
\end{aligned}
\label{fluxes-bef}
\end{equation}
\begin{equation}
\begin{aligned}
& \frac{1}{L^{4}}F_{5} =  4e^{2A+2B-V-4U}dy_0 \wedge dy_1 \wedge d\alpha \wedge d\beta \wedge d r+
\frac{1}{9}\eta \wedge {\textrm{Vol}}_1 \wedge {\textrm{Vol}}_2		
\\[5pt]
& \qquad \quad  {+\frac{\lambda^{2}}{12} \Big[ d\alpha \wedge d\beta \wedge \eta \wedge ( \textrm{Vol}_1+{\textrm{Vol}}_2) + e^{2A-2B-V} dy_0 \wedge dy_1 \wedge d r \wedge ({\textrm{Vol}}_1+{\textrm{Vol}}_2)\Big]}, 
\\[5pt]
&  {\frac{1}{L^{2}} F_3=\frac{\lambda}{6} d\beta \wedge ({\textrm{Vol}}_1-{\textrm{Vol}}_2), \qquad \frac{1}{L^{2}}  B_2 = -\frac{\lambda }{6}\alpha\ ({\textrm{Vol}}_1-{\textrm{Vol}}_2)},
\end{aligned}
\label{fluxes-bef}
\end{equation}
where $\lambda$ is a constant that encodes the deformation of the space
(and the corresponding operator in the 4d CFT). 
The BPS equations for the above system are given by,
\begin{equation}
\label{BPS}
\begin{aligned}
& A' =	\frac{1}{4} \lambda ^2 e^{-2 B-2 U-V}+e^{-4 U-V}, 
\qquad B' =		e^{-4 U-V}-\frac{1}{4} \lambda ^2 e^{-2 B-2 U-V},	 
\\[5pt]
& U' =	e^{V-2 U}-e^{-4 U-V}, \qquad V' = 	-\frac{1}{4} \lambda ^2 e^{-2 B-2 U-V}-e^{-4 U-V}-2 e^{V-2 U}+3 e^{-V}.
\end{aligned}
\end{equation}
It is possible to recover the AdS$_5\times$T$^{1,1}$ solution by setting $\lambda=0$ and
\begin{equation}
A=B=r, \qquad U=V=0.
\label{AdST11sol}
\end{equation}

Further we can recover the AdS$_3$ solution 
by setting $\lambda=2$ (this is just a conventional value adopted in
\cite{Donos:2014eua}) {and}
\begin{equation}
A=\frac{3^{3/4}}{\sqrt{2}}r, \qquad B=\frac{1}{4}\ln\left(\frac{4}{3}\right), \qquad U=\frac{1}{4}\ln\left(\frac{4}{3}\right), \qquad V=-\frac{1}{4}\ln\left(\frac{4}{3}\right) .
\label{AdS3sol}
\end{equation}
{A flow
	(triggered by the deformation parameterized by  $\lambda$) between the asymptotic AdS$_5$ and
	the AdS$_3$ fixed point}
can be {found numerically}.

%


Up to this point, we have set the stage for our study, 
but most of the solutions discussed above have already been stated in the literature. {Here} we just added some new backgrounds and technical elaborations on the known ones.
Below, we will present genuinely new type IIA/B solutions. 
The technical tool {that we have used is} non-Abelian T-duality. 
This technique, when applied to an $\mathrm{SU(2)}$
isometry of the previously discussed solutions will generate new type IIA configurations. {In our examples these new solutions are nonsingular}. Their lift to eleven dimensions will 
produce new, smooth, AdS$_3$ configurations in M-theory. {Moreover, performing an additional Abelian T-duality transformation
	we generate new type IIB  backgrounds} with all fluxes turned on and an AdS$_3$ fixed point at the IR.
We move on to describe these.

\section{New backgrounds in type IIA: uses of non-Abelian T-duality} \label{natdduals}

In this section we apply the rules \eqref{natdintronatdqrule}, \eqref{natdintronatddilatonrule}, \eqref{natdintronatdrrrules} of \natd on the type IIB backgrounds presented above. We have checked explicitly that the generated \bgs  are indeed IIA \sugra solutions.   

\subsection{The non-Abelian T-dual of the twisted solutions}
\label{natdtwistedduals}

We will start by applying non-Abelian T-duality (NATD) to the 
background obtained via a twisted compactification
in section \ref{natdprevioussolutions}. The configuration we will focus on
is a particular case of that in eq. \eqref{NN02}. 
Specifically, in what follows we consider $U = V = 0$. These values for the functions $U$ and $V$ are compatible with the BPS system \eqref{O17}. In this case the background \eqref{NN02} simplifies to, 
\begin{equation}
\label{NN01}
\begin{aligned}
& \frac{ds^{2}}{L^2}=e^{2 A} \left(- dy_0^2+d y_1^2\right)+e^{2 B} ds^2_{\Sigma_{{2}}}+ dr^2+\frac{1}{6} \left(\sigma_{1}^2+\sigma_{2}^2\right)+\frac{1}{6} \left(\omega_1^2+\omega_2^2\right) + \left( \eta + z A_1 \right)^2 \ ,
\\[5pt]
& \frac{F_5}{L^4}=4 \textrm{Vol}_5+2 J \wedge J \wedge \left( \eta + z A_1 \right) - z \textrm{Vol}_{\Sigma_2} \wedge J \wedge \left( \eta + z A_1 \right) - ze^{-2B} \textrm{Vol}_\mathrm{AdS_3} \wedge J \ .
\end{aligned}
\end{equation}

As before, all other RR and NS fields are taken to vanish. Also, the 1-form $A_1$, the line element of the Riemann surface $\Sigma_2$ and the corresponding volume form, for each of the three cases that we consider here, are given in eq. \eqref{NN0}.

We will now present the {details} for the background after NATD
has been applied on the $\mathrm{SU(2)}$ {isometry} described by the coordinates $(\theta_2,\phi_2,\psi)$. As is well-known
a gauge fixing has to be implemented {during the NATD procedure}. This leads to a  choice of three 'new coordinates' among the 
Lagrange multipliers {$(x_1,x_2,x_3)$} used in the NATD procedure  
and the 'old coordinates' ($\theta_2,\phi_2,\psi$) 
\footnote{The process of NATD and the needed gauge fixing was 
	described in detail in \cite{Itsios:2013wd,Macpherson:2014eza}.}. 
In all of our examples we consider a gauge fixing of the form,
\begin{equation}
\theta_2 = \phi_2 = \psi = 0.
\label{gaugefixing012}
\end{equation}
As a result, the Lagrange multipliers $x_1, x_2$ and $x_3$ play the r\^ole of the dual coordinates in the new background.
To display the natural symmetries of the background, 
we will quote the results using  spherical coordinates (the expressions in cylindrical and cartesian coordinates are written in appendix C of \cite{Bea:2015fja}). 
We define,
\begin{equation}
\begin{aligned}
& x_1=\rho \cos \xi \sin \chi \ , ~~~~~~ x_2=\rho \sin \xi \sin \chi \ , ~~~~~~ x_3=\rho \cos \chi \ ,
\\
& \Delta =L^4 +54 \alpha'^2 \rho^2 \sin^2 \chi +36 \alpha'^2 \rho^2 \cos^2 \chi \ ,  ~~~~~ \tilde{\sigma}_{3}= \cos \theta_1 d \phi_1 + 3 z A_1.
\label{NN10}
\end{aligned}
\end{equation}
%
%
The NSNS sector of the transformed IIA background reads,
\begin{equation}
\label{NN11}
\begin{aligned}
& e^{-2 \widehat{\Phi}}=\frac{L^2}{324 \alpha'^3} \Delta, 
\\
& \frac{d\hat{s}^{2}}{L^2}=e^{2 A} \left(- dy_0^2+d y_1^2\right)+e^{2 B} ds^2_{\Sigma_2}+ dr^2+\frac{1}{6} \left(\sigma_{1}^2+\sigma_{2}^2\right) + \frac{\alpha'^2}{\Delta} \Bigg[ 6 \Big( \sin^2 \chi \big( d \rho^2 + \rho^2 (d \xi + \tilde{\sigma}_{3})^2 \big) 
\\
& \qquad + \rho \sin (2 \chi) d\rho d\chi+\rho^2 \cos^2 \chi d \chi^2 \Big) + 9 \Big( \cos \chi d \rho - \rho \sin \chi d \chi \Big)^2 +\frac{324 \alpha'^2}{L^4} \rho^2 d\rho^2 \Bigg] ,
\\
& \widehat{B}_2=\frac{\alpha'^3}{\Delta} \Bigg[ 36 \rho \cos \chi \Big(\rho \tilde{\sigma}_{3}\wedge d\rho +\rho \sin \chi d\xi \wedge \big(\sin \chi d\rho+\rho \cos \chi d \chi \big)\Big)
\\
& \qquad + \Big(  \frac{L^4}{\alpha'^2} \tilde{\sigma}_{3} - 54 \rho^2 \sin^2 \chi d \xi  \Big)\wedge \big( \cos \chi d \rho - \rho \sin \chi d \chi \big) \Bigg] \ ,
\end{aligned}
\end{equation}

while the RR sector is, 
\footnote{
	According to the democratic formalism, the higher rank RR forms are related to those of lower rank through the relation $ F_p = (-1)^{[\frac{p}{2}]} * F_{10-p}$.
}
\begin{align}
\label{NN12}
\widehat{F}_0 &=0, \qquad \widehat{F}_2=\frac{L^4}{54\alpha'^{\frac{3}{2}}} \left( 2 {\sigma_{1}\wedge \sigma_{2}} + 3 z \textrm{Vol}_{\Sigma_2} \right) \ ,
\nonumber\\[5pt]
\widehat{F}_4 &=\frac{L^4}{18 \sqrt{\alpha'}} \Bigg[   3 z e^{-2B}  \big(d\rho \cos \chi - \rho \sin \chi d\chi \big)  \wedge \textrm{Vol}_\mathrm{AdS_3} + z \ \rho \ \cos \chi \textrm{Vol}_{\Sigma_2} \wedge {\sigma_{1} \wedge \sigma_{2}}
\nonumber\\[5pt]
& -\frac{18 \alpha'^2 }{\Delta} \rho^2 \sin  \chi \Big( z \textrm{Vol}_{\Sigma_2}  + \frac{2}{3} {\sigma_{1} \wedge \sigma_{2}} \Big) \wedge \Big( 2  \cos \chi \big(\sin \chi d\rho+\rho \cos \chi d \chi \big)
\\
&  +3   \sin  \chi \big(\rho \sin \chi d \chi -\cos \chi d \rho \big) \Big)\wedge \big(d \xi +  \tilde{\sigma}_{3} \big) \Bigg]. 
\nonumber
\nonumber
\end{align}

The SUSY preserved by this background is discussed in 
appendix \ref{natdappendixsusy2}. We have checked that the equations of motion are solved by this background.
\subsection{The non-Abelian T-dual of the Donos-Gauntlett solution}
\label{natddualDG}
In this section we will briefly present the result of applying NATD to the Donos-Gauntlett solution \cite{Donos:2014eua} that we described in section \ref{natdDG}.

Like above, we will perform the NATD choosing a gauge such that the new coordinates are $(x_1,x_2,x_3)$.
We will quote the result in spherical coordinates (the expressions in cylindrical and cartesian coordinates are written in appendix C of \cite{Bea:2015fja}).
The expressions below are naturally more involved than those in section \ref{natdtwistedduals}.
{This is due to the fact that now there is a non-trivial NS 2-form that enters in the procedure (explicitly, in the string sigma model) 
	which makes things more complicated.}
As above, we start with some definitions,
\begin{equation}
\mathcal{B}_{\pm}=\rho \cos \chi\pm\frac{L^2 \lambda}{6 \alpha'} \alpha \ , \qquad \mathcal{B}=\mathcal{B}_{+} \ ,  \qquad
\Delta =L^4 e^{4 U+2 V}+54 \alpha'^2 e^{2 U} \rho^2 \sin^2 \chi +36 \alpha'^2 \mathcal{B}^2 e^{2V} \ .
\label{s1}
\end{equation}
The NSNS sector is given by,
%
\bea
e^{-2 \widehat{\Phi}} & =&\frac{L^2}{324 \ \alpha'^3} \Delta,   \label{s5} 
\nonumber\\
\frac{d\hat{s}^{2}}{L^2} & =&e^{2 A} \left(- dy_0^2+d y_1^2\right)+e^{2 B} \left( d \alpha^2+ d \beta^2\right)+ dr^2+\frac{e^{2 U}}{6} \left(\sigma_{1}^2+\sigma_{2}^2\right)+
\nonumber\\
&&  +\frac{\alpha'^2}{\Delta} \Bigg[ 6 e^{2 U+2 V} \Bigg( \sin^2 \chi \Big( d \rho^2 + \rho^2 \big(d \xi + \sigma_{3} \big)^2 \Big)+ \rho \sin (2 \chi) d\rho d\chi+\rho^2 \cos^2 \chi d \chi^2 \Bigg) 
\nonumber\\
&&  + 9 e^{4 U} \big( \cos \chi d \rho - \rho \sin \chi d \chi \big)^2 + \frac{324 \alpha'^2}{L^4} \Big( \big(\mathcal{B} \cos \chi+ \rho \sin^2 \chi \big) d\rho+ \rho \big(\rho \cos \chi - \mathcal{B} \big)\sin \chi d\chi \Big)^2 \Bigg],
\eea
\bea
\widehat{B}_2 & =&L^2\frac{\lambda}{6} \ \alpha \ {\sigma_{1}\wedge \sigma_{2}}+\frac{\alpha'^3}{\Delta} \Bigg[ 36 \ \mathcal{B} \  e^{2V} \Bigg(\sigma_{3}\wedge \Big( \big(\mathcal{B} \cos \chi+ \rho \sin^2 \chi \big) d\rho+ \rho \big(\rho \cos \chi - \mathcal{B} \big) \sin \chi d\chi \Big) 
\\
&&\hspace{-6mm}  +\rho \sin \chi d\xi \wedge \big(\sin \chi d\rho+\rho \cos \chi d \chi \big) \Bigg)+ e^{2U} \Big( e^{2V+2U} \frac{L^4}{\alpha'^2} \sigma_{3} - 54 \rho^2 \sin^2 \chi d \xi  \Big)\wedge \big( \cos \chi d \rho - \rho \sin \chi d \chi \big) \Bigg].
\nonumber
\eea
The RR sector reads,
\bea
\label{s11}
 \widehat{F}_0&=&0, \qquad\\
\widehat{F}_2&=&\frac{L^2}{6\alpha'^{\frac{3}{2}}} \Big( \lambda \alpha' d\mathcal{B}_{-}\wedge d\beta + \frac{2}{9} L^2 {\sigma_{1}\wedge \sigma_{2}} \Big), 
\eea

\bea
 \widehat{F}_4&=&\frac{L^4 \lambda}{36 \sqrt{\alpha'}} \Bigg[ e^{V}\Big( \frac{2 L^2}{\alpha'}d\alpha -3 e^{-2B-2V} \lambda \big( \cos \chi d \rho - \rho \sin \chi d \chi \big) \Big) \wedge \textrm{Vol}_\mathrm{AdS_3} \\
&&  \quad \; -\frac{6 \alpha'}{L^2} \Big( \rho \sin \chi \big( \sin \chi d\rho + \rho \cos \chi d \chi \big) \mathcal{B} d \mathcal{B}\Big) \wedge d \beta \wedge {\sigma_{1} \wedge \sigma_{2}}\nn\\
& &\quad \; +\frac{36 \alpha'^2 }{\Delta} \rho \sin  \chi \Big( \frac{3 \alpha'}{L^2} d \beta \wedge d\mathcal{B}_{-}  -\frac{2}{3 \lambda} {\sigma_{1} \wedge \sigma_{2}} \Big) \wedge \Big( 2 e^{2V}\mathcal{B}  \big(\sin \chi d\rho+\rho \cos \chi d \chi \big)+ 
\nn\\
&& \quad \; +3 e^{2U}  \rho \sin  \chi \big(\rho \sin \chi d \chi -\cos \chi d \rho \big) \Big)\wedge \big(d \xi +  \sigma_{3} \big) \Bigg] \ .\nn
\eea

Here again, aspects of the SUSY preserved by this background are
relegated to appendix \ref{natdappendixsusy2}. We have checked that the equations of motion are solved by this background.

\section{T-dualizing back from type IIA to IIB }
\label{natdt}
In this section, we will construct {\it new} type IIB 
Supergravity backgrounds with an AdS$_3$ factor at the IR. The idea is to obtain
these new solutions by performing an (Abelian) T-duality
on the configurations  described 
by eqs. \eqref{NN11}-\eqref{NN12} which, in turn,
were obtained by performing NATD on the backgrounds 
of eq. \eqref{NN01}.
The full chain of dualities is NATD -T. 
The new solutions present an AdS$_3$ fixed point and
all RR and NS fields are switched on. 

It should be interesting to study if the AdS$_3$ fixed point of this geometry falls  within known classifications \cite{Gauntlett:2006ux}. If not,
to use them as inspiring ansatz to extend these taxonomic efforts.

In order to perform the T-duality, we will choose a Killing vector that has no fixed points, in such a way that the dual geometry has no singularities. An adapted system of coordinates for that Killing vector is obtained through the change of variables,
\begin{equation}
\alpha = \textrm{arccosh} \left( \cosh a \cosh b \right) \ , \qquad \beta = \arctan \left(  \frac{\sinh b}{\tanh a}  \right) \ ,
\label{NN2TTwewewe1}
\end{equation}
obtaining \footnote{
	In fact using the coordinate transformation eq. \eqref{NN2TTwewewe1} we obtain $A_1 = \sinh a \ db + \textrm{total derivative}. $
},
\begin{equation}
A_1= \sinh a \ db \ ,  \qquad  \textrm{Vol}_{\Sigma_2}= \cosh a\  da \wedge db \ , \qquad  ds^2_{\Sigma_2}=da^2 + \cosh^2 a \ db^2 \ .
\label{NN0000021}
\end{equation}

The Killing vector that we choose is the one given by translations along the $b$ direction. Its modulus is proportional to the quantity,
\begin{equation}
\Delta_T = 54 \alpha'^2 z^2 \rho^2 \sin^2 \chi  \sinh^2 a + e^{2B} \cosh^2 a \  \Delta \ ,
\label{moduluskillingvector} 
\end{equation}
where $\Delta$ is defined in eq.(\ref{NN10}). Since $\Delta_T$  is never vanishing the isometry has no fixed points. 

To describe these new configurations, we define,
\begin{equation}
A_3=3 z \sinh a \ \big( \cos \chi d\rho - 
\rho \sin \chi d \chi \big) - db \ .
\label{NN1TT}
\end{equation}
We will write with a tilde the \bg produced in this section; its NSNS sector is: 

\bea
\label{NN2TT}
e^{-2 \widetilde{\Phi}} & =&\frac{L^4}{324 \alpha'^4} \Delta_T \ , \\
\frac{d\tilde{s}^{2}}{L^2} & =& e^{2 A} \left(- dy_0^2+d y_1^2\right)+e^{2 B} da^2+ dr^2+\frac{1}{6} \left(\sigma_{1}^2+\sigma_{2}^2\right)
\nonumber\\
&& +\frac{\alpha'^2}{\Delta_T} \Bigg[  \frac{\Delta}{L^4} db^2 + 6 e^{2B} \cosh^2 a  \Big(  \rho^2 \sin^2 \chi      \big(  \sigma_{3}+ d\xi \big)^2 + \big(d \rho \  \sin \chi +\rho \ \cos \chi \  d \chi \big)^2 \Big)
\nonumber\\
&&+ 9 \big( z^2 \sinh^2 a+e^{2B} \cosh^2 a  \big) \Big( \big(d\rho \ \cos \chi -\rho \  \sin \chi \ d\chi \big)^2 + \frac{36 \alpha'^2}{L^4} \rho^2 d\rho^2 \Big) 
\nonumber\\
&&   - 6 z \sinh a \ d b  \Bigg(  \Big(\frac{36 \alpha'^2}{L^4} \rho^2+1\Big)  \cos \chi \  d\rho - \rho \ \sin \chi \ d \chi \Bigg) \Bigg] \ ,
\\[5pt]
\widetilde{B}_2 & =&\frac{\alpha'^3}{\Delta_T}     \Bigg[  e^{2B} \cosh^2 a   \Bigg( 36 \rho \cos \chi \Big[ \rho \sigma_{3}\wedge d\rho + \rho  \sin \chi  d\xi \wedge \Big(\sin \chi   d\rho +\rho  \cos \chi   d\chi \Big)\Big]
\nonumber\\[5pt]
&&\hspace{-6mm} + \Big( \frac{L^4}{\alpha'^2} \sigma_{3} - 54 \rho ^2 d\xi  \sin ^2 \chi  \Big)\wedge \Big(d \rho  \cos \chi - \rho  \sin \chi d \chi  \Big) \Bigg)  
- 18 \ z \ \rho^2 \ \sinh a \sin ^2 \chi \Big(      d \xi  \wedge A_3 + d b \wedge \sigma_{3}   \Big) \Bigg] \ ,
\nonumber
\eea
%
and the RR sector reads:
%
\bea
\label{NN3TT}
&\widetilde{F}_1& =\frac{z L^4}{18 \alpha'^2} \cosh a \ d a \ ,\\
\widetilde{F}_3 &=& \frac{L^4}{54\alpha'} \big(  2 A_3 + 3 z \rho \cos \chi \cosh a \ d a  \big)    \wedge         \sigma_{1}\wedge \sigma_{2},\\
&&\qquad +\frac{ z L^4 \alpha'  \cosh a }{\Delta_T} \rho^2 \sin \chi \big(  \sigma_{3} +  d \xi \big) \wedge \Big[    e^{2B} \cosh^2 a  \Big( \big ( 2 + \sin^2 \chi \big) d\chi - \sin \chi \cos \chi d \rho \Big) 
\nonumber\\[5pt]
&& \qquad - z \sinh a \sin \chi A_3 \Big] \wedge d a \ ,
\nonumber
\eea
\bea
 \widetilde{F}_5 &=& \frac{L^4}{6} e^{-2B}  \Big( 24 e^{4B} \rho \cosh a  da \wedge d \rho + z \big( d\rho  \cos \chi -\rho d\chi \sin \chi \big) \wedge d b \Big) \wedge \textrm{Vol}_\mathrm{AdS_3} 
\\
& & + \frac{L^4 \alpha'^2 \cosh a}{18 \Delta_T} \Bigg[   e^{2B} \cosh a \rho  \sin \chi  d \xi \wedge  \Bigg(  \big( d\rho  \sin \chi +\rho d \chi  \cos \chi \big)  \wedge \Big[ 24 \rho  \cos \chi A_3
\nonumber\\[5pt]
&&  - z \Big( \frac{L^4}{\alpha'^2} + 54 \rho ^2 \sin ^2 \chi  \Big) \cosh a da \Big]
+ 18 \ \rho \  \sin  \chi  \big( 3 z \rho \cos \chi \cosh a d a -2 d\beta \big) \wedge \big(d \rho  \cos \chi -\rho  d \chi  \sin \chi \big) \Bigg)
\nonumber\\[5pt]
& & - 18 z^2 \  \rho ^3 \sinh a \ \sin ^2 \chi \  d \xi \wedge \Big( 3 z \sinh a \big(\sin^2 \chi d \rho + \rho \sin \chi \cos \chi d \chi \big) + \cos \chi A_3 \Big)  \wedge d a  \Bigg] \wedge  \sigma_{1}\wedge \sigma_{2} \ .    
\nonumber
\eea
%
We have checked that the equations of motion are solved by this background, either assuming the BPS equations (\ref{BPSH2}) or the non-SUSY solution (\ref{O15}). We have also checked that the Kosmann derivative vanishes without the need to impose further projections. These facts point to the conclusion that this new and smooth solution is also SUSY preserving.

We will now present new backgrounds of eleven-dimensional supergravity.

\section{ Lift to M-theory } \label{natdlift}

Here, we lift the solutions of sections \ref{natdtwistedduals} and \ref{natdDG} to eleven dimensions using the relations \eqref{natdintrosugraiiametriclift}, \eqref{natdintrosugraiiaf4lift} of subsection \ref{nadtintrosugraiia}. This way, they become new and smooth backgrounds of M-theory that describe 
the strong dynamics of a SUSY 2d CFT.
It is well known that given a massless solution of the 
type IIA supergravity can be uplifted to eleven dimensions solution, with the following relations:
\begin{equation}
ds^2_{11} = e^{-\frac{2}{3} \widehat{\Phi}} \ ds^2_{IIA} \ + \ e^{\frac{4}{3} \widehat{\Phi}} \ \big(  dx_{11} + \widehat{C}_1  \big)^2 \ ,
\end{equation}
where $\widehat{\Phi}$ is the dilaton of the 10-dimensional 
solution of the Type IIA SUGRA and $\widehat{C}_1$ is 
the 1-form potential that corresponds to the 
RR 2-form of the Type IIA background. Also, by
$x_{11}$ we denote the $11^{\textrm{th}}$ coordinate which corresponds 
to a $\mathrm{U(1)}$ isometry as neither the metric tensor or flux  
explicitly depend on it.

The 11-dimensional geometry is supported by a 3-form potential $C^M_3$ 
which gives rise to a 4-form $F^M_4 = dC^M_3$. 
This 3-form potential can be written in terms 
of the 10-dimensional forms and the differential 
of the $11^{\textrm{th}}$ coordinate as,
\begin{equation}
C^M_3 = \widehat{C}_3 + \widehat{B}_2 \wedge dx_{11} \ .
\end{equation}
The 3-form $\widehat{C}_3$ corresponds to the closed part of 
the 10-dimensional RR form 
$\widehat{F}_4 = d\widehat{C}_3 -\widehat{H}_3 \wedge \widehat{C}_1$.
Here, $\widehat{B}_2$ is the NS 2-form of the 10-dimensional 
type-IIA theory and $\widehat{H}_3 = d\widehat{B}_2$.
Hence we see that in order to describe the 11-dimensional 
solution we need the following ingredients,
\begin{equation}
ds^2_{IIA} \ , \quad \widehat{\Phi} \ , \quad \widehat{B}_2 \ , \quad \widehat{C}_1 \ , \quad\widehat{C}_3 \,\,\,\, \textrm{or} \,\,\,\, \widehat{F}_4 \ .
\end{equation}
Let us now present these quantities for the cases of interest.
\subsection{Uplift of the NATD of the twisted solutions}
\label{uplifteds2h2natd}
As we mentioned above, in order to specify 
the M-theory background we need to read the field 
content of the 10-dimensional solution. For the case at hand we 
wrote the metric $ds^2_{IIA}$, 
the dilaton $\widehat{\Phi}$
and the NS 2-form $\widehat{B}_2$ in eq. \eqref{NN11}. 
Moreover, from the expression of the RR 2-form 
in eq. \eqref{NN12} we can immediately extract the 
1-form potential $\widehat{C}_1$, 
\begin{equation}
\widehat{C}_1 = \frac{L^4}{54 \ a'^{\frac{3}{2}}} \ 
\big(  3 \ z \ A_1 - 2 \ \s_3  \big) \ .
\end{equation}
The 3-form potential--$\widehat{C}_3$--can be obtained 
from the RR 4-form of eq. \eqref{NN12}. After some algebra one finds,
\begin{equation}
\begin{aligned}
\widehat{C}_3 & = \frac{L^4 \ \ e^{-2B}}{6 \ a'^{\frac{1}{2}}} \ z \ \r \ \cos\chi \ \textrm{Vol}_\mathrm{AdS_3} + \frac{L^4 \ z \ \big(L^4 + 36 a'^2 \r^2\big) \ \cos\chi}{6 \ \a'^{\frac{1}{2}} \ \Delta} A_1 \wedge \sigma_{3} \wedge d\r
\\[5pt]
& + \frac{L^4 \ \a'^{\frac{3}{2}} \ \r^2 \sin\chi}{\Delta} \Big( z \ A_1 - \frac{2}{3} \sigma_{3}  \Big) \wedge d\xi \wedge \Big( \r \ \big( 2 + \sin^2\chi  \big) \ d\chi -\sin\chi \ \cos\chi \ d\r  \Big)
\\[5pt]
& + \frac{L^4 \ z \ \r}{18 \ \a'^{\frac{1}{2}} \ \Delta} \Big( \Delta \ \cos\chi \ \big(  \sigma_{3} \wedge \textrm{Vol}_{\Sigma_2} - 2 A_1 \wedge d \sigma_{3} \big) - 3 L^4 \sin\chi \ A_1 \wedge \sigma_{3} \wedge d\chi   \Big) \ .
\end{aligned}
\end{equation}
We close this section by observing that, if the coordinate $\rho$
takes values in a finite interval, then the radius of the 
M-theory circle, $e^{\frac{4}{3} \widehat{\Phi}}$
never blows up, because the function $\Delta$ that appears in the expression 
of the dilaton is positive definite. We have checked that the equations of motion of the 11-dimensional Supergravity are solved by this background.
\subsection{Uplift of the NATD of the Donos-Gauntlett solution}
\label{upliftnatdDG}
Here, the NSNS fields of the 10-dimensional theory have 
been written in detail in eq. \eqref{s5}. 
In order to complete the description of the M-theory background 
we need also to consider the potentials 
$\widehat{C}_1$ and $\widehat{C}_3$ that are encoded in the RR fields of 
eq. \eqref{s11}. Hence from the RR 2-form potential 
we can easily read $\widehat{C}_1$,
\begin{equation}
\widehat{C}_1 = \frac{L^2 \l \, \mathcal{B}_{-}}{6 \a'^{\frac{1}{2}}} \, d\beta - \frac{L^4}{27 \a'^{\frac{3}{2}}} \, \s_3 \ ,
\end{equation}
where the function $\mathcal{B}_{-}$ has been defined in 
eq. \eqref{s1}. Also, from the RR 4-form we can obtain 
the potential $\widehat{C}_3$ which in this case is,
\begin{equation}
\begin{aligned}
\widehat{C}_3 & = \frac{e^{-2B-V} L^4 \l}{36 \a'^{\frac{3}{2}}} \Big( 2 e^{2B + 2V} L^2 \alpha - 3 \l \a' \r \cos\chi \Big) \textrm{Vol}_\mathrm{AdS_3} 
\\[5pt]
& - \frac{\a'^{\frac{1}{2}} L^2 \r \sin\chi}{18 \Delta} \Big( 18 \a' \l \mathcal{B}_{-} \big( \s_3 + d\xi \big) \wedge d\beta + 4 L^2 \s_{3} \wedge d\xi \Big) \wedge \Sigma_1
\\[5pt]
& + \frac{L^2 \l \a'^{\frac{1}{2}}}{12 }  \Big(\mathcal{B}_{-}^2 + \mathcal{B}^2 + \r^2 \sin^2\chi \Big) d\beta \wedge d\s_{3} \ .
\end{aligned}
\end{equation}
Here for brevity we have defined the 1-form $\Sigma_1$ in the following way:
\begin{equation}
\Sigma_1 = 6 \a' e^{2V} \mathcal{B} \ \big(  \sin\chi d\r +\r \cos\chi d\chi  \big) - 9 \a' e^{2U} \r \sin\chi \ \big(  \cos\chi d\r -\r \sin\chi d\chi  \big)\ .
\end{equation}
Finally, we observe that the radius of the 
M-theory circle is finite, for reasons similar to those discussed in
the  previous example. We have checked that the equations of motion of the 11-dimensional Supergravity are solved by this background.

This completes our presentation of this set of new and exact solutions. A summary of all the solutions can be found in Table \ref{natdsolutionstable}. The expressions for these backgrounds
in cartesian and cylindrical coordinates are written in appendix C of \cite{Bea:2015fja} .

We will now move on to the second part of this chapter. We will study 
aspects of the field theories that our new and smooth backgrounds 
are defining.



\section{General comments on the Quantum Field Theory}\label{natdcomments}
Let us start our study of the correspondence between our new metrics 
with their respective field theory  dual. We will state some general points that these field theories will fulfill.

In the case of the backgrounds corresponding to the compactifications described in section \ref{natdtwistedsolutions}, our field theories are obtained by a twisted KK-compactification on a two dimensional manifold---that can be $\mathrm H_2, \  \mathrm{S^2}$ or $\mathrm{T^2}$. The original field theory is, as we mentioned, the Klebanov-Witten quiver, that controls the  high energy dynamics of our system. The bosonic part of the global symmetries for this  QFT in the UV are 
\begin{equation}
SO(1,3)\times \mathrm{SU(2)}\times \mathrm{SU(2)}\times \mathrm{U(1)}_R\times \mathrm{U(1)}_B ,
\end{equation}
where, as we know the $SO(1,3)$ is enhanced to $SO(2,4)$.
The theory contains two vector multiplets ${\cal W}^i=(\lambda^i, A_\mu^i)$, for $i=1,2$, together with four chiral multiplets 
${\cal A}_\alpha=(A_\alpha,\psi_{\alpha})$ for $\alpha=1,2$ and ${\cal B}_{\dot{\alpha}} =(B_{\dot{\alpha} } 
,\chi_{\dot{\alpha} }   )$ with $\dot{\alpha}=1,2$.

These fields transform as vector, spinors and scalars under $SO(1,3)$---that is $A_\alpha, B_{\dot{\alpha}}$ are singlets, the fermions transform in the $\bf(\frac{1}{2}, 0) \oplus (0,\frac{1}{2})$ and the 
vectors in the $\bf (\frac{1}{2},\frac{1}{2})$. The transformation  under the 'flavor' quantum numbers 
$\mathrm{SU(2)}\times \mathrm{SU(2)}\times \mathrm{U(1)}_R\times \mathrm{U(1)}_B $ is,
\begin{eqnarray}
& & A_\alpha=(2,1,\frac{1}{2}, 1),\;\;\;~~  B_{\dot{\alpha}}=(1,2,\frac{1}{2},-1),\nonumber\\
& & \psi_\alpha=(2,1,-\frac{1}{2},1),\;\;\; \chi_{\dot{\alpha}}=(1,2,-\frac{1}{2},-1),\\
& & \lambda^i=(1,1,1,0),\;\;\; ~~~~ A_\mu^i=(1,1,0,0).
\nonumber
\label{transflawsxx}
\end{eqnarray}

The backgrounds in section \ref{natdtwistedsolutions},
are describing the strong coupling regime of the field theory above, in the case in which we compactify the  D3-branes
on $\Sigma_2$ twisting the theory. This means, mixing the R-symmetry $\mathrm{U(1)}_R$ with the $SO(2)$ isometry of $\Sigma_2$.
This twisting is reflected in the metric fibration between 
the $\eta$-direction (the Reeb vector) and the $\Sigma_2$.
The fibration is implemented by a vector field $A_1$ 
in eq. \eqref{NN02}. The twisting mixes the R-symmetry of the QFT, 
represented by $A_1$ in the dual description,  
with (part of) the Lorentz group.
In purely field theoretical terms, we are modifying the covariant derivative of different fields that under the  combined action of 
the spin connection and the R-symmetry (on the curved part of the space) will read $D_\mu\sim \partial_\mu+\omega_\mu+ A_\mu$.

In performing this twisting, the fields decompose under 
$SO(1,3)\to SO(1,1)\times SO(2)$. 
The decomposition is straightforward for the bosonic fields. 
For the fermions, we have that $\bf(\frac{1}{2}, 0) $ 
decomposes as ${\bf (+,\pm)}$ and similarly 
for the $(0,\frac{1}{2})$ spinors. 

The twisting itself is the 'mixing' between the $\pm$ charges of the spinor
and its R-symmetry charge. There is an analog operation for the vector and 
scalar fields. Some fields are scalars under the diagonal group in
$\mathrm{U(1)}_R \times SO(2)_{\Sigma_2}$. Some are spinors and some are vectors.
Only the scalars under the diagonal group are massless. These determine the 
SUSY content of the QFT.
This particular example 
amounts to preserving two supercharges. 
There are two massless vector multiplets and two massless matter multiplets. The rest of the fields get a mass whose set by the inverse size of the compact manifold. In other words, the field theory at low energies is a two dimensional  CFT (as indicated by the AdS$_3$ factor), preserving $(0,2)$ SUSY and  obtained by a twisted compactification
of the Klebanov-Witten CFT. The QFT is deformed in the UV by a relevant operator of dimension two, as we can read from eq. \eqref{O20x}. 

An alternative way to think about this QFT is as the one describing
the low energy excitations of a stack of $N_c$ D3-branes wrapping  
a calibrated space $\Sigma_2$
inside a Calabi-Yau 4-fold.

The situation with the metrics in section \ref{natdDG} is 
more subtle. In that case there is also a flow from the Klebanov-Witten quiver to a two-dimensional CFT preserving (0,2) SUSY. The difference is that this second QFT is not apparently obtained via a twisting procedure. As emphasized by the authors of \cite{Donos:2014eua}, 
the partial breaking of SUSY is due (from a five-dimensional 
supergravity perspective) to 'axion' fields depending on the torus directions. These axion fields are proportional to a  
deformation parameter---that we called $\lambda$ in eq. \eqref{fluxes-bef}. 
The deformation in the UV QFT is driven 
by an operator of dimension four that was identified to be $Tr(W_1^2-W_2^2)$ and a  dimension six operator that acquires a VEV, as discussed in \cite{Donos:2014eua}. 

To understand the dual field theory to the  IIA backgrounds obtained after non-Abelian T-duality and presented in section \ref{natdtwistedduals} involves more intricacy. Indeed, it is at present unclear what is the analog field theoretical operation of non-Abelian T-duality. There are, nevertheless, important hints. 
Indeed, the foundational paper of Sfetsos and Thompson \cite{Sfetsos:2010uq}, that sparked the interest of the uses of non-Abelian T-duality in quantum field theory duals, showed that if one starts with a background of the form AdS$_5\times$S$^5/ \mathbb{Z}_2$, a particular solution of the Gaiotto-Maldacena system (after lifting to M-theory) is generated \cite{Gaiotto:2009gz}.
This is hardly surprising, as the backgrounds of eleven-dimensional supergravity with an AdS$_5$ factor and preserving $\mathcal{N}=2$ SUSY in four dimensions, have been classified. What is interesting is that the solution generated by Sfetsos and Thompson
appears as a zoom-in on the particular class of solutions in \cite{Maldacena:2000mw}. This was extended in 
\cite{Itsios:2013wd} that computed the action of non-Abelian T-duality on the end-point of  the flow from the $\mathcal{N}=2$ conformal quiver with adjoints to the Klebanov-Witten CFT. Again, not surprisingly, the  backgrounds obtained correspond to the $\mathcal{N}=1$ version of the Gaiotto $T_N$ theories---
these were called {\it Sicilian} field theories by Benini, Tachikawa and Wecht
in  \cite{Benini:2009mz}, see Fig. \ref{RoadMap}.
It is noticeable, that while the Sicilian theories can be 
obtained by a twisted compactification of M5-branes on $\mathrm H_2, \mathrm S^2,  \mathrm{T^2}$, 
the case obtained using non-Abelian T-duality corresponds only 
to M5-branes compactified on $\mathrm S^2$ and preserving minimal SUSY in four dimensions. What we propose in this chapter is that the twisted compactification on $ \Sigma_2$ of a Sicilian gauge theory
can be studied by using the backgrounds we discussed in section \ref{natdtwistedduals} and their M-theory counterparts.
We will elaborate more about the 2-d CFTs and 
their flows in the coming sections.

In the following, we will calculate different observables of  these QFT's by using the backgrounds as a 'definition' of the 2d SCFT. The backgrounds are smooth and thus the observables have trustable results. 
Hence, we are defining a QFT by its observables, 
calculated in a consistent way by the dual solutions. 
The hope is that these calculations together with 
other efforts can help map the space of 
these new families of CFTs. To the study of these observables we turn now.

\section{Quantized charges}\label{natdpagecharges}
In this section, we will study the quantized charges on the string side.
This analysis appears in the field theory part of the chapter because these charges will, as in the
canonical case of AdS$_5\times$S$^5$, translate into the ranks of the gauge theory local symmetry groups.

The NATD produced local solutions to the 10-dim SUGRA equations of motion. Nevertheless, it is still not known how to obtain the global properties of these new geometries. 
Some quantities associated to a particular solution, like the Page charges below, are only well-defined when the global properties of the background are known. Since we have only a local description of our solution, we will propose very reasonable global results for the Page charges, mostly based on physical intuition.

Let us start by analyzing a quantity that is proposed to be periodic in the string theory.
We follow the ideas introduced in \cite{Lozano:2014ata} and further elaborated in \cite{Macpherson:2014eza}. 
To begin with, we focus on the NATD version of the twisted solutions; described in section \ref{natdtwistedduals}. Let us define the quantity,
\begin{equation}
b_0= \frac{1}{4 \pi^2 \alpha'} \int_{\Pi_2} B_2  ~~ \in  \ [0,1] \ ,
\label{page9b}
\end{equation}
where the cycle $\Pi_2$ is defined as,
\begin{equation}
\Pi_2=\mathrm{S^2}   ~~~  \Big\{  \chi , \xi  , \alpha=0, \rho=\textrm{const} , d\beta=-\frac{1}{3z} d\xi    \Big\} \ .
\label{page9ceeett}
\end{equation}
As the topology of the NATD theory is not known, we propose that this cycle is present in the geometry. This cycle will have a globally defined volume form, which in a local description can be written as  $\textrm{Vol}_{\Pi_2}=\sin \chi \ d \xi \wedge d\chi$. 
We then find,
\begin{equation}
b_0=\frac{1}{4 \pi^2 \alpha'} \int_{\Pi_2} \alpha' \rho \sin \chi d\xi \wedge d \chi= \frac{\rho}{\pi}  ~~ \in  \ [0,1].
\label{page9deee}
\end{equation}
Again, we emphasize that this is a {\it proposal} made in \cite{Lozano:2014ata} and used in \cite{Macpherson:2014eza}. Moving further than $\pi$ along the variable $\rho$ can be 'compensated' by performing a large gauge transformation on the $B_2$-field,
\begin{equation}
B_2 \rightarrow B_2'=B_2 - \alpha '  n \pi \sin \chi d \xi \wedge d\chi.
\label{page10scsc}
\end{equation}
We will make extensive use of this below.

Let us now focus on the conserved magnetic charges defined for our backgrounds. We will start the analysis for the case of the solutions in section \ref{natdtwistedsolutions}.

\subsection{Page charges for the twisted solutions}
We will perform this study for the solutions before and 
after the NATD, and we will obtain how the Page charges transform under the NATD process. Page charges  (in contrast to Maxwell charges) have  proven fundamental in understanding aspects of the dynamics of field theories---see for example \cite{Benini:2007gx}.

In particular, for D3-branes we have, \eqref{natdintrosugraiiapagecharges}:
\begin{equation}
N_{D3} \big{|}_{\Pi_5}=\frac{1}{2\kappa_{10}^2 T_{D3}} \int_{\Pi_5} \left( F_5 - B_2 \wedge F_3 + \frac{1}{2}B_2 \wedge B_2 \wedge F_1 \right).\;\;
\nonumber
\label{page201tgrtg}
\end{equation}

The topology of the internal space is 
$\Sigma_2 \times \mathrm{S^2} \times \mathrm{S^3}$. We consider the following cycles:
\begin{equation}
\begin{aligned}
\Pi_5^{(1)}&=&\mathrm{S^2} \times \mathrm{S^3}  ~\Big\{\theta_1,\phi_1,\theta_2,\phi_2, \psi \Big\},~~~~  \Pi_5^{(2)}&=\Sigma_2 \times \mathrm{S_1^3}                ~  \Big\{\alpha,\beta, \theta_1,\phi_1, \psi \Big\}, \\
\Pi_5^{(3)}&=& \Sigma_2 \times \mathrm{S_2^3}   ~ \Big\{ \alpha,\beta ,\theta_2,\phi_2, \psi \Big\}\ .  ~~~~~~~~~
\label{page6tt}
\end{aligned}
\end{equation}
The background fields $B_2$, $F_1$ and $F_3$ are vanishing, and only $F_5$ contributes. 
The specific components of $F_5$ in eq. \eqref{NN01} that have non vanishing pullback on these cycles are,
\begin{equation}
F_5= \frac{L^4}{9} \ \textrm{Vol}_1 \wedge \textrm{Vol}_2 \wedge \eta + \frac{L^4}{6} z \textrm{Vol}_{\Sigma_2} \wedge \left( \textrm{Vol}_1 + \textrm{Vol}_2 \right) \wedge \eta + ...
\label{page6a}
\end{equation}
We explicitly see that it is a globally defined form, as all the involved quantities ($\eta$, $\textrm{Vol}_{\Sigma_2}$, $\textrm{Vol}_1$, $\textrm{Vol}_2$) are well defined globally.
The associated Page charges of D3-branes for the background  around 
eq. \eqref{NN01} are,
\begin{equation}
N_{D3} \big{|}_{\Pi_5^{(1)}}=\frac{4 L^4}{27 \alpha'^2 \pi} \ ,~~~~~~~~~~~~~~ \widehat{N}_{D3} \big{|}_{\Pi_5^{(2)}}=\widetilde{N}_{D3} \big{|}_{\Pi_5^{(3)}}=\frac{L^4}{\alpha'^2} \frac{z \  vol \left( \Sigma_2 \right)}{18 \pi^2} \ ,
\label{page9dvdfv}
\end{equation}
where $vol\left( \Sigma_2 \right)$ is the total volume of the two-manifold $\Sigma_2$.\footnote{Notice that we use Vol for volume elements (differential forms) and $vol$ for the actual volumes of the manifolds (real numbers).}
As usual, the first relation quantizes the size of the space,
\begin{equation}
L^4=\frac{27 \pi}{4} \alpha'^2 N_{D3}.
\label{page10xdf}
\end{equation}
We have then defined three D3-charges. The one associated with $N_{D3}$ is the usual one appearing
also in the AdS$_5\times$S$^5$ case. The other two can be thought as charges  'induced'  by the wrapping of the D3-branes on the Riemann surface. The reader may wonder  whether these charges are present in the backgrounds found after NATD. We turn to this now.

We use now the Page charges expressions for IIA SUGRA (\ref{natdintrosugraiiapagecharges}. 
We label the radius of the space of the geometry in eq. \eqref{NN11} by $\widehat{L}$, to distinguish it from $L$, the quantized radius before the NATD. In order to properly define the cycles to be considered, we should know the topology of this NATD solution. However, we have obtained only a local expression for this solution, and we do not know the global properties. As we explained above, we will  present a proposal to define the Page charges that would explain the transmutation of branes through the NATD. We propose the relevant cycles to be, \footnote{Intuitively, we can think that the branes transform under NATD as 3 consecutive T-dualities. For example, in the first 5-cycle of eq. (\ref{page6tt}), the NATD is performed along 3 of the coordinates, ($\theta_2, \phi_2, \psi$), in such a way that we end up with a 2-cycle, the first cycle in eq.(\ref{page7r}), associated with D6-branes. In the second 5-cycle of eq.(\ref{page6tt}) the NATD only affects the $\psi$-direction, so it disappears, and two more directions are added in order to complete the 3 T-dualities, ending up with a 6-cycle in eq.(\ref{page7r}), associated with D2-branes.}
\begin{equation}
\Pi_2^{(1)}  \ \  \Big\{\theta_1,\phi_1 \Big\} ,  ~~~~ 
\Pi_6^{(2)}= \Big\{\alpha,\beta,\theta_1, \phi_1, \rho, \xi,\chi=\frac{\pi}{2} \Big\} ,  ~~~~  \Pi_2^{(3)} \  \  \Big\{\alpha,\beta \Big\} .
\label{page7r}
\end{equation}
The associated charges are,
\begin{equation}
\label{page8zr}
\begin{aligned}
& N_{D6} \big{|}_{\Pi_2^{(1)}}=\frac{2 \widehat{L}^4}{27 \alpha'^2 } \ ,    \qquad \quad  \widetilde{N}_{D6}\big{|}_{\Pi_2^{(3)}}=\frac{\widehat{L}^4}{\alpha'^2} \frac{z \ vol \left( \Sigma_2 \right)}{36 \pi} \ ,
\\[10pt]
& \widehat{N}_{D2} \big{|}_{\Pi_6^{(2)}}=\frac{\widehat{L}^4}{\alpha'^2} \frac{z \  vol \left( \Sigma_2 \right)  vol \left(\rho, \xi \right)}{ 144 \pi^4}=\frac{n^2}{4}\frac{\widehat{L}^4}{\alpha'^2} \frac{z \  vol \left( \Sigma_2 \right)}{36 \pi}.
\end{aligned}
\end{equation}
%
%
In the last expression, we performed the integral over the $\rho$-coordinate in the interval $[0, n\pi]$.
These three charges are in correspondence with the ones before the NATD in eq. \eqref{page9dvdfv}.
Indeed, we can compute the quotients,
\begin{align}
& & \frac{\widehat{N}_{D3} }{N_{D3}}=\frac{\widetilde{N}_{D3}}{N_{D3}}=\frac{3}{8\pi}z \  vol \left( \Sigma_2 \right), \qquad \frac{4 \widehat{N}_{D2} }{n^2 N_{D6}}=\frac{\widetilde{N}_{D6}}{N_{D6} }=\frac{3}{8\pi}z \  vol \left( \Sigma_2 \right).
\label{manaxx}
\end{align}
These quotients indicate a nice correspondence between charges before
and after the duality.

Using the first relation in eq. \eqref{page8zr}  we quantize the size $\widehat{L}$ of the space after NATD to be
$\widehat{L}^4=\frac{27}{2} \alpha'^2 N_{D6}$.

A small puzzle is presented by the possible existence of charge for D4-branes, as there would be no quantized number before the NATD to make them correspond to. To solve this puzzle, we propose that there should be a globally defined closed non-exact form that allows us to perform a large gauge transformation for the $\widehat{B}_2$, in such a way that all the D4 Page charges vanish. In local coordinates, we have a gauge transformation,
\begin{equation}
\widehat{B}_2 \rightarrow \widehat{B}'_2 = \widehat{B}_2 + \alpha' d\left[ \rho \cos \chi \ \tilde{\sigma}_3 \right],
\label{page9ss9}
\end{equation}
written in such a way that the integrand has at least one leg along a non-compact
coordinate,
\begin{equation}
\widehat{F}_4-\widehat{B}'_2 \wedge \widehat{F}_2=\frac{z \widehat{L}^4}{6 \sqrt{\alpha'}} e^{-2B} \left(\cos \chi d\rho - \rho \sin \chi d \chi \right) \wedge \textrm{Vol}_\mathrm{AdS_3}.
\label{page9adtrtg}
\end{equation}
Hence, any Page charge for D4-branes is vanishing. To be precise, the  D2 charge $\widehat{N}_{D2}$ must be computed after choosing this gauge, as it depends on $\widehat{B}_2$, but it turns out to be the same as calculated in eq. \eqref{page8zr}.

The motion in the $\rho$-coordinate, as we discussed above, can be related to large gauge transformations of the $\widehat{B}_2$-field.
The  large transformation that 'compensates' for motions in $\rho$, namely
$
\widehat{B}_2 \rightarrow \widehat{B}'_2=\widehat{B}_2 - \alpha '  n \pi \sin \chi d \xi \wedge d\chi,
$
has the effect of changing the Page charges associated with D4-branes, that were initially vanishing. Indeed, if we calculate  for the following four cycles,
\begin{equation}
\Pi_4^{(1)}  ~~  \Big\{\theta_1,\phi_1, \chi, \xi \Big\} \ , ~~~~~~ ~~
\Pi_4^{(2)} ~~  \Big\{\alpha,\beta , \chi, \xi \Big\} \ , ~~~~
\label{page12y}
\end{equation}
the Page charge of D4-branes varies according to,
\begin{equation}
\begin{aligned}
\Delta N_{D4} \big{|}_{\Pi_4^{(1)}}&=\frac{1}{2\kappa_{10}^2 T_{D4}} \int_{\Pi_4^{(1)}} \left(- \Delta  \widehat{B}_2 \wedge \widehat{F}_2  \right) =- n \frac{L'^4}{\alpha'^2}\frac{2}{27} =-n N_{D6},   
\\[10pt]
\Delta N_{D4} \big{|}_{\Pi_4^{(2)}}&=\frac{1}{2\kappa_{10}^2 T_{D4}} \int_{\Pi_4^{(2)}} \left(- \Delta  \widehat{B}_2 \wedge \widehat{F}_2  \right) = -n \frac{L'^4}{\alpha'^2}\frac{z}{36 \pi}  vol \left( \Sigma_2 \right)=-n \widetilde{N}_{D6}  .
\label{page11n}
\end{aligned}
\end{equation}
We can interpret these findings in the following way. Our QFT  (after the NATD) can be thought as living on the world-volume of a superposition of D2- and D6-branes. Motions in the $\rho$-coordinate induce charge of D4-branes, which can be interpreted as new gauge groups appearing. This suggest that we are working with a linear quiver, with many gauge groups. Moving $n \pi$-units in $\rho$ generates or 'un-higgses'  new gauge groups of rank $ n N_{D6}$ and  $n \widetilde{N}_{D6}$. Computing volumes or other observables that involve integration on the $\rho$-coordinate amounts to working with a  QFT with different gauge group, depending on the range in $\rho$ we decide to integrate over. Notice that $\rho$ is not a holographic coordinate. Motions in $\rho$ are not changing the energy in the dual QFT. For the AdS-fixed points the theory is conformal and movements in $\rho$ do not change that.

In the paper \cite{Macpherson:2014eza}, the motion in the $\rho$-coordinate was argued to be related to a form of 'duality' (a Seiberg-type of duality was argued to take place, in analogy with the mechanism of the Klebanov-Strassler duality cascade, but in a CFT context). That can not be the whole story as other observables, like for example the number of degrees of freedom in the QFT, change according to the range of the $\rho$-integration. Hence the motion in $\rho$ can not be just a duality. 
We are proposing here that moving in the $\rho$-coordinate amounts to changing the quiver, adding gauge groups, represented by the increasing D4 charge.

We will now present the same analysis we have performed above, but for the case of the background in section \ref{natdDG}.

\subsection{Page charges for the Donos-Gauntlett solution}
Part of the analysis that follows was carefully done  in the original work of \cite{Donos:2014eua} and here we will extend the study for the solution after the NATD that we presented in section \ref{natdDG}. We will give an outline of the results as the general structure is similar to the one displayed by the twisted solutions of the previous section.

We focus on the original Donos-Gauntlett background first. We denote by $d_{\alpha}$, $d_{\beta}$ the two radii of the torus (the cycles of the torus are then of size $2\pi d_\alpha$ and $2\pi d_\beta$ respectively) and consider
five different cycles in the geometry,

\begin{equation}
\label{page20ab}
\begin{aligned}
& \Pi_5^{(1)}=\mathrm{S^2} \times \mathrm{S^3}  ~~  \Big\{\theta_1,\phi_1,\theta_2,\phi_2, \psi \Big\} ,  ~~
\Pi_5^{(2)}=\mathrm{T^2} \times \mathrm{S_1^3}                ~~   \Big\{\alpha,\beta, \theta_1,\phi_1, \psi \Big\},     ~~\\
&\Pi_5^{(3)}=\mathrm{T^2} \times \mathrm{S_2^3}   ~~ \Big\{ \alpha,\beta ,\theta_2,\phi_2, \psi \Big\} , \\
& \Pi_3^{(4)}=S_1^1 \times s(S)  ~~ \Big\{ \alpha, \theta_1= \theta_2, \phi_1= - \phi_2 , \psi=\textrm{const} \Big\},   ~~ \\
&\Pi_3^{(5)}=S_2^1 \times s(S)  ~~ \Big\{ \beta, \theta_1= \theta_2, \phi_1= - \phi_2 , \psi=\textrm{const} \Big\} .
\end{aligned}
\end{equation}

The  Page charges associated with D3-, D5- and NS5-branes are,
\begin{equation}
\label{page21a}
\begin{aligned}
& N_{D3} \big{|}_{\Pi_5^{(1)}}=\frac{4 L^4}{27 \alpha'^2 \pi} \ ,  \qquad \widehat{N}_{D3} \big{|}_{\Pi_5^{(2)}}= - \widetilde{N}_{D3} \big{|}_{\Pi_5^{(3)}}=  2 \ \frac{L^4}{\alpha'^2} \frac{\lambda^2 \ d_{\alpha} d_{\beta}}{9} \ ,
\\[10pt]
& N_{NS5} \big{|}_{\Pi_3^{(4)}}=- 2 \ \frac{L^2 \lambda \ d_{\alpha}}{3 \alpha'} \ , \qquad 
N'_{D5} \big{|}_{\Pi_3^{(5)}}=2 \ \frac{L^2 \lambda \ d_{\beta}}{3 \alpha'} \ .
\end{aligned}
\end{equation}
After the NATD, we focus on the background around eqs. (\ref{s5}) and (\ref{s11}). 
We consider the  following cycles in the geometry,
%
\begin{equation}
\label{page273}
\begin{aligned}
& \Pi_2^{(1)}  ~~   \Big\{ \theta_1,\phi_1 \Big\}, ~~~~
\Pi_6^{(2)}         ~~   \Big\{ \alpha, \beta, \theta_1,\phi_1, \rho, \xi , \chi=\frac{\pi}{2}  \Big\}, ~~~~
\Pi_2^{(3)}    ~~ \Big\{ \alpha,\beta \Big\}, ~~~~
\\[10pt]
& \Pi_6^{(4)} ~~ \Big\{ \beta, \theta_1, \phi_1, \rho, \chi, \xi \Big\}, ~~~~
\Pi_2^{(5)}  ~~ \Big\{ \beta, \chi, \rho=\rho_0 \Big \}.   ~~~~~~~~~~~~~~~~~~~~~~~~~~~~~~~
\end{aligned}
\end{equation}
The  correspondent Page charges defined on them (the $\rho$-coordinate is taken in the $[0, n\pi]$ interval),
%
%
\begin{equation}
\label{page8khg}
\begin{aligned}
& N_{D6} \big{|}_{\Pi_2^{(1)}}=\frac{2 \widehat{L}^4}{27 \alpha'^2 } \ ,    \qquad     \widetilde{N}_{D6}\big{|}_{\Pi_2^{(3)}}=  - \frac{\widehat{L}^4}{\alpha'^2} \frac{ \lambda^2  \pi}{18} d_{\alpha} d_{\beta} \ , \qquad  N'_{D6} \big{|}_{\Pi_2^{(5)}}=\frac{ L'^2 }{ \alpha' } \frac{\lambda}{3} \rho_0 \ d_{\beta} \ ,
\\[10pt]
& \widehat{N}_{D2} \big{|}_{\Pi_6^{(2)}}=-\frac{\widehat{L}^4}{\alpha'^2} \frac{  \lambda^2 \   vol \left( \mathrm{T^2} \right)  vol  \left(\rho , \xi \right)}{288 \pi^4}  =- \frac{\widehat{L}^4}{\alpha'^2} \lambda^2 \frac{n^2 \pi}{72} d_{\alpha} d_{\beta} ,
\\[10pt]
&  N_{D2} \big{|}_{\Pi_6^{(4)}}= \frac{\widehat{L}^2}{\alpha'} \frac{\lambda}{24 \pi^3} d_{\beta} \  vol \left( \rho, \chi, \xi \right)= \frac{\widehat{L}^2}{\alpha'} \lambda \frac{n^3 \pi}{18} d_{\beta}.  ~~~~~~~~~~~
\end{aligned}
\end{equation}
From the first relation we obtain,
$
\widehat{L}^4=\frac{27}{2} \alpha'^2 N_{D6}$.  Like in the case of the twisted solutions, we can choose a gauge for the $\widehat{B}_2$ field
%
%
\begin{equation}
\label{page9joijo}
\begin{aligned}
& \widehat{B}_2 \rightarrow \widehat{B}'_2=\widehat{B}_2 + \delta \widehat{B}_2 \ ,
\\[5pt]
& \delta \widehat{B}_2=\frac{9 \alpha'^2 \lambda}{2 L^2} \ d \beta \wedge \Big( \rho \sin \chi \ d \big( \rho \sin \chi \big)+ \mathcal{B} \ d \mathcal{B} \Big)+ L^2 \frac{ \lambda}{6} \big( d \alpha \wedge \sigma_{3} - \alpha \ \sigma_{1}\wedge \sigma_{2} \big), 
\end{aligned}
\end{equation}
such that the Page charge of D4-branes, when computed on  every possible compact  4-cycle, is vanishing. 
Indeed, after the gauge transformation, we have
\begin{equation}
\widehat{F}_4-\widehat{B}'_2 \wedge \widehat{F}_2  =\frac{L^4 \lambda}{36 \sqrt{\alpha'}} e^{V}\left( \frac{2 L^2}{\alpha'}d\alpha -3 e^{-2B-2V} \lambda \ d(\rho \cos \chi) \right) \wedge \textrm{Vol}_\mathrm{AdS_3}.
\label{page9hu}
\end{equation}
Any integral over compact manifolds is vanishing. Like in the case of the twisted solutions, $\widehat{N}_{D2}$ and $N_{D2}$ should be recalculated after choosing this gauge;  but their value turns out to be unchanged.

We can apply the same  string theory considerations on the quantity $b_0$ that now is defined as an integral over the cycle,
\begin{equation}
\Pi_2=\mathrm{S^2} \ ,  \qquad \{  \chi , \xi , \alpha = \textrm{const},\rho= \textrm{const}    \}.
\label{page9c}
\end{equation}
We then calculate,
\begin{equation}
b_0=\frac{1}{4 \pi^2 \alpha'} \int_{\Pi_2} \alpha' \rho \sin \chi d\xi \wedge d \chi= \frac{\rho}{\pi}  ~ \in  \ [0,1].
\label{page9d0}
\end{equation}
If we move further than $\pi$ along the variable $\rho$, we can compensate this by performing the large gauge transformation
$
\widehat{B}_2 \rightarrow \widehat{B}'_2=\widehat{B}_2 - \alpha '  n \pi \sin \chi d \xi \wedge d\chi .
$
We now consider the correspondent variation of Page charges for D4-branes, that can be calculated using the following cycles,
\begin{equation}
\Pi_4^{(1)} ~~ \{ \theta_1,\phi_1, \chi, \xi \} \ , ~~~~~~~~ 
\Pi_4^{(2)} ~~ \{\alpha,\beta , \chi, \xi \}, ~~~~
\label{page1200}
\end{equation}
to be,
\begin{equation}
\begin{aligned}
\Delta Q^P_{D4} \big{|}_{\Pi_4^{(1)}}&=\frac{1}{2\kappa_{10}^2 T_{D4}} \int_{\Pi_4^{(1)}} \left(- \Delta  \widehat{B}_2 \wedge \widehat{F}_2  \right) =- n \frac{L^4}{\alpha'^2}\frac{2}{27}   =- n N_{D6} \ ,
\\[5pt]
\Delta Q^P_{D4} \big{|}_{\Pi_4^{(2)}}&=\frac{1}{2\kappa_{10}^2 T_{D4}} \int_{\Pi_4^{(2)}} \left(- \Delta  \widehat{B}_2 \wedge \widehat{F}_2  \right) =-  n \frac{L^4}{\alpha'^2}\frac{\pi \lambda^2}{18 }  d_{\alpha} d_{\beta}=n \widetilde{N}_{D6} \ .
\label{page1100}
\end{aligned}
\end{equation}
The variation of the Page charges of D2-branes vanishes under this large gauge transformation. For the Donos-Gauntlett solution we observe a structure very similar to the one discussed for the twisted solutions. 
Again, here we would propose that the NATD background 'un-higgses'
gauge groups of rank $n N_{D6}$ as we move in units of $n\pi$ in the $\rho$-coordinate.

We move now to the study of another important observable of 
our dual 2-d and 4-d CFTs.

\section{Central charges and c-theorem} \label{natdcentralcharge}

In the following sections, 
we will quote the results for central charges 
according to eq. \eqref{natdintrocfunctiongeneraldef}
for the conformal field theories in two and four dimensions. Following that,
we will write the result that it yields for the flows between theories.

\subsection{Central charge at conformal points} \label{natdccconformal}

As anticipated, we will quote here the results for eq. \eqref{natdintrocfunctiongeneraldef} for the different AdS$_3$ and AdS$_5$ fixed points. We start with the twisted solutions of section \ref{natdtwistedsolutions}. We will use that the volume of the $T^{1,1}$ space is $ vol(T^{1,1})=16\pi^3/27$.

\noindent{\underline{Twisted geometries}}

For the IR AdS$_3$ fixed point, the volume of the internal compact manifold is $
vol (M_7)= \frac{1}{3}   L^7  vol (\Sigma_2) vol (T^{1,1})$, and the central charge is,
\begin{equation}
c= 9 \big{|} N_{D3} \widehat{N}_{D3} \big{|}  =  \frac{L^8}{\alpha'^4} \frac{ vol (\Sigma_2) vol (T^{1,1})}{24 \pi^6}  .
\label{Central42}
\end{equation}
At the UV AdS$_5$ fixed point, the volume of the internal compact manifold is $vol (M_5)= L^5  vol (T^{1,1})$, and the result for the central charge is,
\begin{equation}
c=\frac{27}{64} N_{D3}^2=\frac{L^8}{\alpha'^4} \frac{ vol (T^{1,1})}{64 \ \pi^5} \ .
\label{Central3aoee}
\end{equation}
After the NATD, we must  consider the new radius of the space $\widehat{L}$ and the volume of the new 5-dim compact space $4 \pi^2 vol (\rho, \chi, \xi)$. The computations turn out to be similar as the previous ones, and we obtain that the central charges before and after NATD, for both the two and four dimensional CFTs, are related by,
\begin{equation}
\frac{\hat{c}}{c}=\frac{\widehat{L}^8}{L^8} \frac{4 \pi^2  vol (\rho, \chi, \xi)}{ vol (T^{1,1})}.
\label{Central5aa}
\end{equation}
Let us comment on the quantity $vol (\rho, \chi, \xi)$, that  appears 
in the calculation of the Page charges in section \ref{natdpagecharges}---see for example, eq. \eqref{page8khg} and also in the computation of the 
entanglement entropy of section \ref{natdEEWilson}.
Indeed, if we calculate,  \footnote{We identify the integral with the volume of the manifold spanned by the new coordinates. This becomes more apparent if we use the expressions in the appendix, in different coordinate systems. }
\begin{equation}
\begin{aligned}
& vol (\rho, \chi, \xi)=\int_{0}^{n\pi} \rho^2 d\rho \int_{0}^{2\pi} d\xi \int_{0}^{\pi}\sin\chi d\chi= \frac{4\pi^4}{3}n^3 \ ,
\\[10pt]
& \frac{\hat{c}}{c}=\frac{\widehat{L}^8}{L^8} \frac{4 \pi^2  vol (\rho, \chi, \xi)}{ vol (T^{1,1})}= \frac{36 \pi N_{D6}^2}{N_{D3}^2} n^3 \ .
\end{aligned}
\label{xxyz}
\end{equation}
We have performed the $\rho$-integral in the interval $[0,n\pi]$. 
The logic behind this choice was spelt out 
in section \ref{natdpagecharges}, see below eq. \eqref{page11n}. 
The proposal is that moving in units of $\pi$ in the $\rho$-coordinate implies 'un-higgsing' a gauge group, hence we would have a linear quiver gauge theory. The central charge captures this un-higgsing procedure, increasing according to how many groups we 'create'. What is interesting is the $n^3$
behavior in eq. \eqref{xxyz}. Indeed, if $n$ were associated with the rank 
of a gauge group, this scaling would be precisely the one obtained 
in Gaiotto-like CFTs (also valid for the $\mathcal{N}=1$ 'Sicilian' theories 
of \cite{Benini:2009mz}). Indeed,   
the NATD procedure when applied to the AdS$_5\times$T$^{1,1}$ 
background creates  metric and fluxes  similar to those characterizing 
the Sicilian CFTs. The backgrounds in sections \ref{natdprevioussolutions} to \ref{natdlift} are dual to a compactification of the Klebanov Witten CFT  
and (using the NATD background) the Sicilian CFT 
on a two space $\Sigma_2$. The two-dimensional 
IR fixed point of these flows is described by our 'twisted AdS$_3$' and its NATD. The central charge  of the Sicilian CFT and its compactified version is presenting a behavior that goes like
a certain rank to a third power $c\sim n^3$. 
This suggest that crossing $\rho=\pi$ amounts to adding D4-branes and 
Neveu-Schwarz five branes and $n$ is the number of  branes 
that were added ---see eqs. (\ref{page11n}) and (\ref{page1100})---  or crossed\footnote{Thanks to Daniel Thompson for a 
	discussion about this.}. Had we integrated on the interval $[n\pi, (n+1)\pi]$, we would have obtained a scaling like
$c\sim N_{D_4}^2$ at leading order in $n$.

\noindent {\underline{Donos-Gauntlett geometry}}

We study here the central charge for the Donos-Gauntlett background in section \ref{sectionDG}.
For the AdS$_3$ fixed point, the volume of the internal compact manifold is

$$vol (M_7)= L^7 \left(4/3 \right)^{5/4} 4 \pi^2 d_{\alpha} d_{\beta} \ vol (T^{1,1}) ~~,$$ 

and the central charge is,
\begin{equation}
c= 3 \ \big{|} N_{D3} \widehat{N}_{D3} \big{|}=\frac{2}{3} \frac{L^8}{\alpha'^4} \frac{d_{\alpha} d_{\beta} vol (T^{1,1})}{\pi^4} \ . 
\label{Central3ao1}
\end{equation}
For the AdS$_5$ fixed point, the volume of the internal compact manifold $vol (M_5) =  L^5  vol (T^{1,1})$, and the central charge results in,
\begin{equation}
c=\frac{27}{64} N_{D3}^2=\frac{L^8}{\alpha'^4} \frac{ vol (T^{1,1})}{64 \ \pi^5} \ .
\label{Central3ao}
\end{equation}

The quotient of central charges before and after the NATD for the Donos-Gauntlett QFT, are given by a similar expression to that in eq. \eqref{Central5aa}.

We move now to study a quantity that gives an idea 
of the degrees of freedom along a flow.

\subsection{Central charge for flows across dimensions}
\label{CCdifdims}
In the previous section, we calculated the central charge for two and four dimensional CFTs dual to the AdS$_3$ and AdS$_5$ fixed points of the flow. In this section, we will use the developments in  \cite{Freedman:1999gp} and \cite{Klebanov:2007ws}, to write a c-function along the flows between these fixed points. We will find various subtleties,
\begin{itemize}
	\item{When considered as a low energy two-dimensional CFT, the definition of the c-function evaluated on the flows will not detect the presence of the four dimensional CFT in the far UV.}
	\item{We attempt to generalize the formula of \cite{Klebanov:2007ws}
		for anisotropic cases (that is for field theories that undergo a spontaneous compactification on $\Sigma_2$). This new definition will detect both the two dimensional and four dimensional conformal points, but will not necessarily be decreasing towards the IR. This is not in contradiction with 'c-theorems' that assume Lorentz invariance.}
\end{itemize}
We move into discussing these different points in our particular examples.
To start, we emphasize that the formulas in \eqref{manaza}-\eqref{natdintrocfunctiongeneraldef}, contain the same information. 
Indeed, the authors of \cite{Klebanov:2007ws}, present a 'spontaneous compactification' of a higher dimensional Supergravity (or String theory) to $d+2$ dimensions, see eq. \eqref{natdintrocfunctionmiddle}. Moving the reduced system to Einstein frame and observing that the $T_{\mu\nu}$ of the  matter in the lower dimension satisfies certain positivity conditions imposed in \cite{Freedman:1999gp}, use of  eq. \eqref{natdintrocfunctionaprime} implies eq. \eqref{natdintrocfunctiongeneraldef}. Hence, we will apply eqs. \eqref{natdintrocfunctionmiddle}-\eqref{natdintrocfunctiongeneraldef} to our different compactifications in sections \ref{natdprevioussolutions} to \ref{natdlift}.

{\underline{Twisted and Donos-Gauntlett solutions.}}
For the purpose of the flows both twisted and Donos-Gauntlett solutions present
a similar qualitative behavior.
We start by considering the family of backgrounds in eq. \eqref{NN02} as dual to field theories in $1+1$ dimensions.
In this case the quantities relevant for the calculation of the central charge are,
\begin{equation}
\begin{aligned}
& d=1,\;\; \alpha_0 =L^2 e^{2A},\;\;\; \beta_0=e^{-2A},\;\;\; e^{\Phi}=1,
\\[5pt]
& \frac{ds_{int}^2}{L^2}=e^{2 B} ds^2_{\Sigma_{{2}}}+e^{2U}ds_{KE} ^2+e^{2V} \left( \eta + z A_1 \right)^2.
\end{aligned}
\end{equation}
We calculate the quantity
\begin{equation}
\widehat{H}= {\cal N}^2 e^{2(B+4U+V+A)},     \qquad      {\cal N}=\frac{(4\pi)^3 vol (\Sigma_2) L^8}{108}.
\label{ccc}
\end{equation}
Then, we obtain
\begin{equation}
c=\frac{{\cal N} e^{2B+4U+V}}{2 \pi G_N^{(10)}(2B'+4U'+V'+A')}.
\end{equation}
Using the BPS equations describing these flows in eq. \eqref{O17} we can get an expression without derivatives. Specializing for the solution with an $\mathrm H_2$ in section \ref{natdtwistedsolutions}, we find
\begin{equation}
c=\frac{{\cal N}}{9 \pi G_N^{(10)}}\frac{(1+e^{2r})^2}{1+2e^{2r}}.
\label{vava}
\end{equation}
We can calculate this for the background we obtained in 
section \ref{natdtwistedduals}, by application of NATD. 
The result and procedure will be straightforward, 
but we will pick a factor of the volume of the space 
parametrized by the new coordinates,
$
vol (\rho, \chi, \xi)$.

For the purposes of the RG-flow, the quotient of the central charges will 
be the same as the quotient in eq. \eqref{xxyz}. This was indeed observed 
in the past \cite{Itsios:2013wd}, \cite{Macpherson:2014eza}
and is just a consequence of the invariance of the quantity 
$\big(e^{-2\Phi}\sqrt{\det[g]}\big)$ under NATD.

Coming back to eq. \eqref{vava}, we find that in the far IR, represented by $r\to -\infty$, 
the central charge is constant. But in the far UV ($r\to \infty$), 
we obtain a result that is not characteristic of a CFT. 
Hence, this suggest that the definition is only capturing the behavior 
of a 2-dim QFT. In other words, the four dimensional QFT may 
be thought as a two dimensional QFT, but  with an infinite number of fields. 

The absence of the four dimensional fixed point in our eq. \eqref{vava} can be accounted if we generalize the prescription to calculate central charges for an anisotropic 4-dim QFT. Holographically
this implies working with a background of the form,\footnote{A natural generalization of eq. \eqref{olaqase} is $
	ds^2 =  -\alpha_0 dt^2 +\alpha_1 dy_1^2+....+\alpha_d dy_d^2
	+\Pi_{i=1}^{d}\alpha_i^{\frac{1}{d}}
	\beta dr^2 + g_{ij}d\theta^i d\theta^j.$}
\begin{equation}
ds_{10}^2 =  -\alpha_0 dy_0^2 +\alpha_1 dy_1^2+\alpha_2 ds_{\Sigma_2}^2
+\left( \alpha_1 \alpha_2^2 \right)^{\frac{1}{3}}
\beta_0 dr^2 + g_{ij}d\theta^i d\theta^j.
\label{olaqase}
\end{equation}
In this case we define,
\begin{equation}
\begin{aligned}
& G_{ij}d\xi_i d\xi_j=\alpha_1 dy_1^2 + \alpha_2 ds_{\Sigma_2}^2
+ g_{ij}d\theta^i d\theta^j,  
\\[5pt]
& \widehat{H}= \left( \int d\theta^i \sqrt{e^{-4\Phi} \det[G_{ij}] }  \right)^2, 
\\[5pt]
& c=d^d\frac{\beta_0^{\frac{d}{2}} \widehat{H}^{\frac{2d+1}{2}}}{\pi G_N^{(10)}(\widehat{H}')^d}.
\end{aligned}
\label{centralanisotropic}
\end{equation}

We can apply this generalized definition to the flow for the twisted 
$\mathrm H_2$ background of section \ref{natdtwistedsolutions}, and Donos-Gauntlett background of section \ref{natdDG} (for more examples the reader is referred to appendix D of \cite{Bea:2015fja} ).
In this case, we consider them as dual to a field theory in $3+1$  
anisotropic dimensions (two of the dimensions are compactified on a $\Sigma_2$).
The quantities relevant for the calculation of the central charge are,
\begin{equation}
\begin{aligned}
& & d=3,\;\; \alpha_1 =L^2 e^{2A},\;\;\; \alpha_2=L^2 e^{2B} , \;\;\; \beta_0=e^{\frac{-2A-4B}{3}},\;\;\; e^{\Phi}=1,
\\[5pt]
& &G_{ij}d\xi_i d\xi_j=L^2 \left( e^{2A} dy_1^2 +e^{2B} ds^2_{\Sigma_2}+e^{2U}ds_{KE} ^2+e^{2V} \left( \eta + z A_1 \right)^2\right).
\end{aligned}
\end{equation}
We calculate 
\begin{equation}
\widehat{H}= {\cal N}^2 e^{2(2B+4U+V+A)},\;\;\; 
{\cal N}=\frac{(4\pi)^3  L^8}{108}.
\label{cccc}
\end{equation}
Then,  we obtain
\begin{equation}
c=\frac{ 27 {\cal N} e^{4U+V}}{8 \pi G_N^{(10)} (2B'+4U'+V'+A')^3}.
\label{c000}
\end{equation}
Focusing on the $\mathrm H_2$ case, if we use the solution that 
describe this flow---see eq. \eqref{O19} we get an analytical expression,
\begin{equation}
c=\frac{\mathcal{N}}{\pi \ G_N^{(10)}} \left( \frac{1+e^{2r}}{1+2e^{2r}} \right)^3 \ .
\end{equation}
Notice that, by definition, this quantity gives the correct central charge 
in the UV (a constant, characterizing the 4-d fixed point). 
In the IR, the quantity turns out to be constant too, 
so it is capturing the presence of a 2-d fixed point. 
Nevertheless, it is probably not an appropriate candidate for 
a 'c-function between dimensions' 
as it is not necessarily increasing towards the UV. 
This is not in contradiction with the logic of 'a-theorems' and proofs like the
ones 
in \cite{Freedman:1999gp} or \cite{Komargodski:2011vj}, 
as the metric does not respect Lorentz invariance. 
Hence, it is not satisfying the assumptions of the theorems.
For the Donos-Gauntlett case analogous things happen. 
It would be very nice to try to apply the recent ideas of 
\cite{Gukov:2015qea}
to this flow. Notice that this feature of a 'wrong monotonicity' for the central
charge was also observed---for theories breaking Lorentz invariance in Higher Spin theories---see the papers \cite{Gutperle:2011kf}.

Let us move now to study other observables defining the 2-d and 4-d
QFTs.

\section{Entanglement entropy and Wilson loops} \label{natdEEWilson}
In this section, we will complement the work done above,
by studying a couple of fundamental observables in the QFTs defined by 
the backgrounds in sections \ref{natdprevioussolutions} to \ref{natdlift}.

Whilst at the conformal points the functional dependence of
results is determined by the symmetries, the interest will be 
in the coefficients accompanying the dependence. Both observables
interpolate smoothly between the fixed points.

\subsection{Entanglement entropy}
\label{EE}

\def\reg{\mathrm{reg}}

The aim of this section is to compute the 
entanglement entropy on a strip, which extends along the 
direction $y_1 \in [-\frac{d}{2},\frac{d}{2}]$, and study 
how this observable transforms under NATD. 
The input backgrounds for our calculations will be the Donos-Gauntlett and the $\mathrm H_2$ flow as well as their non-Abelian T-duals. Since the procedure is the same in all of the cases and due to the similarity of the geometries we will present the results in a uniform way.

This prescription states that the holographic entanglement entropy between the strip and its complement is given by the minimal-area static surface that hangs inside the bulk, and whose boundary coincides with the boundary of the strip. The general form of the entanglement entropy for the non-conformal case is,

\begin{equation}
\label{prescription} S_{EE} = \frac{1}{4G_N^{(10)} } \int d^8 \sigma e^{-2
	\Phi} \sqrt{G^{(8)}_{\rm ind}}~.
\end{equation}

For the strip, we chose the embedding functions to be $y_0 = \textrm{const}$ and $r=r(y_1)$ and then using the conservation of the 
Hamiltonian  we arrive at an expression for $r(y_1)$ that makes the area minimal under that embedding. With that we compute,

\begin{equation}
\label{Ryu-Takayanagi}
S_{EE}= \frac{\tilde{L}^8}{54}\frac{\pi}{G^{(10)}_N} vol (\Sigma_2)V_3 \int\limits_{r_*}^{\infty} dr \ e^{-A} \frac{G^2}{\sqrt{G^2 - G^2_*}} \ ,
\end{equation}
where   the form of the function $G$ depends on the geometry of the background 
that we consider, 
\begin{equation}
G = \left\{ \begin{array}{ll}
e^{A+2B} & \textrm{for the twisted solutions}
\\[5pt]
e^{A+2B+4U+V} & \textrm{for the DG solution}
\end{array}
\right. \ .
\end{equation}
Above, $r_*$ is the radial position of the hanging surface tip and we define $G_*=G(r_*)$. Also, with $vol(\Sigma_2)$ we denote the volume of 
the Riemann surface $\Sigma_2$. Notice that the form of the 
function $G$ is the same before and after NATD. 
Moreover we consider,
\begin{equation}
\label{Radii}
\tilde{L} = \left\{ \begin{array}{ll}
L                     & \textrm{before NATD}
\\[5pt]
\widehat{L}& \textrm{after NATD}
\end{array}
\right. \ ,
\end{equation}
%
The quantity $V_3$ is defined as,
\begin{equation}
\label{3dimVol}
V_3 = \left\{ \begin{array}{ll}
16 \ \pi^2 & \textrm{before NATD}
\\[5pt]
\int d \chi \ d \xi \ d \r \ \r^2 \sin\chi & \textrm{after NATD}
\end{array}
\right. \ .
\end{equation}
Before the NATD transformation $V_3$ comes from the 3-dimensional 
submanifold that is spanned by the coordinates 
$(\th_2,\phi_2,\psi)$, while after the NATD it comes 
from the submanifold that is spanned by the dual 
coordinates $(\r,\chi,\xi)$. This 
implies that the entropies before and after NATD are proportional all along the flow, 
for any strip length. 
A discussion on possible values of the  quantity 
$\int d \chi \ d \xi \ d \r \ \r^2 \sin\chi$ 
can be found in the quantized charges 
section \ref{natdpagecharges} and in the discussion 
on central charges in section \ref{natdcentralcharge}.

When computed by the Ryu-Takayanagi formula eq. \eqref{Ryu-Takayanagi} the entanglement entropy 
is UV divergent. In order to solve this we compute the 
regularized entanglement entropy ($S^{\reg}$) by subtracting 
the divergent part of the integrand of  eq. \eqref{Ryu-Takayanagi}. 
The regularized entanglement entropy is given by,
\begin{equation}
\label{RegEE}
S^{\reg} = \frac{\tilde{L}^8}{54}\frac{\pi}{G^{(10)}_N} vol (\Sigma_2)V_3 \Bigg\{ \int\limits_{r_*}^{\infty} dr \   \Big( \frac{e^{-A} G^2}{\sqrt{G^2 - G^2_*}} - F_{UV}  \Big)    - \int\limits^{r_*} dr \ F_{UV}  \Bigg\} \ ,
\end{equation}
where the last integral is an indefinite integral with the result being evaluated at $r=r_*$ and 
\begin{equation}
F_{UV}= \left\{ \begin{array}{ll}
\frac{1+e^{2r}}{3} & \textrm{for the $\mathrm H_2$ twisted solution}
\\[5pt]
e^{2r}+\frac{\lambda^2}{3}& \textrm{for the Donos-Gauntlett solution}
\end{array}
\right. \ .
\end{equation}
From the formulas \eqref{Radii}, \eqref{3dimVol} and \eqref{RegEE} it is obvious that the regularized entanglement entropies before and after the NATD transformation differ by the factor
\begin{equation}
\label{EEquotient}
\frac{\widehat{S_{EE}}^{\reg}}{S_{EE}^{\reg}} = \frac{\widehat{L}^8}{L^8}\frac{\int d \chi \ d \xi \ d \r \ \r^2 \sin\chi}{16 \ \pi^2} \ .
\end{equation}
In the formula above we denoted by $\widehat{S_{EE}}^{\reg}$ the value of the
entanglement entropy after the NATD transformation. 
As discussed below eq. (\ref{vava}), the quantity $\big(e^{-2\Phi}\sqrt{\det[g]}\big)$ is invariant under NATD, and this explains why the ratio (\ref{EEquotient}) is constant along the flow.

At this point let us normalize the regularized 
entanglement entropy by defining the quantity,
\begin{equation}
\label{SPrimeDef}
S'_{EE} = \frac{54}{\tilde{L}^8} \frac{G^{(10)}_N}{\pi \ vol(\Sigma_2) V_3} \ S^{\reg} \ .
\end{equation}
In what follows we present the behavior of $S'_{EE}$ in the UV and the IR for the geometries of interest. We express the results in terms of the width of the strip $d$,
%
\begin{equation}
d = 2 G_* \int\limits_{r_*}^{\infty} dr \ \frac{e^{-A}}{\sqrt{G^2 - G_{*}^2}} \ .
\end{equation}
The UV/IR behavior written in 
eqs \eqref{TwistedUVEE}, \eqref{DGUVEE}, \eqref{TwistedIREE} 
and \eqref{DGIREE} below, 
are just consequences of the fact that in far UV and far 
IR the dual QFT is conformal. 
The functional forms are universal, so our main interest 
is the constant appearing in them, 
and also as a cross-check of numerical results.

\subsubsection*{Behavior in the UV}

\underline{Twisted geometries}

In the case of the twisted $\mathrm H_2$ geometry  we find that the width of the strip is,
\begin{equation}
d = e^{-r_*} \int\limits_{1}^{\infty} \frac{du}{u^2} \frac{2}{\sqrt{u^6 -1}} = \frac{2 \sqrt{\pi} \Gamma\big(  \frac{2}{3} \big)}{\Gamma\big(  \frac{1}{6} \big)} \ e^{-r_*} \ .
\end{equation}
Here in the integration we changed the variable $r$ by $u = \frac{e^r}{e^{r_*}}$. From the calculation of the normalized entropy $S'_{EE}$ we observe that in the UV this behaves like $\frac{1}{d^2}$, namely
\begin{equation}
\label{TwistedUVEE}
S'_{EE} = - \frac{\pi^{3/2}}{6} \Bigg(  \frac{\Gamma\big(  \frac{2}{3} \big)}{\Gamma\big(  \frac{1}{6} \big)}  \Bigg)^3 \frac{1}{d^2} + \frac{1}{3} \ln d + \frac{1}{3} \ln \Bigg(  \frac{\Gamma\big(  \frac{1}{6}  \big)}{2 \sqrt{\pi}} \Gamma\Big(  \frac{2}{3}  \Big)  \Bigg) \ ,
\end{equation}
where we also included subleading and next-to subleading terms.

\noindent \underline{Donos-Gauntlett geometry}

Similarly in the case of the Donos-Gauntlett geometry we find that the width of the strip in terms of $r_*$ (considering also a subheading term) is,
\begin{equation}
d = \frac{2\sqrt{\pi} \Gamma\big(  \frac{2}{3} \big)}{\Gamma\big(  \frac{1}{6}  \big)} e^{-r_*} + \l^2 \big(  \frac{5}{72} + \frac{11}{24} I_1  \big) e^{-3 r_*} \ ,
\end{equation}
where
\begin{equation}
I_1 = \int\limits_{1}^{\infty} dz \ \frac{z^2}{(z^4 + z^2 + 1)\sqrt{z^6-1}} = 0.1896 \ldots
\end{equation}
Here as well we end up with a $\frac{1}{d^2}$ behavior, a logarithmic subleading contribution and a constant $c_1$ for the regularized entropy,
\begin{equation}
\label{DGUVEE}
S'_{EE} = - 2 \pi^{3/2} \Bigg(  \frac{\Gamma\big(  \frac{2}{3} \big)}{\Gamma\big(  \frac{1}{6} \big)}  \Bigg)^3 \frac{1}{d^2} + \frac{\l^2}{3} \ln d + c_1 \ ,
\end{equation}
where $c_1$ ,
\begin{equation}
c_1 = \lambda^{2}\left(-\frac{1}{3}\ln\left(\frac{2\sqrt{\pi}\Gamma\left(\frac{2}{3}\right)}{\Gamma\left(\frac{1}{6}\right)}\right)+\left(\frac{\ln2}{9}+\frac{11}{48}I_{1}\right)+\frac{\Gamma\left(-\frac{1}{3}\right)}{6\Gamma\left(\frac{2}{3}\right)}\left(\frac{5}{72}+\frac{11}{24}I_{1}\right)\right) \ .
\end{equation}

\subsubsection*{Behavior in the IR}

\underline{Twisted geometries}

The calculation for the IR limit is more tricky. The origin of the subtlety is that the integrals we have to evaluate now run all along the flow and we do not know the analytical properties of the integrands. In order to address this issue we split the integration into the intervals $[r_*,a]$ and $[a,+\infty)$ choosing $a$ to be in the deep IR but always greater than $r_*$. 

Following this prescription in the calculation, for the 
width of the strip we find,
\begin{equation}
d = 
\frac{4}{3} \ e^{-\frac{3 r_{*}}{2}}.
\end{equation}
%
If we do the same analysis when we calculate the normalized entropy we find,
\begin{equation}
\label{TwistedIREE}
S'_{EE} = \frac{2}{9} \ln d + \frac{2}{9} \ln \frac{3}{2} \ .
\end{equation}
%
%
The logarithmic dependence  of the leading term on $d$ is the expected for a $1+1$ theory.

\noindent \underline{Donos-Gauntlett geometry}

We close this section presenting the corresponding results for the Donos-Gauntlett geometry. As in the case of the twisted geometries we split the integrations in the same way. Then for the width of the strip we find,
\begin{equation}
d = e^{-a_0} \frac{2\sqrt{2}}{3^\frac{3}{4}} e^{-\frac{3^\frac{3}{4}}{\sqrt{2}}r_*} \ .
\end{equation}
The normalized entropy displays again a logarithmic behavior in terms of the width of the strip,
\begin{equation}
\label{DGIREE}
S'_{EE} = \frac{8}{9} \ln d + \frac{8}{9} \ln \Big(  e^{a_0} \frac{3^{\frac{3}{4}}}{\sqrt{2}}  \Big) + c_2 \ ,
\end{equation}
where the constant $c_2$ has the value,
\begin{equation}
c_2= \int_{0}^{\infty} \left( e^{-A} G-  e^{2r}-\frac{4}{3} \right)dr + \int_{-\infty}^{0} \left( e^{-A} G- e^{2r}-\left(\frac{4}{3}\right)^{\frac{5}{4}} \right)dr = -0.0312 \ldots
\end{equation}

\begin{figure}
	\centering
	\label{fig: S'} 
	\begin{subfigure}[b]{0.62\textwidth}
		\centering
		\includegraphics[width=\textwidth]{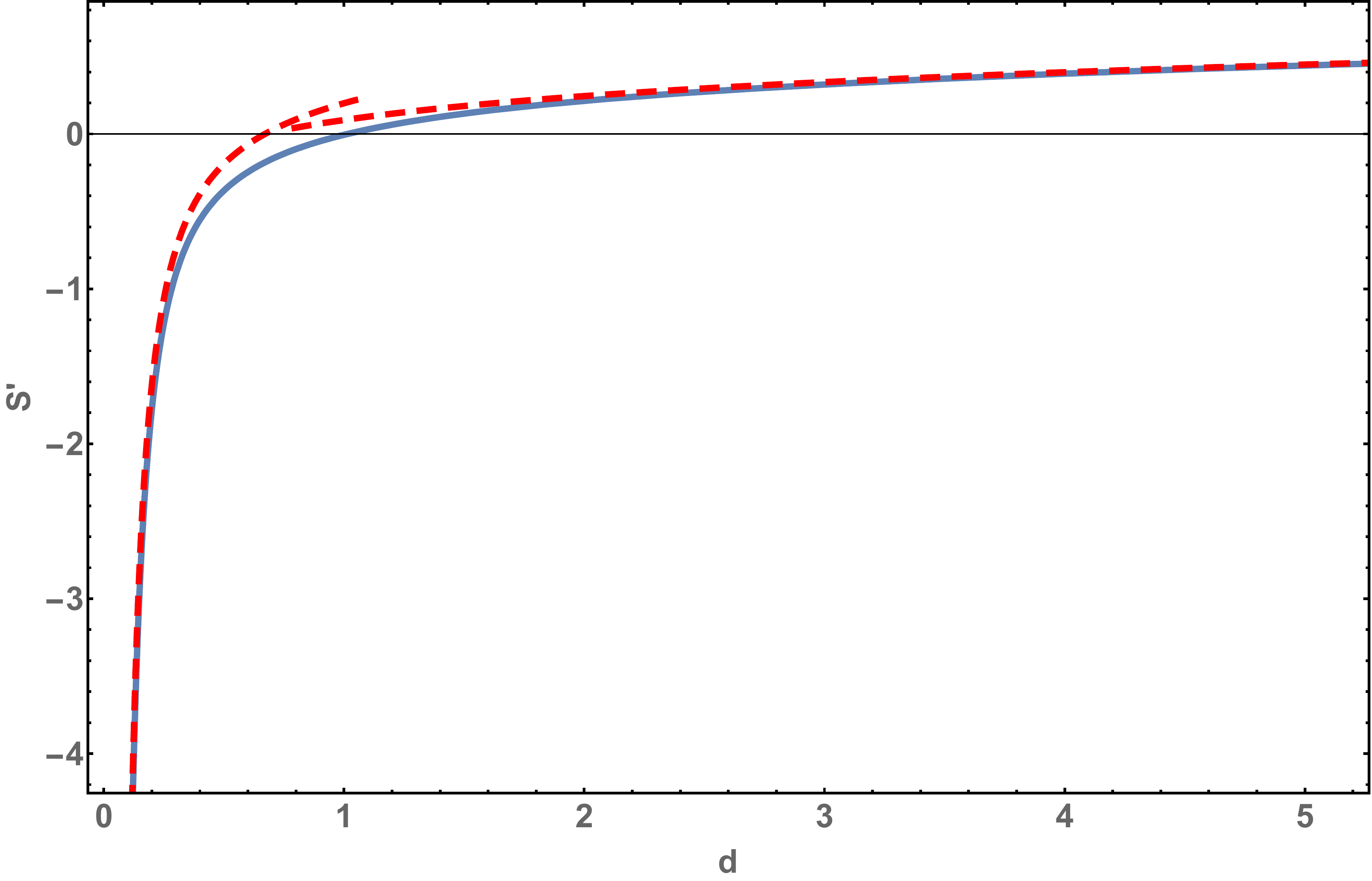}
	\end{subfigure}
	~
	\begin{subfigure}[b]{0.62\textwidth}
		\centering
		\includegraphics[width=\textwidth]{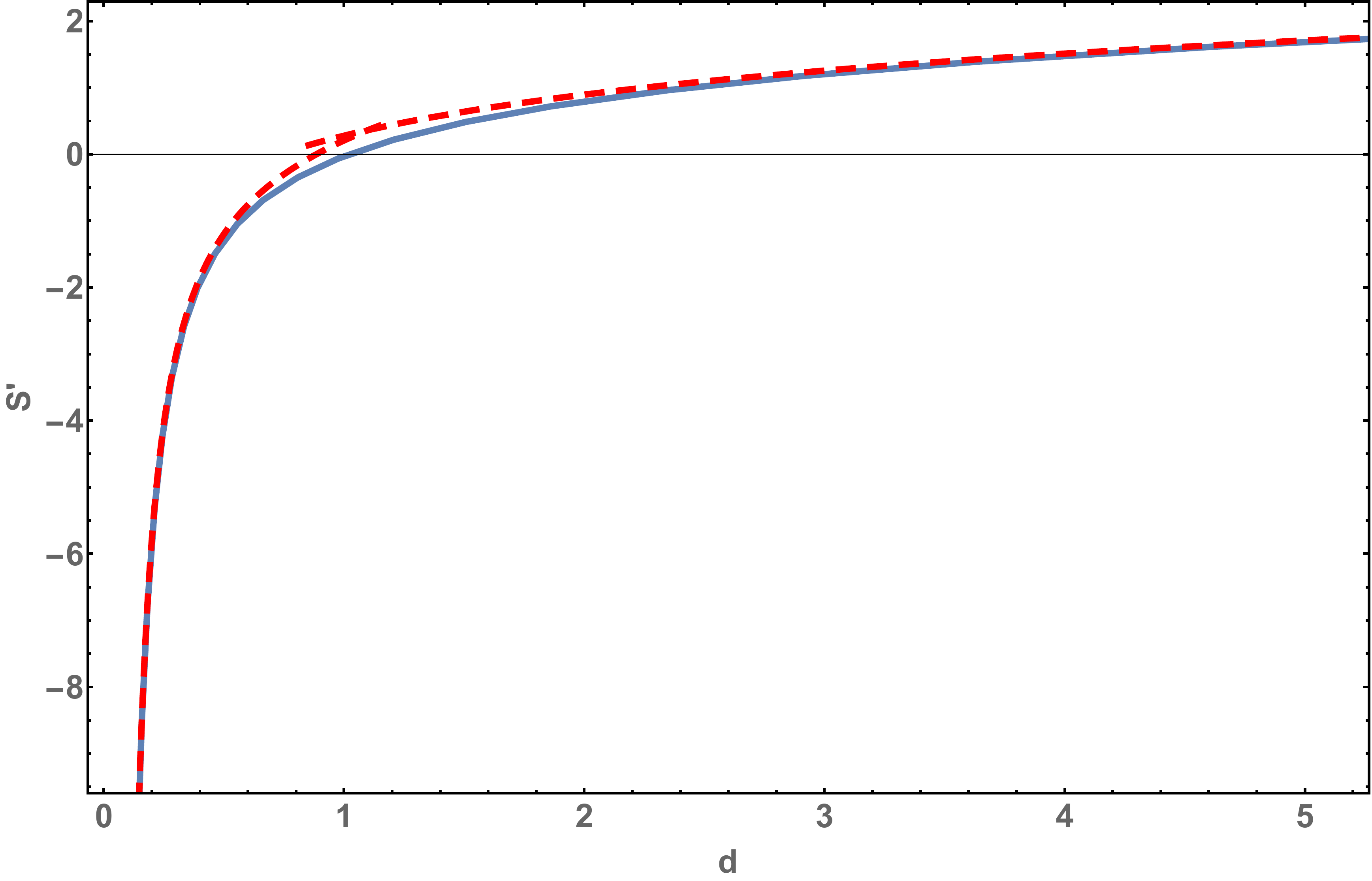}
	\end{subfigure}
	\caption{$S'_{EE}$ as a function of $d$ for twisted $\mathrm H_2$ (left) and Donos-Gauntlett solution (right).
		The continuous curves correspond to the numerical value, while the dashed red ones to the UV and IR limits.
	}
	\label{EEnatduality0}
\end{figure}

The results for the entanglement entropy are shown in Fig. \ref{EEnatduality0}. We will now perform a similar analysis for Wilson loops.

\subsection{Wilson loop} \label{natdwilsonloop}

In our calculations we consider 
an embedding of the form $r = r(y_1)$ for the string. Such an embedding gives rise to the following induced metric on the string,
\begin{equation}
\label{InducedStringMetric}
ds_{\textrm{st}}^2 = \tilde{L}^2 \Big(  e^{-2A} \ dy_0^2 + \big(  e^{2A} + r'^2  \big) \ dy_1^2  \Big) \ ,
\end{equation}
where $\tilde{L}$ is defined in \eqref{Radii}. It is obvious that the above induced metric is the same for 
all of the geometries that we have discussed in this chapter so far (even for the duals). For this reason we believe that it is not necessary to make any distinction with respect to these geometries for the moment. Moreover, this means that the observable is 'uncharged' under NATD and thus it has the same functional form when computed in the initial and dualized geometries. The interest will be 
in the numerical coefficients our calculation will give.

The Nambu-Goto lagrangian density for the string takes the form,
\begin{equation}
\mathcal{L} = \frac{1}{2\pi\alpha'} \sqrt{-\det(g_{\textrm{ind}})} = \frac{\tilde{L}^2}{2\pi \alpha'} \ e^{A} \ \sqrt{e^{2A} + r'^2} \ ,
\end{equation}
where $g_{\textrm{ind}}$ stands for the induced metric \eqref{InducedStringMetric}.

An important observable that can be computed holographically is the potential energy of two non-dynamical sources added to the QFT. As explained in the subsection \ref{natdintrowilsonloop} ,the conservation of the Hamiltonian implies that,
\begin{equation}
\frac{e^{3A}}{\sqrt{e^{2A} + r'^2}} = e^{2 A_*} \ ,
\end{equation}
where $A_*$ is the value of the function $A(r)$ at the tip of the hanging string $r=r_*$. We can solve the last equation for $r'$ and use the result to calculate the distance between the endpoints of the string. If we do this we can express $d$ in terms of $r_*$
\begin{equation}
\label{QuarkDistance}
d = e^{2 A_*} \int\limits_{r_*}^{\infty} dr \ \frac{e^{-A}}{\sqrt{e^{4A} - e^{4A_*}}} \ .
\end{equation}

The Nambu-Goto action now reads
\begin{equation}
\label{NambuGoto}
S_{NG} = \frac{T \tilde{L}^2}{\pi \alpha'} \int\limits_{r_*}^{\infty} dr \ \frac{e^{3A}}{\sqrt{e^{4A} - e^{4A_*}}} \ ,
\end{equation}
where $T = \int dt$. The integral in eq. \eqref{NambuGoto} is divergent since we are considering quarks of infinite mass sitting at the endpoints of the string. We can regularize  this integral by subtracting the mass of the two quarks and dividing by $T$ as it is shown below
\begin{equation}
\label{QuarkAntiquarkEnergy}
\frac{E}{\tilde{L}^2} \alpha' = \frac{1}{\pi} \int\limits_{r_*}^{\infty} dr e^A \ \Big(  \frac{e^{2A}}{\sqrt{e^{4A} - e^{4A_*}}} - 1  \Big) - \frac{1}{\pi} \int\limits_{-\infty}^{r_*} dr \ e^{A} \ .
\end{equation}
This formula gives us the quark-antiquark energy. In order to calculate the same observable starting
with the NATD geometries one must take into account that the AdS radius $L$ is different from that of the original geometries. In fact both results are related in the following way
\begin{equation}
\frac{\widehat{E}}{E} = \frac{\widehat{L}^2}{L^2} \ .
\end{equation}
In the last expression the hats refer to the dual quantities.

At this point we will explore the UV and IR limits of the quark-antiquark energy both for the twisted and the  Donos-Gauntlett geometries.

\subsubsection*{Behavior in the UV}

\underline{Twisted and Donos Gauntlett geometry}

First we focus on the twisted solution where the Riemann surface is the hyperbolic space, i.e. $\Sigma_2 = \mathrm H_2$. In section \ref{natdtwistedsolutions} we saw that in this case the function $A(r)$ behaves like $A(r) \sim r$. Taking this into account we can compute the distance between the quarks from the formula \eqref{QuarkDistance}. The result of this is
\begin{equation}
d = \frac{2 \sqrt{2} \ \pi^\frac{3}{2}}{\Gamma \Big(  \frac{1}{4} \Big)^2} e^{-r_*} \ .
\end{equation}
Solving this equation for $r_*$ we can substitute into the result coming from the formula \eqref{QuarkAntiquarkEnergy}. This will give the quark-antiquark energy in terms of $d$ which in our case is
\begin{equation}
\label{QQb1}
E = - \frac{\tilde{L}^2}{\alpha'} \ \frac{4 \ \pi^2}{ \Gamma  \Big(  \frac{1}{4} \Big)^4} \frac{1}{d} \ ,
\end{equation}
as expected for a CFT. The main point of interest in the previous formula is in the numerical coefficient.


Similar considerations for the case of the Donos-Gauntlett geometry give the same results as in the twisted case above. This is because the asymptotic behavior of the function $A(r)$ in the UV is the same in both cases.

\subsubsection*{Behavior in the IR}

\underline{Twisted geometry}

Again in the IR region we address again the same difficulty that we found in the computation of the entanglement entropy. We use the same trick to overcome it, that is we split the integrations into the intervals $[r_*,a]$ and $[a, +\infty)$ where $a$ has value in the deep IR but always greater than $r_*$. 


In section \ref{natdtwistedsolutions} we saw that in the case where $\Sigma_2 = \mathrm H_2$, the IR behavior of the function $A(r)$ is $A(r) \sim \frac{3}{2} \ r$. Applying this into the formula \eqref{QuarkDistance} we obtain the following result,
\begin{equation}
d = \frac{4 \sqrt{2} \ \pi^\frac{3}{2}}{3 \ \Gamma \Big(  \frac{1}{4} \Big)^2} e^{-\frac{3}{2} r_*} \ .
\end{equation}
As before we solve the previous result for $r_*$ and we substitute it into the expression that we find from the calculation of the quark-antiquark potential. This way we express the energy as a function of the distance between the quarks,
\begin{equation}
\label{QQb2}
E = - \frac{\tilde{L}^2}{\alpha'} \ \frac{16 \ \pi^2}{ 9 \ \Gamma \Big(\frac{1}{4} \Big)^4} \frac{1}{d} \ .
\end{equation}

\noindent \underline{Donos-Gauntlett geometry}

Repeating the same steps for the case of the Donos-Gauntlett geometry we find that the distance between the quarks is,
\begin{equation}
d = \frac{4 \pi^\frac{3}{2}}{ 3^\frac{3}{2} \Gamma \Big( \frac{1}{4} \Big)^2} e^{-a_0 - \frac{3^{\frac{3}{4}}}{\sqrt{2}} r_*} \ .
\end{equation}
Then, expressing the energy in terms of the distance $d$ we find again a dependence proportional to $\frac{1}{d}$,
\begin{equation}
\label{QQb3}
E = - \frac{\tilde{L}^2}{\alpha'} \frac{8 \pi^2}{3^{\frac{3}{2}} \Gamma \Big(  \frac{1}{4}  \Big)^4} \frac{1}{d} \ .
\end{equation}

Let us point out that the behavior in eqs \eqref{QQb1}, \eqref{QQb2} and \eqref{QQb3} are just consequences of the fact that far in the UV and far in the IR the QFT is conformal.

\begin{figure}[h]
	\begin{subfigure}[h]{0.49\textwidth}
		\centering
		\includegraphics[width=\textwidth]{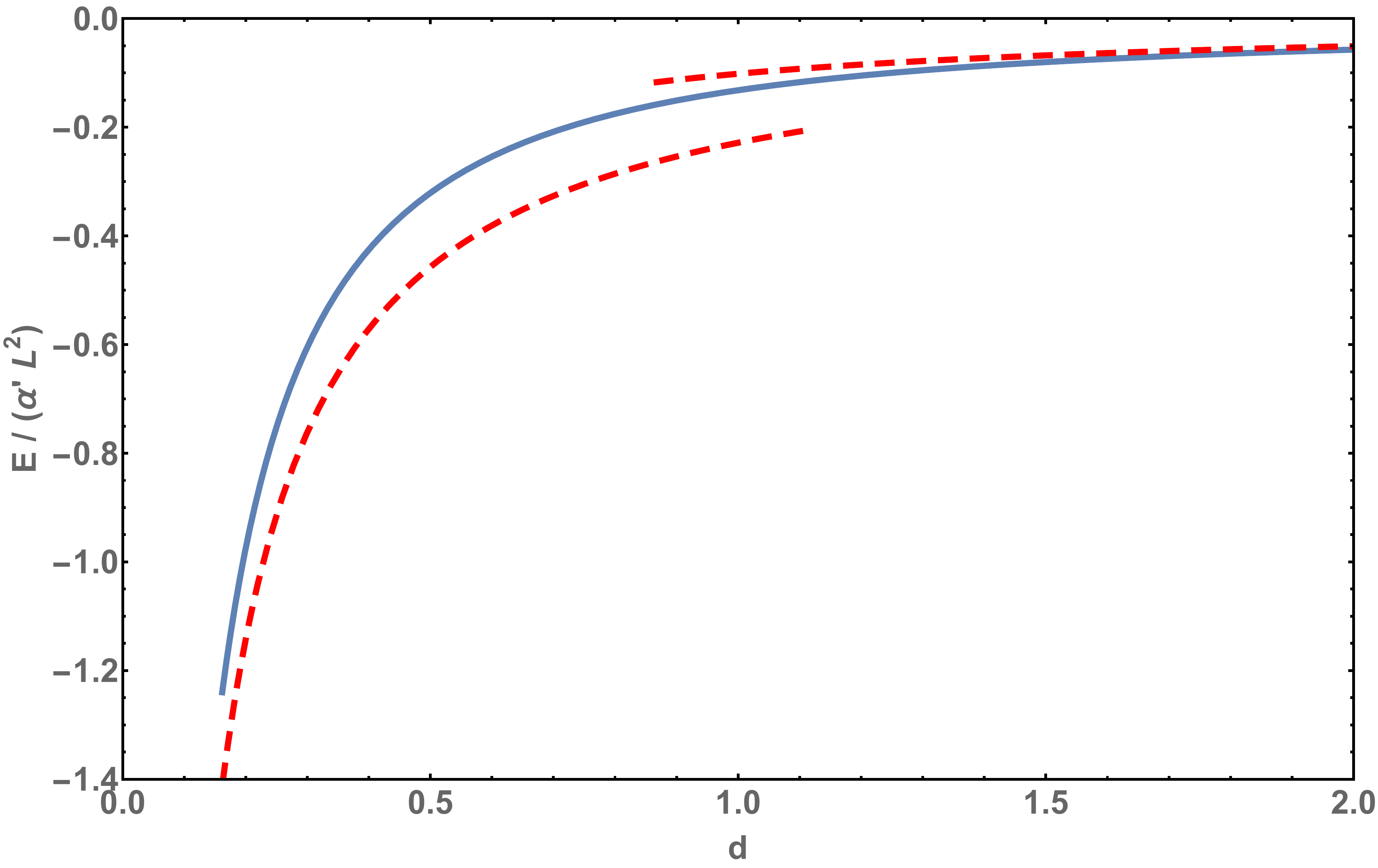}
	\end{subfigure}
	\begin{subfigure}[h]{0.49\textwidth}
		\centering
		\includegraphics[width=\textwidth]{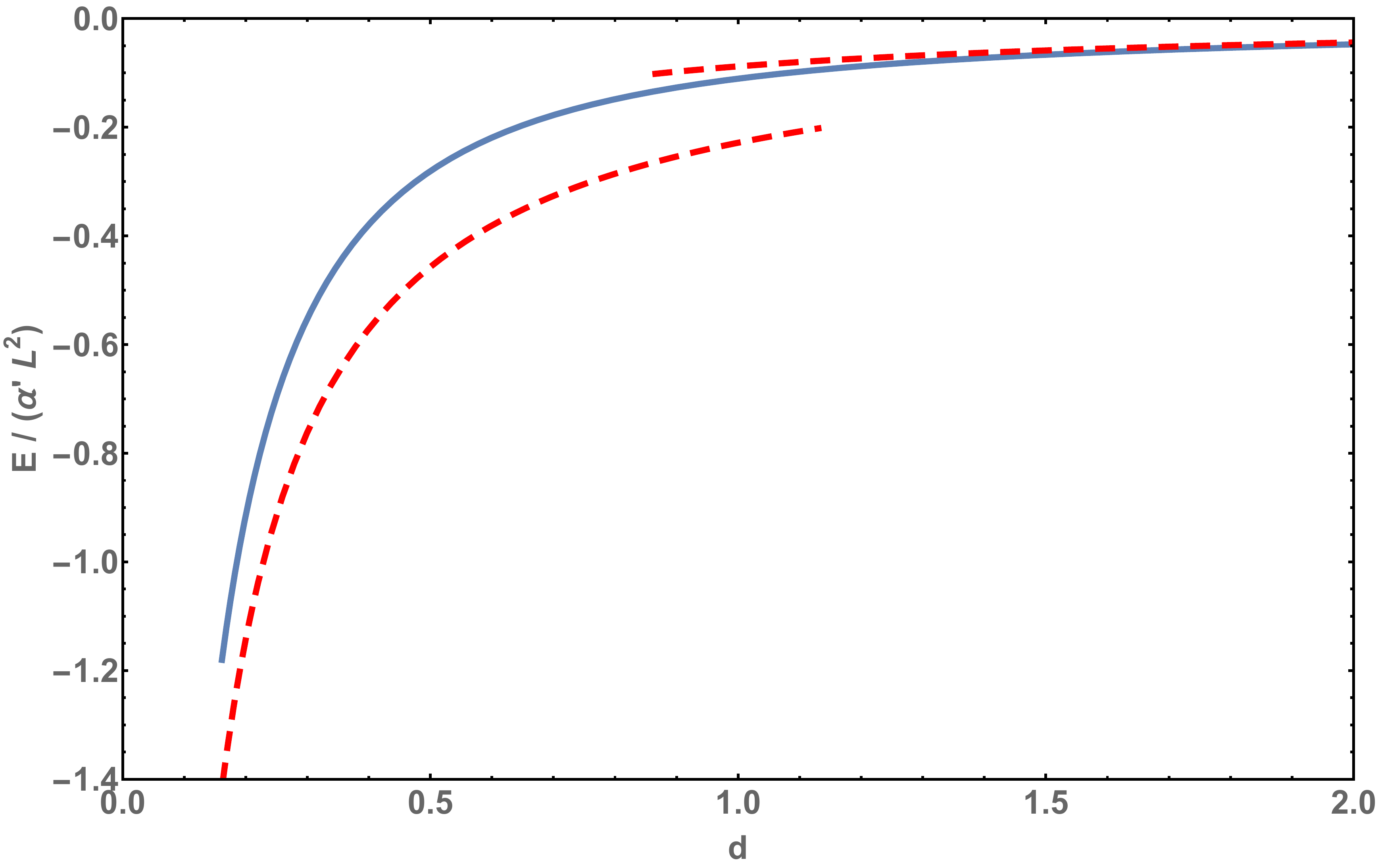}
	\end{subfigure}
	
	\caption{The quark-antiquark potential $\frac{E}{\alpha' L^{2}}$ as a function of the distance $d$ in the cases of the twisted $\mathrm H_2$ (left) and Donos-Gauntlett (right) solutions. The continuous curves correspond to the numerical results and the dashed ones to the UV and IR limits, which are respectively four and two-dimensional. } 
	\label{WLnatduality00}
\end{figure}

The results for the quark-antiquark potential are shown in Fig. \ref{WLnatduality00}.

%
%
%
%
%
%
%
%

\section{Discussion}\label{natdconclusions}

We started by studying backgrounds dual to 
two-dimensional SUSY CFTs. The 2-d CFTs were obtained
by compactification of the four-dimensional Klebanov-Witten CFT
on a torus or on a compact hyperbolic plane. The 2-d CFT preserves (0,2)
SUSY.

On those type IIB backgrounds we performed a NATD transformation,
using an $\mathrm{SU(2)}$-isometry of the 'internal space' (or, equivalently, 
a global symmetry of the dual CFT). As a result, we constructed 
{\it new, smooth} and SUSY preserving backgrounds in type IIA and 11-dimensional supergravity
with an AdS$_3$ fixed point. A further T-duality was used to construct new, smooth and SUSY
type IIB background whose IR is of the form AdS$_3\times$M$_7$ and all fluxes are
active. 

We analyzed the dual QFT by computing its observables, using
the smooth backgrounds mentioned above. By studying the Page charges, we observed that there is a correspondence between the branes of the starting type IIB solution and those of the type IIA solution after NATD.

The behavior of the central charge in the original CFT 
(the compactification of the Klebanov-Witten theory to 2-d) 
is $c\sim N_{D3}^2$, while
after the NATD goes like $c\sim N_{D6}^2 n^3$. This new (cubic)
dependence suggest a relation with long linear quivers, 
which would imply that $n$ is measuring the number of  D4- and NS5-branes.
The picture that emerges is that of a 2-d CFT living on the intersection
of D2- , D6- and NS5-branes, with induced D4 charge every time 
an NS-brane is crossed. Quantized charges support this interpretation. \label{natdqftguess}

Entanglement entropy and Wilson loops
had the expected universal dependence on $d$ (the quark-antiquark separation or
the length of  strip separating the two regions) at the fixed points. The interest of the expressions
is on the coefficients, not determined by conformal invariance. Interestingly, along the flow the observables smoothly interpolate between the IR and the UV behaviors, which are fixed points of different dimensionality. Both for the entanglement entropy and the Wilson loops we found that the quotient of their values before and after NATD is constant along the flow, as it is expected.
It would be interesting to connect these studies with previous calculations done for 
AdS$_3$, either at the sigma model or the supergravity level. We are presenting new backgrounds, hence new 2-d CFTs on which 
studies done in the past could be interesting to revisit. It would be also interesting to make more precise the QFT dual to these backgrounds. The key point needed for this purpose is to understand the global properties of the supergravity background after the NATD. Besides, it could be interesting to check whether our solutions fall within the existent classifications of, for example \cite{Gauntlett:2006ux,Beck:2015hpa}. Probably, for this purpose the global properties of the solution are needed. Another interesting point would be to further study new observables that select (or explore) the values of $\rho=n\pi$ argued in this work to be special values of the $\rho$-coordinate, in the line of  \cite{Lozano:2016kum}.


\begin{subappendices}
	
	\section{Appendix}
	\subsection{Supergravity}
	\label{natdintrosugra}
	In the previous chapter we introduced in some detail the Lovelock and LGB gravity theories; now we discuss another family of gravitational actions that play a very distinguished role in  AdS/CFT : Supergravities, that are low energy limit of string theories. 

	Supersymmetry (SUSY) is a proposed symmetry that relates bosonic and fermionic fields. It has been very explored since the early 1970's, with many remarkable properties both formal and phenomenological \footnote{So far no clear experimental evidence of it is available, though.}. As mentioned in the motivation, it was applied to string theories where it reduced the necessary number of spacetime dimensions and eliminated a tachyon among other implications.

	The SUSY applications to string theories and holography are many and very important: microstate counting for BH entropy, strong-weak dualities, AdS/CFT discovery, identification of the dual CFT in cases of high SUSY, exploration of highly coupled regimes of field theories, classifications of possible fields theories, exact solvability ...; the list goes on and on. It also generically allows to find stable analytic solutions by solving first order equations (frequently called BPS solutions); this has become a mainstream technique that we also have used. 

	The SUSY of superstring theories is also inherited by their low energy limits. The extension of gravitational theories with SUSY was also studied for its own sake, and it is called Supergravity (SUGRA). The SUSY in these cases is a gauge symmetry, namely the transformations parameters are position dependent and new fields appear, very much like in electromagnetism.

	These SUGRA theories are particularly useful in the context of AdS/CFT. In the top-down approach, the duality is normally used in the low energy limit of the string theory side that is described by classical SUGRA. At the beginning of AdS/CFT, only solutions of string theory effective actions were used and they are often found relying on SUSY. Surprisingly, purely bosonic solutions are almost always considered, although they normally preserve some of the supersymmetries of the theory they belong to.   

	In general, SUGRA theories can be constructed in different dimensions, gauge symmetries and different amounts of SUSY \footnote{Meaning that several linearly independent symmetry transformations exist.}. For our purposes we will introduce type IIA/B and 11D SUGRAs, which are the bosonic low energy limits of type IIA/B superstrings and of M-theory respectively. The 11D case is fairly special, because for $D>11$ nontrivial consistent field theories cannot have massless particles with spins greater than two.  

	\subsection{11D and type IIA, IIB SUGRAs} \label{natdintrosugras}
	In this subsection the main concepts of the bosonic sectors of SUGRAs (11d, type IIA and type IIB) used in this thesis are briefly reviewed. Let's start with the type eleven dimensional case. 
	\subsubsection{11D SUGRA} \label{natdintro11dsugra}
	The bosonic sector of 11D SUGRA is given by the vierbein $g_{\m\n}$ and a 4-form $F_4$ with dynamics given by:
	\bea 
	S= \frac{1}{2 \kappa^2} \int_{M} \dd x^{11} \sqrt{-g}  \left( R -\frac{1}{48}   F_{\sigma_1 \sigma_2 \sigma_3 \sigma_4} F^{\sigma_1 \sigma_2 \sigma_3 \sigma_4 }\right)  +   \frac{1}{2 \kappa^2}\int_{M} \frac{1}{6} A \wedge F \wedge F, \label{natdintrosug11daction}
	\eea
	where $A$ is a three form verifying $F=\dd A$ (at least locally). 

	The bosonic equations of motion of eleven-dimensional supergravity are a generalization of the Einstein:
	\begin{equation}
	R_{\mu \nu}= \frac{1}{12} F_{\mu \sigma_1 \sigma_2 \sigma_3} F_{\nu}^{ ~ \sigma_1 \sigma_2 \sigma_3} - \frac{1}{144} g_{\mu \nu}  F_{\sigma_1 \sigma_2 \sigma_3 \sigma_4} F^{\sigma_1 \sigma_2 \sigma_3 \sigma_4 } ~~ .
	\label{natdintrosugra11deinstein}
	\end{equation}
	and Maxwell equations is:
	\begin{equation}
	d * F = \frac{1}{2} F \wedge F ~~ .
	\label{natdintrosugra11dmaxwell}
	\end{equation}
	We will call $g_{\m\n}$ and $F$ called a background in 11D. When it verifies equations (\ref{natdintrosugra11deinstein}) and (\ref{natdintrosugra11dmaxwell}) the background is a classical bosonic solution to 11D SUGRA. 

	The supersymmetric character of the theory is given by a list of rules that transform bosons and fermions and leave invariant the equations of motion, the so called SUSY transformations. For bosonic solutions, where the fermionic fields are set to zero, the supersymmetric variation of bosonic fields is automatically zero, and the supersymmetric variation of the fermionic field called gravitino is given by:
	\begin{equation}
	\delta_{\epsilon} \psi_\m = ~~  D_{\mu} \epsilon  \coloneqq \nabla_{\mu} \epsilon  + \frac{1}{36} \Gamma^{\sigma_1 \sigma_2 \sigma_3} F_{\mu_{} \sigma_1 \sigma_2 \sigma_3}  \epsilon  - \frac{1}{288}  \Gamma_{\mu}^{~ \sigma_1 \sigma_2 \sigma_3 \sigma_4 }  F_{\sigma_1 \sigma_2 \sigma_3 \sigma_4}  \epsilon ,
	\label{natdintrosugra11dpsieq}
	\end{equation}
	where $\nabla_\m \epsilon$ given by the geometrical covariant derivative of the spinor:
	\bea
	\nabla_\m \epsilon := \partial_\m \epsilon+ \frac{1}{4} \omega_{\m a b} \Gamma^{ab} \epsilon. \label{natdintrosugra11dcovd}
	\eea
	We say that a classical bosonic solution of eleven-dimensional supergravity is supersymmetric if there is a nonzero spinor $\varepsilon$ such that $\delta_\varepsilon \psi_\m=0$. That spinor is called a Killing spinor. As equation ($\ref{natdintrosugra11dcovd}$) is linear, the solutions form a vector space of fields \footnote{In very much the same way Killing vectors do.}. For 11D SUGRA, the dimension of this vector space can range from 0 to 32, and it is common to use the parameter $\nu \coloneqq \text{dim} \{ \text{Killing spinors}\}/32$ to refer to the supersymmetry of a given background. It is very common to refer as SUSY solutions as $\nu$ BPS solutions. In this manner 1/2 BPS solution of 11D SUGRA will have $\nu=1/2$, 1/4 BPS means $\nu=1/4$, etcetera.

	The BPS solutions have two key properties that make them extremely useful. (\ref{natdintrosugra11dpsieq}) is a first order equation that surprisingly implies that the background is solution of the much more difficult to solve (\ref{natdintrosugra11deinstein}),(\ref{natdintrosugra11dmaxwell}). But, additionally, the solutions found in this way are automatically stable. This explains why the technique has become so popular.

	
	These are the key notions of 11D SUGRA that appear often in the AdS/CFT literature. Now we will review similar ones for type IIA and IIB SUGRAs, and the relation to the former. 
	\subsubsection{Type IIA supergravity} \label{nadtintrosugraiia}
	It is a non-chiral maximally supersymmetric supergravity in ten dimensions. The bosonic fields are $g,\Phi,B_2,C_1,C_3$:  $g$ is the metric, $\Phi$ is the dilaton, a real scalar, $B_2$ is the Kalb-Ramond 2-form, $C_1,C_3$ are 1,3-forms respectively and called the potentials. Let us define their field strengths:
	\begin{eqnarray}
	H_3 &=&d B_2~, \\
	F_2 &=& d C_1 ~,   \\
	F_4 &=&d C_3 + H_3 \wedge C_1 ~.
	\label{natdintrosugraiiahffdefs}
	\end{eqnarray}
	satisfying the following Bianchi identities:
	\bea
	d H \eq 0,\\
	d F_2 \eq m H_3,\\
	d F_4 \eq H_3 \wg F_2.
	\eea
	The equations of motion are obtained from the action (in string frame \footnote{This is the form one obtains directly from the low energy limit of string theory, hence the name. Yet, the field redefinition ${g}_{\m\n}=e^{\frac{\Phi}{2}} \tilde g_{\m\n}$ eliminates the prefactor $e^{-2\Phi}$ from \eqref{natdintrosugraiiaaction}. That field choice is called the Einstein frame because of the minimal coupling of $\Phi$ that it produces. }):
	$$S_{IIA}= \frac{1}{2 \kappa_{10}^2} \int d x^{10} \sqrt{-g} \Big\{ e^{-2 \Phi} \left( R + 4 \partial_{\mu} \Phi  \partial^{\mu} \Phi - \frac{1}{2 \cdot 3!} H_3^2 \right) - \frac{1}{2 } m^2  - \frac{1}{2 \cdot 2!} F_2^2 - \frac{1}{2\cdot 4!} F_4^2 \Big\}  $$
	\begin{equation}
	-\frac{1}{4 \kappa_{10}^2} \int B_2 \wedge dC_3 \wedge dC_3  ~,
	\label{natdintrosugraiiaaction}
	\end{equation}
	where $2 \kappa_{10}^2=(2 \pi)^7 \alpha'^4 g_s^2$, and $g_s$ is the string coupling constant (often taken equal to 1). The truncation $m=0$ is frequently considered and referred as massless IIA SUGRA.

	The equations of motions for $g_{\m\n}$ and $\Phi$ are:
	\bea
	\label{natdintrosugraiiaeinstein}
	&& R_{\m\n}+2 D_\m D_\n\Phi -
	{1\ov 4} \mathrm{H_2}_{\m\n}
	=e^{2\Phi}\Bigg[ (F_2^2)_{\m\n} + {1\ov 12} (F_4^2)_{\m\n}
	\nonumber- {1\ov 4} g_{\m\n}
	\left( F_2^2 +{1\ov 24} F_4^2 + m^2\right)\Bigg]\ ,\\
	\label{natdintrosugraiiadilaton}
	&&R + 4 D^2 \Phi - 4 (\del \Phi)^2 - {1\ov 12} H_3^2 = 0 \ .
	\eea
	The fluxes (a.k.a. Maxwell) equations read:
	\bea
	&& d\left(e^{-2\Phi}\star H_3\right) - F_2\wedge \star F_4 -  F_4\wedge F_4 = m \star F_2\ ,\\
	&& d \star F_2 + H_3\wedge \star F_4 =0 \ , \nonumber \\
	&&
	d \star F_4 + H_3\wedge  F_4 =0\ .
	\nonumber
	\label{natdintrosugraiiamaxwell}
	\eea
	
	There are some key quantities $N_{D_p}$ and $N_{NS5}$ associated to the forms $F_2,F_4$ and $H_3$:
	\bea \label{natdintrosugraiiapagecharges}
	\begin{aligned}
		&  N_{Dp} \big{|}_{\Pi_{8-p}}=\frac{1}{2\kappa_{10}^2 T_{Dp}} \int_{\Pi_{8-p}}  \left( \sum\limits_{i} F_i  \right) \wedge  e^{-B_2},      ~~~~~~         \left. N_{NS5} \right|_{\Pi_3} =  \frac{1}{2\kappa_{10}^2 T_{NS5}} \int_{\Pi_3} H_3 \ ,
		\\[5pt]
		& 2\kappa_{10}^2 =(2\pi)^{7} \alpha'^{4}  g_s^2,      ~~~~~~   T_{Dp} =\frac{1}{(2\pi)^{p} \alpha'^{\frac{p+1}{2}}  },  ~~~~~~ T_{NS5} = T_{D5} \ .
	\end{aligned}
	\eea
	where every $\Pi_d$ is a d-cycle in the geometry. These quantities are called Page charges and they measure the amount of D$-p$ or $NS5$ branes associated with the geometry, therefore providing vital information about the field theory dual \cite{Benini:2007gx}. \eqref{natdintrosugraiiapagecharges} are also valid for the IIB SUGRA fields.

	As explained in the 11D case, we are only interested in bosonic solutions, and most of them will be supersymmetric. To find the latter we introduce the IIA SUSY transformations for the fermions: dilatino $\lambda$ and gravitino $\psi_m$: 
	\begin{eqnarray}
	\delta_{\epsilon}\lambda &=& \left[ \frac{1}{2}\Gamma^m \partial_m \Phi + \frac{1}{4\cdot3!}H_{mnp}\Gamma^{mnp} \Gamma_{11} + \frac{e^{\Phi} }{8} \left( \frac{3}{2!} F_{mn} \Gamma^{mn} \Gamma_{11} -\frac{1}{4!} F_{mnpq}\Gamma^{mnpq} \right)  \right] \epsilon \ ,  \nonumber  \\
	\delta_{\epsilon}\psi_m &=& \left[  \nabla_m+ \frac{1}{4\cdot2!}H_{mnp}\Gamma^{np} \Gamma_{11}  - \frac{e^{\Phi} }{8} \left( \frac{1}{2} F_{np} \Gamma^{np} \Gamma_{11} +\frac{1}{4!} F_{npqr}\Gamma^{npqr} \right) \Gamma_m  \right] \epsilon \ ,
	\label{natdintrosugraiiavariations}
	\end{eqnarray}
	where $\Gamma_{11}$ is the product of all gamma matrices, and $m,n,p,q,r \in \{1,...,10\}$.\\
	Finally, it is possible to get massless IIA SUGRA as a consistent truncation of the 11D SUGRA; equivalently any massless IIA solution can be uplifted to eleven dimensions. The uplifted metric is given by: \footnote{We said before that M-theory contains IIA superstring. The present embedding of IIA supergravity into 11d is the corresponding fact for their low energy limits.}
	\begin{equation} \label{natdintrosugraiiametriclift}
	ds^2_{11} = e^{-\frac{2}{3} {\Phi}} \ ds^2_{IIA} \ + \ e^{\frac{4}{3} {\Phi}} \ \big(  dx_{11} + {C}_1  \big)^2 \ ,
	\end{equation}
	where by $x_{11}$ we denote the $11^{\textrm{th}}$ coordinate which corresponds to a $\mathrm{U(1)}$ isometry in the 11D theory upon which we compactify.

	The 11-dimensional geometry is supported by a 3-form potential $C^M_3$ which gives rise to the 4-form $F^M_4 = dC^M_3$. 
	This 3-form potential can be written in terms 
	of the 10-dimensional forms and the differential 
	of the $11^{\textrm{th}}$ coordinate as, 
	\begin{equation}
	C^M_3 = {C}_3 + {B}_2 \wedge d x_{11} \ .  \label{natdintrosugraiiaf4lift}
	\end{equation}
	The 3-form ${C}_3$ corresponds to the closed part of 
	the 10-dimensional RR form 
	${F}_4 = d{C}_3 -{H_3} \wedge {C}_1$.
	\subsubsection{Type IIB supergravity} \label{natdintroiib}
	There exists also a chiral maximal supergravity in ten dimensions, type IIB supergravity. It can be related with IIA using T-duality, that will be also explained in section \ref{natdintrotduality}.

	The bosonic fields are similar to those of IIA: $g_{\m\n}$, $\Phi$, $B_2$, $\chi,A_2, A_4$, where $g$ is a ten-dimensional metric, $B_2$ is the Kalb Ramond 2-form, $\Phi$ is the dilaton (again a real scalar),  $\chi, A_2, A_4$ are 0,2,4-differential forms serving as potentials of the following field strengths:
	\begin{eqnarray}
	H_3 &=&dB_2~, \\
	F_1 &=&d\chi ~,   \\
	F_3 &=&dA_2+ \chi H_3  ~, \\
	F_5 &=&dA_4 + H_3 \wedge A_2 ~.
	\label{natdintrosugraiibfieldstrengths}
	\end{eqnarray} 
	The Bianchi identities for the RR forms are given by:
	\bea
	d H_3 = 0 \ , \quad  dF_1 =  0 \ , \quad dF_3 = H_3\wedge F_1 \ ,  \quad dF_5 = H_3 \wedge F_3. \label{natdintrosugraiibbianchis}
	\eea
	The EoM can be derived from the following action:
	$$S_{IIB}= \frac{1}{2 \kappa_{10}^2} \int d\xi^{10} \sqrt{-g} \Big\{ e^{-2 \Phi} \left( R + 4 \partial_{\mu} \Phi  \partial^{\mu} \Phi - \frac{1}{2 \cdot 3!} H_3^2 \right) - \frac{1}{2} \partial_{\mu} \chi  \partial^{\mu} \chi - \frac{1}{2 \cdot 3!} F_3^2- \frac{1}{4 \cdot 5!} F_5^2 \Big\}  $$
	\begin{equation}
	+\frac{1}{4 \kappa_{10}^2} \int dA_2 \wedge H_3 \wedge ( A_4 + \frac{1}{2} B_2 \wedge A_2)  ~,
	\label{natdintrosugraiibaction}
	\end{equation}
	where $2 \kappa_{10}^2=(2 \pi)^7 \alpha'^4 g_s^2$. The equations of motion obtained from this action are supplemented by a further condition, which states that the $F_5$ form is self-dual:
	\begin{equation}
	F_5=*F_5 ~.
	\label{natdintrosugraiibselfdualityf5}
	\end{equation}
	Like in type IIA, the Page charges are given by eq. (\ref{natdintrosugraiiapagecharges}). The difference in this case is that the $N_{D_p}$ will have odd $p$.

	The SUSY transformations for the dilatino $\lambda$ and the gravitino $\psi_m$ for type IIB supergravity in string frame are \cite{Itsios:2013wd}:
	\begin{eqnarray}
	\delta_{\epsilon}\lambda &=& \left[ \frac{1}{2}\Gamma^m \partial_m \Phi + \frac{1}{4\cdot3!}H_{mnp}\Gamma^{mnp} \tau_3 - \frac{e^{\Phi} }{2}  F_m \Gamma^m(i\tau_2) -\frac{e^{\Phi}}{4\cdot3!} F_{mnp}\Gamma^{mnp} \tau_1 \right] \epsilon \ ,  \\
	\delta_{\epsilon}\psi_m &=& \left[  \nabla_m+ \frac{1}{4\cdot2!}H_{mnp}\Gamma^{np} \tau_3 + \frac{e^{\Phi}}{8} \left( F_n \Gamma^n (i\tau_2)+\frac{1}{3!}F_{npq}\Gamma^{npq} \tau_1 + \frac{1}{2\cdot5!}F_{npqrt}\Gamma^{npqrt} (i\tau_2)  \right) \Gamma_m \right] \epsilon \ ,
	\nonumber
	\label{natdintrosugraiibvariations}
	\end{eqnarray}
	where $\tau_i \ , \; i = 1,2,3$, are the Pauli matrices.

	\subsection{SUSY analysis}
	\label{natdappendixsusy2}
	
	\subsubsection{SUSY preserved by the twisted solutions}\label{susyvar}
	In this part of the appendix we write explicitly the variations of the dilatino and gravitino for the ansatz (2.1-2.4), for the 3 cases $\mathrm{H_2}$, $\mathrm{S^2}$ and $\mathrm{T^2}$.  The SUSY transformations for the dilatino $\lambda$ and the gravitino $\psi_m$ for type IIB supergravity in string frame are given by \eqref{natdintrosugraiibvariations}.

	Let us consider the $\mathrm{H_2}$ case in detail (the $\mathrm{S^2}$ case is obtained analogously). Recall that the vielbein is written in (\ref{vielbein00}).
	
	The dilatino variation vanishes identically, as the fields involved are vanishing. The $m=0$ component of the gravitino reads,
	\begin{equation}
	\delta_{\epsilon} \psi_0=\left[  \frac{A'}{2L} \Gamma_{04} - \frac{e^{-4U-V}}{2L} \Gamma_{04} \Gamma_{0123} i \tau_2 + \frac{e^{-2B-2U-V}}{16L} z \big(\Gamma_{014} - \Gamma_{239} \big) \big(   \Gamma_{78} - \Gamma_{56} \big) \Gamma_0 i\tau_2 \right] \epsilon \ .
	\label{gravitino0}
	\end{equation}
	First, we use the chiral projection of type IIB,
	\begin{equation}
	\Gamma_{11}\epsilon=\epsilon \ ,
	\label{gravitino00}
	\end{equation}
	where we define $\Gamma_{11}=\Gamma_{0123456789}$. We also impose the following projections (K\"ahler projections),
	\begin{equation}
	\Gamma_{56}\epsilon= - \Gamma_{78}\epsilon= - \Gamma_{49}\epsilon \ .
	\label{gravitino01}
	\end{equation}
	Then, expression (\ref{gravitino0}) simplifies to,
	\begin{equation}
	\delta_{\epsilon} \psi_0= \Gamma_{04}\left[ \frac{A'}{2L}  - \frac{e^{-V-4U}}{2L}+\frac{e^{-2B-2U-V}}{4L}z \Gamma_{0178} i \tau_2 \right]\epsilon \ .
	\label{gravitino02}
	\end{equation}
	We now impose the usual projection for the D3-brane,
	\begin{equation}
	\Gamma_{0123} \ i \tau_2 \epsilon= \epsilon \ ,
	\label{gravitino03}
	\end{equation}
	and also a further projection related to the presence of the twisting,
	\begin{equation}
	\Gamma_{23}\epsilon= \Gamma_{78}\epsilon \ .
	\label{gravitino04}
	\end{equation}
	Then, imposing that expression (\ref{gravitino02}) vanishes we obtain,
	\begin{equation}
	A'-e^{-V-4U}+\frac{z}{2} e^{-2B-2U-V}=0 \ .
	\label{gravitino05}
	\end{equation}
	For the component $m=1$ of the gravitino equation, we obtain that it is zero when we impose the projections  and equation (\ref{gravitino05}). For the component  $m=2$ we have,
	\begin{equation}
	\delta_{\epsilon} \psi_2=\left[  \frac{B'}{2L} \Gamma_{24} +\frac{e^{-2B+V}}{4L} z \Gamma_{39} - \frac{e^{-V-4U}}{2L} \Gamma_{24}\Gamma_{0123} i \tau_2 - \frac{ e^{-2B-2U-V}}{4L} z \Gamma_{24} \Gamma_{0178} i\tau_2 \right] \epsilon \ .
	\label{gravitino06}
	\end{equation}
	Combining projections (\ref{gravitino01}) and (\ref{gravitino04}) we get  $\Gamma_{39} \epsilon=-\Gamma_{24} \epsilon$. Then, (\ref{gravitino06}) gives the condition,
	\begin{equation}
	B'-e^{-V-4U}- \frac{z}{2}  e^{-2B-2U-V} -\frac{z}{2} e^{-2B+V}=0 \ .
	\label{gravitino07}
	\end{equation}
	For $m=3$, after imposing the projections and equation \eqref{gravitino07} we arrive at,
	\begin{equation}
	\delta_{\epsilon} \psi_3= - \frac{e^{-B}}{2L} \cot \alpha \ (1+3z) \Gamma_{23} \epsilon \ .
	\label{gravitino08}
	\end{equation}
	There are two contributions to this term, one coming from the curvature of the $\mathrm{H_2}$ (through the spin connection) and another coming from the twisting $A_1$. That is, here we explicitly see that the twisting is introduced to compensate the presence of the curvature, in such  a way that some SUSY can be still preserved. Then, we impose,
	\begin{equation}
	z=-\frac{1}{3} \ .
	\label{gravitino09}
	\end{equation}
	For $m=4$, the variation is,
	\begin{equation}
	\delta_{\epsilon} \psi_4= \frac{1}{L}\partial_r \epsilon - \frac{1}{2L} \left[ e^{-4U} + 2e^{-2B -2U-V} \right]  \epsilon \ .
	\end{equation}
	From the condition $\delta_{\epsilon} \psi_4=0$, we obtain a differential equation for $\epsilon$. Solving for it we arrive at the following form for the Killing spinor,
	\begin{equation}
	\epsilon= e^{1/2 \int (e^{-4U} + 2e^{-2B -2U-V} )  dr} \ \epsilon_0 \ ,
	\end{equation}
	where $\epsilon_0$ is spinor which is independent of the coordinate $r$. For $m=5,6,7,8$ the variations vanish as long as,
	\begin{equation}
	U'+e^{-V-4U}-e^{V-2U}=0 \ .
	\label{gravitino011}
	\end{equation}
	Finally, for $m=9$ the graviton variation vanishes if,
	\begin{equation}
	V'-3e^{-V}+2e^{V-2U}+e^{-V-4U}-\frac{z}{2} e^{-2B-2U-V} +\frac{z}{2}e^{-2B+V}=0 \ .
	\label{gravitino012}
	\end{equation}
	Summarizing, the variations of the dilatino and gravitino vanish if we impose the following projections on the Killing spinor,
	\begin{equation}
	\Gamma_{11} \epsilon = \epsilon \ , ~~~~~~ \Gamma_{56} \epsilon = - \Gamma_{78} \epsilon= - \Gamma_{49} \epsilon \ ,  ~~~~~~ \Gamma_{0123} \ i \tau_2 \epsilon = \epsilon \ , ~~~~~~  \Gamma_{23} \epsilon =\Gamma_{49} \epsilon \ ,
	\label{gravitino013}
	\end{equation}
	and the BPS equations (\ref{O17}), together with the condition $z=-1/3$.
	%
	%
	For the case of the 2-torus, if we focus on the $m=3$ component,
	\begin{equation}
	\delta_{\epsilon} \psi_3= \left[   -\frac{z e^{-2B+V}}{4L}\Gamma_{29}  -\frac{3 z e^{-B}}{2L} \ \alpha \ \Gamma_{78} + \frac{B'}{2L} \Gamma_{34}  - \frac{e^{-V-4U}}{2L} \Gamma_{34} - \frac{z e^{-2B-2U-V}}{4L} \Gamma_{34} \right] \epsilon \ ,
	\label{gravitino014}
	\end{equation}
	we see that there is one term depending on $\alpha$, due to the twisting. Contrary to the $\mathrm{H_2}$ and $\mathrm{S^2}$ cases, here there is no curvature term that could cancel it. This will force $z=0$, obtaining $A'=B'$, which does not permit an AdS$_3$ solution. 
	
	Finally, after all this analysis we deduce that the Killing spinor does not depend explicitly on the coordinates $(\theta_2,\phi_2,\psi)$ on which we perform the NATD transformation. In fact it only has a dependence on the coordinate $r$.
	
	\subsubsection{SUSY preserved by the NATD solutions}
	
	In the above subsection we calculated the amount of SUSY that is preserved by the type IIB supergravity solutions of the section \ref{natdtwistedsolutions} by examining the dilatino and the gravitino variations. Here we compute the portion of SUSY that is preserved by a supergravity solution after a NATD transformation following the argument of \cite{Sfetsos:2010uq}, which has been proven in \cite{Kelekci:2014ima}. According to this, one just has to check the vanishing of the Lie-Lorentz (or Kosmann) derivative \cite{Kosmann:1972kd} of the Killing spinor along the Killing vector that generates the isometry of the NATD transformation. More concretely, suppose that we want to transform a supergravity solution by performing a NATD transformation with respect to some isometry of the background that is generated by the Killing vector $k^\mu$. Then there is a simple criterion which states that if the Lie-Lorentz derivative of the Killing spinor along $k^\mu$ vanishes, then the transformed solution preserves the same amount of SUSY as the original solution. In the opposite scenario one has to impose more projection conditions on the Killing spinor in order to make the Lie-Lorentz derivative vanish. Thus in that case the dual background preserves less supersymmetry than the original one.
	
	We recall that given a Killing vector $k^\mu$ the Lie-Lorentz derivative on a spinor $\epsilon$ along $k^\mu$ maps the spinor $\epsilon$ to another spinor and is defined as,
	\begin{equation}
	\label{KosmannDer}
	\mathcal{L}_{k} \epsilon =  k^\mu D_\mu \epsilon + \frac{1}{4} \big( \nabla_\mu k_\nu \big) \Gamma^{\mu\nu} \epsilon =        k^\m D_\m \epsilon + \frac{1}{8} (dk)_{\m\n} \Gamma^{\m\n} \epsilon \ ,
	\end{equation}
	where $D_\m \epsilon = \partial_\m \epsilon + \frac{1}{4} \omega_{\m\r\s} \Gamma^{\r\s} \epsilon$. For further details about the Lie-Lorentz derivative we urge the interested reader to consult \cite{Ortin:2002qb}.  
	
	In this chapter we constructed type IIA supergravity solutions by applying a NATD transformation with respect to the $\mathrm{SU(2)}$ isometry of the original backgrounds that corresponds to the directions $(\th_2,\phi_2,\psi)$. The non-vanishing components of the associated Killing vectors are,
	\begin{equation}
	\label{KillingV}
	\begin{array}{lll}
	k_{(1)}^{\th_2} = \sin\phi_2 \ , & k_{(1)}^{\phi_2} = \cot\th_2 \cos\phi_2 \ , & k_{(1)}^{\psi} = - \frac{\cos\phi_2}{\sin\th_2} \ ,
	\\[10pt]
	k_{(2)}^{\th_2} = \cos\phi_2 \ , & k_{(2)}^{\phi_2} = -\cot\th_2 \sin\phi_2 \ , & k_{(2)}^{\psi} = \frac{\sin\phi_2}{\sin\th_2} \ ,
	\\[10pt]
	k_{(3)}^{\phi_2} = 1 \ .  & \textrm{}   & \textrm{}
	\end{array}
	\end{equation}
	In what follows we will compute the Lie-Lorentz derivative along the three Killing vectors $(k_{(1)},k_{(2)},k_{(3)})$ using the geometries of the sections \ref{natdtwistedsolutions} and \ref{natdDG}. It turns out that in all cases the Lie-Lorentz derivative vanishes without the requirement of imposing further projections on the Killing spinor. This means that the new solutions that we found using the technique of NATD preserve the same SUSY as the original solutions.
	\newline
	
	Let us now compute the Lie-Lorentz derivative along the Killing vector \eqref{KillingV} for the twisted geometries that are described by the formulas \eqref{NN02}-\eqref{vielbein00}. In the previous section, which deals with the supersymmetry of the starting solutions, we mentioned that the Killing spinor does not depend on the isometry coordinates $(\th_2,\phi_2,\psi)$. This means that the first term in \eqref{KosmannDer} reduces to,
	\begin{equation}
	k_{(i)}^\m D_\m \epsilon = \frac{1}{4} \ \omega_{\m\r\s} \ k_{(i)}^\m \ \Gamma^{\r\s} \epsilon \ , \quad i=1,2,3 \ .
	\end{equation}
	Hence for each of the three Killing vectors we find,

	\bea
		\label{KosmannTwisted1}
	 k_{(1)}^\m D_\m \epsilon \eq \frac{z}{12} \ e^{2V-2B} \ \cos\phi_2 \sin\theta_2 \Gamma^{23} \epsilon - \frac{e^{2V-2U}}{6} \ \cos\phi_2 \sin\theta_2 \big(\Gamma^{56} - \Gamma'^{78} \big)  \epsilon	\\
	& &- \frac{1}{2} \ \cos\phi_2 \sin\theta_2 \Gamma'^{78} \epsilon + \frac{e^{V-U}}{2 \sqrt{6}} \ \big(\cos\th_2 \cos\phi_2 \Gamma'^8 - \sin\phi_2 \Gamma'^7 \big) \Gamma^9 \epsilon 	\nn\\
	& &+ \frac{e^U \ U'}{2 \sqrt{6}} \big( \cos\theta_2 \cos\phi_2 \Gamma'^7 + \sin\phi_2 \Gamma'^8 \big) \Gamma^4 \epsilon + \frac{e^V \ V'}{6} \sin\theta_2 \cos\phi_2 \Gamma^{49} \epsilon \ ,\nn
	\eea
\bea
	 k_{(2)}^\m D_\m \epsilon &=& -\frac{z}{12} \ e^{2V-2B} \ \sin\phi_2 \sin\theta_2 \Gamma^{23} \epsilon + \frac{e^{2V-2U}}{6} \ \sin\phi_2 \sin\theta_2 \big( \Gamma^{56} - \Gamma'^{78} \big) \epsilon \\ 
	& &+ \frac{1}{2} \ \sin\phi_2 \sin\theta_2 \Gamma'^{78} \epsilon - \frac{e^{V-U}}{2 \sqrt{6}} \ \big( \cos\th_2 \sin\phi_2 \Gamma'^8 + \cos\phi_2 \Gamma'^7 \big) \Gamma^9 \epsilon \nn \\
	&&  - \frac{e^U \ U'}{2 \sqrt{6}} \big( \cos\theta_2 \sin\phi_2 \Gamma'^7 - \cos\phi_2 \Gamma'^8 \big) \Gamma^4 \epsilon - \frac{e^V \ V'}{6} \sin\theta_2 \sin\phi_2 \Gamma^{49} \epsilon \ , \nn
\eea	
\bea  k_{(3)}^\m D_\m \epsilon \eq -\frac{z}{12} \ e^{2V-2B} \ \cos\theta_2 \Gamma^{23} \epsilon + \frac{e^{2V-2U}}{6} \ \cos\theta_2 \big( \Gamma^{56} - \Gamma'^{78} \big) \epsilon \\
&& + \frac{1}{2} \ \cos\theta_2 \Gamma'^{78} \epsilon + \frac{e^{V-U}}{2 \sqrt{6}} \ \sin\th_2 \Gamma'^8 \Gamma^9 \epsilon - \frac{e^U \ U'}{2 \sqrt{6}} \sin\theta_2 \Gamma^4 \Gamma'^7 \epsilon  - \frac{e^V \ V'}{6} \cos\theta_2 \Gamma^{49} \epsilon \ .\nn
\eea
For convenience we have defined the rotated $\Gamma$-matrices,
	\begin{equation}
	\label{GammaRot}
	\Gamma'^7 = \cos\psi \ \Gamma^7 + \sin\psi \ \Gamma^8 \ , \quad \Gamma'^8 = -\sin\psi \ \Gamma^7 + \cos\psi \ \Gamma^8 \ .
	\end{equation}
	Let us now compute the 1-forms that are dual to the Killing vectors. What one has to do is to lower the index of the Killing vectors \eqref{KillingV} using the metric \eqref{NN02} which gives the following result,
	\begin{equation}
	\label{KosmannTwisted2}
	\begin{aligned}
	& k_{(1)} = - \frac{L^2}{3} e^{2V} \sin\theta_2 \cos\phi_2 \ \big(\eta + z A_1\big) + \frac{L^2}{6} e^{2U} \big(  \sin\phi_2 d\theta_2 + \sin\theta_2 \cos\theta_2 \cos\phi_2 d\phi_2  \big) \ ,
	\\[5pt]
	& k_{(2)} = \frac{L^2}{3} e^{2V} \sin\theta_2 \sin\phi_2 \ \big(\eta + z A_1\big) + \frac{L^2}{6} e^{2U} \big(  \cos\phi_2 d\theta_2 - \sin\theta_2 \cos\theta_2 \sin\phi_2 d\phi_2  \big) \ ,
	\\[5pt]
	& k_{(3)} = \frac{L^2}{3} e^{2V} \cos\theta_2 \ \big(\eta + z A_1\big) + \frac{L^2}{6} e^{2U} \sin^2\theta_2 d\phi_2 \ .
	\end{aligned}
	\end{equation}
	The second term of \eqref{KosmannDer} can be computed by acting with the exterior derivative on the above 1-forms and contracting the result with $\Gamma$-matrices. Notice that in order to compare with \eqref{KosmannTwisted1} one has to express the components of $dk_{(i)}, \,\, i=1,2,3$ using the flat frame \eqref{vielbein00}. Finally for the second term of \eqref{KosmannDer} we find,
	\begin{equation}
	\label{KosmannCheck}
	\frac{1}{8} (dk_{(i)})_{\m\n} \Gamma^{\m\n} \epsilon = - k_{(i)}^\m D_\m \epsilon, \,\, i=1,2,3,
	\end{equation}
	which means that the Lie-Lorentz derivative along the Killing vectors $k_{(i)}, \,\, i=1,2,3$ vanishes.
	\newline

	In the case of the Donos-Gauntlett geometry \eqref{metric-bef} we notice that all the necessary expressions are quite similar to those computed in the previous subsection. This is because the only significant difference between the line element of the twisted geometries and that of the Donos-Gauntlett geometry is just a fiber term. As in the previous case, the Killing spinor does not depend on the isometry coordinates $(\theta_2,\phi_2,\psi)$. This implies that the derivative term $k^\m \partial_\m \epsilon$ in eq. \eqref{KosmannDer} has no contribution to the result. Then the first term of eq. \eqref{KosmannDer}, for each of the three Killing vectors, can be easily obtained from eq. \eqref{KosmannTwisted1} by setting $z=0$. Similarly, if we set $z=0$ into eq. \eqref{KosmannTwisted2} we find the 1-forms $k_{(1)}, k_{(2)}, k_{(3)}$ for the Donos-Gauntlett case. Once we know these 1-forms we can follow the same prescription as in the previous subsection and compute the second term of \eqref{KosmannDer} for each Killing vector. It happens again that this term, when computed for every Killing vector, is related to the first term by a minus sign and thus the Lie-Lorentz derivative vanishes without imposing further projections on the Killing spinor.
	\newpage
	
\end{subappendices}

\chapter{Holographic Ward identities in 1+1 QFT} \label{ward}
\newpage
The goal of the present chapter is the holographic description of Ward Identities in 1+1 QFTs. We start with a review of the concepts of symmetry breaking, Ward identities and Goldstone bosons in the field theory side \ref{introwardsbwi}. The peculiarities of 1+1 dimensional case are covered in section \ref{introward1p1}. In \ref{introwardscalarfield}, we will introduce the basics of holographic renormalization for a scalar, while the AdS/CFT description of symmetry breaking will be given in section \ref{wardintrohsb}. Finally, we start the presentation of the original results of the chapter, reaching the desired \hocy \wi in section \ref{wardward} and discussing them in \ref{warddiscussion}.

\section[\texorpdfstring{Symmetry breaking and Ward identities in QFT}{Symmetry breaking and Ward identities in QFT}]{Symmetry breaking and Ward identities in QFT. Goldstone theorem}  \label{introwardsbwi}
Symmetries play an absolutely fundamental role in High Energy Physics. Nevertheless, this symmetries might be broken below a certain energy scale, and therefore be absent in the experimental data. In this situations, we say that the symmetry is broken. Concerning our mathematical descriptions, the symmetry breaking can be explicit (in the action, Hamiltonian or EoM), spontaneous (violated by the vacuum) or both at the same time (concomitant). There are very important phenomenological consequences. 
In High Energy Physics, a prominent example is the Higgs mechanism\footnote{Indeed discovered nearly simultaneously by Higgs, Brout and Englert, and Guralnik, Hagen and Kibble.} \cite{Englert:1964et,Guralnik:1964eu,Higgs:1964pj,Higgs:1964ia,Higgs:1966ev} (actually inspired by a condensed matter related proposal \cite{Anderson:1963pc}). The discovery \cite{Aad:2012tfa,Chatrchyan:2012xdj} of the Higgs boson at CERN was recently awarded the Nobel Prize. The Higgs field gives mass to the gauge bosons $W$ and $Z$ without explicitly breaking gauge symmetry. Another very important example the is chiral symmetry breaking of QCD, which endows nucleons with mass \cite{Nambu:1961fr,Nambu:1961tp}. This work was also worthy of the Nobel prize. Another very investigated but still unobserved case is SUSY breaking. In condensed matter systems, the list of symmetry breaking instances is very large, and some examples include every crystal (spatial translation is broken), (anti)ferromagnets, superconductors and superfluids, to name a few. Indeed, the BCS theory of superconditivity, that relies on symmetry breaking, has also been worthy of Nobel Prize \cite{BCS}. The relevance and applicability of the concept is therefore clear. In the following we give an introduction of the very elementary concepts from the field theory side.

Consider an action invariant under some gauge group $G$. In classical Physics the Noether theorem asserts that every continuous symmetry will be associated with a conserved current. In the quantum setting, the analog result is given by the Ward identities, that will play also a central role when the symmetry is broken. We begin by briefly discussing the different kinds of symmetry breaking. The simplest possibility is the explicit breaking, in which we add a term to the action that is not invariant under $G$. For concreteness and also because is our case of interest, let us take $ G=\mathrm{U(1)} $:
\bea
S_{tot} \eq S_{inv} + \int d^dx\, \half m \o + c.c. ,
\eea
where $S_{inv}$ is invariant under $ \mathrm{U(1)} $ and for simplicity we consider $ \o $ to have unity charge, therefore it breaks the symmetry. $ m $ is the explicit breaking parameter.

The second possibility is called spontaneous symmetry breaking (SSB). It happens when the vacuum state $ |0\rangle $ is not invariant under all elements of $G$, or alternatively, it is not annihilated by all generators $ T_i $  of $ G $ :
\bea
T_i |0\rangle & \neq & 0 \qquad \mathrm{for\; some}\,i.
\eea 
For $ G=\mathrm{U(1)} $ this generically implies that:
\bea \label{introwardvdef}
\esp{\o} \eq v \neq 0,
\eea 
and $ v $ is called the vacuum expectation value of $ \o $, VEV for short. Despite $ \o $ being charged, we are going to take always real $ v $.\footnote{This can always be achieved with a gauge transformation} These two quantities $ m,v $ will appear extremely often: what we mean by explicit breaking is $ m \neq 0, v=0 $ and by spontaneous $ m=0,v \neq 0 $. Holographically, they will be related to source and VEV of the \sc field dual to $ \o $. \footnote{This is so for standard quantization of the scaar. In alternative quantization of the scalar, their roles are exchanged.}  The third possibility is when both $ m\neq 0, v \neq 0 $ and receives the name of  \csb.   

Now that we have introduced the first definitions of \sb, let us explain its effect on the particle spectrum. This is indeed the content of the Goldstone theorem from the 1960's. \cite{Goldstone:1962es}. It states that when there is spontaneous \sb, massless scalar particles will appear; they are called Goldstone bosons (\gb). There are as many GB as symmetry generators are broken. It must be emphasized that \textit{this result holds in $d$-dimensional QFT for $d\geq 2$}. The peculiarities of $1+1$ dimensional \sb\ are reviewed in section \ref{introward1p1}, as we want to describe them \hocy. When the breaking is concomitant with dominant spontaneous component, Goldstone scalars become slightly massive and then they are referred as pseudo-Goldstone bosons.

Let us relate the \sb\ and \gb\ with the \wi as promised above. The Ward identity is, by definition:
\bea \label{introwarddefinition}
\esp{\djx\, \im0} \eq m \esp{\imx\, \im0} + i \esp{\reo} \d^d(x),
\eea
where $ J^\mu $ is the current operator. The relations:
\begin{align}
	\begin{aligned}
	\label{introinterestingwi}
	\imoimo &= -i f(\Box)        \, \delta^d(x),\\
	\djimo  &= -i (m f(\Box) - v)\, \delta^d(x),
	\end{aligned}
\end{align}
follow from the Ward identity, and we will refer to them informally as the Ward identities. $ f(\Box) $ is an unknown function, but the same for both correlators. \footnote{Functions of $\Box$ can be understood as operatorial relations between functions, for example in one dimension: $A^{(4)}(x)=\Box^2 A(x)$ corresponds to $ f(x)=x^2$, and from it we infer that $\frac{A^{(4)}(x)}{A(x)}= \Box^2$. However, these notation becomes more understandable in Fourier space, where $(A^{(4)})(k)= k^4 A(k)$. The Fourier space is usually preferred for explicit computations of correlators.} We will sometimes keep the delta function $\delta^d (x)$ implicit. Furthermore, in two point functions the operator on the left is evaluated in $ x $ and the one on the right in $0$, unless stated otherwise, and often we will not write them. 
The main objective of the present chapter is to \hocy obtain \eqref{introinterestingwi} for 1+1 dimensions; the result can be found in section \ref{wardward}, eqs. \ref{wardwardeq}. 

The so called GMOR relations establish the mass of pseudo-\gb\ in the presence of small \esb. Although we will not use them, its (quantitative) reproduction by a bottom-up model is a remarkable achievement of \holo\ presented in \cite{Argurio:2015wgr}. We will heavily rely on their methods, and they will be reviewed in section \ref{introwardpreviouspaper}. 
\subsection{Symmetry breaking in 1+1 QFT. Absence of \ssb at finite  N } \label{introward1p1}
The Goldstone theorem is valid only for spacetime dimension $ d>2 $. In $ d=2 $ (1+1 CFT), it is crucial to distinguish finite $ N $ and (strictly infinite) \lN behaviour. At finite $N$, there are no Goldstone bosons \cite{Coleman:1973ci}. A similar situation happens in $ 2+1 $ dimensions with non-vanishing temperature \cite{Mermin:1966fe,Hohenberg:1967zz}. These two facts are collectively known as the Coleman-Mermin-Wagner theorem.

A simple physical intuition of the absence of \ssb can be found in \cite{Ma:1974tp}, and we quickly review it here. Consider a complex scalar field $\phi$, with cartesian components $ \sigma,\pi $:
\bea
\sigma \eq \rho \cos \th,\\
\pi    \eq \rho \sin \th. 
\eea
The action will be given by:
\bea
L = \half (\partial_\m \sigma)^2 + \half(\partial_\m \pi)^2 - \lambda^2 \th{^2}.
\eea
The last terms is introduced by hand to break the gauge symmetry and $ \l $ is taken to be small. This parameter $ \lambda $ will make more clear the limit of purely \ssb, and we will take it to zero at the end. Furthermore,  the spatial length of the system  is restricted with an IR cutoff $ L $, that will also be taken to infinity at the end of the computation.\footnote{Otherwise the results present IR divergences.}

Taking into account these limits, we say that the $\mathrm{U(1)}$ symmetry is spontaneously broken iff:
\bea
\lim\limits_{\lambda \rightarrow 0} \lim\limits_{L \rightarrow 0}\esp{\sigma(x)} &\neq& 0.
\eea 
The order of limits is important, and if taken in the opposite order the VEV will always vanish.

It is convenient to introduce the polar coordinates of the field $ \sigma = \rho \cos \th , \pi = \rho \sin \th$ to focus on the phase. With that purpose we fix the value of the modulus $ \rho=\rho_0 $:  \footnote{The validity of such approximation is also discussed in \cite{Ma:1974tp}.}
\bea \label{introwardmaaction}
L \eq \half \rho_0^2\, (\partial_\m \th)^2 -\half \lambda^2 \theta^2. 
\eea 
We are therefore left with a free field $ \th $. The corresponding mode expansion is well known:
\bea
\th(x,t) \eq \frac{1}{\rho_0 \sqrt{L}} \sum_k \frac{1}{2 \sqrt{\omega_k}} \prt{a_k e^{i(\omega t - k x)}+ a^\dagger_k e^{-i(\omega t - k x)}},
\eea

As the field is free, we can explicitly compute the 2pt. \corr:
\bea \label{introward2ptdivergence}
c(x)&:=& \esp{\th(x) \th(0)} =   \frac{1}{2\pi\rho_0^2} \int_0^\infty \frac{d ke^{ikx}}{2(k^2+\lambda^2)^\half} 
\eea
that diverges in the IR as $ \lambda \rightarrow 0 $, the limit of vanishing explicit breaking. We have implicitly taken the $ L\rightarrow \infty $ limit. \footnote{The integral is also UV divergent, but the Coleman theorem is essentially related to IR behaviour (\textit{cf.} \cite{Ma:1974tp}) .}  Therefore the fluctuations become infinitely large in 2d for the free field. With it, we arrive at our desired result for the VEV:
\bea
\esp{e^{i\th(x)}} \rvert_{t=0} \eq \lim\limits_{\lambda\rightarrow 0} e^{-c(0)/2} = 0
\eea
Using \eqref{introward2ptdivergence}, $ \langle e^{i \th(x)}\rangle = 0 $ in the limit of $ \lambda \rightarrow 0 $, i.e. no explicit breaking. As a consequence, both $ \esp{\cos\th}, \esp{\sin \th} $ vanish, so do $ \esp{\rho_0 \cos\th},\esp{\rho_0 \sin\th} $ and therefore the VEV of $ \phi $, $ \esp{\phi} $, vanishes too. This is essentially worded saying that quantum fluctuations erase the VEV in 2d. Notice that it only happens in $ 2d $ as in higher dimensions higher powers of $ k $ will appear in the numerator of \eqref{introward2ptdivergence} and the IR contribution is finite. A similar phenomenon happens in 2+1 thermal field theories.\cite{Mermin:1966fe,Hohenberg:1967zz}. 
 
Nevertheless, symmetry can be\textit{ spontaneously broken (\cite{Coleman:1974jh,Gross:1974jv,Witten:1978qu})  in the \lN limit }, because of the fluctuation suppression it produces. Actually, with non-zero temperature the situation is the same in $ d=3 $, as implied by the Mermin-Wagner theorem. The breaking of symmetry is possible only for strict \lN. This result was actually re-obtained \hocy in \cite{Anninos:2010sq}, using non-classical calculations on the gravity side to include $ 1/N $ effects. We shall not tackle such difficult computation, and instead we will find \hocy the presence of \gb in eq. \eqref{wardgbfound}, as expected for \lN. 

\section[\texorpdfstring{Introduction to A\MakeLowercase{d}S/CFT of symmetry breaking}{Introduction to AdS/CFT of symmetry breaking}]{Introduction to AdS/CFT of symmetry breaking} \label{introwardpreviouspaper}
In this section we will introduce a few notions of holographic renormalization and the AdS/CFT description symmetry breaking. Our original work is totally based on them.
\subsection{Holographic renormalization of a \sc field (standard  quantization)} \label{introwardscalarfield}  
In this subsection we quickly review the formalism of \hren using the example of a complex \sc field to compute 2 points \fns of an operator in the boundary theory. After that, we use the scalar to show an example of alternative quantization. This simple case contains most of the key concepts that will appear in our original work: correspondence between scalar and operator, gauge transformations, counterterms, near boundary expansion and source choice through alternative quantization.
The general goal of holographic renormalization is to compute n-point functions in the gauge theory side. It is based on the precise formulation of  AdS/CFT  relating the partition functions of the boundary and bulk theories. This is the so-called Witten prescription, that we will use in the saddle point approximation for gravity \cite{Witten:1998qj,Gubser:1998bc}:
\bea \label{introwardwittenprescription}
\exp \prt{i S_{ren}}[\r_0,\pi_0] \eq \Bigg \langle \int_{\partial AdS} \exp \prt{-\r_0 \reo - \pi_0 \imo}  \Bigg \rangle,\\
\implies  i S_{ren}[\ro,\po] \eq W[\ro,\po].  
\eea 
where $ W $ is the QFT generating functional. $ \ro, \pi_0 $ are the sources (to be more clearly defined later), and $ \partial {AdS} $ is the boundary of the asymptotically AdS space in the gravity side. \footnote{The on-shell action can be computed using Euclidean or Lorentzian signature; we proceed with the second.} Notice that both sides of the correspondence have the same sources $ \ro,\pi_0$.  If one computes $ S_{ren}[\r_0,\pi_0] $ with the appropriate renormalization, the $ O $ correlators can be computed taking functional derivatives on it:
\bea \label{introwardfunctionalderivatives}
\esp{\reo(x_1)} \eq {\frac{ \delta i S_{ren}}{\d i \rho_0(x_1)}}\bigg\rvert_{\r_0=\pi_0=0},\\
\esp{\reo(x_1)\, \reo(x_2)} \eq  {\frac{ \delta^2  i S_{ren}}{\d i \rho_0(x_1)\, \d i \rho_0(x_2)}}\bigg\rvert_{\r_0=\pi_0=0},\nn \\
\esp{\reo(x_1)...\, \reo(x_n)}  \eq  {\frac{ \delta^n i S_{ren}}{\d i \r_0(x_1)  \ ... \d i\r_0(x_n)}}\bigg\rvert_{\r_0=\pi_0=0}, \nn\\
\esp{\reo(x_1)\,\imo(x_2)} \eq  {\frac{ \delta^2 i S_{ren}}{\d i \r_0(x_1)  \  \d i\pi_0(x_2)}}\bigg\rvert_{\r_0=\pi_0=0}, \nn
\eea
In similar fashion one computes correlators involving only $ \imo $ and the ones involving both $ \reo$ and $ \imo $.

The main issue is how to compute the on-shell \ac to use \eqref{introwardwittenprescription}. We are going to ignore the backreaction of $ \phi $ on the geometry \footnote{This can be done in certain situations, check \cite{Skenderis:2002wp} for more details.}. The dimension will be fixed to $ d=3 $ field theory, therefore 4-dimensional bulk. 
The \ac then is just the free scalar in AdS. 
\bea  \label{introwardscalarac}
S \eq \int d^4x\ \sqrt{-g} \, \prt{-\partial_M \phi^*\, \partial^M \phi  + m_\phi^2 \phi^* \phi}
\eea
The $ m_\phi^2 $ can be negative, but not arbitrarily. There is a minimal physically acceptable value known as the BF bound \cite{Breitenlohner:1982jf}:
\bea
m^2_{BF} = -\frac{d^2}{4} < m_\phi^2 . 
\eea
The mass will determine the dimension of the operator $ \o $ dual to $ \phi $ as:
\bea
\Delta_{O_\phi} \eq \Delta_+ := \frac{d}{2} + \nu,\\
\nu \eq \sqrt{\frac{d^2}{4}+m_\phi^2}.
\eea
Besides, there is an upper bound given by unitarity of the \dft (unitarity bound), that also further restricts the possible values of $ m_\phi^2 $. Taking into account both bounds, we obtain a neat restriction on $ \nu $: 
\bea
\Delta_{\o} > \frac{d-2}{2} \; \mathrm{and}\; m_{BF}^2 < m_\phi^2 \implies 0< \nu < 1 . 
\eea
We are going to fix $ m_\phi^2 $ such that $ \nu=\half $. This value is well inside the BF and the unitarity bounds, and corresponds to the explicit calculations in \cite{Argurio:2015wgr}. 
It is easy to solve the EoM of $ \phi $ in pure AdS, as it turns out to be Bessel equation. 
For values of $ z $ close to 0, the result is the well-known near boundary expansion of an scalar:
\bea \label{introwardscalarexp}
\r \eq  \r_0\, z + \r_1\, z^2+ ...,\\
\pi \eq\pi_0\, z +\pi_1\, z^2+ ...,
\eea
where $ z $ is the radial coordinate we are going to use, and $ z \rightarrow 0 $ is the near boundary region. Notice that linear and quadratic powers of $ z $ appear only because we chose $ \nu=1/2 $. In general they will be $ \Delta_\pm = \frac{d}{2} \pm \n= 1,2$ where $ m $ is the mass of the scalar. $ \ro, \ru, \pi_0,\pi_1 $ are called the modes of $ \rho,\pi $ respectively. 
The next step is to evaluate the on-shell action using the near-\bdry expansion \eqref{introwardscalarexp}. As mentioned before, it is \uv divergent and needs a \uv cutoff. 
In order to make it finite, we need to add some counterterms $ S_{ct} $ that are \bt in the bulk theory and gauge invariant. With it the renormalized \ac is:
\bea \label{introwardsacalarct}
S_{ct} \eq \int_{z=\eps} d^3 x\, \sqrt{-\g} \phi \phi^* = \int_{z=\eps} d^3 x\, \frac{-1}{z^3}\prt{\rho^2+\pi^2},\\
S_{ren} \eq S_{reg}+S_{ct}= \int d^3x \, \half \prt{ \ru\ro + \pu\po } ,
\eea
where  $ \g $ is the pullback of the metric on the boundary.  
Using the renormalized \ac , we can compute the variations:  
\bea \label{introwardscalarstandardvariation}
\delta S_{ren} \eq \int d^3 x\, \, \prt{\ru \d \ro + \pu \d \po }.
\eea
We see that it demands $ \delta \ro = \delta \po = 0 $ on the boundary to have vanishing variation. When \qns fixed by the \vp are constant, all the sources should be constant too. When this happens we will say that the sources are consistent with the \vp. Furthermore, it is the only requirement we will impose on the sources. As $ \ro=\po $ constant in the boundary is consistent with the \vp, we can use them as a legitimate sources.
The higher order modes $ \ru,\pu $ have to be written as functions of the sources $ \ro,\po $. The EoMs and normalizability in the bulk impose relations between the modes, that we write generically as: 
\bea \label{introwardscalarfs}
\ru \eq f_\rho(\Box)\ro,\\
\pu \eq f_\pi (\Box) \po.
\eea
We can now rewrite the action $ S_{ren} $ as:
\bea
S_{ren} \eq \int d^3 x\, \half \prt{\ro\, f_\rho(\Box) \ro  + \po\, f_\pi (\Box) \po }.
\eea 
We finally compute the desired \hoc correlators with functional derivatives \eqref{introwardfunctionalderivatives}: 
\bea
\esp{\reo} \eq 0,\\
\esp{\reo\, \reo } \eq -i f_\rho(\Box),\\
\esp{\imo\, \imo } \eq -i f_\pi (\Box),\\
\esp{\reo\, \imo } \eq 0.
\eea
The first follows as no linear term in $ \ro $ appears, and the same happens in the last one. 

The only matter that remains  is the precise form of $ f_\rho, f_\pi $. The EoM have already been used; only remains to fix the integrations constant in such a way they solution is normalizable in the bulk. The result is:
\bea
\esp{\reo\,\reo} \eq i k,\\
\esp{\imo\,\imo} \eq i k,
\eea
where we are using the Fourier space representation of the correlators, and $ k= \sqrt{k^2} $. This is precisely the conformal result. 


\subsection*{Alternative quantization  of the scalar}

Eq. \eqref{introwardscalarstandardvariation} imposes to choose $ \ro,\po $ as a source. But we can make $ \ru,\pu $ sources if we add an appropriate \bt $S_{bt}$ to the \ac. Indeed:
\bea \label{introwardaltscalar}
S_{bt} \eq \int_{z=\eps} d^3x \sqrt{-\gamma}\ \prt{\phi^* z \partial_z \phi + h.c.},\\
\delta (S_{ren}+S_{bt}) \eq \int d^3 x \ \ro \d \ru + \po \d \pu, 
\eea
where we now must impose $ \d \ru  = \d \pu = 0 $ instead of $ \d \ro = \d \po = 0$ as we did in the case of \eqref{introwardscalarstandardvariation}. Therefore $ \rho_1,\pu $ will be our new sources. $ S_{bt} $ is sometimes called a Legendre term\footnote{In analogy with the Legendre transform applied for example in Mechanics or Thermodynamics. The necessity of a well defined variational principle and the addition of the necessary terms is totally similar to the discussion of section \ref{gbintrogbbt}. }.  \textit{Choosing subleading expansion modes as sources is called alternative quantization}, in opposition to the standard quantization of the previous section, that uses the leading modes as sources.

Different boundary terms can drastically change the holographic meaning of a given gravity solution. For example, in \ref{wardaltscalar} it changes from representing purely  explicit \sb to purely spontaneous. Furthermore, let's mention a relevant application introduced in \cite{Klebanov:1999tb}. When one uses standard quantization, the lowest dual operator dimension described by holography is $ \Delta_{O_\phi}= \frac{d}{2}$. Nevertheless, the unitarity bound of CFT is lower than $ d/2 $ and given by $ \Delta = \frac{d}{2} -1 $. It turns out that the range of dimensions $ \frac{d}{2}-1 < \Delta < \frac{d}{2}$ can be covered using alternative quantization.

A similar kind of \bt will change the source for the vector field in section \ref{wardalternative}, and this will be needed to describe \cc holographically in 2d.

These examples contain key elements that will appear in the original results of the chapter and that are also common to many holographic renormalizations:
\begin{enumerate}
	\item Scalar-operator correspondence,
	\item Relation of \sc mass and operator dimension, allowed dimensions and bounds,
	\item Necessity of gauge invariant combinations of modes,
	\item Counterterms,
	\item Determination of sources. 
\end{enumerate}
Of course there are several aspects left out. Among the most important are the role of backreaction (which is necessary for \hoc stress-energy tensor \cite{deHaro:2000vlm} and correlators involving $ T_{\mu\nu} $) and the systematic determination of counterterms \cite{Papadimitriou:2004ap}. 

\subsection[\texorpdfstring{Holographic U(1) \sb and analytic Goldstone bosons in AdS$_4$}{Holographic U(1) \sb and analytic Goldstone bosons in AdS_4}]{Holographic U(1) \sb and analytic Goldstone bosons in AdS$_\textbf 4$} \label{wardintrohsb}
The purpose of this section is to review how to obtain the \wi \eqref{introinterestingwi}  in $ d>2 $ via AdS/CFT. We will use the methods presented in \cite{Argurio:2015wgr}. \footnote{Some early AdS/CFT work on description of symmetries and their breaking can be found in  \cite{Klebanov:1999tb,Bianchi:2001de,Bianchi:2001kw}.}
The first step is to introduce the \hoc bottom-up action \footnote{Although it has been proven that it can be embedded in \sugra for some values of the parameters.}:
\bea
S=\int d^4\,x \sqrt{-g}\; \prt{\frac{-1}{4} F^{MN}F_{MN} - D_M \phi^* D^M \phi + 2 \phi^* \phi},
\eea
where $ F_{MN}=\partial_M A_N - \partial_N A_M $, $ D_M \phi = \partial_M \phi - i A_M \phi $. We are not going to consider the backreaction of $\phi,A_M$ on the geometry.\footnote{This is justified in \cite{Argurio:2015wgr}, page 9.}
This model is well known in the literature by the name of \hoc superconductor \cite{Hartnoll:2008vx,Hartnoll:2008kx}. We have chosen mass $ m_\phi^2 = -2 $, that corresponds to $ \nu =1/2 $.  It contains the simplest necessary fields to describe \ssb : A $\mathrm{U(1)}$ vector field $ A^M $ minimally coupled to a charged \sc $ \phi $. 

%

The explicit and spontaneous symmetry breaking are introduced through a finite background $ \phi_B $ \cite{Hartnoll:2008kx} in the scalar, around which the field fluctuates. We also decompose the vector in transversal and longitudinal part:
\bea
\phi   \eq \frac{1}{\sqrt{2}} \prt{\phi_B+\r +\pi},\\
\phi_B \eq m z + v z^2,\\
A_\m   \eq T_\m + \partial_\m L .
\eea
We make the standard gauge fixing $ A_z =0 $ (radial gauge). The parameters $ m,v $ in this background will be the same as the $ m,v $ introduced in \eqref{introwarddefinition} and \eqref{introwardvdef}. Using the EoMs for $ \rho,\pi,L $, we find the near UV expansions: \footnote{As stressed in \eqref{introwardscalarexp}, the powers appearing in $ \rho, \pi $ expansion are determined by dimension and scalar mass choice $ m_\phi^2 =-2$. In particular, other choices can lead to non-entire powers or even logarithms. }
\bea \label{introwardfieldsexp}
\rho \eq \ro z + \ru z^2 + ...,\\
\pi  \eq \po z + \pu z^2 + ...,\\
L    \eq \lo z + \lu z^2 + ... \,. 
\eea
We do not consider explicitly the transverse part $ T_\m $ as its EoM are completely decoupled and it is not the source of the desired \wi. The  counterterms and in general all the boundary terms that are added to the action are required to:
\begin{enumerate}
	\item Cancel all divergences. 
	\item Respect the symmetries of the \ac (for example gauge symmetry.)\footnote{If the renormalization can only be done with non-invariant terms, then the quantum theory will not preserve the symmetry. This is called an anomaly. } 
	\item Have a well defined variational principle, consistent with the desired sources.
\end{enumerate}
In this case, we can meet those criteria with the same counterterm \eqref{introwardsacalarct} as the free scalar. To be able to compute correlators we have to write expansion modes in terms of the sources. Notice that gauge transformations mix $ \rho,\pi $ at first order in fluctuations through $ \phi_B $:
\bea
\d \rho \eq - \alpha \pi ,\\
\d \pi  \eq   \alpha \phi_B + \alpha \rho,
\eea
where $ \lo \rightarrow \lo + \alpha $ is the infinitesimal gauge transformation. It is therefore necessary to mix $ \pi $ and $ A_\m $ to write gauge-invariant relations, namely: 
\bea
\pu \eq v l_0 + f(\Box) (\po - m l_0),
\eea
instead of $ \pu $ in terms only of $ \po $ like we did in \eqref{introwardscalarfs}.

To reproduce the \wi \eqref{introinterestingwi} in terms of one unknown function it is not necessary to solve the EoM of the \fl. We only need to solve them if we want to know the form of that unknown function. The only remaining ingredient is the guess of the coupling between sources and operators:
\bea
\int d^3 x\, \prt{\ro \reo + \po \imo - l_0\, \dmu J^\mu}.
\eea
Using it, we \hocy derive exactly the VEV and the desired \wi \eqref{introinterestingwi}:
\bea
\esp{\reo} \eq \frac{\d i S_{ren}}{\d i \rho_0} = v,\\
\imoimo \eq -i f(\Box),\\
\djimo  \eq -i m f(\Box) + i v,
\eea
Once the \fn $ f(\Box) $ is known analytically, which is possible in this model with the $ m_\phi^2 =-2 \iff \nu = 1/2  $ choice, one can study $ \langle \imo\, \imo \rangle$ and search for the Goldstone boson as the IR poles of it. For example, taking the results for $m=0$ \cite{Argurio:2015wgr} (purely spontaneous breaking):
\bea
\imoimo &\sim& -2 i v^{3/2} \frac{\Gamma[\frac{5}{4}]}{\Gamma[\frac{3}{4}]} \frac{1}{k^2}
\eea
signals the \gb.
The case of interest in 1+1 will follow the procedure of this section, except for one crucial difference: the constant term $ l_0 $ will no longer be the dominant term in the expansion of $L$, eq. \eqref{introwardfieldsexp}. For $ l_0 $ to become source, we need especial boundary terms as explained in section \ref{wardalternative}. 



\section[ Renormalization of scalar and U(1) gauge field in  \texorpdfstring{$\mathrm{A}\MakeLowercase{\mathrm{d}}\mathrm S_\mathrm{3}$}{Holographic renormalization of scalar and U(1) gauge field in AdS$_\textbf{3}$}]{Holographic renormalization of scalar and U(1) gauge field in AdS$_\textbf{3}$} 
 \label{wardrenormchargedscalar}
From now on until the end of the chapter we present our original results. We will consider a fixed AdS$_3$ background with a vector field and a complex scalar. This is the minimal bulk field content to describe a symmetry current and an operator charged under it. Later, holographic renormalization is performed in order to find the correct Ward identities describing symmetry breaking. This has a subtlety related to the fact that a vector in a three-dimensional bulk can have different boundary conditions~\cite{Marolf:2006nd}. We will see that coupling it to a charged scalar with a non-trivial profile singles out one boundary condition, the one that correctly corresponds to a conserved current in the boundary theory. Besides reproducing the Ward identities, including the situation where also explicit breaking is present, an analytically Goldstone boson is found in a specific toy example. Finally the implications and context of the result are discussed. 
\subsection[Preamble: a free vector alone in AdS$_3$]{Preamble: a free vector alone in AdS$_\textbf 3$} \label{wardfreevector}
The procedure of holographic renormalization exhibits many subtleties {for a vector} in 2+1 bulk dimensions. Most of them are related to the fact that a vector in AdS$_3$ has properties analogous to a scalar at the Breitenlohner-Freedman (BF) bound \cite{Breitenlohner:1982jf}. We then start the study of the peculiarities of holographic renormalization in two dimensions with a preliminary discussion of a free gauge field in AdS$_3$, before coupling it to matter and analyzing the physics of symmetry breaking.

As a precedent, the case of a vector in AdS$_3$ was discussed briefly in \cite{Marolf:2006nd} along with higher dimensions. It was stated that only a particular boundary condition led to normalizable fluctuations. It was also noted there that it can be useful to dualize the bulk vector into a bulk massless scalar. This approach was also followed in \cite{Faulkner:2012gt}, where different boundary conditions were selected, in the dual frame. 

{Here, we will stick to a vector in bulk AdS$_3$ and discuss how one has to perform renormalization when imposing 
	the boundary condition of \cite{Marolf:2006nd}. In the next sections we will show that a different boundary condition is needed for consistency with the derivation of the Ward identities.}

We take the following bulk action for a free Abelian gauge field:\footnote{{Being in three dimensions, one could include a Chern-Simons term for the vector (see for instance \cite{Andrade:2011sx} for a careful discussion in a similar perspective). Since our aim is to stay as close as possible to the higher dimensional cases, where such a term is not present, we will take here the minimalistic approach and set it to zero. This choice is of course protected by parity.}}
\begin{equation}	\label{Svec}
S=\int\!d^3x \sqrt{-g} \left( -\dfrac{1}{4}F^{MN}F_{MN} \right) \ ,
\end{equation}
where $F_{MN}$ is the usual electromagnetic field strength, and $g_{MN}$ is the AdS$_3$ metric in the Poincar\'e patch,
\begin{equation*}
g_{MN}dx^Mdx^N = \frac{1}{z^2}\left(dz^2-dt^2+dx^2\right) \ .
\end{equation*}
We choose the radial gauge~$A_z\!=\!0$, and we divide the remainder into transversal and longitudinal components,
\begin{equation}	\label{AT+iL}
A_{\mu}=T_{\mu}+\partial_{\mu}L \ , \quad \text{with }\ \partial_{\mu}T^{\mu}=0 \ ,
\end{equation}
so that the action becomes
\begin{equation}	\label{Strv}
S=-\int\!d^3x\; \frac{z}{2}\Big[-\partial_z L\Box\partial_z L +\partial_zT^{\mu}\partial_zT_{\mu} - T^\mu\Box T_\mu \Big] \ .
\end{equation}
{The action \eqref{Svec} leads to the following equations of motion:}
\begin{align}
&	\Box\partial_zL =0 \ , \phantom{\big]} \label{eqvecAz}\\
&	z\partial_z(z\partial_zL) =0  \ , \phantom{\big]} \label{eqvecL}\\
& 	z\partial_z(z\partial_zT_{\mu}) +z^2\Box T_{\mu} =0 \ . \phantom{\big]} \label{eqvecT}
\end{align}
From the last two equations we derive the asymptotic behaviors of the fields,
\begin{equation}	\label{asymvec}
\begin{aligned}
L = \ln\!z\,\tilde{l}_{0} +l_{0} +\:\ldots \ ,	\qquad
T^{\mu} = \ln\!z\,\tilde{t}^\mu_{0} +t^\mu_{0} +\:\ldots \ ,	
\end{aligned}
\end{equation}
while from the first one we can drop the term in the action \eqref{Strv} involving the longitudinal component
. {Note that the presence of the logarithmic terms entails that the constant terms can suffer from an ambiguity. We will deal later on with this issue.}

Then we can integrate by parts and use the equation of motion for $T_\mu$ to express the action as a boundary term:
\begin{equation}
S_{reg}=\frac{1}{2}\int_{z=\epsilon} d^{2}x\; T^{\mu}z\partial_zT_{\mu} = 
\frac{1}{2}\int_{z=\epsilon} d^{2}x\; \big(\ln\!z\,\tilde{t}_0 +t_0\big)\cdot\tilde{t}_0 \ ,
\end{equation}
which needs to be renormalized because of the logarithmic divergence.

If we want to write a counterterm that removes this divergence and is gauge invariant, we may build it out of the field strength, but we soon realize that we then have to make it non local. This turns out to be equivalent to a mass term, which is absolutely local, but gauge invariant only on-shell, by equation~\eqref{eqvecAz}. Indeed,\footnote{{A counterterm with a $1/\ln\!z$ prefactor is typically needed for scalars at the BF bound, see e.g.~\cite{Bianchi:2001kw}.}}
\begin{align}	
\label{Svec_ct}
S_{ct} &= 
-\dfrac{1}{4}\int\!d^2x\; \frac{\sqrt{-\gamma\,}}{\ln\!z\,\gamma^{\rho\sigma}\partial_\rho\partial_\sigma}\, \gamma^{\kappa\mu}\gamma^{\lambda\nu} F_{\kappa\lambda}F_{\mu\nu} = \nn 
&	= \nn
\dfrac{1}{2}\int\!d^2x\;\frac{1}{\ln\!z}\, T_\mu T^\mu = \dfrac{1}{2}\int\!d^2x\;\frac{\sqrt{-\gamma\,}}{\ln\!z}\ \gamma^{\mu\nu}A_\mu A_\nu = \\
&	=
\frac{1}{2}\int_{z=\epsilon}\!d^{2}x\; \big(\ln\!z\,\tilde{t}_0 +2t_0\big)\cdot\tilde{t}_0 \ , 
\end{align}
where $\gamma_{\mu\nu}$ is the induced metric on the two-dimensional boundary, and the identity in the second line holds indeed thanks to the constraint~\eqref{eqvecAz}.

With such counterterm the renormalized action, $S_{ren}\!=\!S_{reg}\!-\!S_{ct}$, reads
\begin{equation}	\label{SrenT}
S_{ren} = -\frac{1}{2}\int_{z=\epsilon} \!d^{2}x\;\; \tilde{t}_0 \cdot t_0 \ .
\end{equation}

The variational principle gives
\begin{equation}
\delta S_{reg} = \int_{z=\epsilon}\!d^{2}x\; \Big[\;
\delta T^{\mu}z\partial_{z}T_{\mu} -\delta L\Box z\partial_zL \Big] 
=\int_{z=\epsilon}\!d^{2}x\; 
\tilde{t}_{0}\cdot\big(\ln\!z\,\delta\tilde{t}_{0}+\delta t_{0}\big)\ ,
\end{equation} 
so that, taking into account the last line of \eqref{Svec_ct}, we have
\begin{equation}
\delta S_{ren}=\delta S_{reg}-\delta S_{ct} = -\int_{z=\epsilon}\!d^{2}x\; 
{t}_{0}\cdot \delta\tilde t_{0}\ .
\end{equation}
The source for the operator dual to $A^\mu$ would thus be $\tilde{t}_0^\mu$, in agreement with \cite{Marolf:2006nd}. Note that as the source is transverse, the dual operator enjoys a gauge symmetry and thus cannot be a conserved current, it would be a pure gauge field.

We stress that the counterterm we have to introduce in two boundary dimensions, which has the form of a mass term, does not have an equivalent in any higher dimensions. This might be reminiscent of the Schwinger model (see e.g.~\cite{Gross:1995bp} for a modern exposition),
{where the photon mass is also generated by exactly the same non-local term.} Note however that here we are dealing with a non-local counterterm, due to a non-local UV divergent term, while in two-dimensional QED the loop-generated mass of the photon is finite. 

In \cite{Marolf:2006nd} it is noted that the above boundary conditions for the vector are equivalent to the usual boundary conditions one would impose on the massless scalar that is equivalent to (the transverse part of) the vector by bulk duality:
\begin{equation}
\partial_L\vartheta\!=\!\sqrt{-g\,}g^{MR}g^{NS}\varepsilon_{LMN}\partial_R A_S\ .
\end{equation}
It is straightforward to see that the usual, local counterterm that one writes for $\vartheta$ corresponds to the non-local one found above. 

Profiting from this dual formulation, the authors of \cite{Faulkner:2012gt} have argued that in order to describe a conserved current in the boundary theory, one should impose mixed boundary conditions on $\vartheta$, which are interpolating between the ordinary and the alternative quantizations \cite{Klebanov:1999tb} (see also \cite{Minces:1999eg}).\footnote{
	We should note that when considering the holographic correspondence between string theory on AdS$_3$ and boundary CFT$_2$ as in \cite{Giveon:1998ns}, there is a natural prescription to describe CFT currents, {which are actually enhanced to Kac-Moody generators.} The techniques are however different from the ones employed here, in particular there is no renormalization.}
Below, we are going to see that coupling the vector to a scalar leaves us with the only choice of the alternative quantization for the vector. 
We consider now a holographic model for spontaneous symmetry breaking in a 1+1~dimensional boundary field theory. We thus consider the following action, of an Abelian gauge field coupled to a complex scalar in AdS$_3$:
\begin{equation}
S=\int\!d^3x \sqrt{-g} \left[ -\dfrac{1}{4}F^{MN}F_{MN}-g^{MN}(D_{M}\phi)^{*}D_{N}\phi-m_{\phi}^{2}\,\phi^{*}\phi \right] \ , 
\label{S-U1}
\end{equation}
where
\begin{equation*}
\begin{aligned}
&	F_{MN} =	\partial_{M}A_{N}-\partial_{N}A_{M} \ ,\\
&	D_{M} =	\partial_{M}-iA_{M} \ .
\end{aligned}
\end{equation*}
From the equations of motion of a free scalar in AdS$_3$, one can infer the following scaling dimensions for the dual boundary operator/source:
\begin{equation}
\Delta_\pm = 1\pm\nu\ , \quad \text{with }\ \nu = \sqrt{1+m_{\phi}^2}\ .
\label{Delta+-}
\end{equation}
We then fluctuate the complex scalar around a fixed background,
\begin{equation}
\phi= \frac{\phi_B+\rho+i\pi}{\sqrt{2}}\ , \quad \text{with }\ \phi_B=m\,z^{1-\nu}+v\,z^{1+\nu} \ .
\end{equation}
We take $m$ and $v$ to be real for definiteness. When the scalar is in the ordinary quantization, the sub-leading piece (proportional to~$v$)  triggers a VEV for the real part of the dual boundary operator, and so  leads to spontaneous symmetry breaking of the global $\mathrm{U}(1)$. The leading piece (proportional to~$m$)  corresponds to explicit breaking of the symmetry, and we keep it different from zero for the moment, in order to use it as a sort of regulator. It will indeed turn out that we will  need it in some intermediary steps. 

Moreover, we fix the radial gauge $A_z=0$ and we conveniently split the gauge field into transverse and longitudinal components as in~\eqref{AT+iL}. We then derive from the variation of the action the following linearized equations of motion for the fluctuated fields:
\begin{align}
&	\Box z^2\partial_zL -\left(\phi_B\partial_z\pi-\pi\partial_z\phi_B\right) =0 \ , \phantom{\frac{}{|}} \label{eqAz}\\
&	z^2\partial_z^2T_{\mu} +z\partial_zT_{\mu} +z^2\Box T_{\mu} -\phi_B^2T_{\mu} =0 \ , \phantom{\frac{}{|}} \label{eqAt}\\
&	z^2\partial_z^2L +z\partial_zL -\phi_B^2 L +\phi_B\pi =0 \phantom{\frac{}{|}} \ , \label{eqL}\\
&	z^2\partial_{z}^{2}\rho -z\partial_{z}\rho -m^{2}\rho +z^2\Box\rho =0 \phantom{\frac{}{|}} \ , \label{eqrho}\\
&	z^2\partial_{z}^{2}\pi -z\partial_{z}\pi -m^{2}\pi +z^2\Box\pi -z^2\phi_B\Box L =0 \ . \label{eqpi}
\end{align}

As we have seen in the previous section, in three dimensions the vector field is at the BF bound, and indeed we have the following asymptotic expansions near the boundary:
\begin{equation} \label{wardl0definition}
T^{\mu}	= \ln\!z\,\tilde{t}_0^\mu +t_0^\mu +\:\ldots \ , \qquad
L = \ln\!z\,\tilde{l}_{0} +l_0 +\:\ldots \ .		
\end{equation}
The asymptotic expansion of the two scalar components depends on the value of the bulk mass. Let us set ourselves in the window between the BF bound~($m^2\!=\!-1$) and the ``massless bound''~($m^2\!=\!0$), and exclude the two extremal values, which would need to be treated separately since they entail logarithms. For all other values in this window, the scalar asymptotic expansions are logarithm-free. We thus have the following expansions,
\begin{equation}
\begin{aligned}
\rho &=	z^{1-\nu}\left(\rho_0+z^2\rho_1+\,\ldots\right) +z^{1+\nu}\left(\tilde{\rho}_0 +z^2\tilde{\rho}_1 +\,\ldots\right) \ , \phantom{\frac{}{|}}\\
\pi  &=	z^{1-\nu}\left(\pi_0+z^2\pi_1+\,\ldots\right) +z^{1+\nu}\left(\tilde{\pi}_0 +z^2\tilde{\pi}_1 +\,\ldots\right)\ ;  \phantom{\frac{|}{}}
\end{aligned}\ \quad \text{for }\ \nu \in\; (0,1) \ .
\end{equation}
We can now, integrating by parts and using the equations of motion, reduce the action to a boundary term, which reads
\begin{align}
S_{reg} & = \int_{z=\epsilon}\!\!d^{2}x\; \frac{1}{2} \bigg[\,
T^{\mu}z\partial_zT_{\mu} -\Box Lz\partial_zL +\frac{1}{z}\left(\pi\partial_z\pi+\rho\partial_z\rho+2\rho\partial_z\phi_B\right) \bigg]\ .
\label{SregGenCase}
\end{align}
By using the asymptotic expansions we obtain
\bea 	\label{regActionAsymptoticGenCase}
S_{reg} \eq  \int_{z=\epsilon}\!\! d^{2}x\; \frac{1}{2} \Big[ \;
\big(\ln\!z\,\tilde{t}_0 +t_0\big)\cdot \tilde{t}_0 -\big(\ln\!z\,\tilde{l}_0 +l_0\big)\,\Box\tilde{l}_0\; + \nn\\
&+& \rho_0\,\Big( (1-\nu)\big(\rho_0+2m\big)z^{-2\nu} +2\tilde{\rho}_0 \Big) +2m(1-\nu)\,\tilde{\rho}_0 +2v(1+\nu)\,\rho_0 + \nn\\
&+ & \pi_0\,\Big((1-\nu)\pi_0\,z^{-2\nu} +2\tilde{\pi}_0\Big) \Big] \ . 
\eea
We see that the divergent pieces of the scalar sector can be removed by the usual counterterm (in which we subtract the background value)
\begin{align} \label{Sctmass}
S^{(m)}_{ct} &= 
\left(1-\nu\right)\int_{z=\epsilon}\! d^{2}x\,\sqrt{-\gamma\,}\;\left(\phi^*\phi-\frac{\phi_B^2}{2}\right) \\
&	= \frac{1}{2}\left(1-\nu\right)\int_{z=\epsilon}\! d^{2}x\,\sqrt{-\gamma\,}\; \Big[\rho^2 +2\phi_B\rho +\pi^2 \,\Big] \ , 	\nonumber
\end{align}\\
leaving only the logarithmic divergences of the vector sector:
\begin{align}	\label{Sreg-Sctm}
S_{reg}-S^{(m)}_{ct} = \frac{1}{2}\int\!d^{2}x\; \Big[&	
\big(\ln\!z\,\tilde{t}_0 +t_0\big)\cdot \tilde{t}_0 -\big(\ln\!z\,\tilde{l}_0 +l_0\big)\,\Box\tilde{l}_0  +2\nu\, \big(\rho_0\tilde{\rho}_0 +2v\,\rho_0 +\pi_0\tilde{\pi}_0\big)\Big]  . 
\end{align}
We would like to remove also these divergences and then express the renormalized action in terms of the sources only. To do so, we need to identify which are the sources. Normally the source is associated with the leading term in the small~$z$ expansion of the field, unless one performs alternative quantization. For the gauge vector field, in any higher dimension the leading piece corresponds to a constant term. In our case instead, due to the fact that the vector is at the BF bound and to the consequent presence of the logarithmic term, the leading term is no longer the constant one. Then we would be naively led to choose the logarithmic term as the source, as we indeed did in the previous section. 
For the transverse part this does not pose any particular problem, but for the longitudinal part, which in the present case does not disappear from the boundary action, it is more problematic. 

Note that the longitudinal part of the vector shifts under gauge transformations, which in the radial gauge~$A_z\!=\!0$ are constant in $z$. It is then the constant term in $L$ that shifts, in any boundary dimensions, including two. Thus in our case it is $l_0$ defined by \eqref{wardl0definition} that shifts under gauge transformations. In other words, it is the constant part of $A_\mu$ that has gauge transformations, and so should be the source of a boundary conserved current. Then it seems reasonable to try to alternatively quantize the vector, and move the source from the coefficient of the logarithm to the $z$-constant, gauge-dependent term.

\subsection{Ordinary quantization for the vector} \label{wardordynary}

In first place, however, let us renormalize the vector in ordinary quantization,  {i.e.~keeping the source to be $\tilde l_0$. The point of this subsection is to show that this approach does however lead to a flawed physical picture, and that a different choice has to be made.}

We then compute first the variation of the regularized action~\eqref{S-U1}, namely
\bea\label{deltaSreg}
\delta S_{reg}\eq \int_{z=\epsilon}\!d^{2}x\; \Big[\;
\delta T^{\mu}z\partial_{z}T_{\mu} -\delta L\Box z\partial_zL +\frac{1}{z}\left(\delta \pi\partial_z \pi+\delta\rho\partial_z\rho+\delta\rho\partial_z\phi_B\right) \Big] \nn \\
\eq \int_{z=\epsilon}\!d^{2}x\; \Big[\;
\tilde{t}_{0}\cdot\big(\ln\!z\,\delta\tilde{t}_{0}+\delta t_{0}\big) -\Box\tilde{l}_0\big(\ln\!z\,\delta\tilde{l}_0 +\delta l_0\big) \,+ \nn \\
\phantom{\Big|} 
&+&(1-\nu)\pi_{0}\big(z^{-2\nu}\delta\pi_0+\delta\tilde{\pi}_0\big) +(1+\nu)\tilde{\pi}_0\delta\pi_0 \,+ \nn \\
& 	+& (1-\nu)\big(\rho_{0}+m\big)\big(z^{-2\nu}\delta\rho_{0}+\delta\tilde{\rho}_{0}\big) +(1+\nu)\big(\tilde{\rho}_{0}+v\big)\delta\rho_{0} \,\Big]\ .
\eea
It is crucial here not to use the constraint~\eqref{eqAz}, which in components yields
\begin{equation}
\Box\tilde{l}_0 = 2\nu\left(m\tilde{\pi}_0-\,v\,\pi_0\right) \ . 		\label{boxtill}
\end{equation} 
This equation is relating the coefficient of the logarithm to the source and VEV of the fluctuating scalar $\pi$. The equations of motion can be used to express VEV's in term of sources, but since we do not know yet whether $\tilde{l}_0$ will be a source or not, we have to remain off-shell to check the variational principle.

We then vary the counterterm for the scalar divergences,
\bea
\delta S_{ct}^{(m)} \eq (1-\nu)\!\int_{z=\epsilon}\!d^{2}x\, \Big[\,
\Big( z^{-2\nu}\big(\rho_{0}+m\big)+\big(\tilde{\rho}_{0}+v\big) \Big)\delta\rho_0 +\big(\rho_{0}+m\big)\delta\tilde{\rho}_{0} \nn\\
&& \qquad \qquad \quad + \Big( z^{-2\nu}\pi_{0} +\tilde{\pi}_{0} \Big)\delta\pi_{0} +\pi_{0}\delta\tilde{\pi}_{0} \Big] \ ,
\eea
so that we get
\bea \label{deltaSreg-Sctm}
\delta(S_{reg}\!-\!S_{ct}^{(m)}) \eq \int_{z=\epsilon}\!d^{2}x\; \Big[\;
\tilde{t}_{0}\cdot\big(\ln \!z\delta\tilde{t}_{0}+\delta t_{0}\big) -\Box\tilde{l}_0\big(\ln\!z\,\delta\tilde{l}_0 +\delta l_0\big) \,+ \nn\\
&+& 2\nu\:\Big( \big(\tilde{\rho}_{0}+v\big)\delta\rho_0 +\tilde{\pi}_0\delta\pi_{0} \Big) \,\Big]\ .
\eea
Thus the scalar sources appear to be well-defined in the ordinary quantization, but of course we still have to renormalize the vector sector. A mass-like counterterm as the one of the previous section \eqref{Svec_ct} will not help in renormalizing the longitudinal component. We propose the following gauge invariant, local counterterm:
\bea\label{Sct0}
S_{ct}^{(0)} \eq \int_{z=\epsilon}\! d^{2}x\,\frac{\sqrt{-\gamma\,}\,\gamma^{\mu\nu}}{\ln\!z\;\phi_B^2}\: (D_\mu\phi)^*D_\nu\phi \\
\eq 
\frac{1}{2}\int_{z=\epsilon}\! d^{2}x\;\Big[ \left(\ln\!z\,\tilde{t}_0 +2t_0\right)\cdot\tilde{t}_0 -\Big(\ln\!z\,\tilde{l}_0 +2l_0 -\frac{2}{m}\pi_0\Big)\Box\tilde{l}_0 \,\Big]\ ,	\nonumber
\eea
whose variation is
\begin{align} \label{varSct0}
\delta S_{ct}^{(0)} = \int_{z=\epsilon}\!d^{2}x\;\Big[\:&
\big(\ln\!z\,\tilde{t}_0 +t_0\big)\cdot\delta\tilde{t}_0 +\tilde{t}_0\cdot\delta{t_0} \\
&
-\Box\tilde{l_0}\,\big(\ln\!z\,\delta\tilde{l}_0 +\delta{l_0}\big) -\Box{l_0}\delta\tilde{l_0} +\frac{1}{m}\left(\Box\pi_0\delta\tilde{l_0}+\Box\tilde{l_0}\delta\pi_0\right)\Big) \,\Big]\ . 	\nonumber
\end{align}
We notice that such counterterm (and its variation as well) is singular in the limit~$m\!\to\!0$, and so in the purely spontaneous case it would not do the job. Keeping instead $m\!\neq\!0$, we obtain
\begin{equation}
\delta\tilde{S}_{ren}=\int_{z=\epsilon}\!d^{2}x\: \Big[ -t_0\cdot\delta\tilde{t}_0 -\frac{\Box}{m}\big(\pi_0-ml_0\big)\delta\tilde{l_0} +\big(2\nu\tilde{\pi}_0-\frac{\Box}{m}\tilde{l_0}\big)\delta{\pi}_0 \; +2\nu\left(\tilde{\rho}_0+v\right)\delta\rho_0 \Big]\ .
\end{equation}
We see that in this way the variational principle is well defined (even if still singular in~$m\!\rightarrow\!0$), and in particular $\tilde{l}_0$ should be considered as the source. Then we had better interpret the constraint~\eqref{boxtill} as an expression for $\tilde{\pi}_0$ in terms of the sources. In this point of view the renormalized action is
\begin{equation}
\tilde{S}_{ren}=\frac{1}{2}\int_{z=\epsilon}\!d^{2}x\: \Big[ -\tilde{t}_0\cdot t_0 -\big(\pi_0-ml_0\big)\frac{\Box}{m}\tilde{l}_0 +2\nu\Big(\rho_0\tilde{\rho}_0+2v\rho_0+\frac{v}{m}\pi^2_0\Big) \Big]\ .
\end{equation}
Again we see that all the terms involving the source of the imaginary part of the dual scalar operator explode for $m\!=\!0$. No theory of spontaneous breaking can be extracted out of this, which is consistent with the fact that the VEV of the longitudinal component of the vector is gauge-dependent in this quantization. Moreover, in the purely spontaneous case the constraint coming from eq.~\eqref{eqAz} becomes
\begin{equation}
\Box\tilde{l}_0=-2\nu\,v\pi_0 \ ,
\end{equation}
which strengthens the idea that $\tilde{l}_0$ cannot be the source of the conserved current. Indeed, $\pi_0$ is the source for the imaginary part of the scalar, and so $\tilde{l}_0$ cannot be another source,\footnote{A similar situation happens in cascading solutions, as for instance in \cite{Bertolini:2015hua}.} and therefore we confirm that we are forced to do the alternative quantization on the vector. 

Let us then show how to alternatively quantize the vector field and put the source back to the gauge-dependent, $z$-constant term~$l_0$, as in any higher dimensions.

\subsection{Alternative quantization for the vector}\label{wardalternative}

We should then try to renormalize in a different way, such that we move the source to the constant term.
This is achieved by considering an additional counterterm, of the Legendre transform kind, such as the following one:
\begin{align}
{S}_{ct}^{(1)} &=
\frac{i}{2\sqrt{2}}\int_{z=\epsilon}\! d^{2}x\, \frac{\sqrt{-\gamma\,}}{\phi_B}\;\gamma^{\mu\nu} z\partial_zA_\mu \Big(D_\nu\phi-(D_\nu\phi)^*\Big)\ = \nn\\
&=
\frac12\int_{z=\epsilon}\! d^{2}x\, \Big[\: \big(\ln\!z\,\tilde{t}_0 +t_0\big)\cdot\tilde{t}_0 -\Big(\ln\!z\,\tilde{l}_0 -\frac{1}{m}\left(\pi_0-ml_0\right)\Big)\,\Box\tilde{l}_0 \,\Big]\ .
\label{Sct1}
\end{align}
Indeed its variation is 
\bea \label{varSct1}
\delta{S}_{ct}^{(1)} \eq \frac{1}{2}\int_{z=\epsilon}\!d^{2}x\;\Big[\:
\big(\ln\!z\,2\tilde{t}_0 +t_0\big)\cdot\delta\tilde{t}_0 +\tilde{t}_0\cdot \delta{t_0} \\
&-& \big(\ln\!z\,2\tilde{l}_0 +l_0\big)\,\delta\Box\tilde{l_0} -\Box\tilde{l_0}\delta{l_0} +\frac{1}{m}\left(\pi_0\delta\Box\tilde{l_0}+\Box\tilde{l_0}\delta\pi_0\right)\Big]\ , 	\nonumber
\eea
and, taking the expression~\eqref{deltaSreg-Sctm}, and the variation of the ordinary counterterm~\eqref{Sct0}, the combination
\bea 
\label{wardvarpl0source}
&& \delta S_{reg}-\delta S_{ct}^{(m)}+\delta S_{ct}^{(0)}-2\delta{S}_{ct}^{(1)} = \nn\\
&& \qquad \qquad \qquad \int_{z=\epsilon}\!d^{2}x\;\Big[\: \tilde{t}_0\cdot\delta{t_0} -\Box\tilde{l}_0\delta{l_0} +2\nu\:\Big( \big(\tilde{\rho}_{0}+v\big)\delta\rho_0 +\tilde{\pi}_0\delta\pi_{0} \Big) \,\Big]\ ,
\eea
yields the desired switch of the sources. Furthermore, the corresponding renormalized action reads
\begin{equation}\label{SrenAQ}
{S}_{ren}=\frac{1}{2}\int_{z=\epsilon} \!d^{2}x\; \Big[ t_0\cdot\tilde{t}_0 +2\nu\, \Big(\, \rho_0\tilde{\rho}_0 +2v\,\rho_0 +\pi_0\tilde{\pi}_0 -\big(m\tilde{\pi}_0-v\pi_0\big)l_0\Big)\Big] \ ,
\end{equation}
where we have used the constraint~\eqref{boxtill} to remove $\tilde{l}_0$, which is not a source anymore. Notice that both~\eqref{wardvarpl0source} and~\eqref{SrenAQ} are finite in the $m\!\to\!0$ limit.

This renormalized action is completely identical to those of higher space-time dimensions and gives the suitable Ward identities for a pseudo-Goldstone boson (see for instance \cite{Argurio:2015wgr} for the three-dimensional case). However, both counterterms we have used \eqref{Sct0} and \eqref{Sct1} are singular for $m\!=\!0$ (even if the final result is not), so our action cannot be renormalized in this way if we set $m\!=\!0$ from the beginning.

We realize however that there is another gauge invariant and local counterterm that yields the same action as $2{S}_{ct}^{(1)}-S_{ct}^{(0)}$ in one step, namely
\bea \label{warduniquectt}
S^{(2)}_{ct} \eq 
\frac{1}{2}\int_{z=\epsilon}\!d^{2}x\,\sqrt{-\gamma\,}\,\gamma^{\mu\nu} \ln\!z\big(z\partial_zA_\mu\big)\big(z\partial_zA_\nu\big) =
\frac{1}{2}\int_{z=\epsilon}\!d^{2}x \ln\!z\left({\tilde{t}_0}\cdot{\tilde{t}_0}-\tilde{l}_0\Box\tilde{l}_0\right) \ .
\eea
This counterterm, subtracted to~\eqref{Sreg-Sctm}, trivially gives the renormalized action~\eqref{SrenAQ}. The counterterm is then equivalent to the combination~$2{S}_{ct}^{(1)}-S_{ct}^{(0)}$, and is not singular when $m\!=\!0$. 

Note that this could be interpreted to mean that the limits $z\rightarrow0$ (high UV) and $m\rightarrow0$ (purely spontaneous breaking) do not commute at large $N$. In order to make explicit this non-commutativity of the two limits, we can write
\begin{equation*}
2S^{(1)}_{ct}-S^{(0)}_{ct} = S^{(2)}_{ct} +\frac{1}{2}\int_{z=\epsilon}\!d^{2}x\:\frac{1}{\ln\!z}\bigg[
2\Box{l_0}\,\frac{\pi_0+\tilde{\pi}_0z^{2\nu}}{m+vz^{2\nu}} 
-\Box\pi_0\,\frac{\pi_0+2\tilde{\pi}_0z^{2\nu}}{m^2+2mvz^{2\nu}+v^2z^{4\nu}} \bigg]\ .
\end{equation*}
We see immediately that if we take first the limit~$z\!\rightarrow\!0$ we get the desired counterterm~${S}_{ct}^{(2)}$, which is independent of~$m$ and so the subsequent $m\!\rightarrow\!0$ limit is ineffective; whereas if we take first the limit~$m\!\rightarrow\!0$, we have no singularities  thanks to $v\!\neq\!0$, but we have surviving divergences in~$z$ when we take the $z\!\rightarrow\!0$ limit afterwards, namely:
\begin{equation}
\Big[2S^{(1)}_{ct}-S^{(0)}_{ct}\Big]_{m=0} = S^{(2)}_{ct} +\frac{1}{2}\int_{z=\epsilon}\!d^{2}x\:\bigg[
\frac{2}{v}\Box{l_0}\,\pi_0z^{-2\nu} -\frac{1}{v^2}\Box\pi_0\,\big(\pi_0z^{-4\nu}+2\tilde{\pi}_0z^{-2\nu}\big) \bigg]\ .
\end{equation}
One could be tempted to interpret this as a signal of the impossibility of taking \mbox{$v\!\neq\!0$} and \mbox{$m\!=\!0$} at the same time, i.e.~no spontaneous symmetry breaking. However the counterterm \eqref{warduniquectt} is perfectly well-behaved on its own, and it can actually be written also for the vector alone, whereas this is not the case for the two intermediate steps~\mbox{(\eqref{Sct0}, \eqref{Sct1})}, which are actually even more singular when all the scalar background is taken to zero. We cannot thus exclude the counterterm~\eqref{warduniquectt}, and with it we have to allow for spontaneous symmetry breaking in two dimensions, confirming the expected holographic evasion of the Coleman theorem.

{The counterterm \eqref{warduniquectt} has an explicit $\ln\!z$ factor, leading to the possibility to add also a finite counterterm with a similar structure and an arbitrary prefactor. This is indeed what was analyzed in \cite{Faulkner:2012gt} in the dual frame, with the interpretation of the introduction of a double-trace current-current deformation and the consequence of a non-trivial RG flow. Here we note that such a finite counterterm would spoil the identification of $t^\mu_{0}$ and $l_0$ as sources, shifting them by an arbitrary amount linear in $t^\mu_{0}$ and $\tilde{l}_0$ respectively. In the following, we take the point of view that the ambiguity in the $\ln\!z$ has been fixed, and we have taken $t^\mu_{0}$ and $l_0$ to be our sources. It is indeed this prescription that allows us to find the expected Ward identities.}

One last comment about alternative quantization for the vector field is that, if one holds $t_0^\mu$ fixed and lets $\tilde{t}_0^\mu$ loose, then according to~\cite{Marolf:2006nd} the fluctuations are not normalizable. This seems the price to pay to describe in the boundary theory a proper conserved current, whose existence we have no reason to exclude for a two-dimensional CFT. In addition, let us say that while for scalars bulk non-normalizability is usually connected to boundary operators with dimension below the unitarity bound, in the present case we do not see what problematic scenario this non-normalizability would correspond to in the dual theory; on the contrary, we have shown that everything works as smoothly as in higher dimensions precisely when we choose the alternative quantization for the vector field. 

\section[\texorpdfstring{Ward identities of symmetry breaking in AdS$_3$/CFT$_2$}{Ward identities of symmetry breaking in AdS_3/CFT_2}]{Ward identities of symmetry breaking in AdS$\boldsymbol{_{3}}$/CFT$\boldsymbol{_{2}}$} \label{wardward}
Once we have obtained the renormalized action \eqref{SrenAQ}, then showing that the Ward identities are realized is systematic. It actually follows from the same arguments as in \cite{Argurio:2015wgr}. First we rewrite the action as
\bea
{S}_{ren} &=& \frac{1}{2}\int_{z=\epsilon}\!d^{2}x\; \Big[ \; + t_0\cdot\tilde{t}_0 
+2\nu\, \Big(\, \rho_0\tilde{\rho}_0 +v\left(2\rho_0+2\pi_0l_0-ml_0l_0\right) +\big(\pi_0-m l_0\big)\big(\tilde{\pi}_0-vl_0\big) \Big)\Big] \ .\nn
\eea
Then we remark that the equations of motions and gauge invariance dictate the relations between VEVs and sources to take the following form:
\begin{equation}\label{wardfpidefinition}
\tilde t_0^\mu = f_t(\Box)  t_0^\mu\ , 	\quad 
\tilde \rho_0 = f_\rho (\Box) \rho_0 \ , \quad 
\tilde{\pi}_0-vl_0 = f_\pi (\Box)\big(\pi_0-m l_0\big)\ ,
\end{equation}
where the $f$'s are typically non-local functions, obtained by solving the EOM with appropriate IR boundary  conditions (i.e.~in the deep bulk). 

Replacing in the action yields the generating functional for one- and two-point functions, depending explicitly on sources only:
\begin{align}\label{wardrenexpansion}
{S}_{ren}=\frac{1}{2}\int_{z=\epsilon}\!d^{2}x\; \Big[ &\; 
t_0\cdot f_t(\Box) t_0 +2\nu\, \Big(\, v\left(2\rho_0+2\pi_0l_0-m\,l_0l_0\right) + \\ 
& 
+\rho_0f_\rho (\Box) \rho_0 +\big(\pi_0-m l_0\big) f_\pi (\Box)\big(\pi_0-m l_0\big)\Big)\Big] \ . \nonumber
\end{align}
Using the usual dictionary, for instance
\begin{equation} \label{warddictionary}
\langle\ImO\rangle = \frac{\delta {S}_{ren} }{\delta \pi_0}\ ,\qquad 
\langle \partial^\mu J_\mu \rangle =- \frac{\delta {S}_{ren} }{\delta l_0}\ ,
\end{equation}
we get for the correlators most relevant  to the Ward identities
\begin{align}
	\begin{aligned} \label{wardwardeq} 
	\big\langle \ImO(x)\,\ImO(x') \big\rangle &= -i\,2\nu\: f_\pi (\Box)\;\delta(x-x')\ , \\
	\big\langle \partial_\mu J^\mu(x)\,\ImO(x') \big\rangle &= -i\,2\nu\: \big( m f_\pi(\Box) -v\big)\; \delta(x-x')\ .
	\end{aligned}
\end{align}
where $f_\pi$ is a function enforced by fluctuations EoM and IR decaying \bc, as defined in eq. \eqref{wardfpidefinition}; $\nu$ is a constant given by $\nu=\sqrt{1+m_\phi^2}$. This is our main result, and we can summarize the whole chapter as:
\begin{enumerate}
	\item  $ l_0 $ must be used as source of a boundary \cc.
	\item  One needs a \bdry term like \eqref{warduniquectt} such that: 1) divergences of the \ac are countered, 2) $ l_0 $ appears as a source  in the variational principle (\eqref{wardvarpl0source}) of the renormalized \ac . 
	\item After the previous two steps, the expected \wi are recovered in \eqref{wardwardeq}.
\end{enumerate} 
As in \cite{Argurio:2015wgr}, we can obtain directly the Goldstone boson pole in the purely spontaneous case from \eqref{wardrenexpansion} and \eqref{warddictionary}. In momentum space, relativistic invariance and the Ward identity force the mixed correlator to be
\begin{equation}
\big\langle J_\mu(k)\,\ImO(-k)\big\rangle = v\,\frac{k_\mu}{k^2}\ ,
\end{equation}
displaying the expected massless pole. Furthermore when turning on $m$, one can argue that $f_\pi$ has to have a pole with a mass square proportional to $m$. Hence also $f_\pi$ has a massless pole in the $m=0$ limit. We will not repeat these steps here because they are clearly independent of the dimension of space-time. The Coleman theorem kicks in only after one considers (perturbative) quantum corrections due to the massless particle. Clearly holography does not capture such quantum corrections, which we then assume to be suppressed by the large $N$ limit. 

In the following section, in order to cover all possibilities (namely, all scalar operator dimensions between $0$ and $2$), we will briefly perform alternative quantization also in the scalar sector. 
Moreover, this will allow us to work out an analytic expression for $f_\pi$ for a specific value of the dimension of the dual boundary operator.


\section[\texorpdfstring{Alternative quantization of the scalar}{Alternative quantization of the scalar 
	and Goldstone Boson}]{Alternative quantization of the scalar. Analytic Goldstone boson} \label{wardaltscalar}
Here, as we did for the vector field, the goal is to move the sources to the subleading terms for the scalar as well. That is, we are interested in considering $\rt, \pt$ as the sources. 

In order to change the boundary conditions we should consider a Legendre transformation of the scalar counterterm~\eqref{Sctmass}, that is
\begin{align}
\hat{S}^{(m)}_{ct}& = 
\frac12\int_{z=\epsilon } d^2x\,\sqrt{-\g\,}\Big(\phi^* z \partial_z \phi + \phi^* z \partial_z \phi -\phi_Bz\partial_z\phi_B\Big) \\
& =
\frac12\int_{z=\epsilon } d^2x\; \Big[(1-\nu)\big(\rho_0\left(\rho_0+2m\right) +\pi_0\pi_0\big)\, z^{-2\nu } +2\,\big(v\rho_0 + m\tilde{\rho}_0 +\rho_0\tilde{\rho}_0 +\pi_0\tilde{\pi}_0\big)\Big] \nonumber .
\end{align}
Then the following combination is free from scalar divergences,
\begin{align}
S_{reg}+S^{(m)}_{ct}-2\hat{S}^{(m)}_{ct} = \frac{1}{2}\int\!d^{2}x\; \Big[\; &
\big(\ln\!z\,\tilde{t}_0 +t_0\big)\cdot\tilde{t}_0 -\big(\ln\!z\,\tilde{l}_0 +l_0\big)\,\Box\tilde{l}_0 + \nn \\
&\qquad\qquad
-2\nu\, \big(\rho_0\tilde{\rho}_0 +2m\,\tilde{\rho}_0 +\pi_0\tilde{\pi}_0\big)\Big] \ ,
\end{align}
and $v$ and $m$ have opposite meanings with respect to~\eqref{Sreg-Sctm}. We can verify by the variational principle that indeed the sources and VEV's are switched. If we take the expression~\eqref{deltaSreg} and subtract the variation of the present counterterm, we obtain
\begin{align}\label{deltaSreg-Sctm2}
\delta(S_{reg}+S^{(m)}_{ct}-2\hat{S}^{(m)}_{ct}) = \int_{z=\epsilon}\!d^{2}x\; \Big[\;&
\tilde{t}_{0}\cdot\big(\ln \!z\delta\tilde{t}_{0}+\delta t_{0}\big) -\big(\ln\!z\,\delta\tilde{l}_0 +\delta l_0\big)\,\Box\tilde{l}_0 \: + \nn\\
&
-2\nu\:\Big( \big(\rho_0+m\big)\delta\tilde{\rho}_{0} +\pi_0\delta\tilde{\pi}_0 \Big) \,\Big]\ ,
\end{align}
as desired.

Then we use the counterterm~\eqref{warduniquectt} to remove the vector divergences as well, and we get the renormalized action where both the vector and the scalar are in the alternative quantization:
\begin{equation}
\hat{S}_{ren} = \frac{1}{2} \int_{z=\epsilon } d^2x\; \Big[\, t_0\cdot\tilde{t}_0 -2\n\,\Big( \rho_0\rt +2m\rt +\pi_0\pt +\left(m\pt-v\pi_0\right) l_0\Big) \Big]\ .
\end{equation}

We remark that, since now the purely spontaneous breaking occurs for $v\!=\!0$, the two counterterms~\mbox{\eqref{Sct0}, \eqref{Sct1}} are now well behaved for the purely spontaneous case, whereas they are singular for the purely explicit one. Since we do not expect any obstruction for explicit symmetry breaking specific to two dimensions, this confirms that the counterterm~\eqref{warduniquectt} is the correct one, while it is the ordinary quantization for the vector which is problematic.

If we now express the VEV's in terms of the gauge-invariant sources in the following way:
\begin{equation}
\tilde{t}_0^\mu = f_t(\Box) t_0^\mu\ , 	\quad 
\rho_0 = \tilde{f}_\rho(\Box) \tilde{\rho}_0 \ , \quad 
\pi_0-ml_0 = \tilde{f}_\pi(\Box)\big(\tilde{\pi}_0-vl_0\big)
\end{equation}
(where the $\tilde{f}$'s are just the reciprocals of the $f$'s), we can rewrite the renormalized action uniquely in terms of the sources,
\begin{align}\label{hatSrenfinal}
\hat{S}_{ren}=\frac{1}{2}\int_{z=\epsilon}\!d^{2}x\; \Big[ &\; 
t_0\cdot f_t(\Box)t_0 -2\nu\, \Big( m\,\big(2\tilde{\rho}_0+2\tilde{\pi}_0l_0-v\,l_0l_0\big) + \\ 
&\qquad\quad
+\tilde{\rho}_0\tilde{f}_\rho(\Box)\tilde{\rho}_0 +\big(\tilde{\pi}_0-vl_0\big)\tilde{f}_\pi(\Box)\big(\tilde{\pi}_0-vl_0\big)\Big)\Big] \ . \nonumber
\end{align}
From this renormalized action we can retrieve Ward identities that are completely equivalent to those in~\eqref{wardwardeq}, with inverted roles for $v$ and $m$ (and $\nu$ going into $-\nu$).

To conclude the discussion, we would like to provide an explicit expression for the two-point correlator of $\ImOt$, where the massless Goldstone pole should be found. For $v=0$, that in alternative quantization corresponds to purely spontaneous breaking, and $\nu=1/2$, corresponding to the dimension of the boundary operator equal to $1/2$, the equation of motion~\eqref{eqL} becomes
\begin{equation}\label{solvablenu}
M''(z) - \left(k^2+ {m^2}{z}^{-1}\right) M(z) =0 \ ,
\end{equation}
where $M=z\partial_zL$. This equation can be analytically solved, and, if we impose boundary conditions such that the solution is not exploding in the deep bulk, we obtain the following well-behaved function
\begin{equation}
M(z)= C\; z\, e^{-\sqrt{k^2\,}z}\: \sfU\bigg[1+\frac{m^2}{2\sqrt{k^2\,}},\,2;\,2\sqrt{k^2\,}\,z\bigg] \ ,
\end{equation}
where $\sfU[a,b;x]$ is the Tricomi's hypergeometric function.

From the constraint~\eqref{eqAz} we get
\begin{equation}
\tilde{\pi}_0= -\frac{1}{m} \left.k^2 M \right|_{z^0} \ ,
\end{equation}
where $\left.M\right|_{z^0}$ is the constant term in the small~$z$ expansion. Similarly, from the equation of motion~\eqref{eqL} we can express the gauge invariant combination involving $\pi_0$ in the following way:
\begin{equation}
\pi_0-ml_0 = -\frac{1}{m} \left. M' \right|_{z^0} \ .
\end{equation}
Then we can derive the final expression for the correlator:
\begin{align}
\big\langle\ImOt(k)\,\ImOt(-k) \big\rangle & = i\,\tilde{f_\pi}(k^2) = i\,\frac{\pi_0-ml_0}{\tilde{\pi}} \\
& =
-\frac{i}{k^2} \left[ \sqrt{k^2\,}-m^2\left( 2\gamma_{EM} +\ln\!\big(2\sqrt{k^2\,}\big) + \psi_{(2)}\bigg[1+\frac{m^2}{2\sqrt{k^2\,}}\bigg]\right)\right]\ , \nonumber
\end{align}
where $\gamma_{EM}$ is the Euler-Mascheroni constant, and $\psi_{(2)}[x]$ is the di-gamma function. Using the expansion $\psi_{(2)}[1+x]\simeq \ln (x) + 1/(2x) + {\mathcal O}(1/x^2)$ for large $x$, one verifies that both the linear term $\sim |k|$ and the logarithmic term $\sim \ln|k|$ in the numerator of the equation above cancel in the $k\to 0$ limit. In this way, the low energy behavior of this correlator exhibits the expected Goldstone massless pole, namely
\begin{equation} \label{wardgbfound}
\big\langle\ImOt\,\ImOt \big\rangle \approx i\, \frac{2m^2}{k^2}\: \big(\gamma_{EM} +\ln m\big) \ .
\end{equation}
{We have thus confirmed the presence of the Goldstone boson, in addition to the constraints imposed by the Ward identities.}

\section{Discussion} \label{warddiscussion}

We have verified from the holographic point of view that in the strict large $N$ limit spontaneous symmetry breaking can occur in two dimensions~\cite{Witten:1978qu}. Indeed, considering the minimal AdS$_3$/CFT$_2$ setup in which symmetry breaking can be produced, we have retrieved the canonical Ward identities \eqref{wardwardeq} as they appear in higher dimensions~\cite{Argurio:2015wgr}. Nevertheless, the way to get this result involves subtleties and peculiarities which are specific to two dimensions.  The most crucial one is that we have to renormalize the gauge field in the alternative quantization, if we want it to properly source a conserved current and then recover the correct Ward identities for the breaking of a global symmetry on the boundary.

{We can consider quantum corrections to the result that we have obtained, taking inspiration from~\cite{Anninos:2010sq}, and compute a bulk tadpole correction to the scalar profile. It would presumably reproduce the infrared divergence which is responsible for preventing the vacuum expectation value, mirroring a similar field theory computation. Such a quantum effect, by the holographic correspondence, is equivalent to a $1/N$ correction in the boundary theory.}

On the other hand, we can think of the question directly in field theory, considering the canonical example of a complex scalar field with a Mexican-hat potential, as in \cite{Ma:1974tp}. In $1\!+\!1$ dimensions, the large quantum fluctuations of the phase would prevent the selection of a specific ground state around the circle at the minimum of the potential. However, if we add an arbitrarily small (but finite) explicit breaking, this would select a particular ground state, and act as a regulator for the infra-red divergence, making such vacuum stable under quantum fluctuations. Hence, for explicit breaking parametrically smaller than the spontaneous one, we expect (even at finite $N$) a mode that is hierarchically lighter than the rest of the spectrum, and whose mass is linear in the explicit breaking parameter, in accordance with the renowned Gell-Mann--Oakes--Renner relation~\cite{GellMann:1968rz}. So, if there are no Goldstone bosons in two dimensions, there definitely {should be} pseudo-Goldstone bosons in two dimensions, {and we have just provided a holographic description of the latter.}

\chapter{Summary and conclusions} \label{sac}

In the present thesis we have the long term motivation of testing and extending the domain of applicability of the AdS/CFT correspondence. This goal has been made concrete in three different problems, which we summarize alongside with the main results in the following pages.

Quantum Theory describes our most essential principles of the very small scales, roughly ranging from molecular and condensed matter to elementary particles. Its success in explaining and predicting physical phenomena is simply extraordinary. The other great revolution of early XX century was the General theory of Relativity, that explains gravity but has profound conceptual implications for the rest of Physics. One can naively apply on it the principles of Quantum Mechanics seeking a Quantum description of Gravity but the resulting theory is not renormalizable. Indeed, finding a consistent Quantum Theory of gravity, and experimentally testing it, is one of the most relevant problems of current Fundamental Physics.

The most developed candidate to reach that goal is string theory. After more than 20 years of extensive work on it, the AdS/CFT correspondence (a.k.a. holography, holographic duality and gauge/gravity correspondence) was discovered by Maldacena in 1997. In its most frequent use, it states the equivalence of QFT at \lN and strong coupling with classical gravity theories in AdS. This means that it is possible to compute strongly coupled QFT quantities (typically very hard) using weakly coupled classical gravity. It is in the context of the correspondence where this thesis' research was done.

\section{AdS to dS phase transition in higher curvature gravity} \label{sacgb}
We have examined the role of higher curvature gravity corrections on thermal phase transitions between AdS and dS geometries. Apart from the natural gravitational interest, this research may be eventually helpful to clarify the holographic correspondence in the case of asymptotically dS geometries. 

Higher curvature corrections to gravity are studied for a variety of reasons: as string theory (or any other high energy gravity theory) low energy effective actions, to test their cosmological effects, from a purely classical gravitation point of view, to explore their holographic meaning, etcetera. Among those theories, Lovelock family have some special properties. We have focused on the simplest non-GR case of the Lovelock family, known as Lanczos-Gauss-Bonnet gravity. The potential holographic relevance of the problem is that we have a \textit{dynamical process} in which the initial state is asymptotically AdS while after some time a dS cosmological horizon is formed. 

Despite our emphasis in higher curvature theories, gravitational phase transitions can also happen in GR; for example, the well known Hawking-Page transition \cite{Hawking:1982dh}. Furthermore, it is certainly possible for a phase transition in GR to change spacetime asymptotics. A simple way is the transition between the different non-degenerated minima of a scalar potential.  \cite{Coleman:1980aw}. Normally, these transitions end in AdS vacuum, while in our case it ends in a dS one.

What is specific of higher curvature corrections is that the transitions can happen between asymptotics of different curvature radius \textit{even in the absence of matter} \cite{Camanho:2013uda}. This process is mediated by bubble nucleation. The interior of such bubbles contains a solution with the geometry of the final vacuum (dS in our case) and it is matched to an exterior solution with the initial asymptotics (AdS in our example) with appropriate junction conditions. When the thermalon -euclidean section of the static bubble- has lower free energy than the initial solution, bubble nucleation is triggered. The new configuration is dynamically unstable, thereby it will either collapse or expand. The last possibility will give rise to the desired change in the asymptotics. Let us remark that the metric is continuous at the junction but the derivatives are not; despite it such solution must be considered as a legitimate competing saddle in the euclidean path integral approximation to free energy \cite{Camanho:2013uda}.

Our main result is a phase transition from thermal AdS to the formation of dS cosmological horizon in Lanczos-Gauss-Bonnet gravity. The transition happens through the proliferation of bubbles hosting dS black holes in their interior. No matter fields are required to match both sides of the bubble due to the Lanczos-Gauss-Bonnet term. For other higher curvature corrections we generically expect a similar phenomenon.  The transition becomes disfavored from a maximum value of the higher curvature coupling $ \lambda $, although thermal fluctuations could trigger them given enough time. In previous studies of AdS to AdS transitions was favored beyond a critical temperature for all $\lambda$.


It is important to remind the limitations of our analysis. The main one is the Boulware-Deser instability of the initial AdS vacuum, meaning that the graviton field presents ghosts around that vacuum. Nevertheless it is easy to find higher order Lovelock theories having two different stable Boulware-Deser vacua, although there are many additional technical complications to compute the bubble free energy. Such analysis was carried out in fourth order Lovelock by the authors of \cite{Camanho:2013uda}, and the transition between two AdS healthy vacua was shown to take place. We also conjecture that the same kind of thermalon mediated phase transitions are generic in higher curvature theories, not exclusively in Lanczos-Gauss-Bonnet or Lovelock. 

There are other limitations deserving some comments. In \cite{Camanho:2014apa}, the quadratic theories of gravity have been found to display causality violation, rendering them inconsistent at least in this respect. Notably, it was found in \cite{DAppollonio:2015fly} that for certain string-brane scattering the massless modes approximation is inconsistent. However, if the \textit{full} stringy structure (in particular the Regge behaviour) is taken into account the causality violating effects found in \cite{Camanho:2014apa} are avoided. Nevertheless, the Lanczos-Gauss-Bonnet theories still remain acceptable models for many purposes, for example the study of hydrodynamics. Another point of concern, related to the previous one and that is actually present in part of our work, is that for small $ \lambda $ the curvature of one of the vacua goes as $ 1/\lambda $, and from a stringy point of view, curvature corrections of order higher than quadratic should be necessary.

The main open question concerning the transitions is clearly their potential holographic meaning, which remains unknown to us. A first step may be to analyze the observables of the static bubble configuration, for example the Wilson loop at large enough distances so that the string penetrates the bubble or even the horizon. A second stage would be the computation of similar observables during the expansion of the bubble, along the lines of, for example, \cite{Li:2013cja}. Finally, one may try to study the dS geometry in which the transition ends, following the research line of dS/CFT correspondence started by Strominger in 2001 \cite{Strominger:2001pn}.

\section[Deformation of KW CFT and new  AdS$_3$ backgrounds via NATD]{Deformation of KW CFT and new  AdS$_\textbf{3}$ backgrounds via non-Abelian T-duality} \label{sacnatd}
We found new solutions of Supergravity using non-Abelian T-duality, a generating technique based on string theory. Such new solutions represent supersymmetric RG flows from four to two dimensions, and we investigate the effects of the Non-Abelian T-duality on the holographic observables. The problem also presents a (super)gravitational motivation, as the generated solutions can fall outside known classifications, as it happens in \cite{Lozano:2015bra}. Apart from being interesting in themselves, the new field theories may help to understand the effect of non-Abelian duality on the string theory sigma model, its supergravity approximation and the interplay with holographic correspondence.

There are deformations of Klebanov-Witten (KW) background AdS$_5\times$T$^{1,1}$ flowing to an AdS$_3$ factor in the IR. We generated new examples applying non-Abelian T-duality on them. After it, we compared the holographic observables of the generated solutions with those of the previously known deformations. The new dual field theories seem to be related to long linear quiver gauge theories. This paragraph summarizes the whole research.

To be more detailed, we have considered two kinds of geometries on which NATD is applied. The first is given by the duals of some 4d QFT placed on a manifold $\Sigma_2$; in particular an $\mathcal N=1$ SUSY flow with $\Sigma_2=\mathrm{H_2}$ and non-SUSY fixed points with $\Sigma_2=\mathrm{H_2}, \mathrm{T^2}, \mathrm{S^2}$; we called them 'twisted solutions'. The second kind is the deformation of KW found by Donos and Gauntlett in \cite{Donos:2014eua}, that we refer to as the Donos-Gauntlett (DG) solution. This last \bg was chosen as seed of the generating technique essentially for three reasons: it is a deformation of KW flowing to AdS$_3$, it is supersymmetric and can be related by two Abelian T-dualities to AdS$_3\times$S$^3\times$S$^3\times$S$^1$. This last background is the near horizon limit of D1-D5 brane intersection and some discussion of its possible field theory dual can be found in \cite{Tong:2014yna}. 

The solutions generated that are T-dual to a flow are also RG flows from AdS$_5$ to an AdS$_3$ factor. They are new type IIA SUGRA backgrounds and preserve \susy and regularity. By construction very non-trivial fields and coordinate dependence appear. On one of the new geometries, we have also performed Abelian T-duality after the non-Abelian one. Furthermore we have lifted the IIA solutions to 11D SUGRA. The main limitation of the NATD procedure is that the generating technique does not give a prescription on the coordinate ranges for the new background, and therefore its global properties are not determined a priori. Nevertheless, recent progress on the subject has given a non-trivial interpretation of the unknown ranges based on the holographic field theory \cite{Lozano:2016kum}.

Concerning the observables dual to the new geometries, we have analyzed Page charges, c-function, entanglement entropy (EE) on a strip and a rectangular Wilson loop (WL) energy before and after applying \natd. The main results have been:
\begin{enumerate}
	\item The new background's dual field theories seems closely related to an infinitely long linear quiver gauge theory. The relation between central charge and Page charges is $c\sim N_{D6}^2 n^3$, instead of the standard quadratic in $ n $. Such dependence suggests a relation with long linear quivers, which would imply that $n$ is measuring the number of  D4- and NS5-branes. The picture that emerges is that of a 2-d CFT living on the intersection of D2- , D6- and NS5-branes, with induced D4 charge every time an NS-brane is crossed.
	
	\item Quark-antiquark Wilson loop energy as a function of distance is trivially invariant. c-function and EE in terms of the energy scale and strip width, respectively, are non-trivially invariant all along the RG flow up to a constant factor which depends on the range of dual coordinates.
\end{enumerate}
The relation to long linear quiver gauge theories was quickly sharpened in \cite{Macpherson:2015tka} and the effect of \natd on the \dft significantly clarified in \cite{Lozano:2016kum, Lozano:2016wrs,Lozano:2017ole}. The change of the dual field theory caused by the duality at the level of SUGRA does not mean automatically that NATD is not a duality of the full (perturbative) string theory, although it points to it, as explained in \cite{Itsios:2016ooc}. Its authors proposed as a motivation of their work to elucidate whether NATD is a string duality or not by the examination of its effects on the meson spectrum. They conjecture that the difference before and after NATD they found may be caused either by the presence of a boundary in the sigma-model field theory or due to finite $ N $ or finite 't Hooft coupling effects. Finally, they propose to directly study the string theory on the non-Abelian T-dual, specifically taking the pp-wave limit and using the formalism initiated in \cite{Berenstein:2002jq}. 

\section{Holographic Ward identities in 1+1 QFT} \label{sacward}
We have studied the extension of the holographic renormalization procedure to reproduce the field theoretic \wi of spontaneous symmetry breaking in a 1+1 holographic superconductor. The result could be of interest for applications to one-dimensional condensed matter systems. Another motivation is to know if the bottom up model we use behaves in physically sensible manner for this low dimension. A final goal is to explore what is specific of three dimensional bulk theory and what are its implications for the QFT side.

Spontaneous symmetry breaking (SSB) is important in High Energy Physics as well as in Condensed Matter. A key result about it is the Goldstone theorem: there is a massless boson (called Goldstone boson) for every given generator whose symmetry is spontaneously broken. The properties of \sb\ are reflected on the Ward identities: the one point function signals SSB and the two point spectrum encodes the Goldstone bosons. The powerful result of Goldstone does not apply to 1+1 QFT: at finite $N$ SSB and Goldstone bosons are not possible (Coleman theorem), although the restriction disappears at \lN. 

Holographically, \sb\ has been studied in several previous works to explore the \hoc\ dictionary and apply it to the less understood case of \sb\ at strong coupling. Our purpose is to \hocy\ derive the \wi\ corresponding to \sb in a 1+1 QFT. The action in the gravity side is that of the \hoc superconductor; we will break its $\mathrm{U(1)}$ gauge symmetry. In 2+1 bulk the \hoc description of \cc\ requires \aq\ for the $\mathrm{U(1)}$ vector field $A_\m$. This happens because a massless vector field saturates the BF bound in 2+1 dimensions; this is the particular way in which the peculiarities of three dimensional bulk are reflected on the Ward identities holographic computation. Once we applied alternative quantization, the \hdal field-theoretic \wi are correctly recovered. Furthermore, we found an analytic Goldstone boson pole for the spontaneous breaking case; this is not in contradiction with Coleman theorem since our use of holography corresponds to large $ N $. 

The result above is an extension of previous ones on Ward Identities and analytic Goldstone bosons to 1+1 CFT \cite{Argurio:2015wgr}. Concerning the goal of learning about the applicability of the correspondence, we can draw two lessons. The first is that we can interpret the \textit{quantitative} holographic reproduction of standard field-theoretic \wi in $ d=2 $ as further evidence of the broad applicability of AdS/CFT beyond the standard dimension $ d=4 $. Our result is far from being the first concerning $ d=2 $, but we have given additional evidence.

The second lesson coming from our study is on whether gravitational bottom-up models do have sensible field theories duals. It must be noted that the holographic superconductor model can be embedded into string theory \textit{for some values of the parameters} in  $d=3$ \cite{Gauntlett:2009dn} and $d=4$ \cite{Gubser:2009qm}. For a particular bottom-up case presented in \cite{Argurio:2015wgr}, it was possible to quantitatively reproduce GMOR relations and Goldstone boson poles. In our case, it is difficult to assure with total certainty that our action cannot be embedded in SUGRA, although it seems unlikely. Despite it, the quantitative predictions for the Goldstone boson pole were quantitatively correct.

We have investigated gravitational phase transitions, non-Abelian T-duality and symmetry breaking in the context of AdS/CFT to test and extend its applicability. All of the previous topics remain worthy of further investigation on the open questions explained above. We also want to mention some additional lines of research. For example the generation of new solutions, in particular time dependent ones, via combination of Abelian and non-Abelian dualities, and possibly other generating techniques. A particularly interesting project is to apply the Abelian transformation to the theories found in \cite{Marques:2015vua}. This work is in progress, and the result about invariance of BH entropy will be published soon \cite{Edelstein:2017apt}. Another possibility for dualities in the SUGRA limit is the study of more field theory observables like baryon vertex, giant gravitons and couplings defined by probe branes, as suggested in \cite{Itsios:2016ooc}. 

Another interesting test of the conjecture would be to extend our result about symmetry breaking to the finite $N$ holography. In \cite{Anninos:2010sq} the thermal field theory analog of Coleman theorem was also tested successfully for a holographic superconductor both for the large $N$ limit and leading $1/N$ corrections. It should be possible to repeat their quantum gravity analysis in our setup to confirm or disproof the vanishing VEV expected for finite $ N $.  

Clearly, many question remain open about what can be described via AdS/CFT and the precise form of the correspondence. Although we have learned many non-trivial lessons in the last 20 years, a more precise map between bulk and boundary theories is only starting to to be developed, for example in the recent efforts about bulk reconstruction. This is expected to clarify in more detail how geometry emerges from field theory. The literature about quantum gravity corrections to the gravity side of the duality is still limited in comparison to the classical limit. The extensions to asymptotically dS and flat spaces are much less common than the AdS case. The same can be said about the correspondence at weak coupling with respect to the standard infinitely strong limit. There are many other relevant questions out of this short list. Clearly, AdS/CFT will remain a very active field of research in the next few years.

\chapter{Resumen}

Los problemas estudiados en la presente tesis tienen como objetivo a largo plazo explorar la validez y extender el dominio de aplicación de la correspondencia AdS/CFT. Este objetivo se ha concretado en tres problemas diferentes que se resumen en este capítulo junto con los resultados principales.

Situemos el presente trabajo en el contexto del desarrollo de la Física en el siglo XX que llevó hacia la citada correspondencia AdS/CFT. La Teoría Cuántica contiene los principios más fundamentales de la materia a la más pequeña escala. Su éxito en explicar y predecir fenómenos físicos en Física de la materia condensada, Molecular, Atómica, Nuclear y de Partículas no tiene parangón. La otra gran revolución es la teoría General de la Relatividad, que explica la gravedad pero afecta profundamente a los fundamentos conceptuales de la Física en todas sus ramas. Cuando se intentan aplicar los principios de la Teoría Cuántica a la Relatividad General de forma estándar, la teoría resultante no es renormalizable. De hecho, el problema de encontrar una descripción cuántica de la gravedad está en los más profundos de la Física Teórica actual. 

Una propuesta que ha sido objeto de estudio muy profundo es la Teoría de Cuerdas. Independientemente de su potencial para explicar fenómenos de la Física Fundamental a muy altas energías, a través de ella se ha descubierto la correspondencia AdS/CFT. Ella establece que ciertas Teorías Cuánticas de Campos en el límite de acoplo fuerte y gran $N$ pueden ser entendidas como teorías de gravedad. Esto hace posible calcular ciertos observables físicos en ellas que normalmente son imposibles usando sólo la Teoría Cuántica de Campos. Es en el contexto de esta correspondencia donde se desarrolla la presente tesis doctoral. Señalemos a continuación los problemas estudiados y sus resultados. 

\section{Transiciones de fase entre AdS y dS en teorías de gravedad con correcciones de order superior en la curvatura }
Se ha examinado el papel de las correcciones de curvatura superior en transiciones de fase térmicas entre geometrías asintóticamente AdS y dS (Anti de Sitter y de Sitter respectivamente). 

Las correcciones de orden superior en la curvatura a la acción gravitatoria de Einstein-Hilbert se estudian por varios motivos: como límite de bajas energías de la Teoría de Cuerdas u otras compleciones ultravioleta, para explorar sus efectos cosmológicos, desde un punto de vista puramente gravitacional, para entender sus implicaciones holográficas, etcétera. En esas teorías, la acciones de Lovelock ocupan un lugar destacado. Se ha estudiado la corrección cuádratica de esas teorías, conocida como gravedad de Lanczos-Gauss-Bonnet. En nuestro caso concreto, el interés holográfico se debe a que el un proceso dinámico hace que se pase de un espacio asintóticamente AdS, en el que la correspondencia se aplica de forma conocida, a otro asintóticamente dS donde su utilización está menos comprendida. 

Pese al énfasis en las teorías con correcciones de curvatura superior, las transiciones de fase gravitacionales también ocurren en la teoría de Einstein-Hilbert, por ejemplo la transición de Hawking-Page\cite{Hawking:1982dh}. Transiciones entre valores efectivos diferentes de la constante cosmológica puede ocurrir en ausencia de correcciones de curvatura superior por fluctuaciones térmicas entre los mínimos del potential de un campo escalar. 

El papel específico de las correciones de orden superior en la curvatura es que la transición es posible incluso en \textit{ausencia de cualquier tipo de materia}. El proceso está mediado por la formación de burbujas, de modo similar a las transiciones de fase en ebullición de los líquidos. El interior de dichas burbujas contiene una solución con geometría dS en su interior. Por otra parte esta solución está yuxtapuesta a otra en la parte exterior de la burbuja, que tiene la asintótica inicial. La burbuja es dinámicamente inestable y se contrae o expande. En este último caso la asintótica llega a cambiar cuando la burbuja alcanza la frontera de AdS. 

Nuestro principal resultado es una transición de fase de AdS térmico a la formación de un horizonte cosmológico en gravedad de Lanczos-Gauss-Bonnet. La transición ocurre por proliferación de burbujas que contienen un agujero negro de geometría de Sitter en su interior. Se espera que el fenómeno sea genérico de otras teorías de gravedad con curvatura superior. La transición es favorecida hasta un valor máximo del acoplo de la curvatura superior $\lambda$, mientras que esta peculiaridad no puede aparecer en estudios anteriores porque en ellos las burbujas contienen en su interior agujeros negros de geometría Anti de Sitter. 

Es importante señalar las limitaciones de nuestro tratamiento. La principal es la inestabilidad de tipo Boulware-Deser en el vacío AdS inicial. Esto significa que el campo del gravitón tiene fantasmas alrededor de esa solución. Sin embargo, es fácil encontrar otras correcciones de order superior en la curvature que tienen vacíos de diferente constante cosmológica y no presentan fantasmas en ellos; encontrar una transición de fase similar entre ellos es técnicamente mucho más díficil. Para transiciones entre asintóticas AdS, se ha comprobado explícitamente que es posible en la referencia \cite{Camanho:2013uda}. Existen buenos motivos para pensar que semejantes entre vacíos estables AdS y dS es también posible. 

No debemos olvidar otros comentarios sobre la validez de nuestro cálculo. En \cite{Camanho:2014apa} se mostró que las teorías cuadráticas de gravedad presentan genéricamente violaciones de causalidad, haciéndolas patológicas al menos en este aspecto. En la referencia \cite{DAppollonio:2015fly} se confirmó explícitamente que tal violación de causalidad aparece para cierto scattering entre branas y cuerdas si se conservan sólo los modos no masivos de la Teoría de Cuerdas. Sin embargo, si se toma en cuenta la torre de estados completa, (en particular la "Reggeización"), la causalidad se preserva. A pesar de esta violación de causalidad, las teorías cuadráticas y en particular LGB siguen siendo modelos efectivos aceptables para otros propósitos, como es el caso de la hidrodinámica holográfica, donde no manifiesta comportamientos físicamente inaceptables.

La cuestión principal que sigue abierta es el significado holográfico de las transiciones de fase encontradas, que sigue siendo esencialmente desconocido. Un primer paso es analizar los observables de la configuración estática de la burbuja, por ejemplo el lazo de Wilson en el infrarrojo. Un segundo paso sería el cálculo de observables similares para la expansión de la burbuja, de un modo paralelo al análisis de \cite{Li:2013cja}. Finalmente, se puede intentar estudiar directamente la geometría dS en que finaliza la transición, siguiendo la línea de investigación iniciada por Strominger en 2001 \cite{Strominger:2001pn}. 

\section[Nuevas soluciones a través de TDNA]{Deformación de la Teoría de Campos conforme de Klebanov-Witten y nuevas soluciones AdS$\boldsymbol{_{3}}$ través de T-dualidad no abeliana}

Hemos encontrado nuevas soluciones de Supergravedad usando T-dualidad no abeliana (TDNA), una técnica de generación de soluciones aparecida en el contexto de Teoría de Cuerdas. Tales nuevas soluciones representan flujos del grupo de renormalización de cuatro a dos dimensiones, y se investigan también los efectos de la TDNA en los observables holográficos. El problema tiene además una motivación de Supergravedad, puesto que es frecuente que las soluciones generadas estén fuera de las clasificaciones conocidas, como ocurre en la referencia \cite{Lozano:2015bra}. Además de ser interesantes en sí mismas, las nuevas teorías de campos pueden ayudar a entender el efecto de TDNA en el modelo sigma de la Teoría de Cuerdas, la aproximación de esta en Supergravedad y su relación con la correspondencia holográfica. 

Existen deformaciones de la solución de Supergravedad de Klebanov-Witten (KW) $\mathrm {AdS}_5 \times \mathrm S^5$ que fluyen hacia un factor $\mathrm {AdS}_3$ en el infrarrojo. Hemos generado nuevos ejemplos mediante la aplicación de TDNA sobre casos anteriormente conocidos. Las nuevas teorías de campos parecen estar relacionadas con teorías de gauge "quiver" largas y lineales. Este párrafo resume toda la investigación realizada en este problema. 

Para ser más detallado, se han considerado dos clases de geometrías en las que se aplica TDNA. La primera está dada por los duales geómetricos de algunas teorías de campos conformes en una variedad $\Sigma_2$; en particular un flujo $\mathcal{N}=1$ con $\Sigma_2=\mathrm H_2$ y puntos fijos no supersimétricos con $\Sigma_2=\mathrm H_2,\mathrm T^2,\mathrm S^2$; las llamamos soluciones con "twist". La segunda clase es la deformación de KW producida por Donos y Gauntlett \cite{Donos:2014eua}, que llamaremos por este motivo solución de Donos-Gauntlett. Este último background fue elegido como semilla de TDNA esencialmente por tres razones: es una deformación que fluye hacia AdS$_3$, es supersimétrica y regular, y está relacionada por dos T-dualidades abelianas con $ \mathrm{AdS}_3\times \mathrm S^3 \times \mathrm S^3 \times \mathrm S^1$. Esta última es el límite cerca del horizonte de una intersección de branas D1-D5; una discusión de su posible teoría dual puede encontrarse en \cite{Tong:2014yna}. 

Las soluciones generadas que son T-duales a un flujo de renormalización también corresponden holográficamente a un flujo de renormalización entre AdS$_5$ y AdS$_3$. Las soluciones encontradas son nuevas geometrías en Supergravedad de tipo IIA y preservan la Supersimetría y regularidad de los flujos iniciales. Debido a la naturaleza intrínseca de TDNA, muchos campos se activan en las nuevas geometrías y las dependencias funcionales son altamente no triviales. En una de las nuevas soluciones hemos aplicado además T-dualidad abeliana. Todos los casos de geometrías IIA se han embebido en Supergravedad de 11 dimensiones. La principal limitación del procedimiento de TDNA es que la técnica no da una prescripción de los rangos de las coordenadas de las soluciones generadas, y por lo tanto las propiedades globales no están totalmente determinadas a priori. Sin embargo, el estudio detallado de este problema ha dado lugar a avances significativos en la comprensión de los efectos holográficos de la TDNA \cite{Lozano:2016kum}. 

Con respecto a los observables duales en las nuevas geometrías, se han calculado las cargas de Page, función c, entropía de entrelazamiento en una banda y lazo de Wilson rectangular antes y después de la aplicación de TDNA. Los resultados principales han sido:
\begin{enumerate}
	\item Las nuevas teorías de campos generadas parecen estar relacionadas con teorías de gauge "quiver" infinitamente largas. La relación entre la carga central y las cargas de Page es de la forma $c \sim N_{D6} n^3$, frente a la relación estándar de tipo cuadrático. Es esta relación la que sugiere la conexión con las teorías "quiver", lo que implicaría que $n$ está midiendo el número de branas D4 y NS5. La imagen que aparece es la de una teoría de campos en dos dimensiones que vive en la intersección de D2, D6 y NS5 branas, y nuevas cargas D4 se añaden cuando se cruza una brana NS5.
	\item La energía del par quark-antiquark calculada a través de lazo de Wilson es trivialmente invariante por efecto de la dualidad. La función c y la entropía de entrelazamiento son no trivialmente invariantes salvo por un factor constante a lo largo del flujo. Tal factor depende de las propiedades globales que la TDNA no establece a priori. 
\end{enumerate} 
La relación con teorías quiver fue rápidamente profundizada en los artículos \cite{Lozano:2016kum,Lozano:2016wrs,Lozano:2017ole}. En \cite{Itsios:2016ooc}, se calculó el espectro de mesones antes y después de TDNA para AdS$_5\times$S$^5$. El cambio encontrado en la teoría de campos dual no significa automáticamente que TDNA no sea una simetría (perturbativa) de la teoría del modelo sigma, aunque lo sugiere, como se explica en la referencia citada. Sus autores proponen que, si no es debido a que TDNA no es simetría de Teoría de Cuerdas, los cambios en la teoría de campos dual podrían ser causados por correcciones en $1/N$ y acoplo finito que la dualidad no captura. 

Otra posibilidad interesante para clarificar el rol de TDNA en Teoría de Cuerdas proviene de una potencial aplicación a correcciones de orden superior en la curvatura para Supergravedad. De hecho, los autores de \cite{Marques:2015vua} encontraron una familia de accciones que es invariante por T-dualidad abeliana e incluye como casos particulares límites de Teoría de Cuerdas. Si las reglas de TDNA corregidas a orden $\alpha'$ no son una simetría de esos casos particulares, entonces es más probable que tampoco sea una simetría perturbativa de la Teoría de Cuerdas. También puede ser interesante comparar los efectos con los casos que no corresponden a Teoría de Cuerdas. 

\section[Identidades de Ward holográficas en 1+1 dimensiones]{Identidades de Ward holográficas en Teoría Cuántica de Campos en 1+1}
Hemos estudiado la extensión de la renormalización holográfica para reproducir las identidades de Ward de la ruptura espontánea de simetría en un modelo de superconductor holográfico en 1+1 dimensiones. El resultado podría ser de interés en aplicaciones a modelos unidimensionales de Materia Condensada. Otra motivación es saber si el modelo "bottom-up" utilizado se comporta de manera físicamente aceptable en dimensiones bajas. Un objetivo adicional es también entender las peculiaridades de gravedad en 2+1 dimensiones y sus implicaciones en la teoría en la frontera. 

La ruptura espontánea de simetría (RES) es importante en Fïsica de Altas Energías así como en Materia Condensada. Un resultado clave sobre ella es el Teorema de Goldstone: existe un bosón sin masa por cada grado de libertad de simetría que esté roto en el vacío elegido. Las propiedades de la ruptura de simetría están reflejadas en las identidades de Ward: la función de un punto señala la RES mientras que la de dos puntos contiene el espectro de bosones de Goldstone. Este poderoso resultado no se aplica a teorías de campos en 1+1 dimensiones (no compactas): a $N$ finito ni RES ni bosones de Goldstone son posibles (Tma. de Coleman), mientras que esta restricción desaparece en el límite de gran $N$. 

Holográficamente, la RES se ha estudiado con anterioridad para entender el diccionario holográfico en este aspecto de la teoría de campos y con posibles aplicaciones en el caso de acoplo fuerte, el cuál es difícil de abordar sin AdS/CFT. La acción en el lado de gravedad viene dada por un modelo de superconductor holográfico, y se rompe su simetría U(1). En la teoría de gravedad en 2+1 la descripción holográfica de corrientes conservadas involucra cuantización alternativa del campo gauge. Esto se debe a que en 2+1 tal campo gauge sin masa está en el límite de Breitenlohner-Freedman. Con la citada cuantización alternativa las identidades de Ward son idénticas a los casos de dimensión más alta. Además, se ha encontrado un bosón de Goldstone en el espectro. Esto significa que el modelo bottom-up que estamos usando proporciona no trivialmente tal resultado cuantitativamente correcto. 

A lo largo de esta tesis, hemos investigado transiciones de fase gravitacionales, T-dualidad no abeliana e identidades de Ward de la ruptura espontánea de simetría en el contexto de AdS/CFT. Los tres problemas anteriores tienen interés suficiente para futuros estudios adicionales. 
Más concretamente, la generación de nuevas soluciones, en particular dependientes del tiempo, a través de la combinación de dualiades abelianas y no abelianas, y posiblemente otras técnicas de generación de soluciones. Un caso más específico es el de las teorías con curvatura superior presentadas en \cite{Marques:2015vua}. Este trabajo se encuentra de hecho en curso, y resultados concernientes a la transformación de la entropía de agujero negro serán publicados en breve \cite{Edelstein:2017apt}. Otra posibilidad interesante es la extensión del tratamiento holográfico de la ruptura espontánea de simetría al caso de correcciones $1/N$, de modo similar a como éstas se incluyen en \cite{Anninos:2010sq}. 

Claramente, muchas preguntas siguen abiertas sobre qué fenómenos de Teoría de Campos se pueden describir gravitatoriamente y cómo hacerlo. Aunque el progreso en los últimos veinte años ha sido extensísimo, el mapa preciso entre ambos lados de la correspondencia aún tiene mucho margen de desarrollo, como se ejemplifica en los recientes esfuerzos para la reconstrucción de la geometría a partir del entrelazamiento en la teoría en la frontera. Comparativamente las correcciones de $1/N$ están mucho menos desarrolladas que el límite de acoplo fuerte. Otras asintóticas distintas de AdS aún no son de uso común (con la excepción de Lifshitz). Hay otros muchos temas de interés que no se han citado en esta breve lista. Sin duda AdS/CFT seguirá siendo un campo de investigación muy activo en los próximos años.

\bibliographystyle{plain}
\bibliography{anibal}

\begin{thebibliography}{100}

\bibitem{Aad:2012tfa}
Georges Aad et~al.
\newblock {Observation of a new particle in the search for the Standard Model
  Higgs boson with the ATLAS detector at the LHC}.
\newblock {\em Phys. Lett.}, B716:1--29, 2012.

\bibitem{TheLIGOScientific:2016wfe}
Benjamin~P. Abbott et~al.
\newblock {Properties of the Binary Black Hole Merger GW150914}.
\newblock {\em Phys. Rev. Lett.}, 116(24):241102, 2016.

\bibitem{Affleck:1980ac}
Ian Affleck.
\newblock {Quantum Statistical Metastability}.
\newblock {\em Phys. Rev. Lett.}, 46:388, 1981.

\bibitem{Akbar:2008vz}
Mohammad Akbar.
\newblock {Generalized Second Law of Thermodynamics in Extended Theories of
  Gravity}.
\newblock {\em Int. J. Theor. Phys.}, 48:2665--2671, 2009.

\bibitem{Altamirano:2013ane}
Natacha Altamirano, David Kubiznak, and Robert~B. Mann.
\newblock {Reentrant phase transitions in rotating anti–de Sitter black
  holes}.
\newblock {\em Phys. Rev.}, D88(10):101502, 2013.

\bibitem{Alvarez:1994dn}
Enrique Alvarez, Luis Alvarez-Gaume, and Yolanda Lozano.
\newblock {An Introduction to T duality in string theory}.
\newblock {\em Nucl. Phys. Proc. Suppl.}, 41:1--20, 1995.

\bibitem{Anderson:1963pc}
Philip~W. Anderson.
\newblock {Plasmons, Gauge Invariance, and Mass}.
\newblock {\em Phys. Rev.}, 130:439--442, 1963.

\bibitem{Andrade:2011sx}
Tomas Andrade, Juan~I. Jottar, and Robert~G. Leigh.
\newblock {Boundary Conditions and Unitarity: the Maxwell-Chern-Simons System
  in $AdS_3/CFT_2$}.
\newblock {\em JHEP}, 05:071, 2012.

\bibitem{Anninos:2010sq}
Dionysios Anninos, Sean~A. Hartnoll, and Nabil Iqbal.
\newblock {Holography and the Coleman-Mermin-Wagner theorem}.
\newblock {\em Phys. Rev.}, D82:066008, 2010.

\bibitem{Argurio:2016xih}
Riccardo Argurio, Gaston Giribet, Andrea Marzolla, Daniel Naegels, and
  J.~Anibal Sierra-Garcia.
\newblock {Holographic Ward identities for symmetry breaking in two
  dimensions}.
\newblock 2016.

\bibitem{Argurio:2015wgr}
Riccardo Argurio, Andrea Marzolla, Andrea Mezzalira, and Daniele Musso.
\newblock {Analytic pseudo-Goldstone bosons}.
\newblock {\em JHEP}, 03:012, 2016.

\bibitem{Attems:2016tby}
Maximilian Attems, Jorge Casalderrey-Solana, David Mateos, Daniel
  Santos-Oliván, Carlos~F. Sopuerta, Miquel Triana, and Miguel Zilhão.
\newblock {Holographic collisions in non-conformal theories}.
\newblock {\em JHEP}, 01:026, 2017.

\bibitem{BCS}
James Bardeen, Leon~N. Cooper, and John~Robert Schrieffer.
\newblock Theory of superconductivity.
\newblock {\em Physical Review}, 108(5):1175--1204, December 1957.

\bibitem{Bardeen:1973gs}
James~M. Bardeen, Brandon Carter, and Stephen~W. Hawking.
\newblock {The Four laws of black hole mechanics}.
\newblock {\em Commun. Math. Phys.}, 31:161--170, 1973.

\bibitem{Barnes:2004jj}
Edwin Barnes, Kenneth~A. Intriligator, Brian Wecht, and Jason Wright.
\newblock {Evidence for the strongest version of the 4d a-theorem, via
  a-maximization along RG flows}.
\newblock {\em Nucl. Phys.}, B702:131--162, 2004.

\bibitem{Bea:2015fja}
Yago Bea, Jose~D. Edelstein, Georgios Itsios, Karta~S. Kooner, Carlos Nunez,
  Daniel Schofield, and J.~Anibal Sierra-Garcia.
\newblock {Compactifications of the Klebanov-Witten CFT and new AdS$_{3}$
  backgrounds}.
\newblock {\em JHEP}, 05:062, 2015.

\bibitem{Beck:2015hpa}
Samuel. Beck, Jan~B. Gutowski, and George Papadopoulos.
\newblock {Supersymmetry of IIA warped flux AdS and flat backgrounds}.
\newblock {\em JHEP}, 09:135, 2015.

\bibitem{Beisert:2010jr}
Niklas Beisert et~al.
\newblock {Review of AdS/CFT Integrability: An Overview}.
\newblock {\em Lett. Math. Phys.}, 99:3--32, 2012.

\bibitem{Bekenstein:1973ur}
Jacob~D. Bekenstein.
\newblock {Black holes and entropy}.
\newblock {\em Phys. Rev.}, D7:2333--2346, 1973.

\bibitem{Benini:2007gx}
Francesco Benini, Felipe Canoura, Stefano Cremonesi, Carlos Nunez, and
  Alfonso~V. Ramallo.
\newblock {Backreacting flavors in the Klebanov-Strassler background}.
\newblock {\em JHEP}, 09:109, 2007.

\bibitem{Benini:2009mz}
Francesco Benini, Yuji Tachikawa, and Brian Wecht.
\newblock {Sicilian gauge theories and N=1 dualities}.
\newblock {\em JHEP}, 01:088, 2010.

\bibitem{Berenstein:2002jq}
David~Eliecer Berenstein, Juan~Martin Maldacena, and Horatiu~Stefan Nastase.
\newblock {Strings in flat space and pp waves from N=4 superYang-Mills}.
\newblock {\em JHEP}, 04:013, 2002.

\bibitem{Bergshoeff:1995as}
Eric Bergshoeff, Christopher~M. Hull, and Tomas Ortin.
\newblock {Duality in the type II superstring effective action}.
\newblock {\em Nucl. Phys.}, B451:547--578, 1995.

\bibitem{Berkowitz:2016muc}
Evan Berkowitz, Masanori Hanada, and Jonathan Maltz.
\newblock {A microscopic description of black hole evaporation via holography}.
\newblock {\em Int. J. Mod. Phys.}, D25(12):1644002, 2016.

\bibitem{Bertolini:2015hua}
Matteo Bertolini, Daniele Musso, Ioannis Papadimitriou, and Himanshu Raj.
\newblock {A goldstino at the bottom of the cascade}.
\newblock {\em JHEP}, 11:184, 2015.

\bibitem{Bianchi:2001de}
Massimo Bianchi, Daniel~Z. Freedman, and Kostas Skenderis.
\newblock {How to go with an RG flow}.
\newblock {\em JHEP}, 08:041, 2001.

\bibitem{Bianchi:2001kw}
Massimo Bianchi, Daniel~Z. Freedman, and Kostas Skenderis.
\newblock {Holographic renormalization}.
\newblock {\em Nucl. Phys.}, B631:159--194, 2002.

\bibitem{Bigazzi:2015bna}
Francesco Bigazzi, Aldo~L. Cotrone, and Roberto Sisca.
\newblock {Notes on Theta Dependence in Holographic Yang-Mills}.
\newblock {\em JHEP}, 08:090, 2015.

\bibitem{Blanco:2013joa}
David~D. Blanco, Horacio Casini, Ling-Yan Hung, and Robert~C. Myers.
\newblock {Relative Entropy and Holography}.
\newblock {\em JHEP}, 08:060, 2013.

\bibitem{Borlaf:1996na}
Javier Borlaf and Yolanda Lozano.
\newblock {Aspects of T duality in open strings}.
\newblock {\em Nucl. Phys.}, B480:239--264, 1996.

\bibitem{Boulware:1985wk}
David~G. Boulware and Stanley Deser.
\newblock {String Generated Gravity Models}.
\newblock {\em Phys. Rev. Lett.}, 55:2656, 1985.

\bibitem{Breitenlohner:1982jf}
Peter Breitenlohner and Daniel~Z. Freedman.
\newblock {Stability in Gauged Extended Supergravity}.
\newblock {\em Annals Phys.}, 144:249, 1982.

\bibitem{Brigante:2007nu}
Mauro Brigante, Hong Liu, Robert~C. Myers, Stephen Shenker, and Sho Yaida.
\newblock {Viscosity Bound Violation in Higher Derivative Gravity}.
\newblock {\em Phys. Rev.}, D77:126006, 2008.

\bibitem{Brown:1986nw}
J.~David Brown and Marc Henneaux.
\newblock {Central Charges in the Canonical Realization of Asymptotic
  Symmetries: An Example from Three-Dimensional Gravity}.
\newblock {\em Commun. Math. Phys.}, 104:207--226, 1986.

\bibitem{Brown:1987dd}
J.~David Brown and Claudio Teitelboim.
\newblock {Dynamical Neutralization of the Cosmological Constant}.
\newblock {\em Phys. Lett.}, B195:177--182, 1987.

\bibitem{Brown:1988kg}
J.~David Brown and Claudio Teitelboim.
\newblock {Neutralization of the Cosmological Constant by Membrane Creation}.
\newblock {\em Nucl. Phys.}, B297:787--836, 1988.

\bibitem{Buchel:2006gb}
Alex Buchel and James~T. Liu.
\newblock {Gauged supergravity from type IIB string theory on Y**p,q
  manifolds}.
\newblock {\em Nucl. Phys.}, B771:93--112, 2007.

\bibitem{Buscher:1987sk}
Thomas~Henry Buscher.
\newblock {A Symmetry of the String Background Field Equations}.
\newblock {\em Phys. Lett.}, B194:59--62, 1987.

\bibitem{Buscher:1987qj}
Thomas~Henry Buscher.
\newblock {Path Integral Derivation of Quantum Duality in Nonlinear Sigma
  Models}.
\newblock {\em Phys. Lett.}, B201:466--472, 1988.

\bibitem{Cai:2003gr}
Rong-Gen Cai and Qi~Guo.
\newblock {Gauss-Bonnet black holes in dS spaces}.
\newblock {\em Phys. Rev.}, D69:104025, 2004.

\bibitem{Camanho:2011rj}
Xian~O. Camanho and Jose~D. Edelstein.
\newblock {A Lovelock black hole bestiary}.
\newblock {\em Class. Quant. Grav.}, 30:035009, 2013.

\bibitem{Camanho:2012da}
Xian~O. Camanho, Jose~D. Edelstein, Gaston Giribet, and Andres Gomberoff.
\newblock {A New type of phase transition in gravitational theories}.
\newblock {\em Phys. Rev.}, D86:124048, 2012.

\bibitem{Camanho:2013uda}
Xian~O. Camanho, Jose~D. Edelstein, Gaston Giribet, and Andres Gomberoff.
\newblock {Generalized phase transitions in Lovelock gravity}.
\newblock {\em Phys. Rev.}, D90(6):064028, 2014.

\bibitem{Camanho:2014apa}
Xian~O. Camanho, Jose~D. Edelstein, Juan Maldacena, and Alexander Zhiboedov.
\newblock {Causality Constraints on Corrections to the Graviton Three-Point
  Coupling}.
\newblock {\em JHEP}, 02:020, 2016.

\bibitem{Camanho:2015ysa}
Xián~O. Camanho.
\newblock {\em {Lovelock gravity, black holes and holography}}.
\newblock PhD thesis, Santiago de Compostela U., 2013.

\bibitem{Camanho:2015zqa}
Xián~O. Camanho, Jose~D. Edelstein, Andrés Gomberoff, and J.~Anıbal
  Sierra-Garcıa.
\newblock {On AdS to dS transitions in higher-curvature gravity}.
\newblock {\em JHEP}, 10:179, 2015.

\bibitem{Camanho:2013pda}
Xián~O. Camanho, José~D. Edelstein, and José~M. Sánchez De~Santos.
\newblock {Lovelock theory and the AdS/CFT correspondence}.
\newblock {\em Gen. Rel. Grav.}, 46:1637, 2014.

\bibitem{Cardy:1988cwa}
John~L. Cardy.
\newblock {Is There a c Theorem in Four-Dimensions?}
\newblock {\em Phys. Lett.}, B215:749--752, 1988.

\bibitem{Carter:1971zc}
Brandon Carter.
\newblock {Axisymmetric Black Hole Has Only Two Degrees of Freedom}.
\newblock {\em Phys. Rev. Lett.}, 26:331--333, 1971.

\bibitem{Casalderrey-Solana:2016hrm}
Jorge Casalderrey-Solana and Andrej Ficnar.
\newblock {Holographic 3-jet events}.
\newblock {\em Nucl. Part. Phys. Proc.}, 276-278:115--116, 2016.

\bibitem{Charmousis:2010zz}
Christos Charmousis, Blaise Gouteraux, Bom~Soo Kim, Elias Kiritsis, and Rene
  Meyer.
\newblock {Effective Holographic Theories for low-temperature condensed matter
  systems}.
\newblock {\em JHEP}, 11:151, 2010.

\bibitem{Chatrchyan:2012xdj}
Serguei Chatrchyan et~al.
\newblock {Observation of a new boson at a mass of 125 GeV with the CMS
  experiment at the LHC}.
\newblock {\em Phys. Lett.}, B716:30--61, 2012.

\bibitem{Chesler:2008hg}
Paul~M. Chesler and Laurence~G. Yaffe.
\newblock {Horizon formation and far-from-equilibrium isotropization in
  supersymmetric Yang-Mills plasma}.
\newblock {\em Phys. Rev. Lett.}, 102:211601, 2009.

\bibitem{Chesler:2010bi}
Paul~M. Chesler and Laurence~G. Yaffe.
\newblock {Holography and colliding gravitational shock waves in asymptotically
  $AdS_5$ spacetime}.
\newblock {\em Phys. Rev. Lett.}, 106:021601, 2011.

\bibitem{Coleman:1973ci}
Sidney~R. Coleman.
\newblock {There are no Goldstone bosons in two-dimensions}.
\newblock {\em Commun. Math. Phys.}, 31:259--264, 1973.

\bibitem{Coleman:1980aw}
Sidney~R. Coleman and Frank De~Luccia.
\newblock {Gravitational Effects on and of Vacuum Decay}.
\newblock {\em Phys. Rev.}, D21:3305, 1980.

\bibitem{Coleman:1974jh}
Sidney~R. Coleman, Roman Jackiw, and H.~David Politzer.
\newblock {Spontaneous Symmetry Breaking in the O(N) Model for Large N*}.
\newblock {\em Phys. Rev.}, D10:2491, 1974.

\bibitem{Cvetic:2001bk}
Mirjam Cvetic, Shin'ichi Nojiri, and Sergei~D. Odintsov.
\newblock {Black hole thermodynamics and negative entropy in de Sitter and
  anti-de Sitter Einstein-Gauss-Bonnet gravity}.
\newblock {\em Nucl. Phys.}, B628:295--330, 2002.

\bibitem{DAppollonio:2015fly}
Giuseppe D'Appollonio, Paolo Di~Vecchia, Rodolfo Russo, and Gabriele Veneziano.
\newblock {Regge behavior saves String Theory from causality violations}.
\newblock {\em JHEP}, 05:144, 2015.

\bibitem{Davies:1976ei}
Paul C.~W. Davies, Stephen~Albert Fulling, and William~George Unruh.
\newblock {Energy Momentum Tensor Near an Evaporating Black Hole}.
\newblock {\em Phys. Rev.}, D13:2720--2723, 1976.

\bibitem{Davis:2002gn}
Stephen~C. Davis.
\newblock {Generalized Israel junction conditions for a Gauss-Bonnet brane
  world}.
\newblock {\em Phys. Rev.}, D67:024030, 2003.

\bibitem{deHaro:2000vlm}
Sebastian de~Haro, Sergey~N. Solodukhin, and Kostas Skenderis.
\newblock {Holographic reconstruction of space-time and renormalization in the
  AdS / CFT correspondence}.
\newblock {\em Commun. Math. Phys.}, 217:595--622, 2001.

\bibitem{delaOssa:1992vci}
Xenia~C. de~la Ossa and Fernando Quevedo.
\newblock {Duality symmetries from nonAbelian isometries in string theory}.
\newblock {\em Nucl. Phys.}, B403:377--394, 1993.

\bibitem{Deruelle:2003ck}
Nathalie Deruelle and John Madore.
\newblock {On the quasilinearity of the Einstein-'Gauss-Bonnet' gravity field
  equations}.
\newblock 2003.

\bibitem{Dong:2013qoa}
Xi~Dong.
\newblock {Holographic Entanglement Entropy for General Higher Derivative
  Gravity}.
\newblock {\em JHEP}, 01:044, 2014.

\bibitem{Donos:2014eua}
Aristomenis Donos and Jerome~P. Gauntlett.
\newblock {Flowing from AdS$_{5}$ to AdS$_{3}$ with T$^{1,1}$}.
\newblock {\em JHEP}, 08:006, 2014.

\bibitem{Donos:2008ug}
Aristomenis Donos, Jerome~P. Gauntlett, and Nakwoo Kim.
\newblock {AdS Solutions Through Transgression}.
\newblock {\em JHEP}, 09:021, 2008.

\bibitem{Duff:1998us}
Michael~J. Duff, Hong Lu, and Christopher~N. Pope.
\newblock {AdS(5) x S**5 untwisted}.
\newblock {\em Nucl. Phys.}, B532:181--209, 1998.

\bibitem{Edelstein:2006kw}
Jose~D. Edelstein and Ruben Portugues.
\newblock {Gauge/string duality in confining theories}.
\newblock {\em Fortsch. Phys.}, 54:525--579, 2006.

\bibitem{Edelstein:2017apt}
Jos\'e~Daniel Edelstein, Konstadinos Sfetsos, and Jesus~Anibal Sierra-Garcia.
\newblock {On black hole entropy and T-duality with $\alpha'$ corrections (in
  preparation)}.

\bibitem{Emparan:2008eg}
Roberto Emparan and Harvey~S. Reall.
\newblock {Black Holes in Higher Dimensions}.
\newblock {\em Living Rev. Rel.}, 11:6, 2008.

\bibitem{Englert:1964et}
Francois Englert and Robert Brout.
\newblock {Broken Symmetry and the Mass of Gauge Vector Mesons}.
\newblock {\em Phys. Rev. Lett.}, 13:321--323, 1964.

\bibitem{Faedo:2015urf}
Antón~F. Faedo, Arnab Kundu, David Mateos, Christiana Pantelidou, and Javier
  Tarrío.
\newblock {Three-dimensional super Yang-Mills with compressible quark matter}.
\newblock {\em JHEP}, 03:154, 2016.

\bibitem{Faulkner:2012gt}
Thomas Faulkner and Nabil Iqbal.
\newblock {Friedel oscillations and horizon charge in 1D holographic liquids}.
\newblock {\em JHEP}, 07:060, 2013.

\bibitem{Faulkner:2009wj}
Thomas Faulkner, Hong Liu, John McGreevy, and David Vegh.
\newblock {Emergent quantum criticality, Fermi surfaces, and AdS(2)}.
\newblock {\em Phys. Rev.}, D83:125002, 2011.

\bibitem{Frassino:2014pha}
Antonia~M. Frassino, David Kubiznak, Robert~B. Mann, and Fil Simovic.
\newblock {Multiple Reentrant Phase Transitions and Triple Points in Lovelock
  Thermodynamics}.
\newblock {\em JHEP}, 09:080, 2014.

\bibitem{Freedman:1999gp}
Daniel~Z. Freedman, Steven~Scott Gubser, Krzysztof Pilch, and Nicholas~P.
  Warner.
\newblock {Renormalization group flows from holography supersymmetry and a c
  theorem}.
\newblock {\em Adv. Theor. Math. Phys.}, 3:363--417, 1999.

\bibitem{Freivogel:2005qh}
Ben Freivogel, Veronika~E. Hubeny, Alexander Maloney, Robert~C. Myers, Mukund
  Rangamani, and Stephen Shenker.
\newblock {Inflation in AdS/CFT}.
\newblock {\em JHEP}, 03:007, 2006.

\bibitem{Fursaev:1995ef}
Dmitri~V. Fursaev and Sergey~N. Solodukhin.
\newblock {On the description of the Riemannian geometry in the presence of
  conical defects}.
\newblock {\em Phys. Rev.}, D52:2133--2143, 1995.

\bibitem{Gaiotto:2009we}
Davide Gaiotto.
\newblock {N=2 dualities}.
\newblock {\em JHEP}, 08:034, 2012.

\bibitem{Gaiotto:2009gz}
Davide Gaiotto and Juan Maldacena.
\newblock {The Gravity duals of N=2 superconformal field theories}.
\newblock {\em JHEP}, 10:189, 2012.

\bibitem{Garraffo:2007fi}
Cecilia Garraffo, Gast\'on Giribet, Elias Gravanis, and Steven Willison.
\newblock {Gravitational solitons and C0 vacuum metrics in five-dimensional
  Lovelock gravity}.
\newblock {\em J. Math. Phys.}, 49:042502, 2008.

\bibitem{Gauntlett:2009zw}
Jerome~P. Gauntlett, Seok Kim, Oscar Varela, and Daniel Waldram.
\newblock {Consistent supersymmetric Kaluza-Klein truncations with massive
  modes}.
\newblock {\em JHEP}, 04:102, 2009.

\bibitem{Gauntlett:2006ux}
Jerome~P. Gauntlett, Oisin A.~P. Mac~Conamhna, Toni Mateos, and Daniel Waldram.
\newblock {AdS spacetimes from wrapped M5 branes}.
\newblock {\em JHEP}, 11:053, 2006.

\bibitem{Gauntlett:2006af}
Jerome~P. Gauntlett, Oisin A.~P. Mac~Conamhna, Toni Mateos, and Daniel Waldram.
\newblock {Supersymmetric AdS(3) solutions of type IIB supergravity}.
\newblock {\em Phys. Rev. Lett.}, 97:171601, 2006.

\bibitem{Gauntlett:2009dn}
Jerome~P. Gauntlett, Julian Sonner, and Toby Wiseman.
\newblock {Holographic superconductivity in M-Theory}.
\newblock {\em Phys. Rev. Lett.}, 103:151601, 2009.

\bibitem{GellMann:1968rz}
Murray Gell-Mann, Robert~J. Oakes, and Brian Renner.
\newblock {Behavior of current divergences under SU(3) x SU(3)}.
\newblock {\em Phys. Rev.}, 175:2195--2199, 1968.

\bibitem{Gibbons:1976ue}
Gary~W. Gibbons and Stephen~W. Hawking.
\newblock {Action Integrals and Partition Functions in Quantum Gravity}.
\newblock {\em Phys. Rev.}, D15:2752--2756, 1977.

\bibitem{Giombi:2009wh}
Simone Giombi and Xi~Yin.
\newblock {Higher Spin Gauge Theory and Holography: The Three-Point Functions}.
\newblock {\em JHEP}, 09:115, 2010.

\bibitem{Giveon:1998ns}
Amit Giveon, David. Kutasov, and Nathan Seiberg.
\newblock {Comments on string theory on AdS(3)}.
\newblock {\em Adv. Theor. Math. Phys.}, 2:733--780, 1998.

\bibitem{Giveon:1993ai}
Amit Giveon and Martin Rocek.
\newblock {On nonAbelian duality}.
\newblock {\em Nucl. Phys.}, B421:173--190, 1994.

\bibitem{Goldstone:1962es}
Jeffrey Goldstone, Abdus Salam, and Steven Weinberg.
\newblock {Broken Symmetries}.
\newblock {\em Phys. Rev.}, 127:965--970, 1962.

\bibitem{Gomberoff:2003zh}
Andres Gomberoff, Marc Henneaux, Claudio Teitelboim, and Frank Wilczek.
\newblock {Thermal decay of the cosmological constant into black holes}.
\newblock {\em Phys. Rev.}, D69:083520, 2004.

\bibitem{Gravanis:2007ei}
Elias Gravanis and Steven Willison.
\newblock {`Mass without mass' from thin shells in Gauss-Bonnet gravity}.
\newblock {\em Phys. Rev.}, D75:084025, 2007.

\bibitem{Gross:1995bp}
David~J. Gross, Igor~R. Klebanov, Andrei~V. Matytsin, and Andrei~V. Smilga.
\newblock {Screening versus confinement in (1+1)-dimensions}.
\newblock {\em Nucl. Phys.}, B461:109--130, 1996.

\bibitem{Gross:1974jv}
David~J. Gross and Andre Neveu.
\newblock {Dynamical Symmetry Breaking in Asymptotically Free Field Theories}.
\newblock {\em Phys. Rev.}, D10:3235, 1974.

\bibitem{Gubser:1998bc}
S.~S. Gubser, Igor~R. Klebanov, and Alexander~M. Polyakov.
\newblock {Gauge theory correlators from noncritical string theory}.
\newblock {\em Phys. Lett.}, B428:105--114, 1998.

\bibitem{Gubser:2009qm}
Steven~S. Gubser, Christopher~P. Herzog, Silviu~S. Pufu, and Tiberiu Tesileanu.
\newblock {Superconductors from Superstrings}.
\newblock {\em Phys. Rev. Lett.}, 103:141601, 2009.

\bibitem{Gukov:2015qea}
Sergei Gukov.
\newblock {Counting RG flows}.
\newblock {\em JHEP}, 01:020, 2016.

\bibitem{Gupt:2013poa}
Brajesh Gupt and Parampreet Singh.
\newblock {Nonsingular AdS-dS transitions in a landscape scenario}.
\newblock {\em Phys. Rev.}, D89(6):063520, 2014.

\bibitem{Guralnik:1964eu}
Gerald~S. Guralnik, C.~R. Hagen, and Thomas W.~B. Kibble.
\newblock {Global Conservation Laws and Massless Particles}.
\newblock {\em Phys. Rev. Lett.}, 13:585--587, 1964.

\bibitem{Gutperle:2011kf}
Michael Gutperle and Per Kraus.
\newblock {Higher Spin Black Holes}.
\newblock {\em JHEP}, 05:022, 2011.

\bibitem{Hartnoll:2008vx}
Sean~A. Hartnoll, Christopher~P. Herzog, and Gary~T. Horowitz.
\newblock {Building a Holographic Superconductor}.
\newblock {\em Phys. Rev. Lett.}, 101:031601, 2008.

\bibitem{Hartnoll:2008kx}
Sean~A. Hartnoll, Christopher~P. Herzog, and Gary~T. Horowitz.
\newblock {Holographic Superconductors}.
\newblock {\em JHEP}, 12:015, 2008.

\bibitem{Hawking:1974sw}
Stephen~W. Hawking.
\newblock {Particle Creation by Black Holes}.
\newblock {\em Commun. Math. Phys.}, 43:199--220, 1975.
\newblock [,167(1975)].

\bibitem{Hawking:1982dh}
Stephen~W. Hawking and Don~N. Page.
\newblock {Thermodynamics of Black Holes in anti-De Sitter Space}.
\newblock {\em Commun. Math. Phys.}, 87:577, 1983.

\bibitem{Hayden:2011ag}
Patrick Hayden, Matthew Headrick, and Alexander Maloney.
\newblock {Holographic Mutual Information is Monogamous}.
\newblock {\em Phys. Rev.}, D87(4):046003, 2013.

\bibitem{Headrick:2010zt}
Matthew Headrick.
\newblock {Entanglement Renyi entropies in holographic theories}.
\newblock {\em Phys. Rev.}, D82:126010, 2010.

\bibitem{Headrick:2007km}
Matthew Headrick and Tadashi Takayanagi.
\newblock {A Holographic proof of the strong subadditivity of entanglement
  entropy}.
\newblock {\em Phys. Rev.}, D76:106013, 2007.

\bibitem{Henneaux:1987zz}
Marc Henneaux, Claudio Teitelboim, and Jorge Zanelli.
\newblock {Quantum mechanics for multivalued Hamiltonians}.
\newblock {\em Phys. Rev.}, A36:4417--4420, 1987.

\bibitem{Henningson:1998gx}
Mans Henningson and Kostas Skenderis.
\newblock {The Holographic Weyl anomaly}.
\newblock {\em JHEP}, 07:023, 1998.

\bibitem{Herzog:2009xv}
Christopher~P. Herzog.
\newblock {Lectures on Holographic Superfluidity and Superconductivity}.
\newblock {\em J. Phys.}, A42:343001, 2009.

\bibitem{Higgs:1964pj}
Peter~W. Higgs.
\newblock {Broken Symmetries and the Masses of Gauge Bosons}.
\newblock {\em Phys. Rev. Lett.}, 13:508--509, 1964.

\bibitem{Higgs:1964ia}
Peter~W. Higgs.
\newblock {Broken symmetries, massless particles and gauge fields}.
\newblock {\em Phys. Lett.}, 12:132--133, 1964.

\bibitem{Higgs:1966ev}
Peter~W. Higgs.
\newblock {Spontaneous Symmetry Breakdown without Massless Bosons}.
\newblock {\em Phys. Rev.}, 145:1156--1163, 1966.

\bibitem{Hofman:2008ar}
Diego~M. Hofman and Juan Maldacena.
\newblock {Conformal collider physics: Energy and charge correlations}.
\newblock {\em JHEP}, 05:012, 2008.

\bibitem{Hohenberg:1967zz}
Pierre~C. Hohenberg.
\newblock {Existence of Long-Range Order in One and Two Dimensions}.
\newblock {\em Phys. Rev.}, 158:383--386, 1967.

\bibitem{Hubeny:2009kz}
Veronika~E. Hubeny, Donald Marolf, and Mukund Rangamani.
\newblock {Black funnels and droplets from the AdS C-metrics}.
\newblock {\em Class. Quant. Grav.}, 27:025001, 2010.

\bibitem{Hubeny:2011hd}
Veronika~E. Hubeny, Shiraz Minwalla, and Mukund Rangamani.
\newblock {The fluid/gravity correspondence}.
\newblock In {\em {Black holes in higher dimensions}}, pages 348--383, 2012.
\newblock [,817(2011)].

\bibitem{Hubeny:2007xt}
Veronika~E. Hubeny, Mukund Rangamani, and Tadashi Takayanagi.
\newblock {A Covariant holographic entanglement entropy proposal}.
\newblock {\em JHEP}, 07:062, 2007.

\bibitem{Israel:1966rt}
Werner Israel.
\newblock {Singular hypersurfaces and thin shells in general relativity}.
\newblock {\em Nuovo Cim.}, B44S10:1, 1966.
\newblock [Nuovo Cim.B44,1(1966)].

\bibitem{Israel:1967wq}
Werner Israel.
\newblock {Event horizons in static vacuum space-times}.
\newblock {\em Phys. Rev.}, 164:1776--1779, 1967.

\bibitem{Israel:1967za}
Werner Israel.
\newblock {Event horizons in static electrovac space-times}.
\newblock {\em Commun. Math. Phys.}, 8:245--260, 1968.

\bibitem{Itsios:2017cew}
Georgios Itsios, Yolanda Lozano, Jesus Montero, and Carlos Nunez.
\newblock {The $AdS_5$ non-Abelian T-dual of Klebanov-Witten as a $\mathcal{N}
  = 1$ linear quiver from M5-branes}.
\newblock 2017.

\bibitem{Itsios:2012dc}
Georgios Itsios, Yolanda Lozano, Eoin. O~Colgain, and Konstadinos Sfetsos.
\newblock {Non-Abelian T-duality and consistent truncations in type-II
  supergravity}.
\newblock {\em JHEP}, 08:132, 2012.

\bibitem{Itsios:2013wd}
Georgios Itsios, Carlos Nunez, Konstadinos Sfetsos, and Daniel~C. Thompson.
\newblock {Non-Abelian T-duality and the AdS/CFT correspondence:new N=1
  backgrounds}.
\newblock {\em Nucl. Phys.}, B873:1--64, 2013.

\bibitem{Itsios:2012zv}
Georgios Itsios, Carlos Nunez, Konstadinos Sfetsos, and Daniel~C. Thompson.
\newblock {On Non-Abelian T-Duality and new N=1 backgrounds}.
\newblock {\em Phys. Lett.}, B721:342--346, 2013.

\bibitem{Itsios:2016ooc}
Georgios Itsios, Carlos Nunez, and Dimitrios Zoakos.
\newblock {Mesons from (non) Abelian T-dual backgrounds}.
\newblock {\em JHEP}, 01:011, 2017.

\bibitem{Iyer:1994ys}
Vivek Iyer and Robert~M. Wald.
\newblock {Some properties of Noether charge and a proposal for dynamical black
  hole entropy}.
\newblock {\em Phys. Rev.}, D50:846--864, 1994.

\bibitem{Jacobson:1993xs}
Ted Jacobson and Robert~C. Myers.
\newblock {Black hole entropy and higher curvature interactions}.
\newblock {\em Phys. Rev. Lett.}, 70:3684--3687, 1993.

\bibitem{Kachru:2008yh}
Shamit Kachru, Xiao Liu, and Michael Mulligan.
\newblock {Gravity duals of Lifshitz-like fixed points}.
\newblock {\em Phys. Rev.}, D78:106005, 2008.

\bibitem{Kehagias:2000dga}
Alex Kehagias and Jorge~G. Russo.
\newblock {Hyperbolic spaces in string and M theory}.
\newblock {\em JHEP}, 07:027, 2000.

\bibitem{Kelekci:2014ima}
Özgür Kelekci, Yolanda Lozano, Niall~T. Macpherson, and Eoin~Ó Colgáin.
\newblock {Supersymmetry and non-Abelian T-duality in type II supergravity}.
\newblock {\em Class. Quant. Grav.}, 32(3):035014, 2015.

\bibitem{Kim:2007ix}
Wontae Kim and Myungseok Yoon.
\newblock {Transition from AdS universe to DS universe in the BPP model}.
\newblock {\em JHEP}, 04:098, 2007.

\bibitem{Klebanov:2007ws}
Igor~R. Klebanov, David Kutasov, and Arvind Murugan.
\newblock {Entanglement as a probe of confinement}.
\newblock {\em Nucl. Phys.}, B796:274--293, 2008.

\bibitem{Klebanov:1998hh}
Igor~R. Klebanov and Edward Witten.
\newblock {Superconformal field theory on three-branes at a Calabi-Yau
  singularity}.
\newblock {\em Nucl. Phys.}, B536:199--218, 1998.

\bibitem{Klebanov:1999tb}
Igor~R. Klebanov and Edward Witten.
\newblock {AdS / CFT correspondence and symmetry breaking}.
\newblock {\em Nucl. Phys.}, B556:89--114, 1999.

\bibitem{Klinkhamer:1984di}
Frans~R. Klinkhamer and Nicholas~S. Manton.
\newblock {A Saddle Point Solution in the Weinberg-Salam Theory}.
\newblock {\em Phys. Rev.}, D30:2212, 1984.

\bibitem{Kolekar:2012tq}
Sanved Kolekar, Thanu Padmanabhan, and Sudipta Sarkar.
\newblock {Entropy Increase during Physical Processes for Black Holes in
  Lanczos-Lovelock Gravity}.
\newblock {\em Phys. Rev.}, D86:021501, 2012.

\bibitem{Komargodski:2011vj}
Zohar Komargodski and Adam Schwimmer.
\newblock {On Renormalization Group Flows in Four Dimensions}.
\newblock {\em JHEP}, 12:099, 2011.

\bibitem{Kosmann:1972kd}
Yvette Kosmann.
\newblock {\em Annali di Mat. Pura Appl.}

\bibitem{Lanczos:1932zz}
Cornelius Lanczos.
\newblock {Electricity as a natural property of Riemannian geometry}.
\newblock {\em Rev. Mod. Phys.}, 39:716--736, 1932.

\bibitem{Lanczos:1938sf}
Cornelius Lanczos.
\newblock {A Remarkable property of the Riemann-Christoffel tensor in four
  dimensions}.
\newblock {\em Annals Math.}, 39:842--850, 1938.

\bibitem{Langer:1969bc}
James~S. Langer.
\newblock {Statistical theory of the decay of metastable states}.
\newblock {\em Annals Phys.}, 54:258--275, 1969.

\bibitem{Lewkowycz:2013nqa}
Aitor Lewkowycz and Juan Maldacena.
\newblock {Generalized gravitational entropy}.
\newblock {\em JHEP}, 08:090, 2013.

\bibitem{Li:2013cja}
Yong-Zhuang Li, Shao-Feng Wu, and Guo-Hong Yang.
\newblock {Gauss-Bonnet correction to Holographic thermalization: two-point
  functions, circular Wilson loops and entanglement entropy}.
\newblock {\em Phys. Rev.}, D88:086006, 2013.

\bibitem{Lin:2004nb}
Hai Lin, Oleg Lunin, and Juan~Martin Maldacena.
\newblock {Bubbling AdS space and 1/2 BPS geometries}.
\newblock {\em JHEP}, 10:025, 2004.

\bibitem{Linde:1977mm}
Andrei~D. Linde.
\newblock {On the Vacuum Instability and the Higgs Meson Mass}.
\newblock {\em Phys. Lett.}, B70:306--308, 1977.

\bibitem{Linde:1981zj}
Andrei~D. Linde.
\newblock {Decay of the False Vacuum at Finite Temperature}.
\newblock {\em Nucl. Phys.}, B216:421, 1983.
\newblock [Erratum: Nucl. Phys.B223,544(1983)].

\bibitem{Liu:2006ug}
Hong Liu, Krishna Rajagopal, and Urs~Achim Wiedemann.
\newblock {Calculating the jet quenching parameter from AdS/CFT}.
\newblock {\em Phys. Rev. Lett.}, 97:182301, 2006.

\bibitem{Lovelock:1971yv}
David Lovelock.
\newblock {The Einstein tensor and its generalizations}.
\newblock {\em J. Math. Phys.}, 12:498--501, 1971.

\bibitem{Lozano:2014ata}
Yolanda Lozano and Niall~T. Macpherson.
\newblock {A new AdS$_{4}$/CFT$_{3}$ dual with extended SUSY and a spectral
  flow}.
\newblock {\em JHEP}, 11:115, 2014.

\bibitem{Lozano:2016wrs}
Yolanda Lozano, Niall~T. Macpherson, Jesus Montero, and Carlos Nunez.
\newblock {Three-dimensional $ \mathcal{N}=4 $ linear quivers and non-Abelian
  T-duals}.
\newblock {\em JHEP}, 11:133, 2016.

\bibitem{Lozano:2015bra}
Yolanda Lozano, Niall~T. Macpherson, Jesús Montero, and Eoin~Ó Colgáin.
\newblock {New $AdS_3 \times S^2$ T-duals with $ \mathcal{N}=\left(0,4\right) $
  supersymmetry}.
\newblock {\em JHEP}, 08:121, 2015.

\bibitem{Lozano:2017ole}
Yolanda Lozano, Carlos Nunez, and Salomon Zacarias.
\newblock {BMN Vacua, Superstars and Non-Abelian T-duality}.
\newblock 2017.

\bibitem{Lozano:2016kum}
Yolanda Lozano and Carlos Núñez.
\newblock {Field theory aspects of non-Abelian T-duality and $ \mathcal{N} =$ 2
  linear quivers}.
\newblock {\em JHEP}, 05:107, 2016.

\bibitem{Lozano:2011kb}
Yolanda Lozano, Eoin. O~Colgain, Konstadinos Sfetsos, and Daniel~C. Thompson.
\newblock {Non-abelian T-duality, Ramond Fields and Coset Geometries}.
\newblock {\em JHEP}, 06:106, 2011.

\bibitem{Ma:1974tp}
Shang-keng Ma and R.~Rajaraman.
\newblock {Comments on the Absence of Spontaneous Symmetry Breaking in Low
  Dimensions}.
\newblock {\em Phys. Rev.}, D11:1701, 1975.

\bibitem{Macpherson:2015tka}
Niall~T. Macpherson, Carlos Nunez, Daniel~C. Thompson, and S.~Zacarias.
\newblock {Holographic Flows in non-Abelian T-dual Geometries}.
\newblock {\em JHEP}, 11:212, 2015.

\bibitem{Macpherson:2014eza}
Niall~T. Macpherson, Carlos Núñez, Leopoldo~A. Pando~Zayas, Vincent G.~J.
  Rodgers, and Catherine~A. Whiting.
\newblock {Type IIB supergravity solutions with AdS$_{5}$ from Abelian and
  non-Abelian T dualities}.
\newblock {\em JHEP}, 02:040, 2015.

\bibitem{Maldacena:1998im}
Juan~Martin Maldacena.
\newblock {Wilson loops in large N field theories}.
\newblock {\em Phys. Rev. Lett.}, 80:4859--4862, 1998.

\bibitem{Maldacena:1997re}
Juan~Martin Maldacena.
\newblock {The Large N limit of superconformal field theories and
  supergravity}.
\newblock {\em Int. J. Theor. Phys.}, 38:1113--1133, 1999.
\newblock [Adv. Theor. Math. Phys.2,231(1998)].

\bibitem{Maldacena:2000mw}
Juan~Martin Maldacena and Carlos Nunez.
\newblock {Supergravity description of field theories on curved manifolds and a
  no go theorem}.
\newblock {\em Int. J. Mod. Phys.}, A16:822--855, 2001.
\newblock [,182(2000)].

\bibitem{Marolf:2006nd}
Donald Marolf and Simon~F. Ross.
\newblock {Boundary Conditions and New Dualities: Vector Fields in AdS/CFT}.
\newblock {\em JHEP}, 11:085, 2006.

\bibitem{Marques:2015vua}
Diego Marques and Carmen~A. Nunez.
\newblock {T-duality and $\alpha$'-corrections}.
\newblock {\em JHEP}, 10:084, 2015.

\bibitem{Mermin:1966fe}
N.~David Mermin and H.~Wagner.
\newblock {Absence of ferromagnetism or antiferromagnetism in one-dimensional
  or two-dimensional isotropic Heisenberg models}.
\newblock {\em Phys. Rev. Lett.}, 17:1133--1136, 1966.

\bibitem{Metsaev:1987zx}
R.~R. Metsaev and Arkady~A. Tseytlin.
\newblock {Order alpha-prime (Two Loop) Equivalence of the String Equations of
  Motion and the Sigma Model Weyl Invariance Conditions: Dependence on the
  Dilaton and the Antisymmetric Tensor}.
\newblock {\em Nucl. Phys.}, B293:385--419, 1987.

\bibitem{Minces:1999eg}
Pablo Minces and Victor~O. Rivelles.
\newblock {Scalar field theory in the AdS / CFT correspondence revisited}.
\newblock {\em Nucl. Phys.}, B572:651--669, 2000.

\bibitem{Myers:1987yn}
Robert~C. Myers.
\newblock {Higher Derivative Gravity, Surface Terms and String Theory}.
\newblock {\em Phys. Rev.}, D36:392, 1987.

\bibitem{Myers:1988ze}
Robert~C. Myers and Jonathan~Z. Simon.
\newblock {Black Hole Thermodynamics in Lovelock Gravity}.
\newblock {\em Phys. Rev.}, D38:2434--2444, 1988.

\bibitem{Nambu:1961tp}
Yoichiro Nambu and G.~Jona-Lasinio.
\newblock {Dynamical Model of Elementary Particles Based on an Analogy with
  Superconductivity. 1.}
\newblock {\em Phys. Rev.}, 122:345--358, 1961.

\bibitem{Nambu:1961fr}
Yoichiro Nambu and G.~Jona-Lasinio.
\newblock {DYNAMICAL MODEL OF ELEMENTARY PARTICLES BASED ON AN ANALOGY WITH
  SUPERCONDUCTIVITY. II}.
\newblock {\em Phys. Rev.}, 124:246--254, 1961.

\bibitem{Nishioka:2009un}
Tatsuma Nishioka, Shinsei Ryu, and Tadashi Takayanagi.
\newblock {Holographic Entanglement Entropy: An Overview}.
\newblock {\em J. Phys.}, A42:504008, 2009.

\bibitem{Nishioka:2006gr}
Tatsuma Nishioka and Tadashi Takayanagi.
\newblock {AdS Bubbles, Entropy and Closed String Tachyons}.
\newblock {\em JHEP}, 01:090, 2007.

\bibitem{Nojiri:2001pm}
Shin'ichi Nojiri and Sergei~D. Odintsov.
\newblock {The de Sitter / anti-de Sitter black holes phase transition?}
\newblock In {\em {1st Mexican Meeting on Mathematical and Experimental Physics
  Mexico City, Mexico, September 10-14, 2001}}, 2001.

\bibitem{Nunez:2001pt}
Carlos Nunez, I.~Y. Park, Martin Schvellinger, and Tuan~A. Tran.
\newblock {Supergravity duals of gauge theories from F(4) gauged supergravity
  in six-dimensions}.
\newblock {\em JHEP}, 04:025, 2001.

\bibitem{Ortin:2002qb}
Tomas Ortin.
\newblock {A Note on Lie-Lorentz derivatives}.
\newblock {\em Class. Quant. Grav.}, 19:L143--L150, 2002.

\bibitem{Papadimitriou:2004ap}
Ioannis Papadimitriou and Kostas Skenderis.
\newblock {AdS / CFT correspondence and geometry}.
\newblock {\em IRMA Lect. Math. Theor. Phys.}, 8:73--101, 2005.

\bibitem{Picos:2009pc}
Marco~Antonio Picos~Sol.
\newblock {Regularization techniques for the dilaton transformation under
  T-duality (MSc. thesis)}.
\newblock 2009.

\bibitem{Pons:2004dk}
Josep~M. Pons, Jorge~G. Russo, and Pedro Talavera.
\newblock {Semiclassical string spectrum in a string model dual to large N
  QCD}.
\newblock {\em Nucl. Phys.}, B700:71--88, 2004.

\bibitem{Quevedo:1997jb}
Fernando Quevedo.
\newblock {Duality and global symmetries}.
\newblock {\em Nucl. Phys. Proc. Suppl.}, 61A:23--41, 1998.

\bibitem{Ramallo:2013bua}
Alfonso~V. Ramallo.
\newblock {Introduction to the AdS/CFT correspondence}.
\newblock {\em Springer Proc. Phys.}, 161:411--474, 2015.

\bibitem{Rangamani:2016dms}
Mukund Rangamani and Tadashi Takayanagi.
\newblock {Holographic Entanglement Entropy}.
\newblock 2016.

\bibitem{Rey:1998ik}
Soo-Jong Rey and Jung-Tay Yee.
\newblock {Macroscopic strings as heavy quarks in large N gauge theory and
  anti-de Sitter supergravity}.
\newblock {\em Eur. Phys. J.}, C22:379--394, 2001.

\bibitem{Rocek:1991ps}
Martin Rocek and Erik~P. Verlinde.
\newblock {Duality, quotients, and currents}.
\newblock {\em Nucl. Phys.}, B373:630--646, 1992.

\bibitem{Ryu:2006ef}
Shinsei Ryu and Tadashi Takayanagi.
\newblock {Aspects of Holographic Entanglement Entropy}.
\newblock {\em JHEP}, 08:045, 2006.

\bibitem{Ryu:2006bv}
Shinsei Ryu and Tadashi Takayanagi.
\newblock {Holographic derivation of entanglement entropy from AdS/CFT}.
\newblock {\em Phys. Rev. Lett.}, 96:181602, 2006.

\bibitem{Sakai:2004cn}
Tadakatsu Sakai and Shigeki Sugimoto.
\newblock {Low energy hadron physics in holographic QCD}.
\newblock {\em Prog. Theor. Phys.}, 113:843--882, 2005.

\bibitem{Sarkar:2012wy}
Sudipta Sarkar and Swastik Bhattacharya.
\newblock {Issue of zeroth law for Killing horizons in Lanczos-Lovelock
  gravity}.
\newblock {\em Phys. Rev.}, D87(4):044023, 2013.

\bibitem{Sarkar:2010xp}
Sudipta Sarkar and Aron~C. Wall.
\newblock {Second Law Violations in Lovelock Gravity for Black Hole Mergers}.
\newblock {\em Phys. Rev.}, D83:124048, 2011.

\bibitem{Sfetsos:2010uq}
Konstadinos Sfetsos and Daniel~C. Thompson.
\newblock {On non-abelian T-dual geometries with Ramond fluxes}.
\newblock {\em Nucl. Phys.}, B846:21--42, 2011.

\bibitem{Skenderis:2002wp}
Kostas Skenderis.
\newblock {Lecture notes on holographic renormalization}.
\newblock {\em Class. Quant. Grav.}, 19:5849--5876, 2002.

\bibitem{Strominger:2001pn}
Andrew Strominger.
\newblock {The dS / CFT correspondence}.
\newblock {\em JHEP}, 10:034, 2001.

\bibitem{Strominger:1996sh}
Andrew Strominger and Cumrun Vafa.
\newblock {Microscopic origin of the Bekenstein-Hawking entropy}.
\newblock {\em Phys. Lett.}, B379:99--104, 1996.

\bibitem{tHooft:1993dmi}
Gerard 't~Hooft.
\newblock {Dimensional reduction in quantum gravity}.
\newblock In {\em {Salamfest 1993:0284-296}}, pages 0284--296, 1993.

\bibitem{Tachikawa:2013kta}
Yuji Tachikawa.
\newblock {\em {N=2 supersymmetric dynamics for pedestrians}}, volume 890.
\newblock 2014.

\bibitem{Taylor:2015glc}
Marika Taylor.
\newblock {Lifshitz holography}.
\newblock {\em Class. Quant. Grav.}, 33(3):033001, 2016.

\bibitem{Tong:2014yna}
David Tong.
\newblock {The holographic dual of $AdS_{3} \times S^{3} \times S^{3} \times
  S^{1}$}.
\newblock {\em JHEP}, 04:193, 2014.

\bibitem{Torii:2006gu}
Takashi Torii.
\newblock {Black holes in higher curvature theory and third law of
  thermodynamics}.
\newblock {\em J. Phys. Conf. Ser.}, 31:175--176, 2006.

\bibitem{Unruh:1976db}
William~George Unruh.
\newblock {Notes on black hole evaporation}.
\newblock {\em Phys. Rev.}, D14:870, 1976.

\bibitem{Vasiliev:1990en}
Mikhail~A. Vasiliev.
\newblock {Consistent equation for interacting gauge fields of all spins in
  (3+1)-dimensions}.
\newblock {\em Phys. Lett.}, B243:378--382, 1990.

\bibitem{Wald:1993nt}
Robert~M. Wald.
\newblock {Black hole entropy is the Noether charge}.
\newblock {\em Phys. Rev.}, D48(8):R3427--R3431, 1993.

\bibitem{Weinberg:1988cp}
Steven Weinberg.
\newblock {The Cosmological Constant Problem}.
\newblock {\em Rev. Mod. Phys.}, 61:1--23, 1989.

\bibitem{Wheeler:1985nh}
James~T. Wheeler.
\newblock {Symmetric Solutions to the Gauss-Bonnet Extended Einstein
  Equations}.
\newblock {\em Nucl. Phys.}, B268:737--746, 1986.

\bibitem{Wheeler:1985qd}
James~T. Wheeler.
\newblock {Symmetric Solutions to the Maximally {Gauss-Bonnet} Extended
  Einstein Equations}.
\newblock {\em Nucl. Phys.}, B273:732--748, 1986.

\bibitem{Wilson:1974sk}
Kenneth~G. Wilson.
\newblock {Confinement of Quarks}.
\newblock {\em Phys. Rev.}, D10:2445--2459, 1974.
\newblock [,45(1974)].

\bibitem{Witten:1978qu}
Edward Witten.
\newblock {Chiral Symmetry, the 1/n Expansion, and the SU(N) Thirring Model}.
\newblock {\em Nucl. Phys.}, B145:110--118, 1978.

\bibitem{Witten:1998qj}
Edward Witten.
\newblock {Anti-de Sitter space and holography}.
\newblock {\em Adv. Theor. Math. Phys.}, 2:253--291, 1998.

\bibitem{Zamolodchikov:1986gt}
Alexander~B. Zamolodchikov.
\newblock {Irreversibility of the Flux of the Renormalization Group in a 2D
  Field Theory}.
\newblock {\em JETP Lett.}, 43:730--732, 1986.
\newblock [Pisma Zh. Eksp. Teor. Fiz.43,565(1986)].

\bibitem{Zumino:1985dp}
Bruno Zumino.
\newblock {Gravity Theories in More Than Four-Dimensions}.
\newblock {\em Phys. Rept.}, 137:109, 1986.

\end{thebibliography}

\end{document}